\documentclass[10pt]{report}    
\usepackage{graphicx}
\usepackage{wrapfig}
\usepackage{amssymb}
\usepackage{amsmath}
\usepackage{mathrsfs}
\usepackage{slashed}
\usepackage{braket}
\usepackage{sectsty}
\usepackage{tocloft}
\usepackage{fancyhdr}
\usepackage{endnotes}
\usepackage{makeidx}
\makeindex

\makeatletter
\def\@removefromreset#1#2{%
\let\@tempb\@elt
\expandafter\let\expandafter\@tempa
\csname c@#1\endcsname
\def\@elt##1{\expandafter\ifx
\csname c@##1\endcsname\@tempa\else
\noexpand\@elt{##1}\fi}%
\expandafter\protected@edef
\csname cl@#2\endcsname
{\csname cl@#2\endcsname}%
\let\@elt\@tempb}

\def\theequation{\arabic{equation}}
\@removefromreset{equation}{chapter}
\makeatother

\begin{document}

\chapterfont{\fontfamily{cmss}\mdseries\upshape \large\selectfont} 
\sectionfont{\fontfamily{cmr}\mdseries\itshape\normalsize\selectfont} 
\subsectionfont{\fontfamily{cmr}\mdseries\itshape\normalsize\selectfont} 

\begin{titlepage}
\vspace*{2in}
\begin{centering}
\textbf{\textsf{F. J. Dyson}} \\
\vspace*{0.3in}
\huge{\textbf{\textsf{ADVANCED}}} \\

\Huge{\textbf{\textsf{QUANTUM MECHANICS}}} \\
\vspace*{0.3in}
\normalsize{\textbf{\textsf{Second Edition}}} \\
\end{centering}
\end{titlepage}
\break

\vspace*{1in}
\begin{centering}
\emph{ADVANCED QUANTUM MECHANICS} \\
\vspace*{0.3in}
Lecture notes by Professor F. J. Dyson for a course in Relativistic Quantum Mechanics \\ given at Cornell University in the Fall of 1951.\\ %
\index{Dyson, Freeman~J.}%
\vspace*{0.2in}
Second Edition
\\
\vspace*{0.1in}
The first edition of these notes was written by Professor Dyson. The second edition was prepared by Michael J. Moravcsik; he is responsible for the changes made in the process of re-editing.\\ %
\index{Moravcsik, Michael J.}%
\end{centering}
\vspace*{3in}
\begin{center}
\emph{Generally used notation:}
\end{center}
\begin{tabbing}
\hspace*{1in}\=\hspace{5ex}\=\hspace{6ex}\=\hspace{4ex}\kill
\>$A^*$\>\,\,: \>complex conjugate transposed (Hermitian conjugate)\\
\>$A^{+}$\>\,\,:\>complex conjugate (not transposed)\\
\>$\overline{A}$\>\,\,: \>$A^*\beta = A^*\gamma_4$ = adjoint\\
\>$A^{-1}$\>=\> inverse\\
\>$A^\text{T}$\>=\>transposed\\
\>$\mathbb{I}$\>=\>identity matrix or operator\\
\end{tabbing}
\break

\renewcommand{\contentsname}{Table of Contents}                                                       
\renewcommand{\cfttoctitlefont}{\fontfamily{cmss}\mdseries\upshape \Large\selectfont}  
\renewcommand{\cftchapfont}{\fontfamily{cmss}\mdseries\upshape \large\selectfont}      

\tableofcontents 

\newpage

\pagenumbering{arabic} 
\thispagestyle{empty} 

\vspace*{2in}
\begin{centering}
\textsc{\Large{Lecture Course 491 --- Advanced Quantum Theory}}\\
\textsc{\Large{F. J. Dyson --- Fall Semester 1951}}\\
\end{centering}
\break

\thispagestyle{empty}

\setcounter{page}{1} 



\chapter*{Introduction}
\addcontentsline{toc}{chapter}{Introduction} 

\pagestyle{fancy}
\fancyhead{}
\lhead{\emph{\MakeUppercase{Introduction}}}
\chead{}
\rhead{\thepage}
\lfoot{}
\cfoot{}
\rfoot{}

\section*{Books}

\hspace*{1em}\hspace*{1ex}W. Pauli, ``Die Allgemeinen Principien der Wellenmechanik''; \emph{Handbuch der Physik}, 2 ed., Vol.  24, \\
\index{Pauli, Wolfgang}%
\hspace*{6ex}Part 1; Edwards reprint, Ann Arbor 1947. (In German) \cite{Pauli58}\\

W. Heitler, \emph{Quantum Theory of Radiation}, 2nd Edition, Oxford. 3rd edition just published. \cite{Heitler54}\\
\index{Heitler, Walter}%

G. Wentzel, \emph{Introduction to the Quantum Theory of Wave-Fields}, Interscience, N.Y. 1949 \cite{Wentzel49} \\
\index{Wentzel, Gregor}%

I shall not expect you to have read any of these, but I shall refer to them as we go along. The later part of the course will be new stuff, taken from papers of Feynman and Schwinger mainly. \cite{Schwinger58}, \cite{Miller94}, \cite{Schweber61}, \cite{Schweber94}, \cite{Kaiser05}

\index{Feynman, Richard~P.}%
\index{Schwinger, Julian}%
\section*{Subject Matter}
\addcontentsline{toc}{section}{Subject Matter} 

You have had a complete course in non-relativistic quantum theory. I assume this known. All the general principles of the NR theory are valid and true under all circumstances, in particular also when the system happens to be relativistic. What you have learned is therefore still good.

You have had a course in classical mechanics and electrodynamics including special relativity. You know what is meant by a system being relativistic; the equations of motion are formally invariant under Lorentz transformations. General relativity we shall not touch.%
\index{Lorentz!transformations}%

This course will be concerned with the development of a \emph{Lorentz--invariant Quantum theory}. That is not a general dynamical method like the NR quantum theory, applicable to all systems. We cannot yet devise a general method of that kind, and it is probably impossible. Instead we have to find out what are the possible systems, the particular equations of motion, which can be handled by the NR quantum dynamics and which are at the same time Lorentz--invariant.%
\index{Lorentz!invariance}%

In the NR theory it was found that almost any classical system could be handled, i.e.quantized. Now on the contrary we find there are very few possibilities for a relativistic quantized system. This is a most important fact. It means that starting only from the principles of relativity and quantization, it is mathematically possible only for very special types of objects to exist. So one can \emph{predict} mathematically some important things about the real world. The most striking examples of this are: \\

\hspace*{2ex}(i)\hspace*{2ex}Dirac from a study of the electron predicted the positron, which was later discovered \cite{Dirac28}. \\ %
\hspace*{0.30in}(ii)\hspace*{2ex}Yukawa from a study of nuclear forces predicted the meson, which was later discovered \cite{Yukawa35}. \\ %
\index{Yukawa, Hideki}%
\index{Dirac, P. A. M.}%
\index{positron}%
\index{meson}%

These two examples are special cases of the general principle, which is the basic success of the relativistic quantum theory, that \emph{A Relativistic Quantum Theory of a Finite Number of Particles is Impossible.} A RQ theory necessarily contains these features: an indefinite number of particles of one or more types, particles of each type being identical and indistinguishable from each other, possibility of creation and annihilation of particles.
\index{creation}%
\index{annihilation}%
Thus the two principles of relativity and quantum theory when combined lead to a world built up out of various types of elementary particles, and so make us feel quite confident that we are on the right way to an understanding of the real world. In addition, various detailed properties of the observed particles are necessary consequences of the general theory. These are for example: \\

\hspace*{2ex}(i)\hspace*{2ex}Magnetic moment of Electron (Dirac) \cite{Dirac28}. \\
\hspace*{0.30in}(ii)\hspace*{2ex}Relation between spin and statistics (Pauli) \cite{Pauli40}.  %
\index{Dirac, P. A. M.}%
\index{electron!magnetic~moment}%
\index{Pauli, Wolfgang}%
\index{Pauli!spin \& statistics}%

\section*{Detailed Program}
\addcontentsline{toc}{section}{Detailed Program} 

We shall not develop straightaway a correct theory including many particles. Instead we follow the historical development. We try to make a relativistic quantum theory of \emph{one} particle, find out how far we can go and where we get into trouble. Then we shall see how to change the theory and get over the trouble by introducing many particles. Incidentally, the one--particle theories are quite useful, being correct to a good approximation in many situations where creation  
of new particles does not occur, and where something better than a NR approximation is needed. 
\index{creation}%
An example is the Dirac theory of the H atom.\endnote{
``is'' substituted for ``in''}
\index{hydrogen~atom}%
\index{Dirac, P. A. M.}%

The NR theory gave levels correctly but no fine-structure. (Accuracy of one part in 10,000). The Dirac one-particle theory gives all the main features of the fine-structure correctly, number of components and separations good to 10\% but not better. (Accuracy one part in 100,000).

The Dirac many-particle theory gives the fine-structure separations (Lamb experiment) correctly to about one part in 10,000. (Overall accuracy 1 in 10$^8$.)
\index{Lamb~shift!experiment}%
\index{Dirac, P. A. M.}%

Probably to get accuracy better than 1 in 10$^8$ even the DMP theory is not enough and one will need to take all kinds of meson effects into account which are not yet treated properly. Experiments are so far only good to about 1 in 10$^8$.
\index{meson}

In this course I will go through the one-particle theories first in detail. Then I will talk about their breaking down. At that point I will make a fresh start and discuss how one can make a relativistic quantum theory in general, using the new methods of Feynman and Schwinger. From this we shall be led to the many-particle theories. I will talk about the general features of these theories. Then I will take the special example of quantum electrodynamics and get as far as I can with it before the end of the course. 
\index{Feynman, Richard~P.}%
\index{Schwinger, Julian}%
\index{quantum~electrodynamics}%

\section*{One-Particle Theories}
\addcontentsline{toc}{section}{One Particle Theories} 

Take the simplest case, one particle with no forces. Then the NR wave-mechanics tells you to take the equation $E = \dfrac{1}{2m} p^2$ of classical mechanics, and write
\begin{equation}
E \rightarrow i\hbar \frac{\partial}{\partial t} \qquad \qquad p_x \rightarrow -i\hbar \frac{\partial}{\partial x}
\end{equation}
to get the wave-equation\endnote{
Both the first and the second editions use $\Delta$ for the Laplacian differential operator; as there are many quantities with $\Delta$ in them to denote a change or shift, the use of $\nabla^{2}$ for the Laplacian seems a safer choice.}
\begin{equation}
i\hbar \frac{\partial}{\partial t}\, \psi = - \frac{\hbar^2}{2m} \left( \frac{\partial^2}{\partial x^2} +  \frac{\partial^2}{\partial y^2} +  \frac{\partial^2}{\partial z^2} \right) \psi = - \frac{\hbar^2}{2m} \nabla^{2}  \psi
\end{equation}
satisfied by the wave-function $\psi$. 

To give a physical meaning to $\psi$, we state that $\rho = \psi^* \psi$ is the probability of finding the particle at the point $x\;y\;z$ at time $t$. And the probability is conserved because\endnote{
The text symbols ``div'', ``grad'' and ``curl'' have been replaced throughout by ``$\nabla \, \boldsymbol{\cdot} $'', ``$\nabla$'', and ``$\nabla \times$'', respectively. For example, the equation (3) originally read 
\[
\frac{\partial \rho}{\partial t} + \, \text{div}\, j = 0
\]
}
\index{probability!conserved}%
\begin{equation}
\frac{\partial \rho}{\partial t} + \nabla \boldsymbol{\cdot} \vec{\boldsymbol{\jmath}} = 0
\end{equation}
where
\begin{equation}
\vec{\boldsymbol{\jmath}} = \frac{\hbar}{2mi}\left(\psi^* \nabla \psi - \psi \nabla \psi^* \right)
\end{equation}
\vspace*{0.1in}
where $\psi^*$ is the complex conjugate of $\psi$.

Now do this relativistically. We have classically
\begin{equation}
E^2 = m^2c^4 + c^2 p^2
\end{equation}
which gives the wave equation
\begin{equation}
\frac{1}{c^2} \frac{\partial^2}{\partial t^2}\, \psi = \nabla^{2} \psi - \frac{m^2c^2}{\hbar^2} \, \psi
\end{equation}
This is an historic equation, the Klein-Gordon equation. Schr\"{o}dinger already in 1926 tried to make a RQ theory out of it. But he failed, and many other people too, until Pauli and Weisskopf gave the many-particle theory in 1934 \cite{PauliWeisskopf34}.  
\index{Pauli, Wolfgang}%
\index{Klein-Gordon~equation}%
\index{Weisskopf, Victor}%
\index{Schr\"{o}dinger, Erwin}%
Why? 

Because in order to interpret the wave-function as a probability we must have a continuity equation. 
\index{probability!and~continuity}%
This can only be got out of the wave-equation if we take $\vec{\boldsymbol{\jmath}}$ as before, and 
\begin{equation}
\rho = \frac{i\hbar}{2mc^2} \left(\psi^* \frac{\partial \psi}{\partial t} - \frac{\partial \psi^*}{\partial t} \psi \right)
\end{equation}
But now since the equation is 2$^{\text{nd}}$ order, $\psi$ and $\dfrac{\partial \psi}{\partial t}$ are arbitrary. Hence $\rho$ need not be positive. We have \emph{Negative Probabilities}. This defeated all attempts to make a sensible one-particle theory. 

The theory can be carried through quite easily, if we make $\psi$ describe an assembly of particles of both positive and negative charge, and $\rho$ is the \emph{net} charge density at any point. This is what Pauli and Weisskopf did, and the theory you get is correct for $\pi$-mesons, the mesons  
\index{pi~meson@$\pi$~meson}%
which are made in the synchrotron downstairs. I will talk about it later. 
\index{Pauli, Wolfgang}%
\index{Weisskopf, Victor}%



\newpage

\chapter*{The Dirac Theory}
\addcontentsline{toc}{chapter}{The Dirac Theory}

\pagestyle{fancy}
\fancyhead{}
\lhead{\emph{\MakeUppercase{The Dirac Theory}}}
\chead{}
\rhead{\thepage}
\lfoot{}
\cfoot{}
\rfoot{}

\section*{The Form of the Dirac Equation}
\addcontentsline{toc}{section}{The Form of the Dirac Equation}

Historically before the RQ theory came the one-particle theory of Dirac. This was so successful in dealing with the electron, that it was for many years the only respectable RQ theory in existence. And its difficulties are a lot less immediate than the difficulties of the one-particle KG theory. 
\index{Dirac!equation}%
\index{Dirac!electron~theory}%
\index{Klein-Gordon~equation}%

Dirac said,
\index{Dirac, P. A. M.}%
suppose the particle can exist in several distinct states with the same momentum (different orientations of spin.) Then the  wave-function $\psi$ satisfying (6) must have several \emph{components}; it is not a scalar but a set of numbers each giving the prob. amplitude 
\index{amplitude}%
to find the particle at a given place and in a given substate. So we write for $\psi$ a column matrix 
\index{momentum}%
\begin{equation*}
\psi = 
	\left[
		\begin{matrix}
			\psi_1 \\
			\psi_2 \\
			\cdot \\
			\cdot \\
			\cdot \\
		\end{matrix}
         \right] \qquad \qquad {\text{for the components}} \; \psi_{\alpha}\,; \quad \alpha = 1, 2, \dots
\end{equation*}
Dirac assumed that the probability density at any point is still given by 
\index{Dirac, P. A. M.}%
\index{probability!density}%
\begin{equation}
\rho = \sum_\alpha \psi^*_{\alpha} \psi_{\alpha}
\end{equation}
which we write 
\begin{equation*}
\rho = \psi^*\psi
\end{equation*}
as in the NR theory. Here $\psi^*$ is a \emph{row} matrix
\begin{equation*}
\left[ \psi^*_1, \, \psi^*_2, \,\dots \right]
\end{equation*}
We must have (3) still satisfied. So $\psi$ must satisfy a wave-equation of \emph{First Order} in $t$. But since the equations are relativistic, the equation has to be also of 1$^\text{st}$ order in $x\; y\; z$. Thus the most general possible wave-equation is 
\begin{equation}
\frac{1}{c} \, \frac{\partial \psi}{\partial t} + \sum_1^3 \alpha^k \, \frac{\partial \psi}{\partial x_k} + i \frac{mc}{\hbar} \, \beta\, \psi = 0
\end{equation}
where $x_1 \; x_2 \; x_3$ are written for $x\; y\; z$ and $\alpha^1 \; \alpha^2 \; \alpha^3 \; \beta$ are square matrices whose elements are numbers. The conjugate of (9) gives
\begin{equation}
\frac{1}{c} \, \frac{\partial \psi^*}{\partial t} + \sum_1^3  \frac{\partial \psi^*}{\partial x_k}\, \alpha^{k*} - i \frac{mc}{\hbar} \, \psi^* \,\beta ^*= 0
\end{equation}
where $\alpha^{k*}$ and $\beta^*$ are Hermitian conjugates.
\index{Hermitian!conjugate}%

Now to get (3) out of (8), (9) and (10) we must have $\alpha^{k*} =  \alpha^{k}$, $\beta^* = \beta$ so $\alpha^{k}$ and $\beta$ are \emph{Hermitian}; and
\index{Hermitian} %
\begin{equation}
j_{k} = c \left(\psi^* \, \alpha^k \psi \right)
\end{equation}

Next what more do we want from equation (9)? Two things. (A) it must be consistent with the 2$^{\text{nd}}$ order equation (6) we started from; (B) the whole theory must be Lorentz invariant. 
\index{Lorentz!invariance}%

First consider (A). If (9) is consistent with (6) it must be possible to get exactly (6) by multiplying (9) by the operator
\begin{equation}
\frac{1}{c} \, \frac{\partial}{\partial t} - \sum_1^3 \alpha^\ell \, \frac{\partial}{\partial x_\ell} - i \frac{mc}{\hbar} \, \beta
\end{equation}
chosen so that the terms with mixed derivatives $\dfrac{\partial}{\partial t} \; \dfrac{\partial}{\partial x_k}$ and $\dfrac{\partial}{\partial t}$ cancel. This gives
\begin{multline*}
\qquad \frac{1}{c^2}\, \frac{\partial^2 \psi}{\partial t^2} = \sum \sum_{k \ne \ell} \frac{1}{2} \left( \alpha^k \alpha^\ell  + \alpha^\ell \alpha^k \right) \frac{\partial^2 \psi}{\partial x_k \, \partial x_\ell}
 + \sum_k \alpha_{k}^2 \frac{\partial^2 \psi}{\partial x_{k}^2}  \\
  \shoveleft{\qquad \qquad \qquad \qquad- \frac{m^{2}c^{2}}{\hbar^2} \beta^2 \, \psi+ i \frac{mc}{\hbar} \sum_k \left( \alpha^k \beta + \beta \alpha^k \right) \frac{\partial \psi}{\partial x_k}} \\
\end{multline*}This agrees with (6) if and only if
\begin{align}
&\alpha^k \alpha^{\ell} + \alpha^\ell \alpha^k = 0 \quad k \ne \ell  \notag\\
&\alpha^k \beta + \beta\, \alpha^k = 0 \\
&\alpha^{k \, 2} = \beta^2 = \mathbb{I}\text{, (identity matrix)}\notag
\end{align}
Thus we could not possibly \emph{factorize} the 2$^\text{nd}$ order equation into two first-order operators involving ordinary numbers. But we can do it with \emph{matrices}.  \\ 
\index{Dirac!matrices}%
\index{Pauli!matrices}%

\noindent Consider the Pauli spin matrices

\begin{equation}
\sigma_1 = \left( \begin{matrix}
0 & 1 \\
1 & 0 \\
\end{matrix} \right)
\qquad \qquad
\sigma_2 = \left( \begin{matrix}
0 & -i \\
i & 0 \\
\end{matrix} \right)
\qquad \qquad
\sigma_3 = \left( \begin{matrix}
1 & 0 \\
0 & -1 \\
\end{matrix} \right)
\end{equation}
\\
\noindent you are familiar with. They satisfy
\[
\sigma_k \sigma_\ell + \sigma_\ell \sigma_k = 2 \delta_{\ell k}
\]
But we cannot make 4 matrices of this type all anti-commuting. They must be \emph{at least} 4 $\times$ 4. \\

One possible set of $\alpha^k$ and $\beta$ is
\begin{equation}
\alpha^k = \left( \begin{matrix}
0 & \sigma_k \\
\sigma_k & 0 \\
\end{matrix} \right)
\qquad
\beta = \left( \begin{matrix}
\tiny{\begin{matrix}
1 & 0 \\
0 & 1 
\end {matrix}} & 0 \\
0 & \tiny{\begin{matrix}
-1 & \;\;0 \\
\;\;0 & -1 
\end {matrix}}  \\
\end{matrix} \right)
\end{equation}
In particular
\[
\alpha^1 = \left( \begin{matrix}
0 & 0 & 0 & 1 \\
0 & 0 & 1 & 0 \\
0 & 1 & 0 & 0 \\
1 & 0 & 0 & 0 \\
\end{matrix} \right)
\qquad
\alpha^2 = \left( \begin{matrix}
0 & 0 & 0 & -i \\
0 & 0 & i & 0 \\
0 & -i & 0 & 0 \\
i & 0 & 0 & 0 \\
\end{matrix} \right)
\qquad
\alpha^3 = \left( \begin{matrix}
0 & 0 & 1 & 0 \\
0 & 0 & 0 & \!\!-1 \\
1 & 0 & 0 & 0 \\
0 &\! \!-1 & 0 & 0 \\
\end{matrix} \right)
\]

\noindent These are hermitian as required. Of course if $\alpha^k$ and $\beta$ are any set satisfying (13) then $S\alpha^k S^{-1}$ and $S\beta S^{-1}$ are another set, where $S$ is any \emph{unitary} matrix $SS^* = 1$. And conversely it can be proved that every possible 4 $\times$ 4 matrices $\alpha^k$ and $\beta$ are of this form with \emph{some} such matrix $S$. We do not prove this here. \\

The Dirac equation is thus a set of 4 simultaneous linear partial differential\endnote{
``differential'' replaces ``diff.''}
equations in the four functions $\psi_{\alpha}$.  \\
\index{Dirac!equation}%

\section*{Lorentz Invariance of the Dirac Equation}
\addcontentsline{toc}{section}{Lorentz Invariance of the Dirac Equation}

What does this mean? Consider a general Lorentz transformation: If $x_{\mu}^{\prime}$ are the new coordinates:
\index{Lorentz!transformations}%
\begin{equation}
x_{\mu}^{\prime} = \sum_{\nu = 0}^{3} a_{\mu \nu} x_{\nu} \qquad (x_o = ct)
\end{equation}
In the new coordinate system the wave-function will be $\psi^{\prime}$. Clearly we do not expect that $\psi^{\prime} = \psi$. Example: in the Maxwell theory  which is relativistic, the magnetic field $H$ is no longer a pure magnetic field in a moving system.
\index{Maxwell!electromagnetic~theory}%
Instead it transforms like a tensor. So we have to find \emph{some} transformation law for the $\psi$ which will leave invariant the physical consequences of the equations. 
\index{Dirac!equation!Lorentz~invariance}%

We need in fact two things: (i) the interpretation of $\psi^{*}\psi$ as a probability density must be preserved, (ii) the validity of the Dirac equation  must be preserved in the new system. 
\index{Dirac!equation}%
\index{probability!density}%

First consider (i). The quantity which can be directly observed and must be invariant is the quantity 
\[
\left(\psi^{*}\psi\right) \times V
\]
where $V$ is a volume. Now in going to a new Lorentz system with relative velocity $v$ the volume $V$ changes by Fitzgerald contraction to the value 
\index{Lorentz!system}%
\index{Fitzgerald~contraction}%
\[
V^{\prime} = V \sqrt{1 - \frac{v^2}{c^2}}
\]
Therefore 
\begin{equation}
\left(\psi^{* \prime}\psi^{\prime}\right) = \frac{\psi^{*} \psi}{\sqrt{1 - \dfrac{v^2}{c^2}}}
\end{equation}
and so $\left(\psi^{*} \psi \right) = \rho$ transforms like an \emph{energy},  i.e.like the fourth component of a vector. This shows incidentally that $\psi^{\prime} \ne \psi$. Since $\rho$ and $\vec{\boldsymbol{\jmath}}$ are related by the equation of continuity, the space-components of the 4-vector are 
\begin{equation}
\left(S_1, S_2, S_3 \right) = \psi^{*} \alpha^{k} \psi = \frac{1}{c}\, j_{k}
\end{equation}
So we require that the 4 quantities
\begin{equation}
\left(S_1, S_2, S_3,S_0 \right) = \left(\psi^{*} \alpha^{k} \psi,  \psi^{*} \psi \right) 
\end{equation}
transform like a \emph{4-vector}. This will be enough to preserve the interpretation of the theory. \\

Assume that 
\begin{equation}
\psi^{\prime} = S \psi
\end{equation}
where $S$ is a \emph{linear} operator. Then
\begin{equation}
\psi^{\prime *} = \psi^{*}S^{*}
\end{equation}
So we require
\begin{align}
\psi^{* \prime} \alpha^{k} \psi^{\prime} &= \psi^{*} S^{*} \alpha^{k} S \psi = \sum_{\nu = 0}^3 a_{k \nu} \psi^{*} \alpha^{\nu} \psi \\ \notag
\psi^{* \prime} \psi^{\prime} &= \psi^{*} S^{*} S \psi = \sum_{\nu = 0}^3 a_{0 \nu} \psi^{*} \alpha^{\nu} \psi 
\end{align}
writing \( \alpha^0 = \mathbb{I} \).  \\

\noindent Thus we need 
\begin{equation}
S^{*}\alpha^{\mu} S = \sum_{\nu = 0}^3 a_{\mu \nu} \alpha^{\nu}, \qquad \mu = 0, 1, 2, 3
\end{equation}

Next consider (ii). The Dirac equation for $\psi^{\prime}$ is 
\index{Dirac!equation}%
\begin{equation} 
\sum_0^3 \alpha^{\nu} \frac{\partial}{\partial x^{\prime}_{\nu}} \psi^{\prime} + i \frac{mc}{\hbar} \beta \psi^{\prime} = 0
\end{equation}
Now the original Dirac equation for $\psi$ expressed in terms of the new coordinates is 
\index{Dirac!equation}%
\begin{equation}
\sum_{\mu = 0}^3 \sum_{\nu = 0}^3 \alpha^{\mu} \frac{\partial}{\partial x^{\prime}_{\nu}} a_{\nu \mu} S^{-1} \psi^{\prime} + i \frac{mc}{\hbar} \beta S^{-1} \psi^{\prime} = 0
\end{equation}
The sets of equations (24) and (25) have to be equivalent, not identical. Thus (25) must be the same as (24) multiplied by $\beta S^{-1} \beta$. The condition for this is 
\begin{equation}
\beta S^{-1} \beta \alpha^{\nu} = \sum_0^3 \alpha^{\lambda} a_{\nu \lambda} S^{-1}
\end{equation}
But (23) and (26) are identical if 
\begin{equation}
\beta S^{-1} \beta = S^{*} \qquad \text{which means} \qquad S^{*} \beta S = \beta
\end{equation}
Thus $\beta$ transforms like a scalar, $\alpha^{\nu}$ like a 4-vector when multiplied by $S^{*}S$. \\

\section*{To find the S}
\addcontentsline{toc}{section}{To Find the S}

Given two coordinate transformations in succession, with matrices already found, the combined transformation will correspond to the product of these matrices. Hence we have to consider only 3 simple types of transformation. 
\index{Dirac!equation!Lorentz~invariance}%
\begin{align*}
\text{1) Pure}&\text{ rotations} \\
x_0^{\prime} &= x_0 \qquad \qquad \qquad \qquad \qquad \qquad \qquad x_3^{\prime} = x_3 \\
x_1^{\prime} &= x_1 \cos \theta + x_2 \sin \theta \\
x_2^{\prime} &= -x_1 \sin \theta + x_2 \cos \theta 
\end{align*}
\begin{align*}
\text{2) Pure}&\text{ Lorentz transformations} \\
x_1^{\prime} &= x_1 \qquad \qquad \qquad \qquad \qquad \qquad \qquad x_2^{\prime} = x_2 \\
x_3^{\prime} &= x_3 \cosh \theta + x_0 \sinh \theta \\
x_0^{\prime} &= x_3 \sinh \theta + x_0 \cosh \theta 
\end{align*}
\begin{align*}
\text{3) Pure}&\text{ reflections} \\
x_1^{\prime} &= -x_1 \qquad  x_2^{\prime} =  -x_2 \qquad  x_3^{\prime} =  -x_3  \qquad x_0^{\prime} =  x_0
\end{align*}
\emph{Case 1} Then 
\begin{equation}
S = \cos \tfrac{1}{2} \theta + i \sigma_3 \sin \tfrac{1}{2} \theta
\end{equation}
Here
\[
\sigma_3 = \left( \begin{matrix}
\sigma_3 & 0 \\
        0 & \sigma_3 \\
\end{matrix} \right)
\]
commutes with $\alpha_3$ and $\beta$. 
\[
\sigma_3 \alpha_1 = i \alpha_2, \qquad \qquad \qquad \sigma_3 \alpha_2 = -i \alpha_1
\]
\[
S^{*}= \cos \tfrac{1}{2} \theta - i \sigma_3 \sin \tfrac{1}{2} \theta
\]
Then 
\begin{align*}
S^{*} \beta S &= \beta \\
S^{*} \alpha^0 S &= \alpha^0 \\
S^{*} \alpha^3 S &= \alpha^3 
\end{align*}
as required.
\begin{align*}
S^{*} \alpha^1 S &= \cos \theta\; \alpha^1 + \sin \theta\; \alpha^2\\
S^{*} \alpha^2 S &= -\sin \theta\; \alpha^1 + \cos \theta\; \alpha^2
\end{align*}
\emph{Case 2} 
\begin{equation}
S = S^{*} = \cosh \tfrac{1}{2} \theta + \alpha_3  \sinh \tfrac{1}{2} \theta
\end{equation}
Here \begin{align*}
S^{*} \beta S &= \beta \\
S^{*} \alpha^1 S &= \alpha^1 \\
S^{*} \alpha^2 S &= \alpha^2 \\
S^{*} \alpha^3 S &= \cosh \theta\; \alpha^3 + \sinh \theta\; \alpha^0\\
S^{*} \alpha^0 S &= \sinh \theta\; \alpha^3 + \cos \theta\; \alpha^0
\end{align*}
\emph{Case 3} 
\begin{equation}
S = S^{*} = \beta
\end{equation}
Note that in all cases S is ambiguous by a factor $\pm 1$. So in case 1 a rotation though 360$^\circ$ gives $S = -1$.

\noindent \rule[0.02in]{1in}{0.01in}\\
\noindent \emph{Problem 1} Find the S corresponding to a general infinitesimal coordinate transformation. Compare and show that it agrees with the exact solutions given here. \\
\rule[0.05in]{1in}{0.01in} \\
\index{problem!\emph{1}}%
The $\psi_{\alpha}$'s transforming with these $S$-transformations are called \emph{spinors}. 
\index{spinors}%
They are a direct extension of the non-relativistic 2-component spin-functions. Mathematical theory of spinors is not very useful. In fact we find always in practice, calculations can be done most easily if one avoids any explicit representation of the spinors. \emph{Use only formal algebra and commutation relations of the matrices.} 

\section*{The covariant notation}
\addcontentsline{toc}{section}{The Covariant Notation}

In order to avoid distinction between covariant and contravariant vectors (which we have also unjustifiably ignored in the previous discussion) it is useful to use the imaginary 4$^\text{th}$ coordinate     
\index{Dirac!equation!covariant notation}%
\begin{equation}
x_4 = ix_0 = ict
\end{equation}
In this coordinate system the four matrices\endnote{
In the second edition, this equation lacked a label.}
\begin{equation}
\gamma_{\,\text{1, 2, 3, 4}} = \left( -i \beta \alpha^{\text{1, 2, 3}}, \beta\, \right) \qquad \qquad \text{i.e.}
\end{equation}
\[
\gamma_1 = \left(\, \begin{matrix}
0 & 
       \tiny{\begin{matrix}
			\; \;0 & -i \\
			-i & \; \:0  \\
		\end {matrix}} \\
\tiny{\begin{matrix}
0 & i \\
i & 0
\end{matrix}} & 0 \\
\end{matrix}\, \right); \quad
\gamma_2 = \left(\, \begin{matrix}
0 & 
       \tiny{\begin{matrix}
			0 & \!\!\!-1 \\
			1 & 0  \\
		\end {matrix}} \\
\tiny{\begin{matrix}
\; \;0 & 1 \\
-1 & 0 \\
\end{matrix}} & 0 \\
\end{matrix}\, \right); \quad
\gamma_3 = \left(\, \begin{matrix}
0 & 
       \tiny{\begin{matrix}
			-i & 0 \\
			0 & i  \\
		\end {matrix}} \\
\tiny{\begin{matrix}
i & 0 \\
0 & -i \\
\end{matrix}} & 0 \\
\end{matrix}\, \right); \quad
\gamma_4 = \left( \begin{matrix}
\tiny{\begin{matrix}
1 & 0 \\
0 & 1 
\end {matrix}} & 0 \\
0 & \tiny{\begin{matrix}
-1 & \;\;0 \\
\;\;0 & -1 
\end {matrix}}  \\
\end{matrix} \right)
\]
\\
are a 4-vector. They are all Hermitian and satisfy
\index{Hermitian}%
\begin{equation}
\gamma_{\mu} \gamma_{\nu} + \gamma_{\nu} \gamma_{\mu} = 2 \delta_{\mu \nu}
\end{equation}
The Dirac equation and its conjugate may now be written 
\index{Dirac!equation!conjugate}%
\begin{align}
\sum_1^4 \gamma_{\mu} \frac{\partial \psi}{\partial x_{\mu}} + \frac{mc}{\hbar} \psi &= 0 \notag \\
\sum_1^4  \frac{\partial \overline{\psi}}{\partial x_{\mu}}\gamma_{\mu} - \frac{mc}{\hbar} \overline{\psi} &= 0
\end{align}
with 
\begin{equation}
\overline{\psi} = \psi^{*} \beta \qquad \text{and}
\end{equation}

\begin{equation}
s_{\mu} = i \left( \overline{\psi}\;\gamma_{\mu}\; \psi \right) = \left( \frac{1}{c}\; \vec{\boldsymbol{\jmath}}, i \rho \right)
\end{equation}\\
These notations are much the most convenient for calculations. 

\section*{Conservation Laws. Existence of Spin.}
\addcontentsline{toc}{section}{Conservation Laws -- Existence of Spin}

The Hamiltonian in this theory is\endnote{
In the first edition, the coefficient of $\boldsymbol{\alpha} \boldsymbol{\cdot} \nabla$ is $+i\hbar c$}
\index{Dirac!equation!Hamiltonian}%
\index{spin,~existence~of}%
\index{conservation~laws}%
\begin{equation}
i \hbar \frac{\partial \psi}{\partial t} = H \psi
\end{equation} 
\begin{equation}
H = -i \hbar c \sum_1^3 \alpha^k \frac{\partial}{\partial x_k} + mc^2 \beta = - i \hbar c\; \boldsymbol{\alpha} \boldsymbol{\cdot} \nabla + mc^2 \beta
\end{equation}
This commutes with the momentum $\boldsymbol{p} = -i \hbar \nabla$.  So the momentum $\boldsymbol{p}$ is a constant of motion.\\
\index{momentum}%
\index{momentum!conservation}%

However the angular momentum operator 
\index{angular~momentum}%
\begin{equation}
\boldsymbol{L} = \boldsymbol{r} \times \boldsymbol{p} = - i \hbar \boldsymbol{r} \times \nabla
\end{equation}
is not a constant. For
\begin{equation}
\left[ H, \boldsymbol{L} \right] = - \hbar^2 c\; \boldsymbol{\alpha} \times \nabla
\end{equation}
But
\[
\left[ H, \boldsymbol{\sigma} \right] = - i \hbar c\; \nabla \boldsymbol{\cdot} \left[\boldsymbol{\alpha}, \boldsymbol{\sigma} \right]   \qquad \text{where} \qquad \boldsymbol{\sigma} = \left(\sigma_1, \sigma_2, \sigma_3 \right)
\]
while
\[
\left[\alpha^1, \sigma_1 \right] = 0, \qquad \left[\alpha^1, \sigma_2 \right] = 2i\alpha^3, \qquad \left[\alpha^1, \sigma_3 \right] = - 2i \alpha^2, \quad \text{etc.}
\]
So
\begin{align}
\left[H, \sigma_3 \right] &= 2 \hbar c \, \left( \alpha^1 \nabla_2 - \alpha^2 \nabla_1 \right) \quad \text{and thus} \notag \\
\left[H, \boldsymbol{\sigma} \right] &= 2 \hbar c \, \boldsymbol{\alpha} \times \nabla
\end{align}
Thus 
\begin{equation}
\boldsymbol{L} + \tfrac{1}{2} \hbar \boldsymbol{\sigma} = \hbar \boldsymbol{J}
\end{equation}
is a constant, the total angular momentum, because by (40), (41) and (42) 
\index{angular~momentum!conservation~of}%
\[
\left[ H, \boldsymbol{J} \right] = 0
\]
$\boldsymbol{L}$ is the orbital a.\ m.\  and $\tfrac{1}{2} \hbar \boldsymbol{\sigma}$ the spin a.\ m.\  This agrees with the N.\ R.\  theory. But in that theory the spin and $L$ of a free particle were \emph{separately} constant. This is no longer the case. \\

When a central force potential $V(r)$ is added to H, the operator $\boldsymbol{J}$ still is constant. \\

\section*{Elementary Solutions}
\addcontentsline{toc}{section}{Elementary Solutions}

For a particle with a particular momentum $\boldsymbol{p}$ and energy $E$, the wave function will be 
\index{momentum}%
\index{electron!states|(}%
\begin{equation}
\psi(x, t) = u \exp \left(i \frac{\boldsymbol{p} \boldsymbol{\cdot} \boldsymbol{x}}{\hbar} - i \frac{Et}{\hbar} \right)
\end{equation}
where $u$ is a constant spinor. The Dirac equation then becomes an equation for $u$ only
\index{Dirac!equation}%
\begin{equation}
Eu = \left(c\, \boldsymbol{\alpha} \boldsymbol{\cdot} \boldsymbol{p} + mc^2 \beta \right) u
\end{equation}
We write now 
\begin{equation}
p_{+} = p_1+ ip_2    \qquad \qquad p_{-} = p_1- ip_2
\end{equation}
Then (44) written out in full becomes
\begin{align}
\left( E - mc^2 \right) u_1 &= c\,\left(p_3 u_3 + p_{-} u_4 \right) \notag \\
\notag \\
\left( E - mc^2 \right) u_2 &= c\,\left(p_{+} u_3 - p_3 u_4 \right) \notag \\
\\
\left( E + mc^2 \right) u_3 &= c\,\left(p_3 u_1 + p_{-} u_2 \right) \notag \\
\notag \\
\left( E + mc^2 \right) u_4 &= c\,\left(p_{+} u_1 - p_3 u_2 \right) \notag
\end{align}
These 4 equations determine $u_3$ and $u_4$ given $u_1$ and $u_2$, or vice-versa. And either $u_1$ and $u_2$, or $u_3$ and $u_4$, can be chosen arbitrarily provided that\endnote{
$E$ lacked the exponent 2 in Eq.\ (47).}
\begin{equation}
E^{2} = m^2c^4 + c^2 p^2
\end{equation}
Thus given $p$ and \(E = + \sqrt{m^2c^4 + c^2p^2} \), there are two independent solutions of (46); these are, in non-normalized form:
\begin{equation}
\left( 
\begin{matrix}
	1 \\
       0 \\
  \dfrac{c\, p_3}{E + mc^2}\vphantom{\dfrac{V^2}{V^2}} \\
  \dfrac{c\, p_{+}}{E + mc^2}\vphantom{\dfrac{V^2}{V^2}}
\end{matrix}
\right)
\qquad
\left( 
\begin{matrix}
	0 \\
       1 \\
  \dfrac{c\, p_{-}}{E + mc^2}\vphantom{\dfrac{V^2}{V^2}} \\
  \dfrac{-c\, p_3}{E + mc^2}\vphantom{\dfrac{V^2}{V^2}}
\end{matrix}
\right)
\end{equation}
This gives the two spin-states of an electron with given momentum, as required physically. \\
\index{electron!states|)}%
 
But there are also solutions with \( E = - \sqrt{m^2c^4 + c^2p^2} \). In fact again two independent solutions, making 4 altogether. These are the famous \emph{negative energy states}. Why cannot we simply agree to ignore these states, say they are physically absent? Because when fields are present the theory gives transitions from positive to negative states. e.g.H atom should decay to negative state in 10$^{-10}$ secs. or less.

Certainly negative energy particles are not allowed physically. They can for example never be stopped by matter at rest, with every collision they move faster and faster. So Dirac was driven to
\index{Dirac, P. A. M.}%

\section*{The Hole Theory}
\addcontentsline{toc}{section}{The Hole Theory} 
\index{Dirac!hole~theory}%

All negative-energy states are normally filled by one electron each. Because of the exclusion principle transitions of ordinary electrons to these states are forbidden. If sometimes a negative energy state of momentum $-p$ energy $-E$ is \emph{empty}, this appears as a particle of momentum $p$ energy $+E$, and the opposite charge to an electron, i.e. an ordinary positron. 
\index{momentum}%

Thus we are led at once to a many-particle theory in order to get sensible results. With spin-0 particles, to get positive probabilities. With spin-$\tfrac{1}{2}$ particles, to get positive energies.

The Dirac theory in its one-particle form cannot describe properly the interaction between several particles. But so long as we are talking only about free particles, we can describe them with one-particle wave-functions. \\
\index{Dirac!electron~theory}%

\section*{Positron States}
\addcontentsline{toc}{section}{Positron States -- charge conjugation} 
\index{positron!states|(}%

So which wave-function will describe a positron with momentum $p$ and energy $E$? Clearly the wave function should be of the form  
\index{positron}%
\index{momentum}%
\begin{equation}
\phi(x, t) = v \exp \left( i \frac{\boldsymbol{p} \boldsymbol{\cdot} \boldsymbol{x}}{\hbar} - i \frac{Et}{\hbar} \right)
\end{equation}
as always in quantum mechanics. But the negative-energy electron whose \emph{absence is} the positron has a wave-function 
\index{electron!negative~energy}%
\begin{equation}
\psi(x, t) = u \exp \left( -i \frac{\boldsymbol{p} \boldsymbol{\cdot} \boldsymbol{x}}{\hbar} + i \frac{Et}{\hbar} \right)
\end{equation}
since it has a momentum $-p$ energy $-E$. \\
Thus we must take 
\begin{equation}
\phi = C \psi^{+} , \qquad \text{i.e.} \qquad v = C u^{+}
\end{equation}
where $\psi^{+}$ is $\psi$ with complex conjugate elements but \emph{not} transposed, and $C$ is a suitable constant matrix;
\[
\psi^{+}(x, t) = u^{+} \exp \left( i \frac{\boldsymbol{p} \boldsymbol{\cdot} \boldsymbol{x}}{\hbar} - i \frac{Et}{\hbar} \right)
\]

We know that $u$ is a solution of 
\begin{equation}
Eu = \left( c\, \boldsymbol{\alpha} \boldsymbol{\cdot} \boldsymbol{p} - mc^2 \beta \right) u
\end{equation}
We want the theory to make no distinction between electrons and positrons, and so $v$ must also satisfy the Dirac equation 
\index{Dirac!equation!positron}%
\begin{align}
Ev &= \left( c\, \boldsymbol{\alpha} \boldsymbol{\cdot} \boldsymbol{p} + mc^2 \beta \right) v \notag \\
ECu^{+} &= \left( c\, \boldsymbol{\alpha} \boldsymbol{\cdot} \boldsymbol{p} + mc^2 \beta \right) Cu^{+} 
\end{align}
But from (52) we have for $u^{+}$ the equation
\begin{equation}
Eu^{+} = \left( c\,\boldsymbol{\alpha}^{+} \! \boldsymbol{\cdot} \boldsymbol{p} - mc^2 \beta^{+} \right) u^{+}
\end{equation}
In order that (53) and (54) be identical we should have
\begin{equation}
C\alpha^{k +} = \alpha^{k} C, \qquad \qquad C\beta^{+} = - \beta C
\end{equation}
Now in fact
\[
\alpha^{1+} = \alpha^{1} \qquad \alpha^{3+} = \alpha^{3} \qquad \alpha^{2+} = - \alpha^{2} \qquad \beta^{+} = \beta
\]
Therefore a suitable C will be
\begin{equation}
C = - i \beta \alpha^{2} = \gamma_{2} = \left(\, \begin{matrix}
0 &  \tiny{\begin{matrix}
		0 & \!\!\!-1 \\
		1 & 0  \\
		\end {matrix}} \\
\tiny{\begin{matrix}
\; \;0 & 1 \\
-1 & 0 \\
\end{matrix}} & 0 \\
\end{matrix}\, \right)
\end{equation}
The relation between $\psi$ and $\phi$ is symmetrical because 
\begin{equation}
C^2 = \mathbb{I} \qquad \qquad \text{Hence} \qquad \qquad \psi = C \phi^{+}
\end{equation}
The $\phi$ is called the \emph{charge-conjugate} wave-function corresponding to the negative-energy electron $\psi$. 
\index{electron!negative~energy}%
Clearly 
\begin{equation}
\phi^{*}\phi = \left(C \psi^{+} \right)^{*} \left(C \psi^{+} \right) = \psi^{T}C^{*}C \psi^{+} = \psi^{*}\left(C^{*}C \right)^{T} \psi = \psi^{*} \psi
\end{equation}
And
\begin{equation}
\phi^{*} \alpha^{k} \phi = \psi^{T}C^{*}\alpha^{k}C \psi^{+} = \psi^{*}C \alpha^{kT} C \psi = \psi^{*} \alpha^{k} \psi
\end{equation}
Thus the probability and flow densities are the same for a positron as for the conjugate negative electron. 
\index{probability!density}%

For many purposes it is easier to represent positrons directly by the $\overline{\psi}$ wave-function, e.g. in computing cross-sections for pair creation   and so forth as we shall do later.
\index{creation}%
\index{cross-section!pair~creation}%
But if you actually want to \emph{see} the positron, e.g. in describing the details of a positronium experiment, it is necessary to use the $\phi$ wave-function to represent e.g. the way the spin is pointing.

This is all we shall say about free electrons and positrons.  \\
\index{positron!states|)}%

\section*{Electromagnetic Properties of the Electron}
\addcontentsline{toc}{section}{Electromagnetic Properties of the Electron}

Given an external (c-number) electromagnetic field defined by the potentials
\[
A_{\mu} \qquad \qquad \mu = 1, 2, 3, 4 \qquad \qquad A_{4} = i \Phi
\]
given functions of space and time. Then the motion of a particle in the field is found by substituting in the free-particle Lagrangian 
\index{Lagrangian,~free-particle}%
\begin{align}
&E + e\,\Phi \qquad \text{for $E$} \notag \\
&\boldsymbol{p} + \frac{e}{c} \boldsymbol{A} \qquad \text{for}\; \boldsymbol{p}
\end{align}
where $(-e)$ is the electron charge. We write the momentum-energy 4-vector 
\index{momentum}%
\begin{equation}
p = \left( p_1, \; p_2, \; p_3, \; p_4 = iE/c \right)
\end{equation}
Then we have to substitute simply
\begin{equation}
p_{\mu} + \frac{e}{c} A_{\mu} \qquad \text{for} \; \; p_{\mu}
\end{equation}
Now in the quantum theory
\begin{equation}
p_{\mu} \rightarrow -i \hbar \frac{\partial}{\partial x_{\mu}}
\end{equation}
Therefore the Dirac equation with fields is 
\index{Dirac!equation!with~electromagnetic~fields}%
\begin{equation}
\sum_1^4 \gamma_{\mu} \left( \frac{\partial}{\partial x_{\mu}} + \frac{ie}{\hbar c} A_{\mu} \right) \psi + \frac{mc}{\hbar}\, \psi = 0
\end{equation}
\begin{equation}
\sum_1^4 \left( \frac{\partial}{\partial x_{\mu}} - \frac{ie}{\hbar c} A_{\mu} \right) \overline{\psi}\, \gamma_{\mu} - \frac{mc}{\hbar}\, \overline{\psi} = 0
\end{equation}
In the non-covariant notations this is 
\begin{equation}
i \hbar \frac{\partial \psi}{\partial t} = \left[ -e\, \Phi + \sum_1^3 \left( -i\hbar c\, \frac{\partial}{\partial x_{k}} + e A_{k} \right) \alpha^k + mc^2 \beta \right] \psi
\end{equation}
since by (57), we have \( \overline{\psi} \gamma_{\mu} = \psi^{*} \beta \gamma_{\mu} = \left(C\phi^{+}\right)^{T} \beta \gamma_{\mu} = \phi^{T}C^{T}\beta \gamma_{\mu} \); the wave function \( \phi = C\psi^{+} \)of a positron satisfies by (65)
\begin{equation}
\sum \left( \frac{\partial}{\partial x_{\mu}} - \frac{ie}{\hbar c} A_{\mu} \right)\gamma_{\mu}^{T}\beta C \phi - \frac{mc}{\hbar}\, \beta C \phi = 0
\end{equation}
Multiplying by $C\beta$ this gives
\begin{equation}
\sum\left( \frac{\partial}{\partial x_{\mu}} - \frac{ie}{\hbar c} A_{\mu} \right)\gamma_{\mu} \phi + \frac{mc}{\hbar}\, \phi = 0
\end{equation}
This is exactly the Dirac equation for a particle of positive charge (+e). We have used
\index{Dirac!equation!positron}%
\begin{equation}
C\beta \gamma_{\mu}^{T} \beta C = - \gamma_{\mu} ,
\end{equation}
which follows from (15), (32), and (55). 

\section*{The Hydrogen Atom}
\addcontentsline{toc}{section}{The Hydrogen Atom}

This is the one problem which it is possible to treat very accurately using the one-electron Dirac theory.  The problem is to find the eigenstates of the equation 
\index{Dirac!equation}%
\index{hydrogen~atom|(}%
\begin{align}
E \psi &= H \psi \notag \\
H &= -i \hbar c \, \boldsymbol{\alpha} \boldsymbol{\cdot} \nabla + mc^2 \beta - \frac{e^2}{r}
\end{align}

As in the NR theory, we have as quantum numbers in addition to $E$ itself the quantities 
\begin{equation}
j_{z} = -i \left[\boldsymbol{r} \times \nabla \right]_{3} + \tfrac{1}{2} \sigma_{3}
\end{equation}
\begin{equation}
j(j+1) = J^2 = \left[-i \left( \boldsymbol{r} \times \nabla \right) + \tfrac{1}{2} \boldsymbol{\sigma} \right]^2
\end{equation}
where $j_z$ and $j$ are now \emph{half-odd integers} by the ordinary theory of angular momenta.
\index{angular~momentum}%
These quantum numbers are not enough to fix the state, because each value of $j$ may correspond to two NR states with $\ell = j \pm \tfrac{1}{2}$. Therefore we need an additional operator which commutes with $H$, which will distinguish between states with $\boldsymbol{\sigma}$ parallel or antiparallel to $\boldsymbol{J}$. The obvious choice is 
\[
Q = \boldsymbol{\sigma} \boldsymbol{\cdot} \boldsymbol{J}
\]
But \( \left[ H, \boldsymbol{\sigma} \right] \) is non-zero and rather complicated. So it is better to try
\begin{equation}
Q = \beta \boldsymbol{\sigma} \boldsymbol{\cdot} \boldsymbol{J}
\end{equation}
which is the same in the NR limit.

\noindent Then we have 
\[
\left[ H, Q \right] = \left[ H, \beta \boldsymbol{\sigma} \boldsymbol{\cdot} \boldsymbol{J} \right]  = \left[ H, \beta \boldsymbol{\sigma}  \right] \boldsymbol{\cdot} \boldsymbol{J} + \beta \boldsymbol{\sigma} \boldsymbol{\cdot} \left[ H,  \boldsymbol{J} \right] 
\]
But \( \left[ H, \boldsymbol{J} \right] = 0 \); furthermore, since
\[
\alpha^{k} \beta \sigma_{\ell} = \beta \sigma_{\ell} \alpha^{k} \quad k \ne \ell \qquad \text{and} \qquad \alpha^{k} \beta \sigma_{k} = - \beta \sigma_{k} \alpha^{k} 
\]
we get
\[
\left[H, \beta \boldsymbol{\sigma} \right] = - i \hbar c \, \left\{ \left(\boldsymbol{\alpha} \boldsymbol{\cdot} \nabla \right) \beta \boldsymbol{\sigma} - \beta \boldsymbol{\sigma} \left( \boldsymbol{\alpha} \boldsymbol{\cdot} \nabla \right) \right\} = -2i \hbar c \sum_{k = 1}^3 \alpha^{k} \sigma_{k}\, \beta \, \nabla_{k}
\]
Therefore
\[
\left[H, \beta \boldsymbol{\sigma} \right] \boldsymbol{\cdot} \boldsymbol{J} = -2 \hbar c \sum_{k = 1}^3 \alpha^{k} \sigma_{k}\, \beta \, \nabla_{k} \left( \boldsymbol{r} \times \nabla \right)_{k} - i \hbar c \left( \boldsymbol{\alpha} \boldsymbol{\cdot} \nabla \right) \beta = - i \hbar c \left( \boldsymbol{\alpha} \boldsymbol{\cdot} \nabla \right) \beta = \left[ H, \tfrac{1}{2} \beta \right]
\]
because
\[
\nabla \boldsymbol{\cdot} \boldsymbol{r} \times \nabla = 0 \qquad \text{and} \qquad \alpha^{k} \sigma_{k} = \left( \begin{matrix} 
0 & 1 \\
1 & 0
\end{matrix} \right) \qquad \text{for all $k$'s.}
\]
Hence the quantity which commutes with $H$ and is a constant of the motion is 
\begin{equation}
K = \beta \boldsymbol{\sigma} \boldsymbol{\cdot} \boldsymbol{J} - \tfrac{1}{2} \beta
\end{equation}
There must be a relation between $K$ and $J$. In fact
\begin{align*}
K^2 = \left( \frac{\boldsymbol{\sigma} \boldsymbol{\cdot} \boldsymbol{L}}{\hbar} + 1 \right)^2 &= \frac{L^2}{\hbar^2} + \frac{\boldsymbol{\sigma} \boldsymbol{\cdot} \boldsymbol{L}}{\hbar} + 1 \\
J^2 = \left( \frac{\boldsymbol{L}}{\hbar} + \tfrac{1}{2} \boldsymbol{\sigma} \right)^2 &= \frac{L^2}{\hbar^2} + \frac{\boldsymbol{\sigma} \boldsymbol{\cdot} \boldsymbol{L}}{\hbar} + \tfrac{3}{4}
\end{align*}
Hence
\begin{equation}
K^2 = J^2 + \tfrac{1}{4} = \left(j + \tfrac{1}{2}\right)^2
\end{equation}
Therefore $K$ has \emph{integer} eigenvalues not zero,
\begin{align}
K &= k = \pm \left(j + \tfrac{1}{2} \right) \\
j &= \left| k \right| - \tfrac{1}{2}, \quad k = \pm 1,\; \pm 2,\; \pm 3,\; \dots
\end{align}

Using the eigenvalue for $K$, we can simplify the Hamiltonian, which we could not do as in the NR theory with the eigenvalue of $L^2$ alone.
\index{hydrogen~atom!Hamiltonian}%
First
\begin{equation}
\boldsymbol{\sigma} \boldsymbol{\cdot} \boldsymbol{r}\; \boldsymbol{\sigma} \boldsymbol{\cdot} \left( \boldsymbol{r} \times \nabla \right) = i \boldsymbol{\sigma} \boldsymbol{\cdot} \left( \boldsymbol{r} \times \left( \boldsymbol{r} \times \nabla \right) \right) = i \left( \boldsymbol{\sigma} \boldsymbol{\cdot} \boldsymbol{r} \right) \left( \boldsymbol{r} \boldsymbol{\cdot} \nabla \right) - i r^2 \boldsymbol{\sigma} \boldsymbol{\cdot} \nabla
\end{equation}
Let now
\begin{equation}
\epsilon = -i \alpha^1 \alpha^2 \alpha^3    \qquad \sigma_k = \epsilon \alpha^k
\end{equation}
Then multiplying (78) by $\epsilon^{-1}$ we get:
\[
-r^2 \, i \boldsymbol{\alpha} \boldsymbol{\cdot} \nabla = \boldsymbol{\alpha} \boldsymbol{\cdot} \boldsymbol{r} \, \boldsymbol{\sigma} \boldsymbol{\cdot} \left(\boldsymbol{r} \times \nabla \right) - i\, \boldsymbol{\alpha} \boldsymbol{\cdot} \boldsymbol{r} \left( r \frac{\partial}{\partial r} \right)
\]
Let \( \alpha_r = \dfrac{1}{r}\, \boldsymbol{\alpha} \boldsymbol{\cdot} \boldsymbol{r} \), then by (39) and (42)
\[
-i\, \boldsymbol{\alpha} \boldsymbol{\cdot} \nabla = \frac{1}{r} \alpha_r \left(i \boldsymbol{\sigma} \boldsymbol{\cdot} \boldsymbol{J} - \tfrac{3}{2} i \right) - i \alpha_r \frac{\partial}{\partial r}= \frac{1}{r}\, \alpha_r \left(i \beta K - i \right) - i \alpha_r \frac{\partial}{\partial r}
\]
Thus finally we can write (70) in the form
\begin{equation}
H = mc^2\beta - \frac{e^2}{r} + i\hbar c\, \alpha_r \left( \frac{\beta K}{r} - \frac{1}{r} - \frac{\partial}{\partial r} \right)
\end{equation}
This gives the Dirac equation as an equation in the single variable $r$, having separated all angular variables.
\index{Dirac!equation}%

For the solution of this equation, see -- Dirac, \emph{Quantum Mechanics}, Third Edition, Sec.\  72, pp 268-271. 
\index{Dirac, P. A. M.}%

\section*{Solution of Radial Equation}
\addcontentsline{toc}{section}{Solution of the radial equation}

We may choose a two-component representation in which 
\index{hydrogen~atom!radial~equation|(}%
\begin{equation}
\beta = \left( \begin{matrix}
             1 & 0 \\
             0 & -1
            \end{matrix} \right) ,
            \qquad
\alpha_r =  \left( \begin{matrix}
             0 & i \\
             -i & 0
            \end{matrix} \right) ,
            \qquad
\psi = \left( \begin{matrix}
             u  \\
             v
            \end{matrix} \right)
\end{equation}
Then
\begin{align}
\left(E - mc^2 \right)u &=  - \frac{e^2}{r} \,u + \hbar c \left(\frac{1 + K}{r} + \frac{\partial}{\partial r} \right) v \notag \\
\\
\left(E + mc^2 \right)v &=  - \frac{e^2}{r} \,v + \hbar c \left(-\frac{1 - K}{r} - \frac{\partial}{\partial r} \right) u \notag
\end{align}
Let now
\begin{equation}
a_1 = \frac{-E + mc^2}{\hbar c}\; ; \qquad a_2 = \frac{E + mc^2}{\hbar c}\; ; \qquad \alpha =  \frac{e^2}{\hbar c}\; ,
\end{equation}
the fine structure constant. 
\index{fine~structure~constant}%
\index{alpha@$\alpha$|see{fine~structure~constant}}%
Then 
\begin{align}
\left( -a_1 + \frac{\alpha}{r} \right) u &= \left( \frac{1 + K}{r} + \frac{\partial}{\partial r} \right) v \notag \\
\\
\left( a_2 + \frac{\alpha}{r} \right) v &= \left( \frac{-1 + K}{r} - \frac{\partial}{\partial r} \right) u \notag 
\end{align}
Next put \( a = \sqrt{a_1 a_2} = \sqrt{m^2c^4 - E^2}/ \hbar c \) which is the magnitude of the imaginary momentum of a free electron of energy $E$. Then \( \psi \sim e^{-ar} \) at infinity. Hence we write 
\index{momentum}%
\begin{align}
u &=  \frac{e^{-ar}}{r} f \\
v &= \frac{e^{-ar}}{r} g \notag
\end{align}
So \begin{align}
\left( \frac {\alpha}{r} - a_1 \right) f  &= \left( \frac{\partial}{\partial r} - a  + \frac{k}{r} \right) g \notag \\
\\
\left( \frac {\alpha}{r} + a_2 \right) g  &= \left( - \frac{\partial}{\partial r} + a  + \frac{k}{r} \right) f \notag
\end{align}
Now we try solutions in series
\begin{equation}
f = \sum c_s r^s \; ,  \qquad g = \sum d_s r^s
\end{equation}
This gives
\begin{align}
\alpha\, c_s - a_1 c_{s-1} &= - a d_{s-1} + (s + k)\, d_s  \notag \\
\\
\alpha\, d_s + a_2 d_{s-1} &= + a c_{s-1} + (-s + k)\, c_s \notag
\end{align}
Putting
\[
e_s = a_1 c_{s-1} - a d_{s -1} \qquad \text{we have} \qquad e_s = \alpha\, c_s - (s + k)\, d_s = \frac{a_1}{a} \left( \alpha \, d_s + (s - k)\, c_s \right)
\]
\[
c_s = \frac{a_1 \alpha + a \,(s + k)}{a_1 \alpha^2 + a_1 (s^2 - k^2)} \, e_s \qquad \qquad d_s = \frac{a \alpha - a_1 \,(s - k)}{a_1 \alpha^2 + a_1 (s^2 - k^2)} \, e_s
\]

\[
e_{s+1} = \frac{\left( a_1^2 - a^2 \right) \alpha + 2 s a a_1}{a_1 \alpha^2 + a_1 \left( s^2 - k^2 \right)}\, e_s
\] \\
\noindent Suppose the series do not terminate. Then for large $s$
\[
\frac{e_{s+1}}{e_s} \approx \frac{c_{s+1}}{c_s} \approx \frac{2a}{s} \qquad \qquad \text{hence} \quad f \approx \exp(2ar)
\]
This is permissible when $a$ is imaginary. Thus there is a continuum of states with 
\begin{equation}
E > mc^2
\end{equation}
For real $a$ the series must terminate at both ends in order not to blow up at infinity. Suppose then\endnote{ 
$e_{s}$ lacked a subscript $s$} 
$e_s$ is non-zero for 
\begin{equation}
s = \epsilon +1, \; \epsilon + 2, \; \dots \; \epsilon + n \qquad n \ge 1
\end{equation}
and otherwise zero. This gives
\[
\alpha^2 + \epsilon^2 - k^2 = 0
\]
\[
\left(a_1^2 - a^2 \right) \alpha + 2 \left( \epsilon + n \right) a a_1 = 0
\]
Now not both $c_{\epsilon}$ and $d_{\epsilon}$ are zero, thus the wave function $r^{-1 + \epsilon}$ must be integrable at zero. This gives $\epsilon > - \tfrac{1}{2}$. But $\epsilon = \pm \sqrt{k^2 - \alpha^2}$. Now $k^2 \ge 1$, hence $\sqrt{k^2 - \alpha^2} > \tfrac{1}{2}$, and 
\begin{equation}
\epsilon = + \sqrt{k^2 - \alpha^2}
\end{equation}
Also \endnote{
The quantity $ \left( \left( mc^2 - E \right)^2 - \left(m^2c^4 - E^2 \right) \right )$ lacked an exponent 2.}
\[
\left( \epsilon + n \right)^2 = \left( \frac{a_1^2 - a^2}{2aa_1} \right)^2 \alpha^2 = \left( \frac{ \left( \left( mc^2 - E \right)^2 - \left(m^2c^4 - E^2 \right) \right )^2}{4 \left( m^2c^4 - E^2 \right) \left(mc^2 - E\right)^2} \right) \alpha^2 = \frac{4E^2 \alpha^2}{4\left(m^2c^4 - E^2 \right)}
\]
\[
E^2 = \frac{m^2c^4}{\left(1 + \dfrac{\alpha^2}{\left( \epsilon + n \right)^2 }\right)}
\]
Hence in this case
\begin{equation}
E = \frac{mc^2}{\sqrt{1 + \dfrac{\alpha^2}{\left(n + \sqrt{k^2 - \alpha^2} \right)^2}}}
\end{equation}
Given this positive $E$, $\left(a_1^2 - a^2 \right)$ is negative (see (83) and (84)) and so it is allowable to square $\left(\epsilon + n \right)$ to find these solutions, without introducing any difficulties. So for each
\begin{align}
k &= \pm 1, \pm 2, \pm 3, \dots  \\
n &= 1, 2, 3, \dots \notag
\end{align}
solutions exist, with $E$ given by (92). 

The alternative possibility is that \emph{all $e_s$ are zero}. Suppose not both of $c_{\epsilon}$ and $d_{\epsilon}$ are zero. Then \( \alpha^2 + \epsilon^2 - k^2 = 0 \) as before and so \( \epsilon = \sqrt{k^2 - \alpha^2} \). But now
\[
a_1 c_{\epsilon} - a d_{\epsilon} = 0
\]
\[
\alpha\,c_{\epsilon} - (\epsilon + k) d_{\epsilon}= 0
\]
Hence \( a \alpha - a_1 (\epsilon + k) = 0 \) and $k$ must be positive to make \( \epsilon + k = \sqrt{k^2 - \alpha^2} + k > 0 \). After this the solution goes as before. So solutions (92) exist for 
\begin{equation}
n = 0, \qquad k = +1, +2, +3, \dots
\end{equation}
The principal quantum number $N$ is 
\[
N = n + |k|
\]
Expanding in powers of $\alpha$
\index{fine~structure~constant}%
\index{hydrogen~atom!energy~levels}%
\begin{equation}
E = mc^2 \left[1 - \frac{1}{2} \frac{\alpha^2}{N^2} + \frac{\alpha^4}{N^3} \left( \frac{3}{8N} - \frac{1}{2 |k|} \right) \right] 
\end{equation}
\vspace{-0.25in}
\[
\hphantom{abcdefghkl}\underbrace{\hphantom{\frac{1}{2} \frac{\alpha^2}{N^2}}}_{\text{NR levels}}\hphantom{a}\underbrace{\hphantom{\frac{\alpha^4}{N^3}\left( \frac{3}{8N} - \frac{1}{2 |k|}\right)}}_{\text{fine structure}}
\]
There is \emph{exact} degeneracy between the two states of a given $|k|$. Nonrelativistic states are given by
\begin{align*}
j &= \ell + \tfrac{1}{2}  \rightarrow k = - (\ell + 1) \\
j &= \ell - \tfrac{1}{2} \rightarrow k = + \ell
\end{align*}
So
\[
\left. \begin{array}{l}
           ^2P_{1/2} \quad \text{is} \quad j = \frac{1}{2}, \quad k = 1 \vphantom{\frac{{M^2}^2}{{M^2}^2}} \\
           ^2S_{1/2} \quad \text{is} \quad j = \frac{1}{2}, \quad k = -1\vphantom{\frac{{M^2}^2}{{M^2}^2}} \\     
         \end{array}
          \right\} \rightarrow \text{degenerate}
\]
\vspace*{-0.1in}
\[          
 ^2S_{3/2}\quad \text{is} \quad j = \tfrac{1}{2}, \quad k = -2 \hphantom{\left. \right\}  \rightarrow \text{degenerate}}
\]
\index{hydrogen~atom!radial~equation|)}%
\index{hydrogen~atom|)}%
\section*{Behaviour of an Electron in a Non-Relativistic Approximation}
\addcontentsline{toc}{section}{Behavior of an Electron in non-relativistic approximation} 
\index{electron!in~electromagnetic~field!nonrelativistic~treatment}%

Multiplying the Dirac equation (64) by \( \sum_{\nu} \gamma_{\nu} \left( \dfrac{ \partial }{\partial x_{\nu}} + i \dfrac {e}{\hbar c} A_{\nu} \right) - \dfrac{mc}{\hbar} \) we have
\index{Dirac!equation}%
\begin{equation}
\sum_{\mu} \sum_{\nu} \gamma_{\mu} \gamma_{\nu} \left( \frac{ \partial }{\partial x_{\mu}} + i \frac {e}{\hbar c} A_{\mu} \right)\left( \frac{ \partial }{\partial x_{\nu}} + i \frac {e}{\hbar c} A_{\nu} \right) \psi - \frac{m^2c^2}{\hbar^2} \psi = 0
\end{equation}

Using \( \gamma_{\mu}^2 = 1, \gamma_{\mu} \gamma_{\nu} + \gamma_{\nu}\gamma_{\mu} = 0 \) this gives
\begin{equation}
\sum_{\mu} \left\{ \left( \frac{\partial}{\partial x_{\mu}} + \frac{ie}{\hbar c} A_{\mu} \right)^2 \right\} \psi - \frac{m^2c^2}{\hbar^2} \psi + \frac{ie}{2\hbar c} \sum_{\mu} \sum_{\nu} \sigma_{\mu \nu} F_{\mu \nu} \psi = 0
\end{equation}
Here 
\[
\sigma_{\mu \nu} = \tfrac{1}{2} \left( \gamma_{\mu} \gamma_{\nu} - \gamma_{\nu} \gamma_{\mu} \right)  \qquad \qquad F_{\mu \nu} = \frac{\partial A_{\nu}}{\partial x_{\mu}} - \frac{\partial A_{\mu}}{\partial x_{\nu}}
\] 
Thus $F_{12} = H_3$ magnetic field component
\begin{align*}
F_{14} = i \frac{\partial \Phi}{\partial x_1} + \frac{i}{c} \frac{\partial A_1}{\partial t} & = -i E_1  \quad \qquad \text{electric field} \\
\sigma_{12} &= i \sigma_3	\qquad \qquad \text{spin component}  \\
\sigma_{14} &= i \alpha_1 \qquad \qquad \text{velocity component}
\end{align*}
Thus (97) becomes 
\begin{equation}
\sum_{\mu} \left\{ \left( \frac{\partial}{\partial x_{\mu}} + \frac{ie}{\hbar c} A_{\mu} \right)^2 \right\} \psi - \frac{m^2c^2}{\hbar^2} \psi -\frac{e}{\hbar c} \left\{ \boldsymbol{\sigma} \boldsymbol{\cdot} \boldsymbol{H} - i \boldsymbol{\alpha} \boldsymbol{\cdot} \boldsymbol{E} \right\} \psi = 0
\end{equation}

This is still exact.\\

\noindent Now in the NR approximation
\begin{align*}
i \hbar \frac{\partial}{\partial  t} &= mc^2 + O(1) \\
\left\{ \left( \frac{\partial}{\partial x_{4}} + \frac{ie}{\hbar c} A_{4} \right)^2 \right\} - \frac{m^2c^2}{\hbar^2}  &=  \frac{1}{\hbar^2 c^2} \left\{ \left( -i \hbar  \frac{\partial }{\partial t} - e \Phi \right)^2 - m^2c^4 \right\} \\
&=  \frac{1}{\hbar^2 c^2} \left\{ \left( -i \hbar  \frac{\partial }{\partial t} - e \Phi  - mc^2 \right) \left( -i \hbar  \frac{\partial }{\partial t} - e \Phi  + mc^2 \right) \right\} \\
&=  \frac{1}{\hbar^2 c^2} \left\{ - 2mc^2 + O(1) \right\} \left( -i \hbar  \frac{\partial }{\partial t} - e \Phi  + mc^2 \right)
\end{align*}
Hence 
\[
\left(- i \hbar \frac{\partial}{\partial t} - e \Phi  + mc^2 \right) \psi -\frac{h^2}{2m} \sum_{k = 1}^3 \left\{ \left( \frac{\partial}{\partial x_{k}} + \frac{ie}{\hbar c} A_{k} \right)^2 \right\} \psi  + \frac{e \hbar}{2mc} \left[ \boldsymbol{\sigma} \boldsymbol{\cdot} \boldsymbol{H} - i \boldsymbol{\alpha} \boldsymbol{\cdot} \boldsymbol{E} \right] \psi  + O\left( \frac{1}{mc^2} \right)= 0
\]
The NR approximation means dropping the terms $O \left(1/mc^2 \right)$. Thus the NR Schr\"{o}dinger equation is 
\index{Schr\"{o}dinger!equation}%
\begin{equation}
 i \hbar \frac{\partial \psi}{\partial t} = \left\{ mc^2 - e \Phi  - \frac{h^2}{2m} \sum_{k = 1}^3 \left( \frac{\partial}{\partial x_{k}} + \frac{ie}{\hbar c} A_{k} \right)^2  + \frac{e \hbar}{2mc} \left( \boldsymbol{\sigma} \boldsymbol{\cdot} \boldsymbol{H} - i \boldsymbol{\alpha} \boldsymbol{\cdot} \boldsymbol{E} \right) \right\} \psi 
\end{equation}
The term $\boldsymbol{\alpha} \boldsymbol{\cdot} \boldsymbol{E}$ is really relativistic, and should be dropped or treated more exactly. Then we have exactly the equation of motion of a NR particle with a spin magnetic moment equal to
\index{electron!magnetic~moment}%
\begin{equation}
M = - \frac{e \hbar}{2mc} \sigma
\end{equation}
This is one of the greatest triumphs of Dirac, that he got this magnetic moment right out of his general assumptions without any arbitrariness.
\index{electron!magnetic~moment!Dirac~prediction}%

It is confirmed by measurements to about one part in 1000. Note that the most recent experiments show a definite discrepancy, and agree with the value
\begin{equation}
M = - \frac{e \hbar}{2mc} \sigma \left\{ 1 + \frac{e^2}{2 \pi \hbar c} \right\}
\end{equation}
calculated by Schwinger using the complete many-particle theory.\\
\index{Schwinger, Julian}%
\noindent \rule[0.02in]{1in}{0.01in}\\
\noindent \emph{Problem 2} Calculate energy values and wave functions of a Dirac particle  moving in a homogeneous infinite magnetic field. Can be done exactly. See F. Sauter, \emph{Zeitschrift f\"{u}r Physik} \textbf{69} (1931) 742.  \\
\index{Dirac!wave~function}%
\index{Sauter, Fritz}%
\index{problem!\emph{2}}%
\noindent \emph{Solution}

Take the field $\boldsymbol{B}$ in the $z$ direction.
\[
A_1 = - \tfrac{1}{2} By\; , \qquad A_2 = \tfrac{1}{2} B x
\]
The second-order Dirac equation (98) gives for a stationary state of energy $\pm E$
\index{Dirac!equation!second~order}%
\[
\left( \frac{E^2}{\hbar^2 c^2} - \frac{m^2c^2}{\hbar^2} \right) \psi \,+ \left( \frac{\partial}{\partial x} - \frac{1}{2} \frac{ieB}{\hbar c} y \right)^2 \! \psi \, +  \left( \frac{\partial}{\partial y} + \frac{1}{2} \frac{ieB}{\hbar c} x \right)^2 \! \psi \, + \frac{\partial^2}{\partial z^2}\, \psi - \frac{eB}{\hbar c} \sigma_z \psi = 0
\]
Taking a representation with $\sigma_z$ diagonal, this splits at once into two states with $\sigma_z = \pm 1$. Also
\[
L_z = -i \hbar \left\{ x \frac{\partial}{\partial y} - y \frac{\partial}{\partial x} \right\}
\]
is a constant of the motion, say $L_z = \ell \hbar$ where $\ell$ is an integer. And \( -i\hbar \dfrac{\partial}{\partial z} = p_z\). Let $\lambda = | eB \hbar c |$. Then
\[
\left\{ E^2 - m^2c^4 - c^2 p_z^2 \pm \left(\ell_z \pm 1 \right) \lambda \right\} \psi = \hbar^2 c^2 \left\{ \tfrac{1}{4} \frac{\lambda^2 r^2}{\hbar^4 c^4} - \left( \frac{\partial^2}{\partial x^2} + \frac{\partial^2}{\partial y^2} \right) \right\} \psi
\]
This is an eigenvalue problem with eigenvalues of a two-dimensional harmonic oscillator. 

\noindent Thus 
\[
E^2 = m^2c^4 + c^2p_z^2 + \lambda \left\{n \pm \left(\ell_z \pm 1 \right)   \right\}
\]
where \( \ell_z = 0, \pm 1, \pm 2, \dots , \pm (n -1) \). 

\noindent So the eigenvalues are 
\[
E = \sqrt{m^2c^4 + c^2p_z^2 + M |eB \hbar c |} \qquad \text{with} \; M = 0, 1, 2, \dots
\]
The lowest state has energy exactly $mc^2$. \\
\rule[0.05in]{1in}{0.01in} 

\newpage

\begin{center} \emph{\textbf{Summary of Matrices in the Dirac Theory in our notation}}\end{center}
\addcontentsline{toc}{section}{Summary of Matrices in the Dirac Theory in our notation}
\index{Dirac!matrices!Dyson~notation}%
\index{Dirac!matrices}%
\[
\alpha^k \alpha^{\ell} + \alpha^{\ell} \alpha^k = 2 \delta_{k \ell} \mathbb{I} \qquad \alpha^k \beta + \beta \alpha^k = 0 \qquad \beta^2 = \mathbb{I} \qquad \sigma_k \sigma_{\ell} + \sigma_{\ell} \sigma_k = 2 \delta_{k \ell} \mathbb{I}
\]
\[
\gamma_k = -i \beta \alpha^k \qquad \alpha^k = i\beta \gamma_k \qquad \gamma_4 = \beta \qquad \gamma_{\mu} \gamma_{\nu} + \gamma_{\nu} \gamma_{\mu} =  2 \delta_{\mu \nu} \mathbb{I} \qquad \left( \gamma_k \right)^{*} = \gamma_k 
\]
\[
\alpha^k \gamma_{\ell} - \gamma_{\ell} \alpha^k = 2i \delta_{\ell k} \beta  \qquad \gamma_5 = \gamma_1 \gamma_2 \gamma_3 \gamma_4 \qquad \gamma_{\mu} \gamma_5 + \gamma_5 \gamma_{\mu} =  0  \qquad \alpha^k \gamma_5 - \gamma_5 \alpha^k = 0  \qquad \gamma_5^2 = \mathbb{I}
\]
\\
\noindent We use the following representation:
\[
\sigma_1 = \left( \begin{matrix}
       0 & 1 \\
       1 & 0 
       \end{matrix} \right)
\qquad
\sigma_2 = \left( \begin{matrix}
       0 & -i \\
       i & 0 
       \end{matrix} \right)
\qquad
\sigma_3 = \left( \begin{matrix}
       1 & 0 \\
       0 & -1 
       \end{matrix} \right)
\]
\[
\alpha^k = \left( \begin{matrix}
      \mathbb{O} & \sigma_k \\
      \sigma_k     & \mathbb{O}
      \end{matrix} \right)
\quad \text{i.e.} \quad 
\alpha^1 =  \left( \begin{matrix}
       0 & 0 & 0 & 1 \\
       0 & 0 & 1 & 0 \\
       0 & 1 & 0 & 0 \\
       1 & 0 & 0 & 0 
       \end{matrix} \right)
       \quad 
\alpha^2 =  \left( \begin{matrix}
       0 & 0 & 0 & -i \\
       0 & 0 & i & 0 \\
       0 & -i & 0 & 0 \\
       i & 0 & 0 & 0 
       \end{matrix} \right)
       \quad 
\alpha^3 =  \left( \begin{matrix}
       0 & 0 & 1 & 0 \\
       0 & 0 & 0 & -1 \\
       1 & 0 & 0 & 0 \\
       0 & -1 & 0 & 0 
       \end{matrix} \right)
\]
\[
\beta = \left( \begin{matrix}
      \mathbb{I} & \mathbb{O} \\
      \mathbb{O} & - \mathbb{I}
      \end{matrix}     \right)
      \quad \text{i.e.} \quad
\beta = \gamma_4 = \left( \begin{matrix}
   	1 & 0 & 0 & 0 \\
       0 & 1 & 0 & 0 \\
       0 & 0 & -1 & 0 \\
       0 & 0 & 0 & -1 
       \end{matrix} \right)   
       	\quad
\gamma_5 = \left( \begin{matrix}
	\mathbb{O} & -\mathbb{I} \\
	-\mathbb{I}   & \mathbb{O}
	\end{matrix} \right)
	= \left( \begin{matrix}
	0 & 0 & -1 & 0 \\
	0 & 0 & 0 & -1 \\
	-1 & 0 & 0 & 0 \\
	0 & -1 & 0 & 0 
	\end{matrix} \right)
\]
\[
\gamma_k = \left( \begin{matrix}
      \mathbb{O} & -i\sigma_k \\
      i\sigma_k     & \mathbb{O}
      \end{matrix} \right)
\quad \text{i.e.} \quad 
\gamma_1 =  \left( \begin{matrix}
       0 & 0 & 0 & -i \\
       0 & 0 & -i & 0 \\
       0 & i & 0 & 0 \\
       i & 0 & 0 & 0 
       \end{matrix} \right)
       \quad 
\gamma_2 =  \left( \begin{matrix}
       0 & 0 & 0 & -1 \\
       0 & 0 & 1 & 0 \\
       0 & 1 & 0 & 0 \\
       -1& 0 & 0 & 0 
       \end{matrix} \right)
       \quad 
\gamma_3 =  \left( \begin{matrix}
       0 & 0 & -i & 0 \\
       0 & 0 & 0 & i \\
       i & 0 & 0 & 0 \\
       0 & -i & 0 & 0 
       \end{matrix} \right)
\]
\[
\sigma_k = \epsilon \alpha^k  \qquad \alpha^k = \epsilon \sigma_k \qquad \eta = i\epsilon \beta \qquad \epsilon = -i\alpha^1 \alpha^2 \alpha^3 \qquad \epsilon^2 = \eta^2 = \mathbb{I} \qquad \gamma_5 = - \epsilon
\]
\[
\sigma_k = \eta \gamma_k \qquad \gamma_k = \eta \sigma_k \qquad \epsilon = -i \eta \beta \qquad \eta = - \alpha^1 \alpha^2 \alpha^3
\]
\[
\epsilon = \left( \begin{matrix}
	\mathbb{O} & \mathbb{I} \\
	\mathbb{I}   & \mathbb{O}
	\end{matrix} \right)
	= \left( \begin{matrix}
	0 & 0 & 1 & 0 \\
	0 & 0 & 0 & 1 \\
	1 & 0 & 0 & 0 \\
	0 & 1 & 0 & 0 
	\end{matrix} \right)
	\qquad
\eta = \left( \begin{matrix}
	\mathbb{O} & -i \mathbb{I} \\
	i\mathbb{I}   & \mathbb{O}
	\end{matrix} \right)
	= \left( \begin{matrix}
	0 & 0 & -i & 0 \\
	0 & 0 & 0 & -i \\
	i & 0 & 0 & 0 \\
	0 & i & 0 & 0 
	\end{matrix} \right)
\]		
\[
\alpha^k \sigma_{\ell} + \sigma_{\ell} \alpha^k = 2 \delta_{\ell k} \epsilon \qquad \gamma_k \sigma_{\ell} + \sigma_{\ell} \gamma_k = 2 \delta_{\ell k} \eta \qquad \beta \sigma_k - \sigma_k \beta = 0 
\]
\[
\sigma_k \sigma_{\ell} = \alpha_k \alpha_{\ell} = \gamma_k \gamma_{\ell} = i \sigma_{m} \qquad k, \ell, m = (1, 2, 3) \; \text{cyclicly permuted}
\]
\[
\alpha^k \epsilon - \epsilon \alpha^k = \gamma_{\mu} \epsilon + \epsilon \gamma_{\mu} = \sigma_k \epsilon - \epsilon \sigma_k = 0
\]
\[
\alpha^k \eta + \eta \alpha^k = \gamma_k \eta - \eta \gamma_k = \sigma_k \eta - \eta \sigma_k = \beta \eta + \eta \beta = 0
\]
\[
\left. 
\begin{array}{l}
	\alpha_k \sigma_{\ell} = i \alpha_m \\
	\sigma_k \gamma_{\ell} = i \gamma_m \\
	\gamma_k \alpha_{\ell} = \beta \sigma_m
\end{array} \right\} \quad k, \ell, m = (1,2,3) \; \text{cyclicly permuted}
\]
\\
\noindent Comparison with the Dirac notation: \(\qquad \rho_1 = \epsilon \qquad \rho_2 =  \eta \qquad \rho_3 = \beta\). \\

\noindent Latin indices: 1, 2, 3. Greek indices: 1, 2,  3, 4.

\newpage

\begin{center} \emph{\textbf{Summary of Matrices in the Dirac Theory in the Feynman notation}} \end{center}
\addcontentsline{toc}{section}{Summary of Matrices in the Dirac Theory in the Feynman notation}
\index{Dirac!matrices!Feynman~notation}
\[
\alpha^k \alpha^{\ell} + \alpha^{\ell} \alpha^k = 2 \delta_{k \ell} \mathbb{I} \qquad \alpha^k \beta + \beta \alpha^k = 0 \qquad g_{00} = +1 \quad  g_{k k} = -1 \quad g_{\mu \nu} = 0, \; \mu \ne  \nu
\]
\[
\sigma_k \sigma_{\ell} + \sigma_{\ell} \sigma_k = 2 \delta_{k \ell} \mathbb{I} \qquad \beta^2 = \mathbb{I} 
\]
\[
\gamma_k =  \beta \alpha^k \qquad \alpha^k = \beta \gamma_k \qquad \gamma_0 = \beta \qquad \gamma_{\mu} \gamma_{\nu} + \gamma_{\nu} \gamma_{\mu} =  2 g_{\mu \nu} \mathbb{I} \qquad \left( \gamma_k \right)^{*} = -\gamma_k 
\]
\[
\alpha^k \gamma_{\ell} - \gamma_{\ell} \alpha^k = -2 \delta_{\ell k} \beta  \qquad \gamma_5 = i \gamma_0 \gamma_1 \gamma_2 \gamma_3 \qquad \gamma_{\mu} \gamma_5 + \gamma_5 \gamma_{\mu} =  0  \qquad \alpha^k \gamma_5 - \gamma_5 \alpha^k = 0  \qquad \gamma_5^2 = -\mathbb{I}
\]
\\
\noindent Representation:
\[
\sigma_1 = \left( \begin{matrix}
       0 & 1 \\
       1 & 0 
       \end{matrix} \right)
\qquad
\sigma_2 = \left( \begin{matrix}
       0 & -i \\
       i & 0 
       \end{matrix} \right)
\qquad
\sigma_3 = \left( \begin{matrix}
       1 & 0 \\
       0 & -1 
       \end{matrix} \right)
\]
\[
\alpha^k = \left( \begin{matrix}
      \mathbb{O} & \sigma_k \\
      \sigma_k     & \mathbb{O}
      \end{matrix} \right)
\quad \text{i.e.} \quad 
\alpha^1 =  \left( \begin{matrix}
       0 & 0 & 0 & 1 \\
       0 & 0 & 1 & 0 \\
       0 & 1 & 0 & 0 \\
       1 & 0 & 0 & 0 
       \end{matrix} \right)
       \quad 
\alpha^2 =  \left( \begin{matrix}
       0 & 0 & 0 & -i \\
       0 & 0 & i & 0 \\
       0 & -i & 0 & 0 \\
       i & 0 & 0 & 0 
       \end{matrix} \right)
       \quad 
\alpha^3 =  \left( \begin{matrix}
       0 & 0 & 1 & 0 \\
       0 & 0 & 0 & -1 \\
       1 & 0 & 0 & 0 \\
       0 & -1 & 0 & 0 
       \end{matrix} \right)
\]
\[
\beta = \left( \begin{matrix}
      \mathbb{I} & \mathbb{O} \\
      \mathbb{O} & - \mathbb{I}
      \end{matrix}     \right)
      \quad \text{i.e.} \quad
\beta = \gamma_0 = \left( \begin{matrix}
   	1 & 0 & 0 & 0 \\
       0 & 1 & 0 & 0 \\
       0 & 0 & -1 & 0 \\
       0 & 0 & 0 & -1 
       \end{matrix} \right)     
\]
\[
\gamma_k = \left( \begin{matrix}
      \mathbb{O} & \sigma_k \\
      -\sigma_k     & \mathbb{O}
      \end{matrix} \right)
\quad \text{i.e.} \quad 
\gamma_1 =  \left( \begin{matrix}
       0 & 0 & 0 & 1 \\
       0 & 0 & 1 & 0 \\
       0 & -1 & 0 & 0 \\
      -1 & 0 & 0 & 0 
       \end{matrix} \right)
       \quad 
\gamma_2 =  \left( \begin{matrix}
       0 & 0 & 0 & -i \\
       0 & 0 & i & 0 \\
       0 & -i & 0 & 0 \\
       i  & 0 & 0 & 0 
       \end{matrix} \right)
       \quad 
\gamma_3 =  \left( \begin{matrix}
       0 & 0 & 1 & 0 \\
       0 & 0 & 0 & -1 \\
       -1 & 0 & 0 & 0 \\
       0 & 1 & 0 & 0 
       \end{matrix} \right)
\]
\[
\rho_1 = \left( \begin{matrix}
	\mathbb{O} & \mathbb{I} \\
	\mathbb{I}   & \mathbb{O}
	\end{matrix} \right)
	= \left( \begin{matrix}
	0 & 0 & 1 & 0 \\
	0 & 0 & 0 & 1 \\
	1 & 0 & 0 & 0 \\
	0 & 1 & 0 & 0 
	\end{matrix} \right)
	= \gamma_5
	\qquad \qquad
\rho_2 = \left( \begin{matrix}
	\mathbb{O} & -i \mathbb{I} \\
	i\mathbb{I}   & \mathbb{O}
	\end{matrix} \right)
	= \left( \begin{matrix}
	0 & 0 & -i & 0 \\
	0 & 0 & 0 & -i \\
	i & 0 & 0 & 0 \\
	0 & i & 0 & 0 
	\end{matrix} \right)
\]		
\\
\[
\sigma_k = \rho_1 \alpha^k  \qquad \alpha^k = \rho_1 \sigma_k \qquad \rho_2 = i\rho_1 \beta \qquad \rho_1 = -i\alpha^1 \alpha^2 \alpha^3 \qquad \rho_1^2 = \rho_2^2 = \mathbb{I}
\]
\[
\sigma_k = -i \rho_2 \gamma_k \qquad \gamma_k = i \rho_2 \sigma_k \qquad \rho_1 = -i \rho_2 \beta \qquad \rho_2 = - \alpha^1 \alpha^2 \alpha^3 \beta
\]
\[
\alpha^k \sigma_{\ell} + \sigma_{\ell} \alpha^k = 2 \delta_{\ell k} \rho_1 \qquad \gamma_k \sigma_{\ell} + \sigma_{\ell} \gamma_k = -2 \delta_{\ell k} \rho_2 \qquad \beta \sigma_k - \sigma_k \beta = 0 
\]
\[
\sigma_k \sigma_{\ell} = \alpha_k \alpha_{\ell} = -\gamma_k \gamma_{\ell} = i \sigma_{m} \qquad k, \ell, m = (1, 2, 3) \; \text{cyclicly permuted}
\]
\[
\alpha^k \rho_1 - \rho_1 \alpha^k = \gamma_{\mu} \rho_1 + \rho_1 \gamma_{\mu} = \sigma_k \rho_1 - \rho_1 \sigma_k = 0
\]
\[
\alpha^k \rho_2 + \rho_2 \alpha^k = \gamma_k \rho_2 - \rho_2 \gamma_k = \sigma_k \rho_2 - \rho_2 \sigma_k = \beta \rho_2 + \rho_2 \beta = 0
\]
\[
\left. 
\begin{array}{l}
	\alpha_k \sigma_{\ell} = i \alpha_m \\
	\sigma_k \gamma_{\ell} = i \gamma_m \\
	\gamma_k \alpha_{\ell} = i\beta \sigma_m
\end{array} \right\} \quad k, \ell, m = (1,2,3) \; \text{cyclicly permuted}
\]
\\
\noindent Latin indices: 1, 2, 3. Greek indices: 0, 1, 2,  3.



\newpage
 
\pagestyle{fancy}
\fancyhead{}
\lhead{\emph{\MakeUppercase{Scattering Problems and Born Approximation}}}
\chead{}
\rhead{\thepage}
\lfoot{}
\cfoot{}
\rfoot{}

\chapter*{Scattering Problems and Born Approximation}
\addcontentsline{toc}{chapter}{Scattering Problems and Born Approximation}

\section*{General Discussion}
\addcontentsline{toc}{section}{General Discussion}

The problem of scattering of a Dirac particle by a potential can be treated exactly by finding the continuum solutions of the Dirac equation. This is a complicated business even for the simplest case of a Coulomb force. It was done by Mott, \emph{Proc.\ Roy.\ Soc.} \textbf{A135} (1932) 429. 
\index{Dirac!equation}%
\index{Mott, Neville~F.}%
\index{scattering!and~Born~approximation}%
\index{Coulomb~potential}%

For most purposes in relativistic problems, and always when the scattering is produced by complicated effects involving radiation theory, one uses the \emph{Born approximation}.  That is, we treat the scattering only to 1$^{\text{st}}$ order in the interaction, or only to some definite order in which we are interested.
\index{Born~approximation}%

The formula for scattering from an initial state $A$ to a final state $B$ lying in a continuum of states is, transition probability per unit time, 
\index{probability!transition}%
\begin{equation}
w = \frac{2 \pi}{\hbar} \rho_{E} \left| V_{BA} \right|^2
\end{equation}
This you ought to know. $\rho_E =$ density of final states per unit energy interval. $V_{BA}$ is the matrix element of the potential $V$ for the transition. Here $V$ may be anything, and may be itself a second-order or higher order effect obtained by using higher-order perturbation theory. 

The difficulties in real calculations usually come from the factors 2 and $\pi$ and the correct \emph{normalization of states}. Always I shall normalize the continuum states not in the usual way (one particle per unit volume) which is non-invariant, but instead
\begin{equation}
\text{One particle per volume} \; \frac{mc^2}{| E |}
\end{equation}
where $| E |$ is the energy of the particles. Then  under a Lorentz transformation the volume of a fixed region transforms like $1/| E |$ and so the definition stays invariant. 
\index{Lorentz!transformations}%

Thus a continuum state given by the spinor \( \psi = u \exp \left\{(i \boldsymbol{p} \boldsymbol{\cdot} \boldsymbol{x} - i E t) / \hbar \right\} \) is normalized so that 
\begin{equation}
u^{*}u = \frac{| E |}{mc^2}
\end{equation}
Now if we multiply the Dirac equation  for a free particle, (44), by $\overline{u}$ on the left, we get \( Eu^{*}\beta u = c u^{*}\beta \boldsymbol{\alpha} \boldsymbol{\cdot} \boldsymbol{p} u + mc^2 u^{*}u \); its complex conjugate is   \( Eu^{*}\beta u =- c u^{*}\beta \boldsymbol{\alpha} \boldsymbol{\cdot} \boldsymbol{p} u + mc^2 u^{*}u \) since $\beta \boldsymbol{\alpha}$ is anti-Hermitian; then by adding we get
\index{Hermitian}%
\index{Dirac!equation}%
\begin{equation}
E\,\overline{u} u = mc^2 u^{*}u
\end{equation}
Therefore the normalization becomes 
\index{electron!states}%
\index{positron!states}%
\begin{equation}
\left.
\begin{array}{ll}
\overline{u}u &= +1 \; \; \text{for electron states} \\
			&= -1 \; \; \text{for positron states}
\end{array} \right\} = \epsilon; \qquad \text{This is the definition of}\; \epsilon .
\end{equation}
With this normalization the density of states in momentum space is one per volume $h^3$ of phase space, that is to say
\index{momentum!space}%
\begin{equation} 
\rho = \frac{1}{h^3} \frac{mc^2}{| E |} \, dp_1 dp_2 dp_3
\end{equation}
per volume $dp_1 dp_2 dp_3$ of momentum space, for \emph{each} direction of spin and each sign of  charge. Now we have again an invariant differential
\begin{equation}
\frac{dp_1 dp_2 dp_3}{ | E |}
\end{equation}

\section*{Projection Operators}
\addcontentsline{toc}{section}{Projection Operators}
\index{projection~operator}%
Usually we are not interested in the spin either of an intermediate or of an initial or final state. Therefore we have to make sums over spin states which are of the form
\begin{equation}
S   = \sum_2 \left( \overline{s} O u \right)\left(\overline{u} P r \right)
\end{equation}
where $O$ and $P$ are some kind of operators, $s$ and $r$ some kind of spin states, and the sum is over the two spin states $u$ of an electron of momentum $\boldsymbol{p}$ and energy $E$. \\
\index{momentum}%

\noindent Let us write 
\begin{equation}
\slashed{p} = \sum_{\mu} p_{\mu} \gamma_{\mu} \; , \qquad p_4 = iE/c
\end{equation}
The Dirac equation satisfied by $u$ is 
\index{Dirac!equation}%
\begin{equation}
\left( \slashed{p} - imc \right) u = 0
\end{equation}
The two spin states with momentum 4-vector $-p$ satisfy
\begin{equation}
\left( \slashed{p} + imc \right) u = 0
\end{equation}
As one can easily show from (48), these 4 states are all orthogonal in the sense that $(\overline{u}^{\,\prime}u) = 0$ for each pair $u^{\prime} u$. Therefore the identity operator may be written in the form
\begin{equation}
\mathbb{I} = \sum_4 \left( u \overline{u} \right) \epsilon
\end{equation}
summed over all 4 states with $\epsilon$ defined as earlier. Hence by (111) and (112) we can write (109) as 
\begin{equation}
S = \sum_4 \left( \overline{s} O \frac{ \slashed{p} + imc}{2imc} \epsilon u \right) \left( \overline{u} P r \right) = \left(\overline{s}O \Lambda_{+} P r \right)
\end{equation}  
by virtue of (113); here the operator
\begin{equation}
\Lambda_{+} = \frac{\slashed{p} + imc}{2imc}
\end{equation}
is a projection operator for \emph{electrons} of momentum $p$. 
\index{electron!states!projection~operator}%

In the same way for a sum over the two positron states $u$ with momentum $p$ energy $E$
\index{positron!states}%
\begin{equation}
S   = \sum_2 \left( \overline{s} O u \right)\left(\overline{u} P r \right) = \left(\overline{s}O \Lambda_{-} P r \right)
\end{equation}
with
\begin{equation}
\Lambda_{-} = \frac{\slashed{p} - imc}{2imc}
\end{equation}
and we have 
\begin{equation}
\Lambda_{+} - \Lambda_{-} = \mathbb{I}
\end{equation}

These projection operators are covariant. In Heitler the business is done in a different way which makes them non-covariant and more difficult to handle.
\index{Heitler, Walter}%

Note that here charge-conjugate wave functions are \emph{not} used. The positrons of momentum $p$ are represented by the electron wave functions $u$ of momentum $-p$ energy $-E$. \\
\index{electron!wave~function}%

\section*{Calculation of Spurs}
\addcontentsline{toc}{section}{Calculation of Spurs}

Suppose we have to calculate an expression such as

\begin{equation*}
\tfrac{1}{2} \sum_{I} \sum_{F} \left( \overline{u}_{F} O u_{I} \right) \left( \overline{u}_I O u_{F} \right)
\end{equation*}
summed over electron states only. This gives
\begin{equation*}
\tfrac{1}{2} \sum \left( \overline{u}_{F} O \Lambda_{+} O \Lambda_{+} u_{F} \right) \epsilon
\end{equation*}
summed over all four spin states $u_F$. To calculate this, let us consider the general expression 
\[
\sum_u \epsilon \left( \overline{u} Q u \right)
\]
summed over all 4 spin states, where $Q$ is any $4 \times 4$ matrix.

Let $Q$ have the eigenvectors $w_1, w_2, w_3, w_4$ with eigenvalues $\lambda_1, \lambda_2, \lambda_3, \lambda_4$. Then
\[
Q = \sum_{k = 1}^4 \lambda_k w_k w_k^{*}
\]
and
\[
\sum_u \epsilon \left( \overline{u} Q u \right) = \sum_u \epsilon \sum_{k =1}^4 \lambda_k \left( \overline{u} w_k \right) \left( w^{*} u \right) = \sum_{k =1}^4 \lambda_k w^{*} \left\{ \sum_u \epsilon \left( u \overline{u} \right) \right\} w_k
\]
Hence by (113)
\begin{equation*}
\sum_u \epsilon \left( \overline{u} Q u \right) = \sum \lambda
\end{equation*}
Now \( \sum \lambda \) = sum of diagonal elements of $Q$ = Spur $Q$. Thus\endnote{
The \emph{spur} of a matrix $A$ is the sum of its diagonal elements, and denoted Spur $A$ or Sp $A$. This term is also, and more commonly, known as the \emph{trace} of a matrix, denoted Tr $A$.}
\[
\sum_u \epsilon \left(\overline{u} O u \right) = \text{Sp}\; Q
\]
and this is always easy to calculate.

\noindent \rule[0.02in]{1in}{0.01in}\\
\noindent \emph{Problem 3:} Given a steady potential $V$ a function of position, and a beam of incident particles, electrons. Solve the Schr\"{o}dinger equation in the Born approximation   \\
\hspace*{5ex}a) By stationary perturbation theory \\
\hspace*{5ex}b) By time-dependent perturbation theory. \\
Show that the results agree, with a transition probability per unit time given by \( w = (2 \pi /\hbar) \rho_E | V_{BA} |^2 \). Evaluate the cross section in the case \( V = - Ze^2/r \), averaging spin over initial state and summing over final state. \\
\hspace*{5ex}c) Repeat the calculation with particles obeying the Klein-Gordon equation, leaving out the $V^2$ term, by either method. Compare the angular distribution in the two cases. \\
\index{problem!\emph{3}}%
\index{Schr\"{o}dinger!equation}%
\index{Born~approximation}%
\index{distribution}%
\index{distribution!angular}%
\index{probability!transition}%
\index{Klein-Gordon~equation}%
\noindent \rule[0.05in]{1in}{0.01in} \\
\noindent \rule[0.02in]{1in}{0.01in}\\
\noindent \emph{Problem 4:} A nucleus (O$^{16}$) has an even\endnote{
``ever'' was written for ``even''}
$j = 0$ ground state and an even $j = 0$ excited state at 6 MeV. Calculate the total rate of emission of pairs, and the angular and momentum distributions.\\
\index{problem!\emph{4}}%
\index{distribution!angular}%
\index{distribution!momentum}%
\index{momentum!distribution}%
\noindent \emph{Solution:} Let $\Delta E$ be the excitation energy, $\rho_N$ and $\boldsymbol{j}_N$ the charge and current density operators of the nucleus. Then for the transition we are interested in $\rho_N$ and $\boldsymbol{j}_N$ are functions of position $r$ with the time-variation of the single matrix element given by $\exp \left\{ -i \Delta E/ \hbar \right\}$. Also 
\begin{equation}
\nabla \boldsymbol{\cdot} \boldsymbol{j}_N = - \frac{\partial \rho_H}{\partial t} = i \frac{\Delta E}{\hbar} \rho_N
\end{equation}
The electrostatic potential $V$ of the nucleus has the matrix element given by
\begin{equation}
\nabla^2 V = - 4\pi \rho_N
\end{equation}
The states being spherically symmetric, $\rho_N$ is a function of $r$ only, and so the general solution of Poisson's equation simplifies to\endnote{
This equation differs from the form given in the first edition by a factor of $-\tfrac{6}{\pi}$, but this form also seems incorrect; perhaps the correct relation is 
\[
V(r) = - 4 \pi \int_0^r \frac{1}{r_1^2} \, dr_1 \int_0^{r_1} {r_2}^2 \rho(r_2)\, dr_2
\]  }
\index{Poisson's~equation}%
\begin{equation}
V(r) = - \frac{6 \pi}{r} \int_0^r r_1^2 \, \rho_N\left(r_1\right) \,dr_1  
\end{equation}

Outside the nucleus $V(r) = Ze^2/r$  is constant in time, and so the matrix element of $V(r)$ for this transition is zero. In fact from (119) and (120) we get by integration
\begin{equation}
V(r) = \frac{\hbar}{i\Delta E} (-4 \pi)(-r) j_{No}(r) = \frac{4 \pi r \hbar}{i \Delta E} j_{No}(r)
\end{equation}
where $j_{No}$  is the outward component of the current.

The interaction which creates pairs is then
\index{creation!pair}%
\begin{equation}
I = \int \frac{4 \pi r \hbar}{i \Delta E} \, j_{No} (r) \left(-e \psi^{*} \psi(r) \right) d \tau
\end{equation}
As an approximation consider the de Broglie wavelengths  of all pairs long compared with the nuclear size. Then
\index{De~Broglie~wavelength}%
\begin{equation}
I = \psi^{*} \psi(0)\,\frac{4 \pi \hbar e i}{\Delta E} \, \int r\,j_{No} (r) \, d \tau
\end{equation}
The constant \(\int r\, j_{No} (r) \, d \tau \) is not known exactly. Suppose as an order of magnitude estimate that the nucleus of charge $Ze$ is uniformly spread over a sphere of radius $r_{o}$ in the ground state, and also in the excited state. Since $\rho_N$ is roughly uniform inside the nucleus we have by integrating (119):
\[
j_N = \frac{i \Delta E}{3 \hbar} \overline{r} \rho_N \; , \quad \text{and thus}
\]
\begin{equation}
I = \psi^{*} \psi(0) \left( \frac{-4 \pi e}{3} \right) \, \int r^2 \rho_N (r) \, d \tau = \psi^{*} \psi (0)\left( \frac{-4 \pi e}{3} \right) \, Q e^{-i \Delta E t/\hbar}
\end{equation}
$Q$ is roughly a measure of the charge-moment of inertia of the nucleus, and is equal to 
\[
\tfrac{3}{5} Zer_{o}^2
\]
Thus
\begin{equation}
I = - \frac{4 \pi Z e^2 r_{o}^2}{5} \left\{ \psi^{*} \psi (0) \right\} e^{-i \Delta E t/\hbar}
\end{equation}

So the problem is just to compute the probabilities of pair-emission with this interaction. Note that real radiation is strictly forbidden in a 0--0 transition, and so these pairs are actually observed in the reaction 
\begin{equation}
p + F_{19} \rightarrow O^{*}_{16} + \alpha \rightarrow  O_{16} + e^{+} + e^{-} + \alpha
\end{equation}
Is it correct to take for the interaction just
\[
\int V(r) \left( -e \psi^{*} \psi \right) d\tau
\]
taking the Coulomb potential  of the nuclear charge and ignoring all electrodynamic effects? Yes. Because in general the interaction would be
\index{Coulomb~potential}%
\begin{equation}
\int \left\{ \varphi \left( -e \psi^{*} \psi \right) - \sum_k A_k \left( -e \psi^{*} \alpha_k \psi \right) \right\} d\tau
\end{equation}
where $\varphi$, $A_k$ are the scalar and vector potentials satisfying the Maxwell equations
\index{Maxwell!equations}%
\[
\nabla^2 \varphi + \frac{1}{c} \nabla \boldsymbol{\cdot} \frac{\partial \boldsymbol{A}}{\partial t} = - 4\pi \rho_N
\]
\[
\nabla^2\! \boldsymbol{A} - \frac{1}{c^2} \frac{\partial^2 \! \boldsymbol{A}}{\partial t^2} - \nabla \left\{ \nabla \boldsymbol{\cdot} \boldsymbol{A} + \frac{1}{c} \frac{\partial \varphi}{\partial t} \right\} = - \frac{4 \pi}{c} \boldsymbol{j}_N
\]
The matrix element of the interaction (128) is unchanged by any gauge transformation of the $( \boldsymbol{A}, \varphi)$. Therefore we may take the gauge in which 
\index{gauge!transformation}%
\index{gauge!condition}%
\[
\nabla \boldsymbol{\cdot} \boldsymbol{A} = 0
\]
Incidentally, since $\varphi = V(r)$, the second Maxwell equation reduces to
\index{Maxwell!equations}%
\[
\nabla^2\! \boldsymbol{A} - \frac{1}{c^2} \frac{\partial^2 \! \boldsymbol{A}}{\partial t^2} - = - \frac{4 \pi}{c} \boldsymbol{j}_N
\]
Now, since there is no free radiation present, also $\nabla \times \boldsymbol{A} = 0$, and hence $\boldsymbol{A} = 0$, in this gauge, and therefore we can indeed ignore all electrodynamic effects. 

Let us calculate then the probability  of pair emission with the interaction (126).  A typical final state has an electron of momentum $p_1$ and a positron of momentum $p_2$, with energies $E_1$, $E_2$ and spins $u_1$, $u_2$ respectively. For the creation of this pair the matrix element of $I$ is
\index{probability!pair~emission}%
\index{electron-positron!creation}%
\index{positron}%
\index{creation!pair}%
\[
I = -C\,\overline{u}_1 \beta u_2 \; , \quad C = \frac{4 \pi Ze^2 r_{o}^2}{5} \tag{128a}
\]
The density of final states is by (107)
\begin{equation}
\frac{1}{(2 \pi \hbar)^6} \frac{m^2c^4}{E_1 E_2}\, p_1^2 dp_1\, d\omega_1\, p_2^2 dp_2\, d\omega_2
\end{equation}
where $d\omega_1$ and $d\omega_2$ are the solid angles  for $p_1$ and $p_2$. The creation probability per unit time is thus by (102)  
\index{angle!solid}%
\index{probability!pair~creation}%
\index{creation}%
\begin{equation}
w = \frac{2\pi}{\hbar} \frac{\rho_E}{dE} | I |^2 = \frac{2\pi}{\hbar} \frac{dp_1 dp_2}{d\left(E_1 +  E_2\right)} C^2 \frac{1}{(2\pi \hbar)^6} \frac{m^2 c^4\, p_1^2\, p_2^2\, d\omega_1\, d\omega_2}{E_1 E_2} \sum_{u_1, u_2} \left| \, \overline{u}_1  \beta u_2 \, \right|^2
\end{equation}
Now fixing $p_1$, 
\[
\frac{dp_2}{d(E_1 + E_2)} = \frac{dp_2}{dE_2} = \frac{E_2}{c^2 p_2}
\]
and 
\begin{align*}
\sum_{u_1, u2} \left| \, \overline{u}_1  \beta u_2\, \right|^2 &= \sum_{u_1, u2}  \left( \overline{u}_1  \beta u_2 \right) \left(\overline{u}_2  \beta u_1\right) = \text{Spur} \left\{\beta\,\frac{ \slashed{p}_2 - imc}{2imc}\, \beta \, \frac{ \slashed{p}_1 + imc}{2imc} \right\} \notag \\
&= -1 + \frac{ \boldsymbol{p}_1 \boldsymbol{\cdot} \boldsymbol{p}_2}{m^2 c^2} + \frac{E_1 E_2}{m^2c^4} = \frac{E_1 E_2 - m^2 c^4 + c^2 p_1 p_2 \cos \theta }{m^2 c^4}
\end{align*}
where $\theta$ is the angle between the pair. Then writing in (130)
\index{angle}%
\[
dE_1 = dp_1 \frac{c^2 p_1}{E_1} , \qquad d\omega_1 = 4 \pi , \qquad d\omega_2 = 2\pi \sin \theta \, d\theta
\]
we obtain the differential probability\endnote{
In the second edition, the factor ``$dE_{1}$'' was missing}
in $E_1$ and $\theta$
\index{probability!differential}%
\begin{equation}
w_o = \frac{4Z^2e^4 r_{o}^4}{25 \pi c^4 \hbar^7} \, p_1 p_2 \, dE_1 \left(E_1 E_2 - m^2c^4 + c^2 p_1 p_2 \cos \theta \right) \sin \theta \, d\theta
\end{equation} 
Since 
\[
\Delta E = 6 \, \text{MeV} = 12\, mc^2
\]
we can to a good approximation treat all particles as extreme relativistic. Thus
\begin{equation}
w_o = \frac{4Z^2e^4 r_{o}^4}{25 \pi c^6 \hbar^7}\, E_1^2\, E_2^2\, dE_1 \left(1 + \cos \theta \right) \sin \theta\, d\theta
\end{equation}
So the pairs have an angular distribution concentrated in the \emph{same} hemisphere, and predominantly \emph{equal} energies. Then, since
\index{distribution!angular}%
\[
\int_0^\pi (1 + \cos \theta ) \sin \theta\, d \theta = 2, \qquad \text{and} \qquad \int_0^{\Delta E} E_1^2 E_2^2\, dE_1 = \int_0^{\Delta E} E_1^2 (E_1  + \Delta E)^2 dE_1 = \frac{1}{15} (\Delta E)^5
\]
the total creation probability per unit time is  
\index{creation}%
\index{probability}%
\index{probability!creation}%
\begin{equation}
w_T = \frac{4Z^2 e^4 r_{o}^4}{25 \pi \hbar^7 c^6}\,\frac{1}{15}\,(\Delta E)^5
\end{equation}
Numerically
\[
\frac{Ze^2}{\hbar c} \approx \frac{1}{17} \qquad \text{and} \qquad \frac{\Delta E r_{o}}{\hbar c} \approx \frac{1}{10} \quad \text{since} \; r_{o} = 4 \times 10^{-13} \; \text{cm}
\]
Hence the lifetime will be 
\begin{equation}
\tau = 15 \times 25 \pi \times 10^{5} \times 17^2 \times \frac{1}{4} \times \frac{r_{o}}{c} = 10^{10}\; \frac{r_{o}}{c} \approx 10^{-13} \, \text{sec.}
\end{equation}
\noindent \rule[0.05in]{1in}{0.01in}

\section*{Scattering of Two Electrons in Born Approximation.The M\o ller Formula.}
\addcontentsline{toc}{section}{Scattering of two electrons in Born Approximation -- The M\o ller Formula} 
\index{Moller@M\o ller~scattering}%
\index{Born~approximation}%
\index{electron-electron~scattering}%
\index{scattering!M\o ller}%

We calculate now the transition scattering matrix element $M$ between an initial state $A$ consisting of 2 electrons with momenta $p_1, p_2$ and spin states $u_1, u_2$ and a final state $B$ consisting of two electrons with momenta $p_1^{\prime}, p_2^{\prime}$ and spin states $u_1^{\prime}, u_2^{\prime}$. Thus $M$ gives the probability amplitude for arriving in state $B$ after a long time when the system is known to be in state $A$ to begin with.
\index{probability!scattering~amplitude}%
Hence $M$ itself should be an \emph{invariant} relativistically.

We treat the interaction in Born approximation  i.e. consider the particles to go directly from the free-particle state $A$ to the free-particle state $B$ by applying the interaction operator once to state $A$.
\index{Born~approximation}%
For electrons at reasonably high or relativistic velocities this will be a very good approximation ($e^2/\hbar v \ll 1$). Also we treat the electromagnetic interaction classically, just as in the O$^{16}$ problem, taking the field produced by particle 1 according to classical Maxwell equations  to act directly on particle 2. 
\index{Maxwell!equations}%
This ignores the fact that the field consists of quanta. We shall see later, after we have developed the quantum field theories, that this introduces \emph{no error} so long as we are in the Born approximation.

For the field produced by particle 1 in a transition from the state $p_1, u_1$ to $p_1^{\prime}, u_1^{\prime}$ we have the matrix elements $\varphi_{(1)}, \boldsymbol{A}_{(1)}$ say. We use now not the gauge in which $\nabla \boldsymbol{\cdot} \boldsymbol{A} = 0$, but the covariant gauge in which\endnote{
In the literature, the gauge condition $\nabla \boldsymbol{\cdot} \boldsymbol{A} = 0$ is now called ``Coulomb gauge''; the choice of the gauge condition $\partial_{\mu} A^{\mu} = 0$ (using the Einstein summation convention) is called ``Lorentz gauge''. (See also Eq.\ (588).) In the first edition, Dyson uses Einstein's convention; in the second edition, Moravcsik does not. See also the parenthetical remark following Eq.\ (234a).}
\index{gauge!condition}%
\begin{equation}
\sum_{\mu} \frac{\partial A_{\mu}}{\partial x_{\mu}} = 0 \qquad \qquad A_4 = i \varphi
\end{equation}
So using covariant notations we have in this gauge
\begin{equation}
\sum_{\nu} \frac{\partial^2}{\partial x_{\nu}^2 }A_{\mu \,\text{(1)}} = +4 \pi e s_{\mu \,\text{(1)}}  \qquad \qquad \text{(charge is}\, -e \text{)}
\end{equation}
\begin{equation}
s_{\mu \,\text{(1)}} = i \left( \overline{u}_{1}^{\,\prime} \gamma_{\mu} u_1 \right) \exp \left\{ \sum_{\nu} \frac{i}{\hbar} \left( p_{1 \nu} - p_{1 \nu}^{\prime} \right) x_{\nu} \right\}
\end{equation}
whence
\begin{equation}
A_{\mu \, \text{(1)}} = -4 \pi i e \hbar^2 
\left[ 
\frac{
\left( \overline{u}_{1}^{\,\prime} \gamma_{\mu} u_1 \right) 
\exp \left\{
 \sum_{\nu} \dfrac{i}{\hbar} \left( p_{1 \nu} - p_{1 \nu}^{\prime} \right) x_{\nu} \right\}}
 {\sum_{\lambda} \left( p_{1 \lambda} - p_{1 \lambda}^{\prime} \right)^2} \right]
\end{equation}
where 
\begin{equation}
\sum_{\nu} \left( p_{1 \nu}^{\vphantom{\prime}} - p_{1 \nu}^{\prime} \right)^2 = \left| \boldsymbol{p}_1 - \boldsymbol{p}_1^{\prime} \right|^2 - \frac{1}{c^2} \left(E_1 - E_1^{\prime} \right)^2
\end{equation}

The effect of the field (138) on particle 2 is given by the interaction term in the Dirac equation  for particle 2
\index{Dirac!equation}%
\begin{equation}
-e \varphi + e \boldsymbol{\alpha}  \boldsymbol{\cdot} \boldsymbol{A} = i e \beta \sum_{\mu} \gamma_{\mu} A_{\mu} 
\end{equation}
This gives for particle 2 for the transition from state $p_2, u_2$ to $p_2^{\prime}, u_2^{\prime}$ a transition matrix element
\begin{equation}
\int d\tau \; \psi_2^{\prime \ast} \left( ie \beta \sum_{\mu} \gamma_{\mu} A_{\mu \, \text{(1)}} \right) \psi_2
\end{equation}
a 3-dimensional integral over space at the time $t$ say. For the total transition matrix element $M$ by first order perturbation method
\begin{equation}
M = - \frac{i}{\hbar} \int_{- \infty}^{\infty} dt \int d\tau \; \overline{\psi}_2^{\,\prime} \smash{\left( ie \sum_{\mu} \gamma_{\mu} A_{\mu \, \text{(1)}} \right)} \psi_2 = - \frac{i}{\hbar c} \int d^{\,4} x \; \overline{\psi}_2^{\,\prime} \left( ie \sum_{\mu} \gamma_{\mu} A_{\mu \, \text{(1)}} \right) \psi_2
\end{equation}
where the 4-fold integral is $dx_1\, dx_2 \, dx_3 \, dx_0, \,x_0 = ct$. Putting in the values of $A_{\mu \, \text{(1)}}, \psi^{\prime}_2$ and $\psi_2$, we get
\begin{align}
M & = - \frac{4\pi e^2 \hbar i}{c} \sum_{\mu} \left(\overline{u}^{\, \prime}_2 \gamma_{\mu} u_2 \right) \left(\overline{u}^{\, \prime}_1 \gamma_{\mu} u_1\right) \frac{1}{\sum_{\nu} \left(p_{1 \nu} - p_{1 \nu}^{\prime}\right)^2} \int d^{\,4}x \; \exp \left\{ \sum_{\lambda} \frac{i}{\hbar} \left(p_{1 \lambda} - p_{1 \lambda}^{\prime} + p_{2 \lambda} - p_{2 \lambda}^{\prime} \right) x_{\lambda} \right\} \notag \\
& = - \frac{4\pi e^2 \hbar i}{c} \sum_{\mu, \, \nu} \frac{\left(\overline{u}^{\, \prime}_2 \gamma_{\mu} u_2 \right)\left(\overline{u}^{\, \prime}_1 \gamma_{\mu} u_1\right)}{\left(p_{1 \nu} - p_{1 \nu}^{\prime}\right)^2} (2 \pi \hbar)^4 \delta^{\,4} \! \left(p_1 + p_2 - p_1^{\prime} - p_2^{\prime} \right)
\end{align}
where $ \delta^{\,4}(x) = \prod_{k =1}^{4} \delta(x_k) $. \\

\noindent There is also the exchange process in which the particle $p_1, u_1$ goes to $p_2^{\prime}, u_2^{\prime}$ and \emph{vice-versa}. This gives a contribution  to $M$ with a minus sign since the wave function ought to be taken antisymmetric between the two particles. Hence the final result is
\begin{equation}
M = - \frac{4\pi e^2 \hbar i}{c} (2 \pi \hbar)^4 \delta^{\,4} \! \left(p_1 + p_2 - p_1^{\prime} - p_2^{\prime} \right)  \sum_{\mu, \, \nu} \left\{ \frac{\left(\overline{u}^{\, \prime}_2 \gamma_{\mu} u_2 \right)\left(\overline{u}^{\, \prime}_1 \gamma_{\mu} u_1\right)}{\left(p_{1 \nu} - p_{1 \nu}^{\prime}\right)^2} - \frac{\left(\overline{u}^{\, \prime}_2 \gamma_{\mu} u_1 \right)\left(\overline{u}^{\, \prime}_1 \gamma_{\mu} u_2\right)}{\left(p_{1 \nu} - p_{2 \nu}^{\prime}\right)^2} \right\}
\end{equation}

This covariant formula is elegant and easy to arrive at. The question now is, how does one go from such a formula to a cross-section? 
\index{cross-section!from~amplitudes|(}%

Generally, suppose in such a 2-particle collision process the transition matrix is
\begin{equation}
M = K (2\pi \hbar)^4 \delta^{\,4} \! \left(p_1 + p_2 - p_1^{\prime} - p_2^{\prime} \right)
\end{equation}
Then what will be the cross-section in terms of $K$? We do this calculation once here so that later we can stop when we have found formulae for $M$ of the type (145) which come for example conveniently in this form out of radiation theory.

\section*{Relation of Cross-sections to Transition Amplitudes}
\addcontentsline{toc}{section}{Relation of Cross Sections to Transition Amplitudes}
\index{amplitude}%

Let $w$ be the transition probability per unit volume and per unit time. This is related to the transition probability for a single final state, which is 
\index{probability!transition}%
\begin{equation}
w_{s} = c |K|^2 (2 \pi \hbar)^4 \delta^{\,4}\!\left(p_1 + p_2 - p_1^{\prime} - p_2^{\prime}\right)
\end{equation}
since in $|M|^2$ one of the two $(2\pi \hbar)^4 \delta^{\,4} \! \left(p_1 + p_2 - p_1^{\prime} - p_2^{\prime}\right)/c$ factors represents merely the volume of space-time in which the interaction can occur. The number of final states is by (107) 
\begin{equation}
\frac{1}{(2 \pi \hbar)^6}  \frac{mc^2}{| E_1^{\prime} |} \frac{mc^2}{| E_2^{\prime} |}\, dp_{11}^{\prime}\, dp_{12}^{\prime}\,dp_{13}^{\prime}\,dp_{21}^{\prime}\,dp_{22}^{\prime}\, dp_{23}^{\prime}
\end{equation}
Multiplying (146) by (147) gives the total transition probability 
\index{probability!transition}%
\begin{equation}
w =|K|^2 \frac{1}{(2 \pi \hbar)^2} \frac{m^2c^4}{E_1^{\prime} E_2^{\prime}}\,c\, \delta^{\,4}  \! \left(p_1 + p_2 - p_1^{\prime} - p_2^{\prime}\right)\, dp_{11}^{\prime}\, dp_{12}^{\prime}\,dp_{13}^{\prime}\,dp_{21}^{\prime}\,dp_{22}^{\prime}\, dp_{23}^{\prime}
\end{equation}
As $\delta(ax) =  \dfrac{1}{a} \delta(x)$, we have
\[
\delta^{\,4} \! \left(p_1 + p_2 - p_1^{\prime} - p_2^{\prime} \right) = \delta^{\,3} \! \left(\boldsymbol{p}_1 + \boldsymbol{p}_2 - \boldsymbol{p}_1^{\prime} - \boldsymbol{p}_2^{\prime} \right) c \, \delta(E_1 + E_2 - E_1^{\prime} - E_2^{\prime})
\]
and the integration over $dp_2$ gives then by the momentum conservation 
\index{momentum!conservation}%
\[
w =|K|^2 \frac{c^2}{(2 \pi \hbar)^2}\, \frac{m^2c^4}{E_1^{\prime} E_2^{\prime}}\, \delta(E_1 + E_2 - E_1^{\prime} - E_2^{\prime})\,  dp_{11}^{\prime}\, dp_{12}^{\prime}\,dp_{13}^{\prime}\tag{148a}
\]
Furthermore,
\[
\text{if} \; f(a) = 0, \; \text{we have} \; f(x) = f(a) + f^{\prime}(a) (x - a) = f^{\prime}(a) (x -a)
\]
and thus 
\[
\delta(f(x)) = \delta \! \left\{ f^{\prime}(a) (x - a)\right\} = \frac{\delta(x - a)}{f^{\prime}(a)}
\]
Applying this to (148a) with $f(x) = f(p_{13}^{\prime}) = E_1 + E_2 - E_1^{\prime} -E_2^{\prime}$ and $a = (p_{13}^{\prime})_c =$ the value of $p_{13}^{\prime}$ giving momentum and energy conservation, we get
\[
\delta(E_1 + E_2 - E_1^{\prime} - E_2^{\prime}) = \frac{1}{\dfrac{d(E_1 + E_2 - E_1^{\prime} - E_2^{\prime})}{dp_{13}^{\prime}}} \,\delta \! \left\{ p_{13}^{\prime} - (p_{13}^{\prime})_c \right\}
\]
Hence we finally obtain
\[
w = |K|^2 \frac{m^2c^4}{E_1^{\prime} E_2^{\prime}} \,\frac{c^2}{(2 \pi \hbar)^2}\,  \frac{dp_{11}^{\prime}\, dp_{12}^{\prime}\,dp_{13}^{\prime}}{d \!\left(E_1^{\prime} + E_2^{\prime}\right)}
\]

Choose a Lorentz-system  in which $p_1$ and $p_2$ are both along the $x_3$-direction and take $p_{11}^{\prime}$ and $p_{12}^{\prime}$ as the variables over which the transition probability is taken.
\index{probability!transition}%
This is necessary for relativistic invariance. Then $p_{11}^{\prime}$ and $p_{12}^{\prime}$ being fixed and having from the momentum conservation $p_{13}^{\prime} = -p_{23}^{\prime}$ we get 
\index{Lorentz!system}%
\index{momentum!conservation}%
\begin{equation}
\frac{d\left(E_1^{\prime} + E_1^{\prime} \right)}{dp_{13}^{\prime}} = \left| \frac{dE_1^{\prime}}{dp_{13}^{\prime}} - \frac{ dE_2^{\prime}}{dp_{23}^{\prime}}\right| = c^2 \frac{\left|E_2^{\prime}\, p_{13}^{\prime} - E_1^{\prime}\,p_{23}^{\prime} \right|}{E_1^{\prime} E_2^{\prime}}
\end{equation}
Then the cross-section $\sigma$ is defined in this system by
\begin{equation}
\sigma = \frac{w V_1 V_2}{\left| \boldsymbol{v}_1 - \boldsymbol{v}_2 \right|}
\end{equation}
where $V_1$ is the normalization volume for particle 1, and $v_1$ its velocity. In fact by (103)
\begin{equation}
V_1 = \frac{mc^2}{E_1} \qquad V_2 = \frac{mc^2}{E_2} \qquad \left(\boldsymbol{v}_1  - \boldsymbol{v}_2 \right) = \frac{c^2 \boldsymbol{p}_1}{E_1} - \frac{c^2 \boldsymbol{p}_2}{E_2}
\end{equation}
Hence the cross-section becomes
\begin{equation}
\sigma = \frac{w \left(mc^2 \right)^2}{c^2\left| \boldsymbol{p}_1E_2 - \boldsymbol{p}_2 E_1\right|} = |K|^2 \frac{\left( mc^2 \right)^4}{c^2 \left| E_2 p_{13} - E_1 p_{23} \right| \left|E_2^{\prime}\, p_{13}^{\prime} - E_1^{\prime}\, p_{23}^{\prime} \right|} \frac{1}{\left(2\pi \hbar\right)^2 } \, dp_{11}^{\prime}\, dp_{12}^{\prime}
\end{equation}
It is worth noting that the factor $\boldsymbol{p}_1 E_2 - \boldsymbol{p}_2 E_1$ is invariant under Lorentz transformations leaving the $x_1$ and $x_2$ components unchanged (e.g.boosts parallel to the $x_3$ axis.)\endnote{
Rewritten. In v.1, Dyson writes ``The factor $\boldsymbol{p}_1 E_2 - \boldsymbol{p}_2 E_1$ is invariant for a Lorentz transformation parallel to the 3 axis.'' In v.2, Moravcsik writes ``It is worth noting that the factor $\boldsymbol{p}_1 E_2 - \boldsymbol{p}_2 E_1$ is invariant under Lorentz transformations leaving the $x_1$ and $x_2$ components unchanged.''}
\index{Lorentz!transformations}%
To prove this, we have to show that $p_{13}E_2 - p_{23}E_1 = \tilde{p}_{13} \tilde{E}_2 - \tilde{p}_{23} \tilde{E}_1$ (where $\tilde{\hphantom{p}}$ denotes the quantities after the Lorentz transformation) because we have chosen a Lorentz system in which the direction of the momentum vector is the $x_3$ axis.
\index{Lorentz!transformations}%
Then 
\begin{align*}
\tilde{E} &= E \cosh \theta - cp \sinh \theta \\
\tilde{p} &= p \cosh \theta - \frac{E}{c} \sinh \theta
\end{align*}
Since $E^2 = p^2 c^2 + m^2 c^4$, we can write
\[
E = mc^2 \cosh \phi \qquad	pc = mc^2 \sinh \phi , \qquad \qquad \text{which makes}
\]
\[
\tilde{E} = mc^2 \cosh (\phi - \theta) \qquad \tilde{p}c = mc^2 \sinh (\phi - \theta) \qquad \qquad \text{and thus}
\]
\begin{align*}
\tilde{E}_2 \tilde{p}_{13} - \tilde{E}_1 \tilde{p}_{23} &= m^2 c^3 \left\{ \cosh \left( \phi_2 - \theta \right) \sinh \left( \phi_1 - \theta \right) -  \cosh \left( \phi_1 - \theta \right) \sinh \left( \phi_2 - \theta \right)  \right\} \\
&= m^2 c^3 \sinh \left( \phi_1 - \phi_2 \right)
\end{align*}
independently of $\theta$. Hence we see that $\sigma$ is invariant under Lorentz transformations parallel to the $x_3$ axis. 
\index{Lorentz!transformations}%
\index{cross-section!from~amplitudes|)}%

\section*{Results for M\o ller Scattering}
\addcontentsline{toc}{section}{Results for M\o ller Scattering}

One electron initially at rest, the other initially with energy $E = \gamma mc^2$; 
\index{angle!scattering}%
\index{Moller@M\o ller~scattering}%
\index{electron-electron~scattering}%
\begin{align*}
\gamma &= \dfrac{1}{\sqrt{1 - (v/c)^2}} \\
\text{scattering angle} &= \theta \; \text{in the lab system} \\
				   &= \theta^{*} \; \text{in the center-of-mass system}
\end{align*}

Then the differential cross-section is (Mott and Massey, \emph{Theory of Atomic Collisions}, 2$^{\text{nd}}$ ed., p.\ 368) 
\index{cross-section!differential,~for~M\o ller~scattering}%
\index{Mott, Neville~F.}%
\index{Massey, H. S. W.}%
\begin{equation}
2 \pi \sigma (\theta) \, d\theta = 4 \pi \left( \frac{e^2}{mv^2} \right)^2 \left( \frac{ \gamma +1}{\gamma^2} \right) dx \left\{ \frac{4}{(1 - x^2)^2} - \frac{3}{1 - x^2} + \left( \frac{\gamma -1}{2 \gamma} \right)^2 \left( 1 + \frac{4}{1 - x^2} \right) \right\}
\end{equation}
with
\[
x = \cos \theta^{*} = \frac{2 - (\gamma + 3) \sin^2 \theta}{2 + (\gamma - 1) \sin^2 \theta}
\]
Without spin you get simply
\[
4 \pi \left( \frac{e^2}{mv^2} \right)^2 \left( \frac{\gamma + 1}{\gamma^2} \right) \,dx \, \left\{ \frac{4}{(1 -x^2)^2} - \frac{3}{1-x^2} \right\}
\]
Effect of spin is a measurable \emph{increase} of scattering over the Mott formula. 
\index{Mott, Neville~F.}%
Effect of exchange is roughly the $\dfrac{3}{1 - x^2}$ term. Positron-electron scattering is very similar. 
\index{electron-positron!scattering}%
Only the exchange effect is different because of annihilation possibility.
\index{annihilation}%

\section*{Note on the Treatment of Exchange Effects}
\addcontentsline{toc}{section}{Note on the Treatment of Exchange Effects}

The correctly normalized initial and final states in this problem are
\begin{align}
& \frac{1}{\sqrt{2}} \left\{ \psi_1(1) \psi_2(2) - \psi_1(2) \psi_2(1) \right\}  \notag \\
& \\
& \frac{1}{\sqrt{2}} \left\{ \psi_1^{\prime}(1) \psi_2^{\prime}(2) - \psi_1^{\prime}(2) \psi_2^{\prime}(1) \right\} \notag
\end{align}
where $\psi_2(1)$ means the particle 2 in the state 1, and so on. With these states the matrix element $M$ is exactly as we have calculated it including the exchange term.

The number of possible final states is only one half of the number of states of two distinguishable particles. But this does not bring a factor \(\frac{1}{2} \) into the differential cross-section, because the density of antisymmetrical states, in which one of the two particles has a momentum in a particular range $dp_1\, dp_2 \, dp_3$ is exactly the same as the density of states of 2 distinguishable particles in which the particle labelled 1 lies in the given range. Hence the general rule: the differential cross-section does \emph{not} have a factor \( \frac{1}{2} \), the total cross section \emph{does} because each final state may only be counted once in integrating over the angles. 
\index{angle}%

\section*{Relativistic Treatment of Several Particles}
\addcontentsline{toc}{section}{Relativistic Treatment of Several Particles}

The M\o ller treatment of the interaction of two electrons succeeds because the field of particle 1 is calculated for all time without taking any account of the effect of particle 2 on particle 1.
\index{Moller@M\o ller~scattering}%
How can one do a better calculation taking such reactions into account? Clearly we must construct  an equation of motion which follows the motions of both particles continuously in time and keeps them in step with each other. So we must have a Dirac equation for 2 electrons, taking exact account of their interaction by including in the equation the behaviour of the Maxwell field too. 
\index{Dirac!equation}%
\index{Maxwell!field}%

This kind of 2-particle Dirac equation  is no longer relativistically invariant, if we give each particle a separate position in space but all the same \emph{time}.
\index{Dirac!equation}%
To avoid this Dirac constructed the \emph{many-time} theory in which each electron has its own private time coordinate, and satisfies its private Dirac equation.
\index{Dirac, P. A. M.}%
\index{Dirac!equation}%
This theory is all right in principle. But it becomes hopelessly complicated when pairs are created and you have equations with new time-coordinates suddenly appearing and disappearing. 
\index{creation!pair}%
In fact the whole program of quantizing the electron theory as a theory of discrete particles each with its private time becomes nonsense when you are dealing with an infinite ``sea'' or an indefinite number of particles. So we have come to the end of what we can do with the relativistic quantum theory of particles.

Where did the theory go wrong? Obviously a lot of the troubles of arose from the fact that a particle was always described by an operator $r$ representing its position at the time $t$, $t$ being a number and not an operator. This made the interpretation of the formalism \emph{essentially} non-relativistic even when the equations were formally invariant. In equations like the K.\ G.\  and the 
Dirac equation,  the space and time coordinates appear symmetrically. Thus we are led to the following new view-point. 
\index{Klein-Gordon~equation}%
\index{Dirac!equation}%

Relativistic quantum theory is the study of quantities $\psi$ which are functions of four coordinates $x_1$, $x_2$, $x_3$, $x_0$, all the coordinates being c-numbers and only the expressions containing $\psi$ being operators describing the dynamical system.

The dynamical system is specified by the quantity $\psi$ existing at all points of space-time, and so consists of a system of fields. Relativistic quantum theory is necessarily a field theory.

The process of reinterpreting a one-particle wave-function like the Dirac $\psi$  as a quantized field operator is called \emph{Second Quantization}.
\index{Dirac!equation}%
\index{second~quantization}%

%


\newpage

\pagestyle{fancy}
\fancyhead{}
\lhead{\emph{\MakeUppercase{Field Theory}}}
\chead{}
\rhead{\thepage}
\lfoot{}
\cfoot{}
\rfoot{}

\chapter*{Field Theory}
\addcontentsline{toc}{chapter}{Field Theory}
Before we can begin on the program of constructing our quantum theory of fields, we must make some remarks about Classical Field Theory.
\section*{Classical Relativistic Field Theory}
\addcontentsline{toc}{section}{Classical Relativistic Field Theory}

We take a field with components (vector, spinor etc.) labeled by a suffix $\alpha$. Let
\begin{equation}
\phi_\mu^\alpha = \frac{\partial \phi^\alpha}{\partial x_\mu}
\end{equation}
The theory is fully described by an invariant function of position called the Lagrangian Density, 
\index{Lagrangian~density}%
\begin{equation}
\mathscr{L} = \mathscr{L}\left( \phi^\alpha(x), \phi^\alpha_\mu(x) \right) ,
\end{equation}
a function of $\phi^\alpha$ and its \emph{first} derivatives at the point $x$. The behaviour of the field is fixed by the \emph{Action Principle}. If $\Omega$ is any finite or infinite region of space-time, then
\index{action}%
\begin{equation}
I(\Omega) = \frac{1}{c} \int_\Omega \mathscr{L} \, d^{\,4}x
\end{equation}
is stationary for the physically possible fields $\phi^\alpha$. Thus the variation $\varphi^\alpha \rightarrow \phi^\alpha + \delta \phi^\alpha$ produces no change in $I$ to first order in $\delta \phi^\alpha$, if $\delta \phi^\alpha$ is an arbitrary variation equal to zero on the boundary of $\Omega$. 

It is always assumed that $\mathscr{L}$ is at most quadratic in the $\phi^\alpha_\mu$ and is in various other respects a well-behaved function.

Let $\Sigma$ be the boundary of $\Omega$, and $d\sigma$ an element of 3--dimensional volume on $\Sigma$, $n_{\mu}$ the outward unit vector normal to $d\sigma$, and 
\begin{equation}
\begin{split}
&d\sigma_{\mu} = n_{\mu} d\sigma, \quad \sum_{\mu} n_{\mu}^2 = -1 \quad \mu = 1, 2, 3, 4  \quad x_0 = ct  \\
&d\sigma_{\mu} = \left(dx_2\, dx_3\,dx_0,\, dx_1\, dx_3\,dx_0,\, dx_1\, dx_2\,dx_0,\, -i\,dx_1\, dx_2\,dx_3 \right)
\end{split}
\end{equation}
Then 
\begin{align}
c\, \delta I(\Omega) &= \int_\Omega \sum_{\alpha} \left( \frac{\partial \mathscr{L}}{\partial \phi^{\alpha}} \delta \phi^{\alpha} + \sum_{\mu} \frac{\partial \mathscr{L}}{\partial \phi^{\alpha}_{\mu}} \delta \phi^{\alpha}_{\mu} \right)\,d^{\,4}x \notag \\
 &= \int_\Omega \sum_{\alpha} \left\{ \frac{\partial \mathscr{L}}{\partial \phi^{\alpha}}  - \sum_{\mu} \frac{\partial}{\partial x_{\mu}} \left( \frac{\partial \mathscr{L}}{\partial \phi^{\alpha}_{\mu}} \right) \right\} \delta \phi^{\alpha}\, d^{\,4}x \,+ \int_{\Sigma} \sum_{\alpha, \mu} n_{\mu} \frac{\partial \mathscr{L}}{\partial \phi_{\mu}^{\alpha}} \delta \phi^{\alpha} \, d\sigma
\end{align}
So the principle of action gives the field equations
\index{action}%
\begin{equation}
\frac{\partial \mathscr{L}}{\partial \phi^{\alpha}}  - \sum_{\mu} \frac{\partial}{\partial x_{\mu}} \left( \frac{\partial \mathscr{L}}{\partial \phi^{\alpha}_{\mu}} \right) = 0
\end{equation}
defining the motion of the fields.

\noindent The quantity 
\begin{equation}
\pi_{\alpha} = \frac{1}{c} \sum_{\mu} n_{\mu} \frac{\partial \mathscr{L}}{\partial \phi_{\mu}^{\alpha}}
\end{equation}
is the momentum conjugate to $\phi^{\alpha}$, defined at $x$ and with respect to the surface $\Sigma$.

A more general type of variation is made by varying not only the $\phi^{\alpha}$ but also the boundary of $\Omega$, each point $x_{\mu}$ being moved to the position $(x_{\mu} + \delta x_{\mu})$ where $\delta x_{\mu}$ is either constant or may vary over the surface. Writing ${}_{N}\phi^{\alpha}$ for the new $\phi^{\alpha}$ and ${}_{O}\phi^{\alpha}$ for the old one, we have
\begin{align}
\delta \phi^{\alpha}(x) &= {}_{N}\phi^{\alpha}(x + \delta x) - {}_{O}\phi^{\alpha}(x) \notag \\
{}_{O}\phi^{\alpha}(x + \delta x) &= {}_{O}\phi^{\alpha}(x) + \sum_{\mu} \delta x_{\mu} \, {}_{O}\phi^{\alpha}_{\mu}(x) \notag \\
\Delta \phi^{\alpha}(x) &=  {}_{N}\phi^{\alpha}(x) - {}_{O}\phi^{\alpha}(x)
\end{align}
Therefore under the joint variation
\begin{align*}
c\, \delta I(\Omega) &= \int_{\Omega{_N}} \mathscr{L}\left({}_{N}\phi^{\alpha}(x), {}_N\phi^{\alpha}_{\mu}(x) \right) d\,^4\!x - \int_{\Omega{_O}} \mathscr{L}\left({}_{O}\phi^{\alpha}(x), {}_O\phi^{\alpha}_{\mu}(x) \right) d\,^4\!x \\
 &= \left\{\int_{\Omega{_N}} - \int_{\Omega{_O}} \right\} \mathscr{L}\left({}_{N}\phi^{\alpha}(x), {}_N\phi^{\alpha}_{\mu}(x) \right) d\,^4\!x - \int_{\Omega{_O}} \left\{ \mathscr{L}\left({}_{N}\phi^{\alpha}(x), {}_N\phi^{\alpha}_{\mu}(x) \right) -  \mathscr{L}\left({}_{O}\phi^{\alpha}(x), {}_O\phi^{\alpha}_{\mu}(x) \right) \right\} d\,^4\!x \\
&= \int_{\Sigma} \sum_{\alpha, \mu} n_{\mu} \delta x_{\mu} \, \mathscr{L}\left({}_{N}\phi^{\alpha}(x), {}_N\phi^{\alpha}_{\mu}(x) \right) \, d\sigma + c\,\int_{\Sigma} \sum_{\alpha}  \pi_{\alpha}(x) \Delta \phi^{\alpha}(x)\, d\sigma
\end{align*}
the latter being true by (159) if we assume (160). 

Now since by (162) 
\[
\delta \phi^{\alpha}(x) = {}_{N}\phi^{\alpha}(x) + \sum_{\mu} \delta x_{\mu} \, {}_{N}\phi^{\alpha}_{\mu}(x) - {}_{O}\phi^{\alpha}(x) = \Delta \phi^{\alpha} + \sum_{\mu} \delta x_{\mu} \, {}_{N}\phi^{\alpha}_{\mu}(x)
\]
hence we get finally
\begin{equation}
\delta I(\Omega) = \int_{\Omega} \sum_{\alpha , \mu} \left\{\pi_{\alpha} \delta \phi^{\alpha} + \left( \frac{1}{c} \, n_{\mu} \mathscr{L} - \phi^{\alpha}_{\mu} \pi^{\alpha} \right)\delta x_{\mu}\right\} d\sigma
\end{equation}
with all the new quantities on the RHS.

In the case which is physically of importance, the actual motion is fixed uniquely by specifying the values of the $\phi^{\alpha}$ everywhere on two space-time surfaces $\sigma_2$ and $\sigma_1$ which are the past and future boundaries of the volume $\Omega$. A space-like surface is one on which every two points are outside each other's light-cones, so that the fields can be fixed independently at every point. 

Special case of non-relativistic theory, both $\sigma_1$ and $\sigma_2$ are just space at the time $t_1$ and $t_2$, and $\delta x_{\mu}$ is $ic$ times a displacement of the time by $\delta t_1$ and $\delta t_2$. Then we may write $n_{\mu} = (0, 0, 0, i)$, $\pi_{\alpha} = \partial \mathscr{L}/\partial \dot{\phi}^{\alpha}$, and then for the Hamiltonian 
\index{Hamiltonian}%
\begin{equation}
H = \int d\tau \, \left( \sum_{\alpha} \pi_{\alpha} \dot{\phi}^{\alpha} - \mathscr{L} \right)
\end{equation}
and thus
\begin{equation}
\delta I(\Omega) = \int d\tau \, \sum_{\alpha} \left\{ \left( \pi_{\alpha} \delta \phi^{\alpha} \right) \left( t_1\right) - \left( \pi_{\alpha} \delta \phi^{\alpha} \right) \left(t_2 \right) \right\} - \left\{ H(t_1)\, \delta t_1 - H(t_2)\, \delta t_2  \right\}
\end{equation}

The essential feature of this classical theory is that the Action Principle is stated only for variations vanishing on the boundary of $\Omega$. 
\index{action}%
From this one can \emph{deduce} as in (163) and (165) the effect on $I(\Omega)$ of variations not vanishing on the boundary. This is possible because each state of motion is defined by fixing \emph{as many} field quantities as can be fixed independently, (e.g. all the fields on two space-like surfaces or all the fields and their time-derivatives on one surface) and then the whole past and future of the motion is determined by the field equations. 

Field equations can be written in the Hamiltonian form 
\index{Hamiltonian!field~equations}%
\[
\dot{\phi}^{\alpha} = \frac{\partial H}{\partial \pi_{\alpha}} , \qquad \dot{\pi}_{\alpha} = - \frac{\partial H}{\partial \phi^{\alpha}}
\]
\emph{Examples:} \\

\noindent 1. Klein-Gordon Field, real 
\index{Klein-Gordon~equation}%
\index{Lagrangian~density!Klein-Gordon,~real}%
\begin{equation}
\mathscr{L}_K = -\tfrac{1}{2} c^2 \left\{ \sum_{\mu} \left(\frac{\partial \psi}{\partial x_{\mu}} \right)^2 +\, \mu^2 \psi^2 \right\}
\end{equation}
\noindent 2. Klein-Gordon Field, complex 
\index{Klein-Gordon~equation}%
\index{Lagrangian~density!Klein-Gordon,~imaginary}%
\begin{equation}
\mathscr{L}^{\,\prime}_K = - c^2 \left\{ \sum_{\mu} \left(\frac{\partial \psi}{\partial x_{\mu}}\, \frac{\partial \psi^{*}}{\partial x_{\mu}} \right) +\, \mu^2 \psi \psi^{*} \right\}
\end{equation}
where we consider $\psi$ and $\psi^{*}$ independent one-component fields.\\

\noindent 3. Maxwell Field, four component $A_{\mu}$, Fermi form, 
\index{Maxwell!field!Lagrangian}%
\index{Fermi!form~of~Maxwell~Lagrangian}%
\index{Lagrangian~density!Maxwell}%
\begin{equation}
\mathscr{L}_M = -\tfrac{1}{4} \sum_{\mu, \nu} \left(\frac{\partial A_{\nu}}{\partial x_{\mu}} - \frac{\partial A_{\mu}}{\partial x_{\nu}}\right)^2 - \,\tfrac{1}{2} \sum_{\mu} \left(\frac{\partial A_{\mu}}{\partial x_{\mu}}\right)^2 
\end{equation}
\noindent 4. Dirac Field 
\index{Dirac!field}%
\index{mu@$\mu$}%
\index{Lagrangian~density!Dirac}%
\begin{equation}
\mathscr{L}_D = -\hbar c \, \overline{\psi} \left(\sum_{\lambda} \gamma_{\lambda} \frac{\partial}{\partial x_{\lambda}} + \mu \right) \psi  \qquad \mu = \frac{mc}{\hbar}
\end{equation}
\noindent 5. Dirac Field interacting with Maxwell Field 
\index{Dirac!field!interacting~with~Maxwell~field}%
\index{Lagrangian~density!Dirac-Maxwell}%
\begin{equation}
\mathscr{L}_Q = \mathscr{L}_D + \mathscr{L}_M  - \sum_{\lambda} i e A_{\lambda} \overline{\psi} \gamma_{\lambda} \psi 
\end{equation}
\index{quantum~electrodynamics}%
here $Q$ stands for quantum electrodynamics. \\
\noindent \rule[0.02in]{1in}{0.01in}\\
\index{problem!\emph{5}}%
\index{Hamiltonian}%
\noindent \emph{Problem 5.} Work out these examples: find the field equations, the momentum conjugate to each component of the field, and the Hamiltonian function, (the momenta and Hamiltonian defined for the case of a flat space $\sigma$ only). Verify that the Hamiltonian gives a correct canonical representation of the field equations as Hamiltonian equations of motion.  \\
\rule[0.05in]{1in}{0.01in}
\section*{Quantum Relativistic Field Theory}
\addcontentsline{toc}{section}{Quantum Relativistic Field Theory}

The classical relativistic field theories were usually quantized by using the Hamiltonian form of the field equations and bringing in the commutation relations between coordinates and momenta taken from non-relativistic quantum mechanics. For this approach see Wentzel's book. 
\index{Hamiltonian!field~equations}%
\index{Wentzel, Gregor}%
It is a very bad method, it is complicated; and it is not at all obvious or even easy to prove that the theory so made is relativistic, because the whole Hamiltonian approach is non-covariant.

Just recently we learnt a much better way of doing it, which I shall now expound in these lectures. It is due to Feynman and Schwinger.\endnote{
These three articles may be found in Schwinger, \emph{Selected Papers on Quantum Electrodynamics}.}
\index{Feynman, Richard~P.}%
\index{Schwinger, Julian}%
\begin{tabbing}
\hspace*{3ex}\=\hspace{0.8in}\=\hspace{1in}\=\hspace{1ex}\kill
\>References:\>R. P. Feynman, \>\emph{Rev.\ Mod.\ Phys.} \textbf{20} (1948) 367\\
\>\>\>\emph{Phys.\ Rev.} \textbf{80} (1950) 440\\
\>\>\>\\
\>\>J. Schwinger,\>\emph{Phys.\ Rev.} \textbf{82} (1951) 914\\
\end{tabbing}
It is relativistic all the way, and it is much simpler than the old methods. It is based directly on the Action Principle form of the classical theory which I have just given you, not the Hamiltonian form. 
\index{Hamiltonian}%

In the quantum theory the $\phi^{\alpha}$ are operators defined at each point of space-time as before. They satisfy the same field equations as before, and this is ensured if we assume that the Action Principle
\begin{align}
\delta I (\Omega) &= 0 \notag \\
I (\Omega) &= \frac{1}{c} \int_{\Omega} \mathscr{L}\left(\phi^{\alpha}, \phi^{\alpha}_{\mu} \right) \, d^{\,4}\!x
\end{align}
holds for all variations $\delta \phi^{\alpha}$ of the operators vanishing on the boundaries of $\Omega$. 

In the quantum theory, because of complementarity relations, it is not possible to give numerical values to all field operators throughout a physical motion. In fact the state of motion is specified by giving numerical values to the $\phi^{\alpha}$ on \emph{one} space-like surface. The future of the state of motion cannot then be determined from the field equations, which are in general second-order differential equations. Therefore the action principle (171) which was enough for the classical theory is no longer enough. We must make some \emph{additional} statement about the behaviour of $\delta I$ for variations $\delta \phi^{\alpha}$ which are not zero on the boundaries of $\Omega$. 

A state of motion is specified by specifying a space-time surface $\sigma$ and a set of numerical values $\phi^{\prime \, \alpha}$ for the eigenvalue which the operators $\phi^{\alpha}$ on $\sigma$ have in this state. The state is denoted by the Dirac ket vector $\ket{\phi^{\prime \, \alpha}, \sigma}$. 
\index{Dirac!ket}%
This is a special kind of state in which the $\phi^{\alpha}$ on $\sigma$ have eigenvalues: the general state is a linear combination of  $\ket{\phi^{\prime \, \alpha}, \sigma}$ with various values of $\phi^{\prime \, \alpha}$. The physically observable quantities are expressions such as the matrix element
\begin{equation}
\Braket{\phi^{\prime \, \alpha}_1, \sigma_1 | \phi^{\beta}(x) | \phi^{\prime \, \alpha}_2, \sigma_2}
\end{equation}
of the field operator $\phi^{\beta}(x)$ between the two states specified by $\phi^{\prime \, \alpha}_1$ on $\sigma_1$ and by $\phi^{\prime \, \alpha}_2$ on $\sigma_2$. In particular, the transition probability amplitude between the two states is 
\index{probability!amplitude!transition}%
\begin{equation}
\Braket{\phi^{\prime \, \alpha}_1, \sigma_1 | \phi^{\prime \, \alpha}_2, \sigma_2}
\end{equation}
The squared modulus of this gives the probability  of finding the values  $\phi^{\prime \, \alpha}_1$ for the fields on $\sigma_1$, in the motion which is defined by the fields being given the definite values  $\phi^{\prime \, \alpha}_2$ on $\sigma_2$.
\index{probability}%

\section*{The Feynman Method of Quantization}
\addcontentsline{toc}{section}{The Feynman Method of Quantization}
\index{Feynman!quantization}%

The Feynman method of quantizing the theory consists in writing down an explicit formula for the transition amplitude (173). Namely 
\index{amplitude!transition}%
\begin{equation}
\Braket{\phi^{\prime \, \alpha}_1, \sigma_1 | \phi^{\prime \, \alpha}_2, \sigma_2} = N \sum_{H} \exp \left\{ \frac{i}{\hbar} I_{H}(\Omega) \right\}
\end{equation}
Here $H$ represents a History of the fields between $\sigma_2$ and $\sigma_1$, i.e. any set of classical functions $\phi^{\alpha}(x)$ which are defined in the region $\Omega$ between $\sigma_2$ and $\sigma_1$ and which take the values $\phi^{\prime \, \alpha}_1$ on $\sigma_1$ and $\phi^{\prime \, \alpha}_2$ on $\sigma_2$. $I_H(\Omega)$ is the value of $I(\Omega)$ calculated with these particular functions. The sum $\sum_H$ is taken over all possible histories, a continuously infinite sum whose exact mathematical definition is not easy to formulate. $N$ is a normalization factor independent of the particular states considered, chosen so as to make the sum of the squares of the amplitudes  from a given state to all other states equal to 1.
\index{amplitude}%
This formula is derived by Feynman from very general considerations, applying a Huyghens principle to the solution of wave-mechanics just as it is done in wave-optics.
\index{Huyghens~principle}%
By this one formula the whole theory is quantized and the answer to any physical problem in principle given. The method applies not only to field theory but to ordinary NR quantum theory too. We do not try to derive or justify the Feynman formula here. We just show that it gives the same results as the usual QM. For a discussion of the difficulties in defining the sum $\sum_H$, and a method of doing it in simple cases, see C. Morette, \emph{Phys.\ Rev.} \textbf{81} (1951) 848. 
\index{Morette, C\'ecile}%

From formula (174) we derive at once the most general \emph{Correspondence Principle} giving us back the classical theory in the limit as $\hbar \rightarrow 0$. For suppose $\hbar \rightarrow 0$ then the exponential factor in (174) becomes an extremely rapidly oscillating function of $H$ for all histories $H$ \emph{except} that one for which $I(\Omega)$ is stationary. Therefore in the limit the sum $\sum_H$ reduces to the contribution from the classical motion leading from $\phi^{\prime \, \alpha}_2$ on $\sigma_2$ to $\phi^{\prime \, \alpha}_1$ on $\sigma_1$, all other contributions interfering destructively. The classical motion is defined by the condition that\endnote{
Deleted ``for''; the original statement read ``condition that for its $\delta I(\Omega) = 0$''}
its $\delta I(\Omega) = 0$ for all small variations of the $\phi^{\alpha}$ between 
$\sigma_2$ and $\sigma_1$. This passage to the classical theory is precisely analogous to the passage from wave-optics to geometrical optics when the wave-length of light is allowed to tend to zero. The WKB approximation is gotten by taking $\hbar$ small but not quite zero.
\index{WKB~approximation}%

To establish a connection between the Feynman method and the ordinary method of quantization, Feynman has to define what he means by an operator in his formulation.
\index{Feynman!quantization}%
This he does as follows: Let $x$ be any space-time point inside $\Omega$. Let $\mathcal{O}(x)$ be any field operator defined at $x$, for example $\phi^{\beta}(x)$ or $\phi^{\beta}_{\mu}(x)$. Then $\mathcal{O}(x)$ is given a meaning by defining its matrix element between the states  $\ket{\phi^{\prime \, \alpha}_2, \sigma_2}$ and  $\ket{\phi^{\prime \, \alpha}_1, \sigma_1}$, where $\sigma_2$ and $\sigma_1$ are any two surfaces to the past and future of $x$. This matrix element is
\begin{equation}
\Braket{\phi^{\prime \, \alpha}_1, \sigma_1 | \mathcal{O}(x) | \phi^{\prime \, \alpha}_2, \sigma_2} = N \sum_{H} \mathcal{O}_H(x) \exp \left\{ \frac{i}{\hbar} I_{H}(\Omega) \right\}
\end{equation}
The number $\mathcal{O}_H$ is just the value which the expression $\mathcal{O}$ takes when the $\phi^{\alpha}$ are given the values which they have in the history $H$. It is easily verified that the definitions (174) and (175) are physically reasonable and give the right formal properties of transition amplitudes and operator matrix elements. 
\index{amplitude}%

The Feynman method has one fatal drawback: we cannot use it until we have some way of calculating or at least using the sums over histories, and so far nobody has suggested a practical way of doing this. But Schwinger has shown how to derive from the Feynman method an Action Principle formulation of the theory which avoids this difficulty.
\index{action}%

\section*{The Schwinger Action Principle}
\addcontentsline{toc}{section}{The Schwinger Action Principle}
\index{Schwinger!action~principle}%

Let the sets of eigenvalues $\phi^{\prime \, \alpha}_1$ and $\phi^{\prime \, \alpha}_2$ in (174) be held fixed. Let the numbers $\phi^{\alpha}_H(x)$ be varied so that $\phi^{\alpha}_H(x)$ is replaced by $\phi^{\alpha}_H(x) + \delta \phi^{\alpha}(x)$ where $\delta \phi^{\alpha}(x)$ is an arbitrary infinitesimal c-number quantity. Let the surfaces $\sigma_1$ and $\sigma_2$ be varied so that the point $x_{\mu}$ moves to $x_{\mu} + \delta x_{\mu}$. And let the function $\mathscr{L}$ also be varied so that it is replaced by $\mathscr{L} + \delta \mathscr{L}$ where $\delta \mathscr{L}$ is any expression involving the $\phi^{\alpha}$ and $\phi^{\alpha}_{\mu}$. Under this triple variation (174) gives
\begin{equation}
\delta \Braket{\phi^{\prime \, \alpha}_1, \sigma_1 | \phi^{\prime \, \alpha}_2, \sigma_2} = N \sum_{H}  \left\{ \frac{i}{\hbar} \delta I_H(\Omega) \exp \left(\frac{i}{\hbar} I_{H}(\Omega) \right) \right\}
\end{equation}
Using (175) this may be written 
\begin{equation}
\delta \Braket{\phi^{\prime \, \alpha}_1, \sigma_1 | \phi^{\prime \, \alpha}_2, \sigma_2} = \frac{i}{\hbar} \Braket{\phi^{\prime \, \alpha}_1, \sigma_1 | \delta I(\Omega) | \phi^{\prime \, \alpha}_2, \sigma_2}.
\end{equation}
Here $\delta I(\Omega)$ is the operator obtained by making the three variations on the operator $I (\Omega)$. Formally $\delta I(\Omega)$ is the same as the variation obtained in the classical theory,
\begin{multline}
\delta I(\Omega) = \frac{1}{c} \int_{\Omega} \left\{ \delta \mathscr{L} + \sum_{\alpha, \mu} \left( \frac{\partial \mathscr{L}}{\partial \phi^{\alpha}} - \frac{\partial}{\partial x_{\mu}} \frac{\partial \mathscr{L}}{\partial \phi^{\alpha}_{\mu}} \right) \delta \phi^{\alpha} \right\} d^{\,4}\!x \\
+ \left\{ \int_{\sigma_1} - \int_{\sigma_2} \right\} \sum_{\alpha, \mu} \left\{ \pi_{\alpha} \delta \phi^{\alpha} + \left( \frac{1}{c} n_{\mu} \mathscr{L} - \phi^{\alpha}_{\mu} \pi_{\alpha}\right) \delta x_{\mu} \right\}\, d\sigma
\end{multline}
Only now everything on the RHS of (178) is an operator.

Now what is the meaning of this triple variation as applied to the left side of (174)? Since the $\phi^{\alpha}_H (x)$ are only variables of summation, the change from $\phi^{\alpha}_H (x)$ to $\phi^{\alpha}_H (x) + \delta \phi^{\alpha}(x)$ only affects the left side in changing the boundary values which $\phi^{\alpha}_H (x)$ must take on $\sigma_1$ and $\sigma_2$. Thus instead of $\phi^{\alpha}_H (x) = \phi^{\alpha \prime \prime}_1 (x)$ on $\sigma_1$ we now have the new summation variable
\[
\phi^{\alpha}_H (x) +  \delta \phi^{\alpha} = \phi^{\alpha \prime \prime}_1 (x) +  \delta \phi^{\alpha} \quad \text{on} \; \sigma_1
\]
Therefore the change in $\phi^{\alpha}_H$ is equivalent simply to changing
\begin{align*}
\phi^{\alpha \prime \prime}_1 \quad &\text{to} \quad \phi^{\alpha \prime \prime}_1 + \delta \phi^{\alpha} \quad \text{on} \; \sigma_1 \\
\phi^{\alpha \prime}_2\quad &\text{to} \quad \phi^{\alpha \prime}_2 + \delta \phi^{\alpha} \quad \text{on} \; \sigma_2. 
\end{align*}
The change in $\mathscr{L}$ and in the position of $\sigma$ produces a change in the left side of (174), by virtue of the change in the operators $\phi^{\alpha}$ on $\sigma_1$ and $\sigma_2$ resulting from the variations $\delta \mathscr{L}$ and $\delta x_{\mu}$ in consequence of the field equations.

Hence the net result of the triple variation on the left side of (174) is to give the change in the matrix element $\Braket{\phi^{\alpha \prime \prime}_1, \sigma_1 | \phi^{ \alpha \prime}_2, \sigma_2}$ if the $\phi^{\alpha \prime \prime}_1$ and $\phi^{\alpha \prime}_2$ are left fixed, the operators $\phi^{\alpha}(x)$ on $\sigma_1$ and $\sigma_2$ being modified in consequence of the variations  $\delta \mathscr{L}$ and $\delta x_{\mu}$ according to the field equations, and in addition $\phi^{\alpha}(x)$ on $\sigma_1$ and $\sigma_2$ being changed to  $\phi^{\alpha}(x) - \delta \phi^{\alpha}(x)$. 

Schwinger takes equation (177) as the fundamental principle for setting up the quantum theory. Thus he gets rid of the unpleasant $\sum_H$. 
Out of this action principle come very simply all the main features of a quantum field theory, A -- G below.
\index{action}%

\subsection*{A. The Field Equations}
\addcontentsline{toc}{subsection}{A. The Field Equations}

If we take the special case of a variation $\delta \phi^{\alpha}$ which vanishes on the boundary of $\Omega$ and $\delta \mathscr{L} = \delta x_{\mu} = 0$, then $\Braket{\phi^{\alpha\prime}_1, \sigma_1 | \phi^{ \alpha \prime}_2, \sigma_2}$ depends only on the operators $\phi^{\alpha}$ on $\sigma_1$ and $\sigma_2$ and is unaffected by the variation. Therefore for all such variations
\[
\delta I(\Omega) = 0 \tag{171}
\]
\begin{equation}
\frac{\partial \mathscr{L}}{\partial \phi^{\alpha}} - \sum_{\mu} \frac{ \partial}{\partial x_{\mu}} \, \frac{\partial \mathscr{L}}{\partial \phi^{\alpha}_{\mu}} = 0
\end{equation}
That is to say, the classical action principle and the classical field equations are valid for the quantum field operators.
\index{action}%

We see that (177) is exactly the kind of generalization we want of the old variation principle (171). It includes the information, necessary for a quantum theory, concerning the effect on $I(\Omega)$ of variations not vanishing at the boundary of $\Omega$.

\subsection*{B. The Schr\"{o}dinger Equation for the State-function}
\addcontentsline{toc}{subsection}{B. The Schr\"{o}dinger Equation for the State-function}

Specialize $\sigma_1$ and $\sigma_2$ to be the whole space at the times $t_1$ and $t_2$. Then
\[
\Braket{\phi^{\prime \,\alpha}_1, \sigma_1 | \phi^{\prime \,\alpha}_2, \sigma_2} = \Braket{\phi^{\prime \,\alpha}_1, t_1 | \phi^{\prime \,\alpha}_2, t_2} =  \Psi \left(\phi^{\prime \,\alpha}_1, t_1 \right)
\]
is a Schr\"{o}dinger wave-function giving the probability amplitude for finding the system in the state $\phi^{\prime \,\alpha}_1$ at the time $t_1$, given the initial conditions $\phi^{\prime \,\alpha}_2$  at $t_2$.
\index{Schr\"{o}dinger!equation}%
\index{probability!amplitude}%
The development of $\Psi \left(\phi^{\prime \,\alpha}_1, t_1 \right)$ with time $t_1$ is thus a description of the development of the state of the system with time in the Schr\"{o}dinger representation.

Take in (177) a variation in which $\delta \phi^{\alpha} = \delta \mathscr{L} = 0$, the surface $\sigma_1$ being just moved through the displacement $\delta t$ in the time direction. Then using (165) and (164)
\[
\delta \Psi \left(\phi^{\prime \,\alpha}_1, t_1 \right) = - \frac{i}{\hbar} \Braket{\phi^{\prime \,\alpha}_1, t_1 | H \left(t_1 \right) | \phi^{\prime \,\alpha}_2, t_2} \delta t_1
\]
or
\begin{equation}
i \hbar \frac{d}{dt} \Braket{\phi^{\prime \,\alpha}_1, t_1 | \phi^{\prime \,\alpha}_2, t_2} = \Braket{\phi^{\prime \,\alpha}_1, t_1 | H \left(t_1 \right) | \phi^{\prime \,\alpha}_2, t_2}
\end{equation}
This is the ordinary Schr\"{o}dinger equation in Dirac's notation.
\index{Schr\"{o}dinger!equation}%
\index{Dirac!notation}%
It shows that the Schwinger action principle contains enough information for predicting the future behaviour of a system given initially in a known quantum state.
\index{Schwinger!action~principle}%

\subsection*{C. Operator Form of the Schwinger Principle}
\addcontentsline{toc}{subsection}{C. Operator Form of the Schwinger Principle}

Feynman defined operators by giving the formula (175) for their matrix elements between states specified on two different surfaces. The initial state had to be specified in the past, the final state in the future, the operator referring to some particular time which is taken as present. 
\index{Feynman!definition of operators}%

The usual and generally more useful way of defining operators is to specify their matrix elements between states defined on the same surface. Thus we are interested in a matrix element 
\begin{equation}
\Braket{\phi^{\prime \, \alpha}, \sigma | \mathcal{O} |  \phi^{\prime \prime \, \alpha}, \sigma}
\end{equation}
where $\phi^{\prime \, \alpha}$ and $\phi^{\prime \prime \, \alpha}$ are given sets of eigenvalues and $\sigma$ is a surface which may be past, present or future in relation to the field-points to which $\mathcal{O}$ refers.

Suppose that a reference surface $\sigma_o$ is chosen in the remote past. Let the $\phi^{\alpha}, \sigma$ and $\mathscr{L}$ be varied in such a way that everything on $\sigma_o$ remains fixed. For such a variation, (178) gives if we assume that (179) holds 
\begin{equation}
\delta I(\Omega) = \frac{1}{c} \int_{\Omega} \delta \mathscr{L} \, d^{\,4}\!x +  \int_{\sigma}  \sum_{\alpha, \mu} \left\{ \pi_{\alpha} \delta \phi^{\alpha} + \left( \frac{1}{c} \, n_{\mu} \mathscr{L} - \phi^{\alpha}_{\mu} \pi_{\alpha}\right) \delta x_{\mu} \right\}\, d\sigma
\end{equation}
where $\Omega$ is the region bounded by $\sigma_o$ and $\sigma$.
Let us now first calculate the variation of (181) arising from the change in the meaning of the states $\ket{\phi^{\prime \, \alpha}, \sigma}$ and $\ket{\phi^{\prime \prime \, \alpha}, \sigma}$. The operator $\mathcal{O}$ itself is at this point fixed and not affected by the variations in $\phi^{\alpha}$, $\sigma$ and $\mathscr{L}$. Then
\begin{equation}
\Braket{\phi^{\prime \, \alpha}, \sigma | \mathcal{O} |  \phi^{\prime \prime \, \alpha}, \sigma} = \sum_{\phi^{\prime}_o} \sum_{\phi^{\prime \prime}_o} \Braket{\phi^{\prime \, \alpha}, \sigma | \phi^{\prime \, \alpha}_o, \sigma_o} \Braket{\phi^{\prime \, \alpha}_o, \sigma_o | \mathcal{O} |  \phi^{\prime \prime \, \alpha}_o, \sigma_o} \Braket{\phi^{\prime \prime \, \alpha}_o, \sigma_o | \phi^{\prime \prime \, \alpha}, \sigma}
\end{equation}
therefore, denoting 
\[
\Braket{\phi^{\prime \, \alpha}, \sigma | \mathcal{O} | \phi^{\prime \prime\, \alpha}, \sigma} = \Braket{\sigma^{\prime} | \mathcal{O} | \sigma^{\prime \prime}} \quad \text{etc., we have}
\]
\[
\delta \Braket{\sigma^{\prime} | \mathcal{O} | \sigma^{\prime \prime}}  = \sum_{\prime} \sum_{\prime \prime} \left( \delta  \Braket{\sigma^{\prime} | \sigma^{\prime}_o} \right)  \Braket{\sigma^{\prime}_o | \mathcal{O} | \sigma^{\prime \prime}_o} \Braket{\sigma^{\prime \prime}_o | \sigma^{\prime \prime}} + \sum_{\prime} \sum_{\prime \prime}  \Braket{\sigma^{\prime} | \sigma^{\prime}_o} \Braket{\sigma^{\prime}_o | \mathcal{O} | \sigma^{\prime \prime}_o} \left( \delta  \Braket{\sigma^{\prime \prime}_o | \sigma^{\prime \prime}} \right)
\]
because $\ket{\phi^{\prime \, \alpha}_o}$ and  $\ket{\phi^{\prime \prime\, \alpha}_o}$ are not changed by the variation, and neither is $\mathcal{O}$. Therefore, using (177) we have 
\[
\delta \Braket{\sigma^{\prime} | \mathcal{O} | \sigma^{\prime \prime}} =  \sum_{\prime} \sum_{\prime \prime} \frac{i}{\hbar} \Braket{\sigma^{\prime} | \delta I_{\sigma - \sigma_o} \mathcal{O} | \sigma^{\prime \prime}} + \sum_{\prime} \sum_{\prime \prime} \frac{i}{\hbar} \Braket{\sigma^{\prime} |  \mathcal{O}\,\delta I_{\sigma_o - \sigma} | \sigma^{\prime \prime}}
\]
where the subscript $\sigma - \sigma_o$ refers to the surface integrals in (178). Since $\delta I_{\sigma - \sigma_o} = - \delta I_{\sigma_o - \sigma}$, we get finally
\begin{equation}
\delta \Braket{\phi^{\prime \, \alpha}, \sigma | \mathcal{O} | \phi^{\prime \prime\, \alpha}, \sigma} = \frac{i}{\hbar} \Braket{\phi^{\prime \, \alpha}, \sigma | \boldsymbol{[} \,\delta I(\Omega), \mathcal{O}\, \boldsymbol{]} | \phi^{\prime \prime\, \alpha}, \sigma}
\end{equation}
where $ \boldsymbol{[}\,P, R\,\boldsymbol{]} = PR - RP $.
This applies for the case when $\mathcal{O}$ is fixed and the states vary.

Now we want to calculate the variation of $\Braket{\phi^{\prime \, \alpha}, \sigma | \mathcal{O} | \phi^{\prime \prime\, \alpha}, \sigma}$ for the case when the states are fixed, and $\mathcal{O} = \mathcal{O}\left( \phi^{\alpha}(\sigma) \right)$ changes. This, however, will be the same as for the previous case, except with the opposite sign, because the variation of the matrix element\endnote{
Eq.\ (185) lacked a label in v.2. The discussion beginning at Eq.\ (182) and continuing to Eq.\ (186) is unusually different between the editions. What is here follows Moravcsik's v.2 with the addition of the phrase ``the matrix element'' at Eq.\ (185).}
\begin{equation}
\Braket{\phi^{\prime \, \alpha}, \sigma | \mathcal{O} | \phi^{\prime \prime\, \alpha}, \sigma}
\end{equation}
if \emph{both} the states and $\mathcal{O}$ change simultaneously is zero. Therefore, if we use a representation in which matrix elements of $\mathcal{O}$ are defined between states \emph{not} subject to variation we get\endnote{
A $\delta$ was missing: the equation read $i \hbar \,\mathcal{O} = \boldsymbol{[} \, \delta I(\Omega), \mathcal{O}(\sigma)\, \boldsymbol{]}$.}
\begin{equation}
i \hbar \, \delta \mathcal{O}(\sigma) = \boldsymbol{[} \, \delta I(\Omega), \mathcal{O}(\sigma)\, \boldsymbol{]}
\end{equation}
This is the Schwinger action principle in operator form. It is related to (177) exactly as the Heisenberg representation is to the Schr\"{o}dinger representation  in elementary quantum mechanics.
\index{Schwinger!action~principle!operator~form}%
\index{Schr\"{o}dinger!representation}%

\subsection*{D. The Canonical Commutation Laws}
\addcontentsline{toc}{subsection}{D. Canonical Commutation Laws}

Taking for $\sigma$ the space at time $t$, for $\mathcal{O}(\sigma)$ the operator $\phi^{\alpha}(r, t)$ at the space-point $r$, and $\delta x_{\mu} = \delta \mathscr{L} = 0$ we have by (182) and (186) for an arbitrary variation $\delta \phi^{\alpha}$
\begin{equation}
-i\hbar\, \delta \phi^{\alpha}(r,t) = \sum_{\beta} \int \boldsymbol{[} \,\pi_{\beta}(r^{\prime}, t) \, \delta \phi^{\beta}(r^{\prime}, t), \, \phi^{\alpha}(r, t) \,\boldsymbol{]} \, d^{\,3}\boldsymbol{r}^{\prime}
\end{equation}
because $d\sigma = - n_{\mu} \, d\sigma_{\mu} = -i (-i\,dx_{1}^{\prime} dx_{2}^{\prime} dx_{3}^{\prime}) = -  d^{\,3}r^{\prime}$ by (158); the unit vector in the \emph{increasing} time direction is $i$, and this is the outward direction since we choose $\sigma_o$ in the past. 
Hence for every $r, r^{\prime}$
\begin{equation}
\boldsymbol{[}\, \phi^{\alpha}(r, t), \, \pi_{\beta}(r^{\prime}, t) \, \boldsymbol{]} = i \hbar \, \delta_{\alpha \beta}\, \delta^{3}(\boldsymbol{r}  - \boldsymbol{r}^{\prime})
\end{equation}
Also since the $\phi^{\alpha}(r)$ on $\sigma$ are assumed independent variables, 
\begin{equation}
\boldsymbol{[} \,\phi^{\alpha}(r, t), \, \phi^{\beta}(r^{\prime}, t) \, \boldsymbol{]} = 0
\end{equation}
So this method gives automatically the correct canonical commutation laws for the fields. There is no need to prove that the commutation rules are consistent with the field equations, as was necessary in the older methods.

\subsection*{E. The Heisenberg Equation of Motion for the Operators}
\addcontentsline{toc}{subsection}{E. The Heisenberg Equation of Motion for the Operators}

Suppose that $\sigma$ is a flat surface at time $t$, and that a variation is made by moving the surface through the small time $\delta t$ as in B above. But now let $\mathcal{O}(t) = \mathcal{O}(\sigma)$ be an operator built up out of the field-operators $\phi^{\alpha}$ on $\sigma$. Then by (165) and (186) the change in $\mathcal{O}(t)$ produced by the variation is given by 
\[
i\hbar \, \delta\mathcal{O}(t) = \boldsymbol{[} - H(t)\,\delta t, \, \mathcal{O}(t) \,\boldsymbol{]}
\]
That is to say, $\mathcal{O}(t)$ satisfies the Heisenberg equation of motion
\begin{equation}
i\hbar \, \frac{d\mathcal{O}(t)}{dt} = \boldsymbol{[}\, \mathcal{O}(t), \, H(t) \, \boldsymbol{]}
\end{equation}
where $H(t)$ is the total Hamiltonian operator.

\subsection*{F. General Covariant Commutation Laws}
\addcontentsline{toc}{subsection}{F. General Covariant Commutation Laws}

From (186) we derive at once the general covariant form of the commutation laws discovered by Peierls in 1950 \cite{Peierls52}. 
\index{Peierls, Rudolf~E.}%
This covariant form is not easy to reach in the Hamiltonian formalism. 
\index{Hamiltonian}%

Let two field points $z$ and $y$ be given, and two operators $\mathcal{R}(z)$ and $\mathcal{Q}(y)$ depending on the field quantities $\phi^{\alpha}$ at $z$ and $y$. Let a reference surface $\sigma_o$ be fixed, past of both $z$ and $y$. Suppose the quantity,
\begin{equation}
\delta_{\mathcal{R}}(\mathscr{L}) = \epsilon\, \delta^{4}(x - z) \,\mathcal{R}(z)
\end{equation}
is added to the Lagrangian density $\mathscr{L}(x)$, where $\epsilon$ is an infinitesimal c-number This will make at most a certain infinitesimal change $\epsilon \, \delta_{\mathcal{R}}\phi^{\alpha}(x)$ in the solutions $\phi^{\alpha}(x)$ of the field equations. Supposing the new $\phi^{\alpha}(x)$ to be identical with the old one on $\sigma_o$ then $\delta_{\mathcal{R}}\phi^{\alpha}(x)$ is different from zero only in the future light-cone of $z$.

Similarly adding 
\begin{equation}
\delta_{\mathcal{Q}}(\mathscr{L}) = \epsilon\, \delta^{4}(x - y) \,\mathcal{Q}(y)
\end{equation}
to  $\mathscr{L}(x)$ produces at most a change $\epsilon \, \delta_{\mathcal{Q}}\phi^{\alpha}(x)$ in the $\phi^{\alpha}(x)$. Let $\epsilon \, \delta_{\mathcal{R}}\mathcal{Q}(y)$ be the change in $\mathcal{Q}(y)$ produced by the addition (191), while $\epsilon \, \delta_{\mathcal{Q}}\mathcal{R}(z)$ be the change in $\mathcal{R}(z)$ produced by (192). Suppose $y$ lies on a surface $\sigma$ lying in the future of $z$. Then we take $\mathcal{Q}(y)$ for $\mathcal{O}(\sigma)$ in (186), and $\delta \mathscr{L}$ given by (191). The $\delta I(\Omega)$ given by (182) reduces then simply to
\[
\delta I(\Omega) = \frac{1}{c} \,\epsilon \, \mathcal{R}(z)
\] 
It is assumed that there is no intrinsic change $\delta \phi^{\alpha}$ of $\phi^{\alpha}$ or $\delta x_{\mu}$ of $\sigma$ apart from the change whose effect is already included in the $\delta \mathscr{L}$ term. Thus (186) gives
\begin{align}
\boldsymbol{[} \, \mathcal{R}(z), \, \mathcal{Q}(y) \, \boldsymbol{]} &= \hphantom{-}i \hbar c \, \delta_{\mathcal{R}} \mathcal{Q}(y) \quad (y_0 > z_0) \notag \\
\boldsymbol{[} \, \mathcal{R}(z), \, \mathcal{Q}(y) \, \boldsymbol{]} &= -i \hbar c \, \delta_{\mathcal{Q}} \mathcal{R}(z) \quad (z_0 > y_0)
\end{align}
When $y$ and $z$ are separated by a space-like interval, the commutator is zero, because the disturbance $\mathcal{R}(z)$ propagates with a velocity at most $c$ and therefore can affect things only in the future lightcone of $z$; this means $\delta_{\mathcal{R}} \mathcal{Q}(y) = 0$ in this case.

Peierls' formula, valid for any pair of field operators, is
\index{Peierls!formula}%
\begin{equation}
\boldsymbol{[} \, \mathcal{R}(z), \, \mathcal{Q}(y) \, \boldsymbol{]} = i \hbar c  \left\{ \delta_{\mathcal{R}} \mathcal{Q}(y) - \delta_{\mathcal{Q}} \mathcal{R}(z) \right\}
\end{equation}
This is a useful formula for calculating commutators in a covariant way.

\subsection*{G. Anticommuting Fields}
\addtocontents{toc}{\protect\enlargethispage{2\baselineskip}}
\addcontentsline{toc}{subsection}{G. Anticommuting Fields}

There is one type of field theory which can be constructed easily by Schwinger's action principle, but which does not come out of Feynman's picture. 
\index{Schwinger!action~principle}%
\index{Feynman!quantization}%
Suppose a classical field theory in which a group of field operators $\psi^{\alpha}$ always occurs in the Lagrangian in bilinear combinations like $\overline{\psi}\!\!\!\!\phantom{\psi}^{\beta}\psi^{\alpha}$ with the group of field operators $\overline{\psi}$. Examples, the Dirac $\mathscr{L}_D$ and the quantum electrodynamics $\mathscr{L}_Q$.  
\index{Dirac!Lagrangian}%
\index{quantum~electrodynamics!Lagrangian}%

Then instead of taking every $\phi^{\alpha}$ on a given surface $\sigma$ to commute as in (189), we may take every pair of $\psi^{\alpha}$ to anticommute, thus\endnote{
The notation originally used for anticommutators was 
$AB + BA = [A, B]_{+}$.The more familiar $\{A, B\}$ has been used instead.}
\index{anticommute}%
\begin{equation}
\{ \, \psi^{\alpha}(r, t), \, \psi^{\beta}(r^{\prime}, t) \,\} = 0	\qquad \qquad \{ \, \mathcal{P}, \, \mathcal{R} \, \} = \mathcal{P} \mathcal{R} + \mathcal{R} \mathcal{P}
\end{equation}

The bilinear combination will still commute, like the $\phi^{\alpha}$'s did before. The $\psi^{\alpha}$ commute as before with any field quantities on $\sigma$ other than the $\psi$ and $\overline{\psi}$. Schwinger then assumes (177) to hold precisely as before, except that in calculating $\delta I(\Omega)$ according to (178), the variation $\delta \psi^{\alpha}$ \emph{anticommutes} with all operators $\psi^{\alpha}$ and $\overline{\psi}\!\!\!\!\phantom{\psi}^{\beta}$. 
\index{anticommute}%
In these theories it turns out that the momentum $\pi_{\alpha}$ conjugate to $\psi^{\alpha}$ is just a linear combination of $\overline{\psi}$, because the Lagrangian is only linear in the derivatives of $\psi$. With the anticommuting fields the field equations (179) are deduced as before, also the Schr\"{o}dinger equation (180), the commutation rules being given by (186) and (187).
\index{Schr\"{o}dinger!equation}%
But now in order to make (187) valid, since $\delta \psi^{\beta}$ anticommutes with the $\psi$ and $\pi$ operators,\endnote{
Here, $\psi$ was substituted for the original $\phi$ (the variable in Eq.\ (187)) for clarity.}
the canonical commutation law must be written 
\begin{equation}
\{ \,\psi^{\alpha}(r, t), \, \pi_{\beta}(r^{\prime}, t) \, \} = - i \hbar\, \delta_{\alpha \beta}\, \delta^{3}(\boldsymbol{r} - \boldsymbol{r}^{\prime}) 
\end{equation}
The general commutation rule (194) is still valid provided that $\mathcal{Q}$ and $\mathcal{R}$ are also expressions bilinear in the $\overline{\psi}$ and $\psi$. 

The interpretation of the operators, and the justification for the Schwinger principle, in the case of anticommuting fields, is not clear. 
\index{anticommute}%
But it is clear that the Schwinger principle in this case gives a consistent and simple formulation of a relativistic quantum field theory. And we may as well take advantage of the method, even if we do not quite understand its conceptual basis. The resulting theory is mathematically unambiguous, and gives results in agreement with experiment; that should be good enough.

%


\newpage
 
\pagestyle{fancy}
\fancyhead{}
\lhead{\emph{\MakeUppercase{Examples of Quantized Field Theories}}}
\chead{}
\rhead{\thepage}
\lfoot{}
\cfoot{}
\rfoot{}

\chapter*{Examples of Quantized Field Theories}
\addcontentsline{toc}{chapter}{Examples of Quantized Field Theories}
\section*{I. The Maxwell Field}
\addcontentsline{toc}{section}{I. The Maxwell Field}

\hspace{3ex}Lagrangian
\[
\mathscr{L}_M = -\tfrac{1}{4} \sum_{\mu, \nu} \left(\frac{\partial A_{\nu}}{\partial x_{\mu}} - \frac{\partial A_{\mu}}{\partial x_{\nu}}\right)^2 - \,\tfrac{1}{2} \sum_{\mu} \left(\frac{\partial A_{\mu}}{\partial x_{\mu}}\right)^2 \tag{168}
\]
\hspace{3ex}Field equations
\begin{equation}
\sum_{\mu} \frac{\partial^2}{\partial x_{\mu}^{2}} A_{\lambda} = \Box^2 A_{\lambda} = 0
\end{equation}
\hspace{3ex}Commutation rules for the $A_{\lambda}$: To find these we uses the Peierls method. 
\index{Peierls!method}%
Take two points $y$ and $z$ with $z_0 > y_0$. Let $\mathcal{Q}(y) = A_{\lambda}(y)$, $\mathcal{R}(z) = A_{\mu}(z)$. Note: in this section $x,\, y,\, z,\, k$ etc. are meant to have components 1, 2, 3 and 0, while in $x_{\mu},\, y_{\mu},\, z_{\mu},\, k_{\mu}$, etc. we mean $\mu = 1, 2, 3\; \text{and}\; 4$. When $\delta_{\mathcal{Q}}(\mathscr{L}) = \epsilon \, \delta^4(x - y) A_{\lambda} (y)$ is added to $\mathscr{L}_M$, the field equation for $A_{\mu}$ becomes
\begin{equation}
\Box^2 A_{\mu} + \delta_{\lambda \mu}\, \epsilon \, \delta^4(x - y) = 0
\end{equation}
This equation is satisfied by $A_{\mu} + \delta_{\mathcal{Q}} A_{\mu}(z)$ (by definition), and hence also by $\delta_{\mathcal{Q}} A_{\mu}(z)$ because of (197). Therefore  $\delta_{\mathcal{Q}} A_{\mu}(z)$ is defined by the conditions
\begin{align}
\Box^2 (\delta_{\mathcal{Q}} A_{\mu}(z)) = - \delta_{\lambda \mu} \, \delta^4(z - y) \notag \\
\delta_{\mathcal{Q}} A_{\mu}(z) = 0 \quad \text{for} \; z_0 < y_0. 
\end{align}
That is to say, $\delta_{\mathcal{Q}} A_{\mu}(z)$ is a c-number and is the retarded potential created by a point source acting instantaneously at the space-time point $y$.
\index{create}%
\begin{align}
\delta_{\mathcal{Q}} A_{\mu}(z) = \delta_{\lambda \mu} D_R(z - y) \notag \\
\Box^2 D_R(z - y) = - \delta^4(z - y)
\end{align}
If $x$ is any 4-vector, using
\[
\delta(x^2 - a^2) = \frac{1}{2a} \left\{ \delta(x - a) + \delta(x + a) \right\}, \quad a > 0
\]
we get 
\begin{equation}
D_R(x) = \frac{1}{2\pi} \Theta(x) \, \delta(x^2) = \frac{1}{4 \pi | \boldsymbol{r} |} \, \delta(x_0 - | \boldsymbol{r} |)
\end{equation}
Here
\[
| \boldsymbol{r} | = \sqrt{x_1^2 + x_2^2 + x_3^2}; \qquad x_o = ct; \qquad x^2 = r^2 - x_o^2 \qquad \Theta(x) = \begin{cases}
+1 &\text{for $x > 0$}\\
\;0 &\text{for $x < 0$}
\end{cases}
\]
In the same way
\begin{equation}
\delta_{\mathcal{R}} A_{\lambda}(y) = \delta_{\lambda \mu} D_A(z - y) = \delta_{\lambda \mu} D_R(y - z) 
\end{equation}
where $D_A$ is the advanced potential of the same source,
\[
D_A(x) =  \frac{1}{4 \pi | \boldsymbol{r} |} \, \delta(x_0 + | \boldsymbol{r} |)
\]
Hence we have the the commutation rule (194)
\begin{align}
\boldsymbol{[} \, A_{\mu}(z), \, A_{\lambda}(y) \, \boldsymbol{]} &= i\hbar c\, \delta_{\lambda \mu} \left[ D_A(z - y) - D_R(z - y) \right] \notag \\
& = i \hbar c \; \delta_{\lambda \mu} D(z - y) \qquad \qquad \text{(definition of $D$)}
\end{align}
This invariant $D$-function satisfies by (200) 
\begin{equation}
\Box^2 D(x) = - \delta^4(x) - (- \delta^4(x)) = 0
\end{equation}
as it must. Also
\begin{align}
D(x) &=  \frac{1}{4 \pi | \boldsymbol{r} |} \, \left[ \delta(x_0 + | \boldsymbol{r} |) -  \delta(x_0 - | \boldsymbol{r} |) \right] \notag \\
&= -\frac{1}{2\pi}\, \epsilon(x)\, \delta(x^2)	\qquad \qquad \epsilon(x) = \text{sign}(x_0)
\end{align}

\subsection*{Momentum Representations}
\addcontentsline{toc}{subsection}{Momentum Representations}
\index{momentum!representation}%

We have 
\begin{equation}
\delta^4(x) = \frac{1}{(2\pi)^4} \int \exp(i k \cdot x) \,d^{\,4} k
\end{equation}
where the integral is fourfold, over $dk_1\,dk_2\,dk_3\,dk_4$.
Therefore
\begin{equation}
D_R(x) = \frac{1}{(2\pi)^4} \int_{+} \exp(i k \cdot x) \frac{1}{k^2} \, d^{\,4} k
\end{equation}
where $k^2 = | \boldsymbol{k} |^2 - k_0^2$. The integration with respect to $k_1\, k_2\, k_3$ is an ordinary real integral. That with respect to $k_0$ is a contour integral going along the real axis and \emph{above} the two poles at $k_0 = \pm | \boldsymbol{k} |$.

\begin{center}
\includegraphics{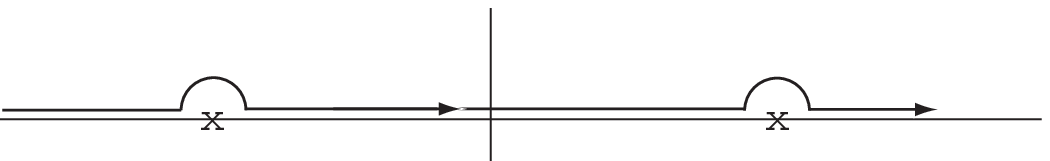}
\end{center}

\noindent For detailed calculations, see the Appendix below.
This gives the correct behaviour of $D_R$ being zero for $x_0 < 0$. Similarly
\begin{equation}
D_A(x) = \frac{1}{(2\pi)^4} \int_{-} \exp(i k \cdot x) \frac{1}{k^2} \, d^{\,4} k
\end{equation}
with a contour going \emph{below} both the poles. Therefore
\begin{equation}
D(x) = \frac{1}{(2\pi)^4} \int_{s} \exp(i k \cdot x) \frac{1}{k^2} \, d^{\,4} k
\end{equation}
with a contour $s$ as shown.
\begin{center}
\includegraphics{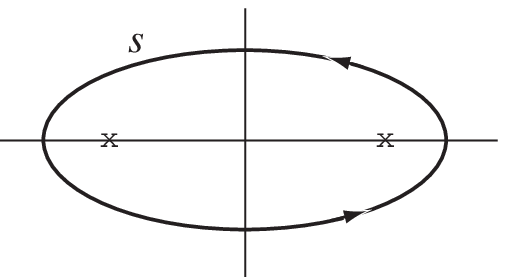}
\end{center}
Evaluating the residues we have 
\begin{equation}
D(x) = -\frac{i}{(2\pi)^3} \int \exp(i k \cdot x) \, \delta(k^2) \, \epsilon(k) \, d^{\,4} k
\end{equation}
this being now an ordinary real integral. \\

\begin{wrapfigure}[9]{r}[10pt]{4cm}
	\centering
\scalebox{0.75}{\includegraphics{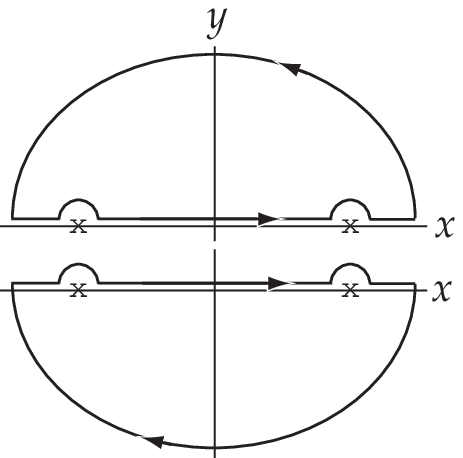}}
\end{wrapfigure}

\noindent \emph{Appendix} \\

\noindent Let us verify, e.g. (207). For $x_0 < 0$, we have to take the top path shown, otherwise the integrand blows up; this gives obviously 0. 

For $x_0 > 0$, we have to take the bottom path; then 
\[
D_R(x) = \frac{1}{(2\pi)^4} \int_{+} \frac{ e^{i \boldsymbol{k} \cdot \boldsymbol{x}} e^{-ik_0x_0}}{\boldsymbol{k}^2 - k_0^2}\, d^{\,3}k \, dk_0
\]
where $\boldsymbol{k}$ and $\boldsymbol{x}$ are 3 dimensional vectors. 

\noindent Now because of the clockwise direction
\begin{align*}
\int_{+} \frac{e^{-ik_0x_0}}{\boldsymbol{k}^2 - k_0^2}\, \, dk_0 &= - \int_{+} \frac{e^{-ik_0x_0}}{(k_0 - | \boldsymbol{k} |)( k_0 + | \boldsymbol{k} |)} \, dk_0 \\
&= 2\pi i(\text{Residue at $k_0 = | \boldsymbol{k} |$  +  Residue at $k_0 = - | \boldsymbol{k} |$}) = 2\pi i \left(\frac{ e^{-i | \boldsymbol{k} | x_0}}{2 | \boldsymbol{k} |} - \frac{ e^{+i | \boldsymbol{k} | x_0}}{2 | \boldsymbol{k} |} \right)
\end{align*}
Hence
\begin{align*}
D_R(x) &= \frac{i}{(2\pi)^3} \int \frac{1}{2 |\boldsymbol{k}|} e^{i \boldsymbol{k} \cdot \boldsymbol{x}} \left\{e^{i | \boldsymbol{k} | x_0} - e^{-i | \boldsymbol{k} | x_0} \right\}d^{\,3} k \vphantom{A_{A_{A_{A_{A_{A_{A_{A_{A_{A}}}}}}}}}}\\
&= \frac{i}{(2\pi)^3}\, 2\pi \iint  \frac{1}{2 |\boldsymbol{k}|} e^{i |\boldsymbol{k}| |\boldsymbol{x}| \cos \theta} \left\{e^{i | \boldsymbol{k} | x_0} - e^{-i | \boldsymbol{k} | x_0} \right\} | \boldsymbol{k}|^2 \, d | \boldsymbol{k}| \, \sin \theta \, d\theta \vphantom{A^{A^{A^{A^{A}}}}}\\
&= -\frac{i}{(2\pi)^2} \int_0^{\infty} \frac{1}{2 |\boldsymbol{k}|}  \frac{1}{i |\boldsymbol{k}| |\boldsymbol{x}|} \left\{e^{i | \boldsymbol{k} | x_0} - e^{-i | \boldsymbol{k} | x_0} \right\} | \boldsymbol{k}|^2 \, \left( \left. e^{i |\boldsymbol{k}| |\boldsymbol{x}| \alpha} \right|_{\alpha = -1}^{\alpha = +1} \right) \, d | \boldsymbol{k}| \vphantom{A^{A^{A^{A^{A^{A^{A}}}}}}}\\
&= - \frac{1}{4\pi^2} \frac{1}{2 | \boldsymbol{x} | } \int_0^{\infty} \left\{ e^{i | \boldsymbol{k} | (x_0 + | \boldsymbol{x} |)} - e^{i | \boldsymbol{k} | (x_0 - | \boldsymbol{x} |)} - e^{-i | \boldsymbol{k} | (x_0 - | \boldsymbol{x} |)} + e^{-i | \boldsymbol{k} | (x_0 + | \boldsymbol{x} |)} \right\} \, d | \boldsymbol{k}| \vphantom{A^{A^{A^{A^{A^{A^{A}}}}}}}\\
&= - \frac{1}{4\pi | \boldsymbol{x} |} \frac{1}{2\pi } \int_{-\infty}^{+\infty} \left\{ e^{i | \boldsymbol{k} | (x_0 + | \boldsymbol{x} |)} - e^{i | \boldsymbol{k} | (x_0 - | \boldsymbol{x} |)} \right\} \, d | \boldsymbol{k}| \vphantom{A^{A^{A^{A^{A^{A^{A}}}}}}} \\
&= - \frac{1}{4\pi | \boldsymbol{x} |} \, \delta(x_0 - | \boldsymbol{x} |) \qquad \text{for $x_0 > 0$.} \vphantom{A^{A^{A^{A^{A^{A^{A}}}}}}} 
\end{align*}

\subsection*{Fourier Analysis of Operators}
\addcontentsline{toc}{subsection}{Fourier Analysis of Operators}

Let us analyze the potential $A_{\mu}$ into Fourier components 
\index{Fourier!components}%
\begin{equation}
A_{\mu}(x) = B \int d^{\,3}k\, | \boldsymbol{k} |^{-1/2} \left\{ a_{k \mu} \exp(i k \cdot x) + \Tilde{a}_{k \mu} \exp(-i k \cdot x) \right\} 
\end{equation}
where the factor $| \boldsymbol{k} |^{-1/2}$ appears only as a matter of convenience; the actual Fourier coefficients are then $| \boldsymbol{k} |^{-1/2} \, a_{k \mu}$ and $| \boldsymbol{k} |^{-1/2}\, \tilde{a}_{k \mu}$.
\index{Fourier!components}%
The integration is over all 4-vectors $(k)$ with $k_0 = + | \boldsymbol{k} |$. $B$ is a normalization factor to be determined later. The $a_{k \mu}$ and $\tilde{a}_{k \mu}$ are operators independent of $x$. \\
Since $A_1, \, A_2, \, A_3 \; \text{and} \; A_0$ are Hermitian, 
\index{Hermitian}%
\begin{align}
\tilde{a}_{k \mu} &= a^{*}_{k \mu} = \; \text{Hermitian conjugate of}\;  a_{k \mu}, \quad \mu = 1, 2, 3, 0 \; \text{and therefore} \notag \\
\tilde{a}_{k 4} &= -a^{*}_{k 4} = \; -\text{Hermitian conjugate of}\;  a_{k 4}
\end{align}

Computing the commutator $\boldsymbol{[} \, A_{\mu}(z), \, A_{\lambda}(y) \, \boldsymbol{]}$ from (211) and comparing the result with (203) in the momentum representation (210), we have first, since the result is a function of $(z - y)$ only\endnote{
In the last commutator, the operator $\tilde{a}_{k^{\prime} \lambda}$ lacked the tilde.}
 \index{momentum!representation}%
\begin{align}
\boldsymbol{[} \, a_{k \mu}, \, a_{k^{\prime} \lambda} \, \boldsymbol{]} &= 0 \notag \\
\boldsymbol{[} \, \tilde{a}_{k \mu}, \, \tilde{a}_{k^{\prime} \lambda} \, \boldsymbol{]} &= 0 \\
\boldsymbol{[} \, a_{k \mu}, \, \tilde{a}_{k^{\prime} \lambda} \, \boldsymbol{]} &= \delta^{3}(\boldsymbol{k} - \boldsymbol{k}^{\prime}) \, \delta_{\mu \lambda} \notag
\end{align}
And the two results for the commutator agree then precisely if we take
\begin{equation}
B = \sqrt{\frac{\hbar c}{16 \pi^3}}
\end{equation}

\subsection*{Emission and Absorption Operators}
\addcontentsline{toc}{subsection}{Emission and Absorption Operators}

The operators $A_{\mu}(x)$ obey the Heisenberg equations of motion for operators (190)
\[
i\hbar \, \frac{\partial A_{\mu}}{\partial t} = \boldsymbol{[}\, A_{\mu}, \, H \,\boldsymbol{]} \tag{190a}
\]
Therefore the operator $a_{k \mu} \exp (i k \cdot x)$ has matrix elements between an initial state of energy $E_1$ and a final state of energy $E_2$, only if 
\begin{equation}
i\hbar(-ic k_0) = E_{1} - E_{2} = \hbar c |\boldsymbol{k}|
\end{equation}
because by (211) and (190a) we have 
\begin{equation}
\psi_1\, i\hbar(-ick_0)a_{k \mu}\, \psi_2 = \psi_1\, \hbar c |\boldsymbol{k} |\, a_{k \mu}\, \psi_2 = \psi_1\,\boldsymbol{[}\, a_{\mu k}, \, H \, \boldsymbol{]} \, \psi_2 = (E_1 - E_2) \psi_1\, a_{\mu k} \, \psi_2
\end{equation}
Now $\hbar c |\boldsymbol{k} |$ is a constant energy, characteristic of the frequency $\omega = ck$ characteristic of the particular Fourier components of the field. 
\index{Fourier!components}%
The operator $a_{k \mu}$ can only operate so as to \emph{reduce} the energy of a system by a lump of energy of this size. In the same way, $\tilde{a}_{k \mu}$ will only operate when 
\[
E_{1} - E_{2} = - \hbar c |\boldsymbol{k} |
\]
to \emph{increase} the energy by the same amount.

This is the fundamental property of the quantized field operators, that they change the energy of a system not continuously but in jumps. This shows that our formalism includes correctly the experimentally known quantum behaviour of radiation.

We call $a_{k \mu}$ the \emph{absorption} operator for the field oscillator with propagation vector $k$ and polarization direction $\mu$. Likewise $\tilde{a}_{k \mu}$ the \emph{emission} operator.

We have thus 4 directions of polarization for a photon of given momentum. There are not all observed in electromagnetic radiation. Free radiation can only consist of transverse waves, and has only 2 possible polarizations. This is because the physically allowable states $\Psi$ are restricted by a supplementary condition
\begin{equation}
\sum_{\mu} \frac{\partial A_{\mu}^{(+)}}{\partial x_{\mu}} \Psi = 0
\end{equation}
where $A_{\mu}^{(+)}$ is the positive frequency part of $A_{\mu}$, i.e. the part containing the absorption operators. In the classical theory we have
\[
\sum_{\mu} \frac{\partial A_{\mu}}{\partial x_{\mu}} = 0
\]
the condition imposed in order to simplify the Maxwell equations  to the simple form $\Box^2 A_{\mu} = 0$.
\index{Maxwell!equations}%
In the quantum theory it was usual to take
\[
\sum_{\mu} \frac{\partial A_{\mu}}{\partial x_{\mu}} \Psi = 0
\]
but this means that photons of a certain kind cannot be \emph{emitted} in a physical state which is physically difficult to understand and brings mathematical inconsistencies into the theory. So we assume only (216) which only says these photons are \emph{not present} and cannot be absorbed from a physical state, which makes good sense. Also in the classical limit $\sum_{\mu} \partial A_{\mu}/\partial x_{\mu}$ is a real quantity, and so  $\sum_{\mu} \partial A_{\mu}/\partial x_{\mu} = 0$ follows correctly from (216) alone.

The method of using (216) as supplementary condition is due to Gupta and Bleuler\endnote{
``Bleuler'' was written ``Bleuber''.};  \\

\hspace*{3ex}S.\ N.\ Gupta, \emph{Proc.\ Roy.\ Soc.\ A} \textbf{63} (1950) 681.

\hspace*{3ex}K.\ Bleuler, \emph{Helv.\ Phys.\ Acta} \textbf{23} (1950) 567. \\
\index{Gupta, S.~N.}%
\index{Bleuler, K.}%

\noindent The older treatment is unnecessary and difficult, so we will not bother about it.

By (211), (216) is equivalent to assuming
\[
\sum_{\mu} \left( k_{\mu} a_{k \mu} \right) \Psi = 0        																	 \tag{216a} 
\]
for each momentum vector $k$ of a photon.

As a result of this work of Gupta and Bleuler, the supplementary conditions do not come into the practical use of the theory at all. We use the theory and get correct results, forgetting about the supplementary conditions.
\index{Gupta-Bleuler~method}%

\subsection*{Gauge-Invariance of the Theory}
\addcontentsline{toc}{subsection}{Gauge Invariance of the Theory}

The theory is gauge-invariant. 
\index{gauge!invariance|(}%
That is to say, adding a gradient\endnote{
Originally, the phrase read $A_{\mu} = \partial \Lambda / \partial x_{\mu}$; this seemed confusing as the original potential is itself $A_{\mu}$.} 
$\Lambda_{\mu} = \partial \Lambda / \partial x_{\mu}$ to the potentials does not change the fields by anything which is physically observable. Therefore all states which differ only by such an addition to the potentials are physically identical. 

If $\Psi$ is any state then
\[
\Psi^{\prime} = \left( 1 + \lambda \sum\nolimits_{\mu} k_{\mu} \tilde{a}_{k \mu} \right) \Psi
\]
is a state obtained from $\Psi$ by emitting a pseudo-photon with potentials proportional to $\partial \Lambda  /\partial x_{\mu}$. Hence $\Psi^{\prime}$ should be indistinguishable from $\Psi$.

Now if $\Psi_{2}$ is any state whatever satisfying the supplementary condition (216a) the matrix element
\begin{align*}
\left(\Psi^{\prime *}, \,\Psi_1 \right) &= \left(\Psi^{*}\left(1 + \lambda \, \sum\nolimits_{\mu} k^{*}_{\mu} \,\tilde{a}_{k \mu}^{*}\right), \, \Psi_2 \right) \notag \\
   &=  \left(\Psi^{*}, \,\left(1 + \lambda \, \sum\nolimits_{\mu} k_{\mu}\, a_{k \mu}^{*}\right) \Psi_2 \right)  = \left(\Psi^{*}, \, \Psi_2 \right)   \tag{216b}
\end{align*}
Hence the matrix elements of $\Psi^{\prime}$ and $\Psi$ to any physical state $\Psi_2$ whatever are equal, and so the results of the theory are all independent of whether the state $\Psi$ is represented by the vector $\Psi$ or by $\Psi^{\prime}$. This is enough to show that the theory is properly gauge-invariant, in spite of the fact that states are specified by the potentials which are not themselves gauge-invariant. 
\index{gauge!invariance|)}%

\subsection*{The Vacuum State}
\addcontentsline{toc}{subsection}{The Vacuum State}

The vacuum state is by definition the state of lowest energy, so that all absorption operators operating on it give zero:
\[
a_{k \mu} \Psi_o = 0																						\tag{217a}
\]
and therefore by (212) 
\[
(a_{k \mu} \Psi_o)^{*} = (\Psi_o^{*}a_{k \mu}^{*}) = \pm (\Psi_o^{*}\tilde{a}_{k \mu}) = 0 									\tag{217b}
\]
Given any operator $\mathcal{Q}$, we are interested in the ``vacuum expectation value'' of $\mathcal{Q}$ defined as 
\begin{equation}
\braket{Q}_o = (\Psi_o^{*}, \mathcal{Q}  \Psi_o)
\end{equation}
Then we have at once
\begin{align}
\braket{a_{k \mu} a_{k^{\prime} \lambda}}_o &=  0 \qquad \text{by (217a)}					\tag{219a} \\
\braket{\tilde{a}_{k \mu} \tilde{a}_{k^{\prime} \lambda}}_o &=  0  \qquad \text{by (217b)}			\tag{219b} \\
\braket{\tilde{a}_{k \mu} a_{k^{\prime} \lambda}}_o &=  0  \qquad \text{by (217a,b)}				\tag{219c} 
\end{align}
\setcounter{equation}{219}
And by the commutation laws (213) and (219c) we have
\begin{equation}
\braket{a_{k \mu} \tilde{a}_{k^{\prime} \lambda}}_o =  \braket{\,\boldsymbol{[}\,a_{k \mu}, \, \tilde{a}_{k^{\prime} \lambda}\,\boldsymbol{]}\,}_o = \delta^{3}(\boldsymbol{k} - \boldsymbol{k}^{\prime}) \, \delta_{\mu \lambda}
\end{equation}

The vacuum expectation value $\Braket{A_{\mu}(z) \, A_{\lambda}(y)}_o$ is thus just the part of the commutator $\boldsymbol{[}\, A_{\mu}(z), \, A_{\lambda}(y) \boldsymbol{]}$ which contains positive frequencies $\exp\{i k \cdot (z - y) \}$ with $k_0 > 0$, as one can see using (211), (219) and (220). Thus\endnote{
Eq.\ (222) lacked the lower limit $k_{0} > 0$ on the integral.}
\begin{equation}
\braket{A_{\mu}(z) \, A_{\lambda}(y)}_o = i\hbar c \, \delta_{\mu \lambda}\, D^{+}(z - y)
\end{equation}
\begin{equation}
D^{+}(x) = -\frac{i}{(2\pi)^3} \int_{k_0 > 0} \exp(i k \cdot x) \, \delta(k^2) \, \Theta(k) \, d^{\,4}k
\end{equation}
We write
\begin{equation}
D(x) = D^{+}(x) + D^{-}(x)
\end{equation}
\begin{equation}
D^{+} = \tfrac{1}{2}\left( D - iD^{(1)}\right)	\qquad D^{-} = \tfrac{1}{2}\left( D + iD^{(1)}\right)
\end{equation}
The even function $D^{(1)}$ is then defined by
\begin{equation}
\Braket{A_{\mu}(z)\, A_{\lambda}(y) + A_{\lambda}(y)\, A_{\mu}(z)}_o = \hbar c \, \delta_{\mu \lambda} \, D^{(1)}(z-y)
\end{equation}
\begin{equation}
D^{(1)}(x) = \frac{1}{(2\pi)^3} \int \exp(i k \cdot x) \, \delta(k^2) \, d^{\, 4}k
\end{equation}
It is then not hard to prove (see the Appendix below) that 
\begin{equation}
D^{(1)}(x) = \frac{1}{2\pi^2 x^2}
\end{equation}

The functions $D$ and $D^{(1)}$ are the two independent solutions of $\Box^2 D = 0$, one odd and the other even. Then we define the function
\begin{align}
\overline{D}(x) &= - \tfrac{1}{2} \epsilon(x) \, D(x) = \tfrac{1}{2}\left(D_R(x) + D_A(x)\right) = \frac{1}{4\pi} \, \delta(x^2) \notag \\
&= \frac{1}{(2\pi)^4} \int \exp(i k \cdot x) \, \frac{1}{k^2}\, d^{\, 4} k
\end{align}
the last being a real principal value integral: This is the even solution of the point-source equation
\begin{equation}
\Box^2 \overline{D}(x) = - \delta^{4}(x)
\end{equation}
\emph{Appendix}
\begin{align*}
D^{(1)}&(x) = \frac{1}{(2\pi)^3} \int e^{i k \cdot x}\, \delta(k^2)\, d^{\,4}k \\
 &= - \frac{1}{(2\pi)^2} \int_{-1}^{+1} d\mu \int_{-\infty}^{\infty} dk_0 \int_{0}^{\infty} d |\boldsymbol{k} | e^{i |\boldsymbol{k} ||\boldsymbol{x} | \mu} e^{-ik_0 x_0} \frac{ \{ \delta(k_0 - | \boldsymbol{k} |) +  \delta(k_0 + | \boldsymbol{k} |) \} | \boldsymbol{k} |^2}{2 | \boldsymbol{k} |} \vphantom{A^{A^{A^{A^{A^{A^{A}}}}}}} \\
&= \frac{1}{(2\pi)^2} \int_0^{\infty} d|\boldsymbol{k}| \frac{1}{i |\boldsymbol{k}| |\boldsymbol{x}|} \left\{e^{i |\boldsymbol{k} ||\boldsymbol{x} |} - e^{-i |\boldsymbol{k} ||\boldsymbol{x} |} \right\}\frac{1}{2|\boldsymbol{k}|} \left\{e^{-i |\boldsymbol{k} | x_0} + e^{i |\boldsymbol{k} | x_0} \right\} |\boldsymbol{k}|^2 \vphantom{A^{A^{A^{A^{A^{A^{A}}}}}}}\\
&= \frac{1}{2 \pi^2} \frac{1}{|\boldsymbol{x} |} \int_0^{\infty} \sin(|\boldsymbol{k} ||\boldsymbol{x} |) \cos(|\boldsymbol{k} |x_0)\,d|\boldsymbol{k}| = \frac{1}{2 \pi^2} \frac{1}{2|\boldsymbol{x} |} \int_0^{\infty} d|\boldsymbol{k}| \left\{ \sin((|\boldsymbol{x} | + x_0)|\boldsymbol{k} |) + \sin((|\boldsymbol{x} | - x_0)|\boldsymbol{k} |) \right\} \vphantom{A^{A^{A^{A^{A^{A^{A}}}}}}}
\end{align*}
Taking the integral in the Abelian sense
\index{Abelian}%
\[
\lim_{\epsilon \rightarrow 0} \int_0^{\infty} e^{-\epsilon x} \sin ax \, dx = \lim_{\epsilon \rightarrow 0} \frac{a}{\epsilon^2 + a^2} = \frac{1}{a}
\]
hence in our case\endnote{
$D^{(1)}(x)$ was added on the left-hand side for clarity}
\[
D^{(1)}(x) = \frac{1}{2 \pi^2} \frac{1}{2 |\boldsymbol{x} |} \left\{\frac{1}{|\boldsymbol{x} | + x_0} + \frac{1}{|\boldsymbol{x} | - x_0} \right\} = \frac{1}{2 \pi^2 x^2}.
\]

\subsection*{The Gupta-Bleuler Method}
\addcontentsline{toc}{subsection}{The Gupta-Bleuler Method} 
\index{Gupta-Bleuler~method|(}%

There is one difficulty in the preceding theory. We assume according to (220) 
\[
\Braket{a_{k \mu} a_{k^{\prime} \lambda}^{*}}_o = \pm \, \delta^3(\boldsymbol{k} - \boldsymbol{k}^{\prime}) \, \delta_{\mu \lambda} \tag{220a}
\]
Here the plus sign holds for $\mu = 1, 2, 3$; the minus sign for $\mu = 4$. Now if the operators $a_{k \mu}, \, a_{k \mu}^{*}$ are represented in the usual way by matrices, as it is done in the elementary theory of the harmonic oscillator (see Wentzel p.\ 33, Eq.\ (6.16)),  the vacuum expectation values of a product $(a_{k \mu} \, a_{k \mu}^{*})$ will always be \emph{positive}, i.e. the plus sign should hold in (220a) for $\mu = 4$ also.
\index{Wentzel, Gregor}%
In fact $(a_{k \mu} \, a_{k \mu}^{*})$ will have a positive expectation value in any state whatever, if the photon oscillators are treated as ordinary elementary oscillators.

Therefore we must distinguish between the scalar product $(\Psi_1^{*},\, \Psi_2)$ as we have defined it by our covariant theory, and the scalar product $(\Psi_1^{*},\, \Psi_2)_{E}$ which one would calculate using the explicit matrix representation of the operators. The product $(\Psi_1^{*},\, \Psi_2)_{E}$ has no physical significance, because the matrix representations of the $a_{k 4}$ refer to states with photons polarized purely in the time dimension, which cannot occur physically. However, it is convenient also to be able to use the matrix representations in practice.

To use the matrix representations, we have only to define an operator $\eta$ by the condition
\[
\eta \Phi = (-1) \Phi 								\tag{220b}
\]
where $\Phi$ is any state in which there is a definite number $N$ of photons polarized in the 4-direction. Then the physical scalar product is given in terms of the explicit matrix representations by
\[
(\Psi_1^{*},\, \Psi_2) = (\Psi_1^{*},\, \eta \Psi_2)_{E}   	\tag{220c}
\]
The definition (220c), introduced by Gupta, makes the matrix representations consistent with all the requirements of the covariant theory, in particular also it gives (220) correctly.
\index{Gupta, S.~N.}%
The physical scalar product is thus an \emph{indefinite metric} regarded from the point of view of the matrix representations. However, we have seen in (216b) that for any \emph{physical} states the scalar product $(\Psi_1^{*},\, \Psi_2)$ is equal to  $(\Psi_{1T}^{*},\,\Psi_{1T})$ where $\Psi_{1T}$ is a state involving transverse photons only, and hence is positive. Thus for physical states the metric is definite and this is all that we require of it.

\subsection*{Example: Spontaneous Emission of Radiation}
\addcontentsline{toc}{subsection}{Example: Spontaneous Emission of Radiation}

This is a purely quantum-mechanical effect. A classical treatment, considering the reaction of the atom to a classical applied Maxwell field, gives a correct account of the absorption of radiation and of stimulated emission, but fails to give the spontaneous emission.
\index{Maxwell!field}%

Let an atom have 2 states, the ground state 1 and an excited state 2 with energy $\hbar c q$. For the transition $2 \rightarrow 1$ let the charge-current density of the atom have the unintegrated matrix elements
\[
j_{\mu A}(x) = j_{\mu A}(r, t) \quad \text{at the point $x = (r, t)$}
\]
The interaction with the Maxwell field has matrix element\endnote{
Eq.\ (230) lacked the sum over $\mu$; \emph{cf.} Eq. (170).}
\index{Maxwell!field}%
\begin{equation}
I = -\frac{1}{c} \int \sum\nolimits_{\mu} j_{\mu A}(r, t) \Braket{A_{\mu}(r, t)}_{\text{emit}} d^{\,3}\boldsymbol{r}
\end{equation}
for making a transition with emission of a photon. The total emission probability per unit time is obtained using time dependent perturbation 
theory:\endnote{
``obtained'' was inserted; statement read  ``per unit time is using \dots ''}
\index{probability!emission}%
\begin{align}
w &= \frac{1}{T} \sum | a_1(T) |^2 = \frac{1}{T} \sum_{\substack{\text{photon} \\ \text{states}}} \left|\,\frac{1}{i \hbar} \int \left\{ - \frac{1}{c} \int \sum\nolimits_{\mu} j_{\mu A}(r, t) \Braket{A_{\mu}(r, t)}_{\text{emit}} d^{\,3}\boldsymbol{r} \right\} dt\, \right|^2 \notag \\
&= \frac{1}{T c^4 \hbar^2} \iint \sum_{\lambda, \mu} j_{\lambda A}^{*}(x^{\prime}) \, j_{\mu A}(x) \Braket{A_{\lambda}^{*}(x^{\prime}) A_{\mu}(x)}_o \, d^{\,4}x \;d^{\,4}x^{\prime}
\end{align}
the integral being over all space for a long time $T$, the sum over the physical photon states only. It is \emph{not} correct to take in (231) the photon states to be the 4 states with polarization in the directions $\mu = 1, 2, 3, 4,$ because these are not physical states.

Using a sum-rule to sum over the states
\[
w= \frac{1}{T c^4 \hbar^2} \iint \sum_{\lambda, \mu} j_{\lambda A}^{*}(x^{\prime}) \, j_{\mu A}(x) \Braket{A_{\lambda}^{*}(x^{\prime}) A_{\mu}(x)}_o \, d^{\,4}x \;d^{\,4}x^{\prime}
\]
Write $\tilde{j}_{\lambda A}(x^{\prime})$ for the matrix element\endnote{
The subscript $A$ was missing in the term $j_{\lambda A}(x^{\prime})$.}
 of $j_{\lambda A}(x^{\prime})$ in the reverse transition $1 \rightarrow 2$. Then
\[
j_{\lambda A}^{*}(x^{\prime}) = \eta_{\lambda}\, \tilde{j}_{\lambda A}(x^{\prime}), \qquad A_{\lambda}^{*}(x^{\prime}) = \eta_{\lambda} A_{\lambda}(x^{\prime})
\]
with $\eta_{\lambda} = +1, \, \lambda = 1, 2, 3; \; \eta_4 = -1$. \\
Hence
\begin{align}
w &= \frac{1}{T c^4 \hbar^2} \iint \sum_{\lambda,\,\mu = 1}^{4} j_{\lambda A}^{*}(x^{\prime}) \, j_{\mu A}(x) \Braket{A_{\lambda}^{*}(x^{\prime}) A_{\mu}(x)}_o \, d^{\,4}x \;d^{\,4}x^{\prime} \notag \\
&= \frac{\hbar c}{(2\pi)^3} \int d^{\,4}k\,\delta(k^{2}) \Theta(k_0) \sum_{\mu = 1}^{4} j_{\mu A}(k) \tilde{j}_{\mu A}(k) \frac{1}{Tc^4 \hbar^2} \int\limits_{0}^{cT} \int\limits_{0}^{cT} dx_0 \, dx^{\prime}_0 \exp\left\{ i(x - x_0)(q - k_o) \right\} \notag \\
&= \frac{1}{(2\pi)^2 \hbar c^2} \sum_{\mu = 1}^{4} \int d^{\,3} \boldsymbol{k} \, \delta(| \boldsymbol{k} |^2 - q^2) j_{\mu A}(\boldsymbol{k}) \tilde{j}_{\mu A}(\boldsymbol{k})
\end{align}
since
\[
\frac{1}{cT} \int\limits_{0}^{cT}  \int\limits_{0}^{cT} dx_0 \, dx^{\prime}_0 \exp\left\{ i(x - x_0)(q - k_o) \right\} = cT\, \frac{\sin^2 \frac{cT}{2} (q - k_0)}{\left(\frac{cT}{2} (q - k_0)\right)^2} \rightarrow \pi \delta(q - k_0) \quad \text{if} \; cT \rightarrow \infty
\]
where\endnote{
The exponential in the first integral lacked $i$; it had the argument $-\boldsymbol{k} \cdot \boldsymbol{r}$.}
\[
j_{\mu A}(\boldsymbol{k}) = \int  j_{\mu A}(r)\, e^{-i \boldsymbol{k} \cdot \boldsymbol{r}} \,d^{\,3}\boldsymbol{r} \qquad \tilde{j}_{\mu A}(\boldsymbol{k}) = \int \tilde{j}_{\mu A}(r)\, e^{i \boldsymbol{k} \cdot \boldsymbol{r}} \, d^{\,3}\boldsymbol{r}
\]

By the charge conservation law 
\[
\sum_{\mu} k_{\mu} j_{\mu A}(k) = 0 \qquad \qquad \sum_{\mu} k_{\mu} \tilde{j}_{\mu A}(k) = 0
\]
and so
\begin{align*}
\sum_{\mu} j_{\mu A}(k) \tilde{j}_{\mu A}(k) &= \frac{1}{q^2} q j_{4A} \, q \tilde{j}_{4A} + \sum_{i}^{3} |j_{iA}(k)|^2
= - \frac{1}{q^2} \sum_{i, \ell = 1}^{3} \left\{ k_{i} j_{iA}(k)\, k_{\ell} \tilde{j}_{\ell A}(k) \right\} + \sum_{i}^{3} |j_{iA}(k)|^2 \notag \\ 
&= - \frac{1}{q^2} | \boldsymbol{k} |^2 |\boldsymbol{j}_{A} |^{2} \cos^{2} \theta + |\boldsymbol{j}_{A} |^{2} = |\boldsymbol{j}_{A} |^{2}(1 - \cos^{2} \theta) = |j_{1A}|^2 + |j_{2A}|^2												\tag{232a}
\end{align*}
where 1 and 2 are the two directions of transverse polarization. This shows how the third and fourth polarization directions do not appear in real emission problems. The same result would be obtained if we used the indefinite metric explicitly, i.e.take the sum in (231) over the 4 polarization states $\mu = 1, 2, 3, 4,$ with the $\mu = 4$ given a minus sign arising from the $\eta$ in (220c). But it is always simpler to work directly with the covariant formalism, than to bother with the non-physical photon states and then have to use $\eta$ to get the right answers. 

Finally, the emission probability in direction of polarization 1 and in direction of propagation given by the solid angle $d\Omega$ is, using (232), (232a), and $\delta(|\boldsymbol{k}|^2 - q^2) = \dfrac{1}{2q} \delta(|\boldsymbol{k}| - q)$ for $q > 0$
\index{probability!emission}%
\index{angle!solid}%
\begin{equation}
w = \frac{q\, d\Omega}{8 \pi^2 \hbar c^2} \left| j_{1A}(x) \right|^2
\end{equation}
For dipole radiation by a one-electron atom with coordinates $(x, y, z)$
\[
j_1 = e \dot{x} = iec q x	\qquad \qquad \boldsymbol{k} \boldsymbol{\cdot} \boldsymbol{r} \ll 1
\]
and
\begin{equation}
w = \frac{e^2 q^3 \, d\Omega}{8 \pi^2 \hbar} \left| \Braket{x}_{12} \right|^2
\end{equation}
This checks with Bethe's \emph{Handbuch} article.\endnote{
Bethe \& Salpeter, ref.\ [20], p.\ 249, Eq.\ 59.7}
\index{Bethe, Hans~A.}%

The example shows how covariant methods will work, even for problems of this elementary sort for which they are not particularly suited. The covariant method avoids the necessity of having to think about the normalization of the photon states, the factors of 2 and $\pi$ etc. being given automatically when one uses (221). 
\index{Gupta-Bleuler~method|)}%

\subsection*{The Hamiltonian Operator}
\addcontentsline{toc}{subsection}{The Hamiltonian Operator}

From the equation
\[
i\hbar \frac{\partial A_{\mu}}{\partial t} = \boldsymbol{[}\, A_{\mu}, \,H\,\boldsymbol{]}
\]
we find
\begin{align*}
\boldsymbol{[}\, a_{k \mu}, \,H\,\boldsymbol{]} &= \hbar c |\boldsymbol{k}| a_{k \mu} \\
\boldsymbol{[}\, \tilde{a}_{k \mu}, \,H\,\boldsymbol{]} &= -\hbar c |\boldsymbol{k}| \tilde{a}_{k \mu}
\end{align*}
Using the commutation rules (213) we can find an operator $H$ which satisfies all these conditions simultaneously. Namely
\[
H = \int d^{\,3}\boldsymbol{k} \, \hbar c |\boldsymbol{k} | \sum_{1}^{4} \tilde{a}_{k \lambda} a_{k \lambda}    \tag{234a}
\]
This operator is in fact unique apart from an arbitrary additive constant. To fix the constant we require\endnote{
There was no subscript  ``o'' on the vacuum expectation value.} 
$\Braket{H}_{o} = 0$ which leads to the result (234a) precisely, as one can see at once from (219). Hence (234a) is the Hamiltonian  of this theory, which is very simple in this momentum representation. 
\index{Hamiltonian}%
\index{momentum!representation}%

To derive $H$ from the Lagrangian is also possible but much more tedious. From (234a) we see that 
\[
N_{k \lambda} = \tilde{a}_{k \lambda} a_{k \lambda}	   \qquad \qquad \text{(not summed)}
\]
is an operator just representing the number of quanta in the frequency $k$ and polarization $\lambda$. It follows at once from the commutation rules (213), from the singular $\delta$-function factor which comes from the continuous spectrum, $N_{k \lambda}$ being in fact the number of quanta per unit frequency range, that $\int N_{k \lambda} d^{\, 3} \boldsymbol{k}$ integrated over \emph{any} region of momentum-space has the integer eigenvalues 0, 1, 2, $\dots$. This is so, because the state with $n_{i}$ particles with momentum $k_{i}$ is $\Psi = \prod_{i = 1}^{\ell} (\tilde{a}_{k_i \lambda})^{n_i} \Psi_o$. Then taking $\int_{\Omega} N_{k \lambda} d^{\, 3} \boldsymbol{k}$ over $\Omega$ including the momenta $k_{1}, k_{2}, \dots, k_{j}$ we get
\begin{align*}
\int_{\Omega}& N_{k \lambda} d^{\, 3} \boldsymbol{k} \, \Psi = \int_{\Omega} \tilde{a}_{k \lambda} a_{k \lambda} \prod_{i = 1}^{\ell} (\tilde{a}_{k_i\lambda})^{n_i} \Psi_o \, d^{\,3} \boldsymbol{k} \\
&= \int_{\Omega} \tilde{a}_{k \lambda} \left\{ \sum_{i=1}^{\ell} n_i \prod_{i = 1}^{\ell} \tilde{a}^{(n_{i} -1)}_{k_i\lambda} \boldsymbol{[} \, a_{k \lambda}, \,\tilde{a}_{k_i \lambda} \,\boldsymbol{]} + \prod_{i = 1}^{\ell} (a_{k_i \lambda})^{n_i} a_{k\lambda} \right\} \Psi_o \, d^{\,3} \boldsymbol{k} \qquad \text{by (213) and (217a)} \\
&= \int_{\Omega} \tilde{a}_{k \lambda} \sum_{i=1}^{\ell} n_i \prod_{i = 1}^{\ell} \tilde{a}^{(n_{i} -1)}_{k_i\ \lambda} \delta^{3}(\boldsymbol{k} - \boldsymbol{k}_i) \Psi_o\, d^{\,3} \boldsymbol{k}  = \sum_{i=1}^{j} n_{i} \prod_{i = 1}^{\ell}\left(\tilde{a}_{k_i \lambda} \right)^{n_i} \Psi_o = \sum_{i =1}^{j} n_i \Psi
\end{align*}

\subsection*{Fluctuations of the Fields}
\addcontentsline{toc}{subsection}{Fluctuations of the Fields}

Since the electromagnetic fields $\boldsymbol{E}$ and $\boldsymbol{H}$ are quantum-mechanical variables, they do not have well-determined values in any state in which energy and momentum are well-defined, for example the vacuum state. A state of the fields can be specified \emph{either} by fixing the values of $\boldsymbol{E}$ and $\boldsymbol{H}$ \emph{or} by specifying the numbers of quanta present with various momenta and energies. The two descriptions are complementary and are both possible only in the classical limit of large numbers of quanta and very strong fields.

An educational discussion of these questions, considering in detail the example of a cavity resonator with one mode of oscillation, has been given by L.\ P.\ Smith, \emph{Phys.\ Rev.} \textbf{69} (1946) 195. 
\index{Smith, Lloyd~P.}%
This is worth reading. Essential is the fact that you cannot fix the time-dependence of the field (phase) with a fixed number of quanta (energy). 

We consider a more general problem. What is the mean-square fluctuation in the vacuum state of a field-quantity? We define
\begin{align}
E_{1}(VT) &= \frac{1}{VT} \int_{VT} E_{1}(x) \, d\tau \, dt \notag \\
H_{1}(VT) &= \frac{1}{VT} \int_{VT} H_{1}(x) \, d\tau \, dt
\end{align}
averaged over some finite space-volume $V$ and also over a time $T$.
\noindent Let $V(\boldsymbol{k}) = \int_{V} e^{-i \boldsymbol{k} \cdot \boldsymbol{r}} \, d\tau$. Then since $\boldsymbol{H} = \nabla \times \boldsymbol{A}$, we have
\begin{align}
\Braket{\left\{H_{1}(VT)\right\}^2}_o &= \frac{1}{V^2 T^2} \iint d\tau \, d\tau^{\prime}\, dt\, dt^{\prime} \, \Braket{\left( \frac{\partial}{\partial x_2} A_{3} - \frac{\partial}{\partial x_3} A_{2} \right)\left( \frac{\partial}{\partial x_2^{\prime}} A_{3}^{\prime} - \frac{\partial}{\partial x_3^{\prime}} A_{2}^{\prime} \right)}_o \notag \\
&= \frac{1}{V^2 T^2} \frac{\hbar c}{2} \iint d\tau \, d\tau^{\prime}\, dt\, dt^{\prime} \, \left(\frac{\partial}{\partial x_{2}} \frac{\partial}{\partial x_{2}^{\prime}} + \frac{\partial}{\partial x_{3}} \frac{\partial}{\partial x_{3}^{\prime}} \right) D^{(1)} (x - x^{\prime}) \quad \text{using (225)} \notag \\
&= \frac{1}{V^2 T^2} \frac{\hbar c}{16 \pi^3} \iint d\tau \, d\tau^{\prime}\, dt\, dt^{\prime} \, \left(\frac{\partial}{\partial x_{2}} \frac{\partial}{\partial x_{2}^{\prime}} + \frac{\partial}{\partial x_{3}} \frac{\partial}{\partial x_{3}^{\prime}} \right)  \int d^{\,4}k \, e^{i k \cdot (x - x^{\prime})} \delta(k^2) \quad \text{using (226)} \notag \\
&= \frac{\hbar c}{16 \pi^3 V^2 T^2} \int_{0}^{T} \int_{0}^{T} \int \frac{d^{\,3} \boldsymbol{k}}{| \boldsymbol{k} |} (k_{2}^{2} + k_{3}^{2})\, |V(k)|^2 \, e^{i |\boldsymbol{k} |x_0}\, e^{-i |\boldsymbol{k} | x_0^{\prime}} \, dt \, dt^{\prime} \notag \\
&= \frac{\hbar c}{16 \pi^3 V^2 T^2} \int \frac{d^{\,3} \boldsymbol{k}}{| \boldsymbol{k} |} (k_{2}^{2} + k_{3}^{2})\, |V(k)|^2 \, \frac{4 \sin^2 (\tfrac{1}{2} c | \boldsymbol{k} T)}{c^2 | \boldsymbol{k} |^2}  \notag \\
\Braket{\left\{E_{1}(VT)\right\}^2}_o &= \frac{1}{V^2 T^2} \frac{\hbar c}{2} \iint d\tau \, d\tau^{\prime}\, dt\, dt^{\prime} \, \left(-\frac{\partial}{\partial x_{4}} \frac{\partial}{\partial x_{4}^{\prime}} - \frac{\partial}{\partial x_{1}} \frac{\partial}{\partial x_{1}^{\prime}} \right) D^{(1)} (x - x^{\prime}) \notag \\
&= \frac{\hbar c}{16 \pi^3 V^2 T^2} \int \frac{d^{\,3} \boldsymbol{k}}{| \boldsymbol{k} |} (| \boldsymbol{k} |^{2} - k_{1}^{2})\, |V(k)|^2 \, \frac{4 \sin^2 (\tfrac{1}{2} c | \boldsymbol{k} T)}{c^2 | \boldsymbol{k} |^2}  = \Braket{\left\{H_{1}(VT)\right\}^2}_o 
\end{align}
Taking for $V$ any finite volume and $T$ a finite time, this mean-square fluctuation is finite. Example: a sphere of radius R gives
\begin{equation}
V(K) = \frac{\,4 \pi}{\, | \boldsymbol{k} |^3}\, (\sin R |\boldsymbol{k}| - R  |\boldsymbol{k}| \cos R |\boldsymbol{k}| )
\end{equation}
But if either $R$ or $T$ tends to zero, the fluctuations tend to $\infty$ and in the limit actually diverge.\endnote{
``if'' substuted for ``it''; and ``gives'' added to the previous sentence.} 
That is to say, only measurements of field-quantities averaged both in space and in time can have any physical reality.

\subsection*{Fluctuation of Position of an Electron in a Quantized Electromagnetic Field. The Lamb Shift.}
\addcontentsline{toc}{subsection}{The Lamb Shift -- Fluctuation of Position of Electron} 
\index{Lamb~shift}%

Consider an electron represented by an extended spherical charge of radius R, lying in a stationary state in the potential $\phi(r)$ of a hydrogen atom. 
\index{hydrogen~atom!Lamb~shift}%
It has a certain wave-function $\psi(r)$. We consider everything non-relativistically except for the quantized radiation field with which the electron interacts. The effect of this fluctuating field is to produce a rapid fluctuation in position of the electron. In fact we have for rapid fluctuations
\[
m \ddot{\boldsymbol{r}} = - e \boldsymbol{E}
\]
Thus a fluctuating component of $\boldsymbol{E}$ with frequency $c | \boldsymbol{K} |$ produces the same fluctuation in $\boldsymbol{r}$ with amplitude multiplied by a factor $\dfrac{e}{m} \dfrac{1}{c^2 | \boldsymbol{K} |^2}$.
\index{amplitude}%
The slow fluctuations of $\boldsymbol{E}$ cannot be followed by the electron if the frequency is less than the atomic frequency $cK_{H}$. Hence we find from (236), making $T \rightarrow 0$
\[
\Braket{r_{1}^2}_o = \frac{e^2}{m^2} \frac{\hbar c}{16 \pi^3 V^2} \int_{K_{H}}^{\infty}  \frac{d^{\,3} \boldsymbol{K}}{| \boldsymbol{K} |} (K_{2}^{2} + K_{3}^{2}) \, |V(K) |^2 \frac{1}{c^4 | \boldsymbol{K} |^4}
\]
because $ \lim\limits_{x \rightarrow 0} \dfrac{\sin^2 x}{x^2} = 1$. 
The integral now converges at $\infty$, because of the finite size of the electron. Since $R$ is very small we may approximate (237) by
\[
V(K) = \begin{cases}
\frac{4}{3}\pi R^3 = V &\text{for $| \boldsymbol{K} | R < 1$}\\
\;0 &\text{for $| \boldsymbol{K} | R > 1$}
\end{cases}
\]
Then, since $(K_{2}^{2} + K_{3}^{2}) = | \boldsymbol{K} |^2 (1 - \cos^{2} \theta)$ and
$\int_{0}^{\pi} \sin^{2} \theta \, \sin \theta \, d\theta  = \frac{4}{3}$, we have\endnote{
The limits on the integral were $-1$ and $+1$, and the value of the integral was given as $\tfrac{2}{3}$.}
\begin{equation}
\Braket{r_{1}^2}_o = \frac{e^2 \hbar}{6 m^{2}c^{3} \pi^{2}} \int_{K_{H}}^{1/R} \frac{d | \boldsymbol{K} |}{|\boldsymbol{K}|} = \frac{e^2 \hbar}{6 m^{2}c^{3} \pi^{2}} \log \left( \frac{1}{RK_{H}} \right)
\end{equation}

This fluctuation in position produces a change in the effective potential acting on the electron. Thus
\begin{align*}
\Braket{V(r + \delta r)} &= V(r) + \Braket{\delta \boldsymbol{r} \cdot \nabla V(r)}_o + \tfrac{1}{2}  \Braket{(\delta \boldsymbol{r})^2}_o \frac{\partial^{2} V}{\partial r^2} + \dots \\
&= V(r) + \tfrac{1}{2} \Braket{r_{1}^{2}}_o \nabla^2 V
\end{align*}
because $\Braket{\delta \boldsymbol{r} \cdot \nabla V(r)}_o = 0$, being odd. 
Now in a hydrogen atom, $\nabla^2 V = e^2 \delta^{3}(\boldsymbol{r})$ (Heaviside units!)
\index{Heaviside~units}%
Hence the change in the energy of the electron due to the fluctuations is\endnote{
The expression for $a_{o}$ is not here in the original, but it appears before Eq.\ (240).} 
($a_{o}$ = Bohr radius)
\begin{align}
\Delta E &= \int \psi^{*} \, \delta V \, \psi \, d\tau = \tfrac{1}{2} \Braket{r_{1}^{2}}_o e^{2} | \psi(0) |^2 \notag \\
              &= \begin{cases}
			\dfrac{e^{4} \hbar}{12 \pi^2 m^2 c^3}  \log \left( \dfrac{1}{RK_{H}} \right) \dfrac{1}{\pi n^3 a_o^3} &\text{for s-states} \\
			\qquad 0 &\text{for all others}
		   \end{cases}
\end{align}
\index{Bohr~radius}
\index{a@$a_{o}$|see{Bohr~radius}}
because for the hydrogen atom\endnote{
The expression for $\rho$ is not in the original.} 
($ \rho = r \sqrt{-8m_{r}E_{n}}/\hbar$)
\[
\psi_{n \ell m}(r, \theta, \varphi) = - \frac{1}{\sqrt{2 \pi}}\, e^{im \varphi} \left\{\frac{(2 \ell + 1)(\ell - | m |)!}{2(\ell + | m |)!} \right\}^{1/2} 
P^{| m |}_{\ell} (\cos \theta) \left[\left(\frac{2}{n a_{o}} \right)^3 \frac{(n - \ell - 1)!}{2n[(n + \ell)!]^{3}} \right]^{1/2} e^{-\rho/2} 
\rho^{\ell} L^{2 \ell + 1}_{n + \ell}(\rho)
\]
and 
\[
\psi_{n 0 0}(0, \theta, \varphi) = \frac{1}{\sqrt{2 \pi}} \frac{\sqrt{2}}{2} \left\{ - \left[ \left( \frac{2}{na_o} \right)^3 \frac{(n - 1)!}{2n(n!)^{3}}  \right]^{1/2} \right\} \left( - \frac{(n!)^{2}}{(n-1)!} \right) = \frac{1}{\pi^{1/2} a_o^{3/2} n^{3/2}}
\]

There will also be a (much bigger) addition to kinetic energy, arising from the fluctuations. We ignore this on the grounds that it will be the same for all atomic states and so will not give any relativistic displacement. Of course this is not a good argument.

Hence we find the first approximation to the Lamb shift;  the $2s$ state is shifted relative to the $2p$ states by
\index{Lamb~shift}%
\[
\Delta E = + \dfrac{e^{4} \hbar}{96 \pi^3 m^2 c^3 a_o^3}  \log \dfrac{1}{RK_{H}}
\]
Now\endnote{
$a_{o} = 0.529177 \times 10^{-8}$ cm; Ry = 13.6056 eV.}
\begin{align*}
a_o &= \frac{4 \pi \hbar^2}{me^2} = \frac{1}{\alpha} \frac{\hbar}{mc} \quad \text{(Bohr radius)} \\
\text{Ry} &= \frac{e^{4}m}{32 \pi^2 \hbar^4} \quad \text{(Rydberg energy unit)} \\
K_{H} &= \frac{\text{Ry}}{4 \hbar c}
\end{align*}
\index{Bohr~radius}%
\index{Rydberg~energy}%
We take $R = (\hbar/mc)$, the electron Compton wave-length since it is at this frequency that the NR treatment becomes completely wrong. 
\index{electron!Compton~wavelength}%
\index{Compton!wavelength}%
Then
\begin{align}
RK_{H} &= \frac{\text{Ry}}{4mc^2} = \frac{1}{8} \alpha^{2} \notag \\
\Delta E &= + \frac{\alpha^3}{3 \pi} \log (8 \times 137^{2}) \; \text{Ry}
\end{align}
Actually $\dfrac{\alpha^{3}}{3\pi}$ Ry = 136 Mc in frequency units. This gives an effect of the right sign and order of magnitude.\\
Method due to Welton \cite{Welton48}. 
\index{Welton, Theodore~A.}%

The size of the log is wrong because the low-frequency cut-off was badly done. We find $\Delta E \sim 1600$ Mc instead of the correct value of 1060 Mc. But physically the origin of the shift is correctly described in this way.

\section*{Ia. Theory of Line Shift and Line Width}
\addcontentsline{toc}{section}{Ia. Theory of Line Shift and Line Width}

To make a better treatment of the effect of the radiation interaction on energy-levels, we must try to solve more exactly the equation of motion for the system atom plus radiation field. The effect of the field is shown not only in a shift of energy levels, but also in a finite \emph{width} of the levels due to real radiation.  Roughly, if the state has a life-time $T$ for decay by radiation, the width $\Gamma$ of the level, or the mean variation in energy of the emitted photons, is given by the uncertainty principle $\Gamma \approx \hbar/T$. The line-shift and line-width are effects of the same kind and cannot be correctly treated except in combination. 

So we make a theory now which treats the atom non-relativistically, but takes account properly of the radiation interaction. This means, we repeat the calculation of spontaneous emission by an atom, but now including the reaction of the radiation on the atom instead of taking the atom as given by a fixed charge-current oscillator.

For this sort of calculation it is always convenient to work in a special representation called the \emph{Interaction Representation}.

\subsection*{The Interaction Representation}
\addcontentsline{toc}{subsection}{The Interaction Representation}

In the Schr\"{o}dinger representation the wave-function $\Psi$ satisfies the equation of motion
\index{Schr\"{o}dinger!representation}%
\begin{equation}
i\hbar \frac{\partial}{\partial t} \Psi = H \Psi
\end{equation}
where $H$ is the Hamiltonian. In the case of an atom interacting with the radiation field, we have 
\index{Hamiltonian}%
\begin{equation}
H = H_A + H_M + H^{S}_I
\end{equation}
where $H_A$ is the Hamiltonian operator for the atom, and $H_M$ that for the Maxwell field  without interaction. $H_M$ is given by (234a), and in quantum electrodynamics according to (170)
\index{Maxwell!field!Hamiltonian}%
\index{quantum~electrodynamics}%
\begin{equation}
H^{S}_I = - \frac{1}{c} \int \sum_{\mu} j_{\mu}^{S}(r) A_{\mu}^{S}(r) d^{\,3}\boldsymbol{r}
\end{equation}
because $j_{\mu}^{S}(r) = ie\, \overline{\psi} \gamma_{\mu} \psi$, and in this case $\mathscr{H} = \sum_{\mu} \pi_{\mu} A_{\mu} - \mathscr{L} = -\mathscr{L}$ because $\pi_{\mu} = \partial \mathscr{L}/\partial \dot{A_{\mu}}$, and $\mathscr{L}_I$ does not contain $\dot{A_{\mu}}$. All the operators in (242), (243) are time-independent\endnote{
``time-independent'' substituted for ``time-dependent'', which describes the Heisenberg representation.},
 Schr\"{o}dinger representation operators, and so are the given the label $S$.
\index{Schr\"{o}dinger!representation}%

Now we choose a new wave function $\Phi(t)$ which is given in terms of $\Psi$ by
\begin{equation}
\Psi(t) = \exp\left\{ -\frac{i}{\hbar} \left( H_{A} + H_{M} \right)t \right\} \Phi(t)
\end{equation}
This $\Phi(t)$ will be a constant for any state representing the atom and the Maxwell field  without interaction.
\index{Maxwell!field}%
Thus the time-variation of $\Phi(t)$ in an actual state describes \emph{just} the effect of the interaction in perturbing the atomic states. From (241) and (244), the time-variation of $\Phi$ is given by the Schr\"{o}dinger equation 
\index{Schr\"{o}dinger!equation}%
\begin{equation}
i \hbar \frac{ \partial \Phi}{\partial t} = H_{I}(t) \Phi
\end{equation}
with 
\begin{equation}
H_{I}(t) = \exp \left\{ \frac{i}{\hbar} (H_{A} + H_{M}) t \right\} H_{I}^{S} \exp \left\{- \frac{i}{\hbar} (H_{A} + H_{M}) t \right\}
\end{equation}
Thus
\begin{equation}
H_{I} = - \frac{1}{c} \int \sum_{\mu} j_{\mu}( r, t) A_{\mu}( r, t ) d^{\,3}\boldsymbol{r}
\end{equation}
with
\begin{equation}
j_{\mu}( r, t) = \exp \left\{ \frac{i}{\hbar} H_{A}  t \right\}  j_{\mu}^{S}( \boldsymbol{r}) \exp \left\{ -\frac{i}{\hbar} H_{A}  t \right\} 
\end{equation}
\begin{equation}
A_{\mu}( r, t ) = \exp \left\{ \frac{i}{\hbar} H_{M}  t \right\}  A_{\mu}^{S}( r ) \exp \left\{ -\frac{i}{\hbar} H_{M}  t \right\} 
\end{equation}
These operators $j_{\mu}( r, t)$ and $A_{\mu}( r, t )$ have precisely the time-dependence of the field-operators in the Heisenberg representation, for the two systems, atom and radiation field, taken separately without interaction. Thus in the Interaction Representation the time-dependence of the Schr\"{o}dinger wave-function is split into two parts, the operators taking the time-dependence of the non-interaction systems, the wave-function taking a time-dependence showing the effects of the interaction only.  
\index{Schr\"{o}dinger!wave~function}%
The operators $A_{\mu}( r, t )$ satisfy the wave-equation $\Box^2 A_{\mu} = 0$ and the covariant commutation laws (203), because we see from (249) that 
\[
\frac{\partial A_{\mu} (r, t)}{\partial t} = i \hbar \, \boldsymbol{[}\, H_M, \, A_{\mu}(r, t) \, \boldsymbol{]}
\]
i.\ e., the time variation of $A_{\mu}(r, t)$ is the same as that of $A_{\mu}(x)$ in the Heisenberg representation without interaction (see (190a)), which in turn leads to the field equations (197). Matrix elements of IR operators given by (246), (248) or (249) between IR wave-functions given by (244) are of course the same as the matrix elements which would be obtained in any other representation.

\subsection*{The Application of the Interaction Representation to the Theory of Line-Shift and Line-Width}
\addcontentsline{toc}{subsection}{Application of Interaction Representation to Line Shift and Line Width}

Consider the solution of Eq. (245) in which the atom is given initially in a stationary unperturbed state $O$ with energy $E_o$, the Maxwell field being in the vacuum state, with no photons present. Let $\Phi_o$ be the interaction representation wave-function representing the atom in state $O$ and the Maxwell field in the vacuum state, without interaction. $\Phi_o$ is independent of time. 
\index{Maxwell!field!vacuum}%

The initial condition $\Phi(t) = \Phi_o$ at time $t = t_o$ is a physically unreal one. It would mean putting the atom into existence at time 0 without any radiation field being excited at this time. This we cannot do physically. In fact the initial condition for an atom in an excited state will depend on how it got into the excited state. This cannot be formulated in a simple way: one needs a complicated model in order to describe the initial excitation of the atom.

We are interested in calculating the variation with time of $\left(\Phi_o^{*} \Phi(t) \right)$, the probability amplitude for finding the atom still in the unperturbed state $\Phi_o$ at time $t$. By (245) we have 
\index{probability!amplitude}%
\begin{equation}
\frac{d}{dt} \left(\Phi_o^{*} \Phi(t) \right) = - \frac{i}{\hbar} \left(\Phi_o^{*} H_{I}(t) \Phi(t) \right)
\end{equation}
Suppose we take the physically unreal initial condition
\[
\Phi(t) = \Phi_o \qquad \text{at $t = t_o$}
\]
Then (250) will give 
\begin{equation}
\frac{d}{dt} \left(\Phi_o^{*} \Phi(t) \right)_{t = t_o} = 0
\end{equation}
from the $H_{I}$ given by (247) has zero expectation value in the Maxwell field vacuum, since $A_{\mu}(r, t)$ also has a zero expectation value in vacuum as one can see from (211) and (217).
\index{Maxwell!field!vacuum}%
Thus $\left(\Phi_o^{*} \Phi(t) \right)$ is momentarily stationary at $t = t_o$. This is however not interesting since the conditions at $t = t_o$ are entirely unphysical.

The physically meaningful quantity is the value of (250) at a time $t$ a long time after $t_o$. Then the atom will have ``settled down'' to a quasi-stationary state of radiative decay, and we may expect that the value we find for (250) is independent of the particular initial condition chosen and will be correct for an atom which has been excited by any reasonable method to the state $\Phi_o$.

We make the calculation so as to include effects of the radiation $H_{I}$ up to second order. This means we include effects of emitting and absorbing one photon only. In fact we know physically that effects from two or more photons are very small, so the approximation is a good one. 

Suppose that $(t - t_o)$ is long compared with all the atomic frequencies. Then a solution of (245), valid to first order in $H_I$, is
\begin{align}
	\Phi_{1}(t) &= \left[1 - \frac{i}{\hbar} \int_{- \infty}^{t} H_{I} (t^{\prime})\, dt^{\prime} \right] a(t)\, \Phi_o \\
		 &+ \; \text{terms involving other atomic states $\Phi_{n}$ with two or more photons present.} \notag
\end{align}
Here $a(t) = \left(\Phi_o^{*}\Phi(t) \right)$ is a slowly varying amplitude, constant to first order in $H_{I}$, representing the slow decay of the atom.
\index{amplitude}%
Note that our treatment is not only a perturbation theory correct to second order in $H_{I}$, but it also must account \emph{exactly} for the big effects produced by the radiative decay over long periods of time. Therefore we do not put $a(t) = 1$ in (252) although this would be correct to first order in $H_{I}$. 

If we put $a(t) = 1$ in (252) we should have just the solution of the radiation emission problem, neglecting all effects of radiation reaction on the atom, which we obtained before from Eg. (230). 

The value of $\dfrac{d}{dt} a(t) = \dfrac{d}{dt} \left(\Phi_o^{*}\Phi(t) \right)$ is obtained correct to second order in $H_{I}$, and including the radiative reaction effects, by substituting (252) into (250). \\
Hence we have 
\begin{equation}
\frac{1}{a(t)} \frac{d}{dt} a(t) = - \frac{1}{\hbar} \int _{-\infty}^{t} dt^{\prime} \, \left\{\Phi_o^{*}H_{I}(t) H_{I}(t^{\prime}) \Phi_o \right\}
\end{equation}
Using (247), (221) this gives\endnote{
Sum over $\mu$ inserted.}
\begin{align}
\frac{1}{a} & \frac{da}{dt} = - \frac{i}{\hbar c} \int _{-\infty}^{t} dt^{\prime} \, \iint d^{\,3}\boldsymbol{r} \,  d^{\,3}\boldsymbol{r}^{\prime} D^{+}(r -r^{\prime}, t - t^{\prime}) \sum_{\mu} \Braket{j_{\mu}(r, t) j_{\mu}(r^{\prime}, t^{\prime})}_{oo} \notag \\
&= -\frac{1}{(2 \pi)^3 \hbar c} \int \frac{d^{\,3} \boldsymbol{k}}{2 | \boldsymbol{k} |} \int _{-\infty}^{t} dt^{\prime} \, \iint d^{\,3}\boldsymbol{r} \,  d^{\,3}\boldsymbol{r}^{\prime}\, \exp\left\{i \boldsymbol{k} \cdot (\boldsymbol{r} - \boldsymbol{r}^{\prime}) - i c| \boldsymbol{k} | (t - t^{\prime}) \right\} \sum_{\mu} \Braket{j_{\mu}(r, t) j_{\mu}(r^{\prime}, t^{\prime})}_{oo}
\end{align}
Let the atomic states be labelled by $n$, the state $n$ having energy $E_{n}$. Let
\begin{equation}
j_{\mu}^{k}(n, m)
\end{equation}
be the matrix element of the operator 
\begin{equation}
\int j_{\mu}^{S} (r) e^{- i \boldsymbol{k} \cdot \boldsymbol{r}}\,d^{\, 3} \boldsymbol{r}
\end{equation}
in the transition $m \rightarrow n$. Then using a matrix product to evaluate $\Braket{j_{\mu}(r, t) j_{\mu}(r^{\prime}, t^{\prime})}_{oo}$,
\begin{equation}
\frac{1}{a} \frac{da}{dt} = -\frac{1}{16 \pi^3 \hbar c} \int \frac{d^{\,3} \boldsymbol{k}}{| \boldsymbol{k} |} \int _{-\infty}^{t} dt^{\prime} \sum_{n} \exp\left\{ \frac{i}{\hbar} (t - t^{\prime})\left( E_o - E_n - \hbar c | \boldsymbol{k} | \right) \right\} \sum_{\mu} | j_{\mu}^{k}(n,0) |^2
\end{equation}
where we made use of (248). \\
As before, the sum extends only over the two transverse polarizations $\mu$, the other two cancelling each other exactly. Now we have to calculate
\begin{equation}
\int_{-\infty}^{0} e^{iax} \, dx = \pi \delta(a) + \frac{1}{ia} = 2 \pi \delta_{+}(a)
\end{equation}
this being the definition of the $\delta_{+}$ function. Thus
\begin{equation}
\frac{1}{a} \frac{da}{dt} = -\frac{1}{8 \pi^2 c} \int \frac{d^{\,3} \boldsymbol{k}}{| \boldsymbol{k} |}  \sum_{n, \mu} | j_{\mu}^{k}(n,0) |^2 \, \delta_{+}\left(E_{n} - E_{0} + \hbar c | \boldsymbol{k} |\right)
\end{equation}
We write
\begin{equation}
\frac{1}{a} \frac{da}{dt} = -\frac{1}{2} \Gamma - \frac{i}{\hbar} \Delta E
\end{equation}
Then $\Delta E$ and $\Gamma$ are real constants given by 
\begin{equation}
\Delta E = - \frac{\hbar}{16 \pi^3 c} \int \frac{d^{\,3} \boldsymbol{k}}{| \boldsymbol{k} |} \sum_{n, \mu} \frac{ | j_{\mu}^{k}(n, 0) |^2 }{E_{n} - E_{0} + \hbar c | \boldsymbol{k} |}
\end{equation}
\begin{equation}
\Gamma = \frac{1}{8 \pi^2 c} \int \frac{d^{\,3} \boldsymbol{k}}{| \boldsymbol{k} |}\sum_{n, \mu}  | j_{\mu}^{k}(n, 0) |^2 \, \delta \left( E_{n} - E_{0} + \hbar c | \boldsymbol{k} | \right)
\end{equation}
These are independent of $t$. Therefore the amplitude of the state $\Phi_o$ in the wave-function $\Phi(t)$ is given for all $t \gg t_o$ by
\index{amplitude}%
\begin{equation}
a(t) = \left(\Phi_o^{*}\Phi(t) \right) = \exp \left\{ - \frac{i}{\hbar} \Delta E\, (t - t_o) - \frac{1}{2} \Gamma \, (t - t_o) \right\}
\end{equation}

The state $\Phi_o$, as a result of the perturbation by the radiation field, has its energy shifted by $\Delta E$, and decays exponentially like
\[
| a(t) |^2 = e^{-\Gamma \, (t - t_o)}																\tag{263/a}
\]
Comparing (232) and (262), we see that $\Gamma$ is exactly the total probability per unit time of radiation from the state $o$ to all other states $n$, calculated neglecting radiation reaction.
\index{probability!radiation!per~unit~time}%
This gives the physical interpretation for the decay law (263/a). When the denominators in (261) have zeroes, the integration over $| \boldsymbol{k} |$ is to be taken as a Cauchy principal value. 
\index{Cauchy~principal~value}%
The energy shift $\Delta E$ is exactly what would be obtained from elementary second-order perturbation theory, if the difficulties arising from vanishing denominators were just ignored.

We calculate now the spectrum of radiation emitted in the transition from the level $o$ to the level $n$, including effects of the level shifts $\Delta E_{o}$ and $\Delta E_{n}$ and the widths $\Gamma_o$ and $\Gamma_n$. Let $b_{nk}$ be the amplitude at time $t$, of the state in which the atom is in state $n$ and the photon is present with propagation vector $k$.
\index{amplitude}%
The equation of motion for $b_{nk}$, including effects of radiation out of the state $n$, is 
\begin{equation}
\frac{db_{nk}}{dt} = \left\{ - \frac{1}{2} \Gamma_{n} - \frac{i}{\hbar} \Delta E_{n} \right\} b_{nk} - Q \exp \left\{ \frac{i}{\hbar} \left( E_{n} - E_{o} + \hbar c | \boldsymbol{k} | \right) t \right\} a(t)
\end{equation}
with $a(t)$ given by (263).\endnote{
Following ``given by (263)'', the second edition has the phrase ``with $o$ suffices''. The first edition lacks this phrase. As the sentence makes more sense without it, it has been deleted.}
Here the last term represents the effects of transitions $o \rightarrow n$, and $Q$ is the space part of a matrix element of $H_{I}^{S}$ which is independent of $t$ and varies only slowly with $k$ so that we can regard $Q$ as a constant for all values of $k$ within the line-width. The exponential is the time part of the matrix element, the exponent being proportional to the energy difference between the atom in state $n$ plus the photon, and the atom in the zero state. The solution of (264) is, taking for convenience $t_o = 0$, 
\begin{equation}
b_{nk} = A \left\{ \exp ( - \beta t) - \exp( - \gamma t) \right\}
\end{equation}
using the initial condition $b_{nk} = 0$ at $t = 0$. Here
\begin{align}
\beta &= \frac{1}{2}\Gamma_o + \frac{i}{\hbar} \left(E_o + \Delta E_{o} - E_{n} - \hbar c | \boldsymbol{k} | \right) \notag \\
\gamma &=  \frac{1}{2}\Gamma_n + \frac{i}{\hbar} \Delta E_{n}
\end{align}
and $A = Q/(\beta - \gamma)$.

The probability that the atom leaves the state $n$ by  a second radiative transition at time $t$, so that a quantum $k$ remains, is 
\index{probability!transition!radiative}%
\[
\Gamma_{n} | b_{nk}(t) |^2
\]
The quantum $k$ remains from the first transition $o \rightarrow n$. After the atom makes the second transition to a continuum of possible states, the final states will no longer be coherent, and so the quanta left behind at different times $t$ will not interfere with each other. The total probability for the emission of a quantum of frequency $k$ in the first transition is thus 
\index{probability!emission}%
\begin{equation}
P(k) = \Gamma_{n} |Q|^2 \frac{1}{ |\beta - \gamma|^2} \int_{0}^{\infty} | e^{-\beta t } - e^{- \gamma t} |^2 \, dt
\end{equation}
Now
\begin{align}
\frac{1}{ |\beta - \gamma|^2} \int_{0}^{\infty} | e^{-\beta t } &- e^{- \gamma t} |^2 \, dt = \frac{1}{(\beta - \gamma)(\beta^{*} - \gamma^{*})} \left\{ \frac{1}{\beta + \beta^{*}} + \frac{1}{\gamma + \gamma^{*}} - \frac{1}{\beta + \gamma^{*}} - \frac{1}{\beta^{*} + \gamma} \right\} \notag \\
&= \frac{1}{\beta - \gamma} \left\{ \frac{1}{(\gamma + \gamma^{*})(\beta^{*} + \gamma)} -  \frac{1}{(\beta + \beta^{*})(\beta + \gamma^{*})} \right\} \notag \\
&= \frac{ \beta + \beta^{*} + \gamma + \gamma^{*} }{(\beta + \beta^{*})(\gamma + \gamma^{*})(\beta + \gamma^{*})(\gamma + \beta^{*})} = \frac{1}{2} \frac{\text{Re} \, (\beta + \gamma)}{\text{Re} \, (\beta) \, \text{Re} \, (\gamma) \, | \beta + \gamma^{*} |^2 }
\end{align}
Hence
\begin{equation}
P(k) = | Q |^2 \frac{\Gamma_{o} + \Gamma_{n}}{\Gamma_{o}} \frac{\hbar^2}{\left(E_{o} + \Delta E_{o} - E_{n} - \Delta E_{n} - \hbar c | \boldsymbol{k} | \right)^{2} + \frac{1}{4} \hbar^2 (\Gamma_{o} + \Gamma_{n})^2 }
\end{equation}
This formula for P(k) gives the natural shape of a spectral line. The maximum intensity occurs at 
\begin{equation}
\hbar c |\boldsymbol{k} | = \left( E_{o} + \Delta E_{o} \right) - \left( E_{n} + \Delta E_{n} \right)
\end{equation}
i.e. at the difference between the energies of the two levels including the radiative level shifts. The width at half-maximum is 
\begin{equation}
\hbar (\Gamma_{o} + \Gamma_{n})
\end{equation}
just the sum of the two widths of the levels given by their reciprocal life-times. 

These formulae (270) and (271) are important in interpreting the modern radio-frequency spectroscopic experiments with their very accurate measurements of line shapes and positions. 

\subsection*{Calculation of Line-Shift, Non-Relativistic Theory}
\addcontentsline{toc}{subsection}{Calculation of Line Shift -- Non-Relativistic Theory}

In all atomic systems, the line widths are finite and easily calculated from the known transition amplitudes. 
\index{amplitude}%
For this, non-relativistic theory is accurate enough for all purposes. The line-shift (261) is much more difficult, and non-relativistic theory is not accurate enough to handle it properly. Still we shall calculate (261) using the NR theory, to see what it gives. It turns out to give quite a lot that is interesting. 

First, in a NR calculation we use the dipole approximation which we also used to derive (234). Supposing a one-electron atom, the electron having mass $m$ and charge $-e$, we put
\begin{equation}
j_{1}^{k}(n 0) = - \frac{e}{m} (p_{1})_{n 0} = - \frac{e}{m} \left\{ \int \psi^{*}_{n} \left( - i \hbar \frac{\partial}{\partial x} \right)\psi_{0} \, d^{\, 3} \boldsymbol{r} \right\}
\end{equation}
The line shift (261) becomes
\[
\Delta E = - \frac{e^2 \hbar}{16 \pi^3 m^2 c} \int \frac{d^{\, 3} \boldsymbol{k}}{| \boldsymbol{k} |} \sum_{n} \frac{ \left| \left( p_{1} \right)_{n0} \right|^2 + \, \left| \left( p_{2} \right)_{n0} \right|^2 }{E_{n} - E_{o} + \hbar c | \boldsymbol{k} |}
\]
and integrating over the direction of $\boldsymbol{k}$, (compare with (238)), 
\begin{equation}
\Delta E = - \frac{e^2 \hbar}{6 \pi^2 m^2 c} \int_{0}^{\infty} | \boldsymbol{k}| \, d  |\boldsymbol{k} | \sum_{n} \frac{ \left| \boldsymbol{p}_{n0} \right|^2}{E_{n} - E_{o} + \hbar c | \boldsymbol{k} |}
\end{equation}
The integral over  $|\boldsymbol{k}|$ is now obviously divergent, even before summing over $n$. Therefore the line-shift is \emph{infinite}. When a complete relativistic theory with positrons is used,  the divergence becomes only logarithmic instead of linear, but it still definitely diverges. 
\index{positron!and~line~shift}%
This was for many years a disaster which destroyed all faith in the theory, and no way of avoiding the difficulty was found until 1947.

\subsection*{The Idea of Mass Renormalization}
\addcontentsline{toc}{subsection}{The Idea of Mass Renormalization}
\index{mass~renormalization|(}%

The line shift (273) is also infinite for a \emph{free} electron of momentum $\boldsymbol{p}$. In this case $\boldsymbol{p}$ is a diagonal operator and the sum over $n$ reduces to the term $n = 0$. Therefore 
\begin{equation}
\Delta E_{F} = - \frac{1}{6 \pi^2} \frac{e^2}{m^2 c^2} \left( \int_0^{\infty} d |\boldsymbol{k} | \right) \boldsymbol{p}^2
\end{equation}
The effect of the radiation interaction is just to give a free electron an additional energy proportional to its kinetic energy $(\boldsymbol{p}^2/2m)$. If the integral in (274) is cut off at an upper limit $K \sim (mc/\hbar)$ in order to allow for the fact that the theory anyway is wrong in the relativistic region; then 
\[
\Delta E_{F} \approx - \frac{1}{6 \pi^2} \frac{e^2}{\hbar c} \frac{\boldsymbol{p}^2}{m} 
\]
is a small correction to the kinetic energy, which would be produced by an increase in the rest-mass of the electron from $m$ to $(m + \delta m)$, 
\begin{equation}
\delta m = \frac{1}{3 \pi^2} \frac{e^2}{c^2} \int_{0}^{\infty} d | \boldsymbol{k} |
\end{equation}

We must now take into account that the observed rest-mass of \emph{any} electron, bound or free, is not $m$ but $m + \delta m$. Therefore in (273) a part
\begin{equation}
- \frac{1}{6 \pi^2} \frac{e^2}{c^2} \left( \int_{0}^{\infty} d | \boldsymbol{k} | \right) \Braket{\boldsymbol{p}^2}_{o o}
\end{equation}
expresses only the effect of the mass-change $\delta m$ on the kinetic energy of the bound electron; this part is already included in the kinetic energy of the electron, when the observed mass ($m + \delta m$) is taken for the mass in the formula $(\boldsymbol{p}^2/2m)$. Therefore the part (276) has to be subtracted from (273), to give the observable line-shift. The subtraction just cancels out the error that was made in identifying the mass $m$ of a ``bare'' electron without electromagnetic interaction with the observed electronic mass.
\index{electron!bare}%

The idea of this mass-renormalization is that, although the ``bare'' mass $m$  appears in the original description of the atom without radiation field, all the final results of the theory should depend only on the physically observable $m + \delta m$. 
\index{bare~mass}%
The idea is originally due to Kramers \cite{Kramers38}, developed by Bethe (\emph{Phys.\ Rev.} \textbf{72} (1947) 339.) 
\index{Kramers, Hendrik~A.}%
\index{mass~renormalization|)}%
\index{Bethe, Hans~A.}%

Subtracting (276) from (273) gives the physically observable line-shift
\begin{equation}
\Delta E = \frac{e^2}{6 \pi^2 m^2 c^2} \int_{0}^{\infty} d | \boldsymbol{k} \, | \sum_{n} \frac{ \left(E_{n} - E_{o} \right) \left| \boldsymbol{p} _{no} \right|^2}{E_{n} - E_{o} + \hbar c | \boldsymbol{k} | }
\end{equation}
The divergence at high $| \boldsymbol{k} |$ is now only logarithmic. Taking an upper limit cut-off for the integral at the point
\[
\hbar c |\boldsymbol{k} | = K
\]
where $K$ is an energy on the order of magnitude of $mc^2$, we have 
\begin{equation}
\Delta E =  \frac{e^2}{6 \pi^2 m^2 c^3 \hbar} \sum_{n}  \left(E_{n} - E_{o} \right) \left| \boldsymbol{p}_{no} \right|^2 \log \frac{K} {E_{n} - E_{o}}
\end{equation}
remembering that the integration over $| \boldsymbol{k} |$ in (277) is to be taken as a Cauchy principal value when $(E_{n} - E_{0})$ is negative.
\index{Cauchy~principal~value}%

From this formula (278) the line-shift for hydrogen  states can be calculated numerically, as was done by Bethe, Brown and Stehn (\emph{Phys. Rev.} \textbf{77} (1950) 370.) 
\index{hydrogen~atom!numerical~calculation~of~energies}%
\index{Bethe, Hans~A.}%
\index{Brown, L. M.}%
\index{Stehn, J. R.}%

Since the log in (278) will be quite large $( \sim 7)$ for states $n$ which are in the non-relativistic range, it is convenient to write
\begin{equation}
\sum_{n}  \left(E_{n} - E_{o} \right)  \left| \boldsymbol{p}_{no} \right|^2 \log  | E_{n} - E_{o} | = \left\{ \sum_{n}  \left(E_{n} - E_{o} \right)  \left| \boldsymbol{p}_{no} \right|^2 \right\} \log \left(E - E_{o}\right)_{\text{av}}
\end{equation}
this being the definition of $\left(E - E_{o}\right)_{\text{av}} $. Then $\left(E - E_{o}\right)_{\text{av}}$ is a non-relativistic energy. Exact calculation gives for the 2s state in hydrogen
\begin{equation}
\left(E - E_{o}\right)_{\text{av}} = 16.6 \,\text{Ry}
\end{equation}
Thus the important transitions are to states which although non-relativistic are continuum states with very high excitation. This is surprising.

Note that in (278) the terms are all positive if $E_{o}$ is the ground state. For higher states there will be both positive and negative contributions. In particular, we shall see that for a Coulomb potential the positive and negative terms cancel almost exactly, for all except $s$ states\endnote{
``$s$ states'' replaces ``$x$ states''.}. 
The cancellation is more or less accidental and seems to have no deeper meaning. 
\index{Coulomb~potential}%

\noindent Now using a sum rule
\begin{equation}
\sum_{n}  \left(E_{n} - E_{o} \right) \left| \boldsymbol{p}_{no} \right|^2 = \Braket{ \, \boldsymbol{p} \boldsymbol{\cdot} \boldsymbol{[} \, H, \, \boldsymbol{p} \, \boldsymbol{]}\, }_{oo}
\end{equation}
where $H$ is the Hamiltonian  for the atom\endnote{
The coefficient of the first integral in Eq.\ (283) was $\hbar$; it has been replaced by $\hbar^{2}$. Also, Eq.\ (283) lacked a label in the second edition.}
\index{Hamiltonian!non-relativistic~atom}%
\begin{equation}
H = \frac{1}{2 m} \boldsymbol{p}^2 + V \qquad \qquad V = - \frac{1}{4 \pi} \frac{e^2}{r}, \qquad \qquad  \boldsymbol{[} \, H, \, \boldsymbol{p} \, \boldsymbol{]} = i \hbar ( \nabla V )
\end{equation}  
\begin{align}
&\Braket{ \, \boldsymbol{p} \cdot \boldsymbol{[} \, H, \, \boldsymbol{p} \, \boldsymbol{]}\, }_{oo} = \hbar^2 \int \psi^{*}_{o} \nabla \cdot \left (\psi_{o} \nabla V \right) \, d\tau \\
&\hphantom{\Braket{ \, \boldsymbol{p} \cdot \boldsymbol{[} \, H, \, \boldsymbol{p} \, \boldsymbol{]}\, }_{oo}\,}= \frac{\hbar^2}{2} \left\{ \int \psi^{*}_{o} \nabla \cdot \left( \psi_{o} \nabla V \right) \, d\tau + \int \psi_{o} \nabla \cdot \left( \psi^{*}_{o} \nabla V \right) \, d\tau \right\} \notag \\
&\hphantom{\Braket{ \, \boldsymbol{p} \cdot \boldsymbol{[} \, H, \, \boldsymbol{p} \, \boldsymbol{]}\, }_{oo}\,}= \frac{\hbar^2}{2} \left\{2 \int \psi^{*}_{o} \psi_{o} \nabla^2 V\, d\tau + \int  \nabla \left( \psi^{*}_{o} \psi_{o} \right) \cdot \nabla V \, d\tau \right\} \notag \\
&\hphantom{\Braket{ \, \boldsymbol{p} \cdot \boldsymbol{[} \, H, \, \boldsymbol{p} \, \boldsymbol{]}\, }_{oo}\,}= \frac{\hbar^2}{2} \int \psi^{*}_{o} \psi_{o} \nabla^2 V\, d\tau  = \tfrac{1}{2} e^{2}\hbar^{2} | \psi_{o}(0) |^2 \notag
\end{align}
where we used Green's vector theorem, the fact that $\Braket{ \, \boldsymbol{p} \cdot \boldsymbol{[} \, H, \, \boldsymbol{p} \, \boldsymbol{]}\, }_{oo}$ is real, and the result that $\nabla^2 V = e^2 \delta^{3}(\boldsymbol{r})$ in Heaviside units. 
\index{Green's~theorem}%
\index{Heaviside~units}%

\noindent Hence\endnote{
Here, the Bohr radius was denoted $a$; it seemed reasonable to use $a_{o}$ instead.}
\begin{align}
\Delta E &= \frac{e^{4} \hbar}{12 \pi^2 m^2 c^3} | \psi_{o}(0) |^2 \log \frac{K}{\left(E - E_{o}\right)_{\text{av}}} \notag \\
&= \frac{e^{4} \hbar}{12 \pi^2 m^2 c^3} \log \frac{K}{\left(E - E_{o}\right)_{\text{av}}} \times \begin{cases}
1/(\pi n^{3} a_{o}^{3})&\text{for $s$ states}\\
\qquad 0 &\text{for others}
\end{cases}
\end{align}
\index{Bohr~radius}%
Compare this with (239). It differs only in having the $\log (K/(E - E_{o})_{\text{av}})$ replacing $\log (1/RK_{H})$. The low frequency photons have now been treated properly instead of being estimated. Only the high-frequency end is still inaccurate because of the vagueness of the cut-off $K$. Taking $K = mc^{2}$, (284) gives for the Lamb shift $2s - 2p$ the value \emph{1040 Megacycles}.  Remarkably close to the experimental value of 1062. 
\index{Lamb~shift}%
\index{Lamb~shift!experiment}%

The success of this calculation of the line shift shows that the correct treatment of the interaction between an electron and the Maxwell field, with the help of the idea of mass renormalization, will give sensible results in agreement with experiment.
\index{Maxwell!field}%
\index{renormalization!mass}%
This calculation could be done non-relativistically because the line shift is mainly a low-frequency and non-relativistic effect.

There are other effects of the radiation interaction, especially the anomalous increase in the observed magnetic moment  of the electron by a factor $\left( 1 + \dfrac{\alpha}{2 \pi} \right)$ over the value given by the Dirac theory,  which are essentially relativistic in character. 
\index{electron!anomalous~magnetic~moment}%
\index{Dirac!electron~magnetic~moment}%
\index{fine~structure~constant}%
For studying these effects, and for calculating the Lamb shift  accurately without an arbitrary cut-off, we need to use a complete relativistic quantum electrodynamics,  in which both electrons and the Maxwell field are handled relativistically.
\index{Lamb~shift}%
\index{quantum~electrodynamics}%
\index{Maxwell!field!relativistic~treatment}%

Therefore we must go back to where we left off the theory of the Dirac electron  on page 31, and start to construct a relativistic field theory of electrons and positrons, similar to the quantized Maxwell field theory. 
\index{Dirac!electron~theory}%
\index{electrons~and~positrons!relativistic~field~theory}%
\index{Maxwell!field!quantized}%

\section*{II. Field Theory of the Dirac Electron, Without Interaction}
\addcontentsline{toc}{section}{II. Field Theory of the Dirac Electron -- Without Interaction}

We apply to the Dirac equation the method of field quantization for anti-commuting fields. The reason why we must do this, and not use commuting fields, we will see later. Write
\index{Dirac!equation!field~quantization}%
\index{mu@$\mu$}
\[
\mu = (mc/\hbar), \qquad m = \,\text{electron mass}
\]
Lagrangian 
\index{Lagrangian~density!Dirac}%
\begin{equation}
\mathscr{L}_{o} = - \hbar c \, \overline{\psi} \left( \sum_{\lambda} \gamma_{\lambda} \frac{\partial}{\partial x_{\lambda}} + \mu \right) \psi
\end{equation}
Note the factor $\hbar$ here. This means that the theory has no classical limit in the sense of the Correspondence Principle. In the classical limit, only charges and currents composed of many particles have any meaning; the $\psi$ field disappears entirely from view. The $\hbar$ has to be put into (285) to make the dimensions right, since $\left( \overline{\psi} \psi \right)$ has dimensions (1/Volume) just as in the 1-particle Dirac theory of which this is an extension.
\index{Dirac!electron~theory}%
\noindent Field equations
\begin{align}
&\sum_{\lambda} \gamma_{\lambda}\, \frac{\partial \psi}{\partial x_{\lambda}} + \mu  \psi = 0 \notag \\
&\sum_{\lambda} \frac{\partial \overline{\psi}}{\partial x_{\lambda}}\, \gamma_{\lambda}  - \mu  \overline{\psi} = 0 
\end{align}
The charge-conjugate field $\phi$ can be defined by 
\[
\phi = C \psi^{+}
\]
according to (51), and it also satisfies
\begin{equation}
\left( \sum_{\lambda} \gamma_{\lambda}\, \frac{\partial}{\partial x_{\lambda}} + \mu  \right) \phi = 0
\end{equation}

\subsection*{Covariant Commutation Rules}
\addcontentsline{toc}{subsection}{Covariant Commutation Rules}

We proceed as for the Maxwell field. 
\index{Maxwell!field}%
Take two points $z$ and $y$ with $z_{0} > y_{0}$. Let
\begin{align}
\mathcal{Q}(y) &= \overline{\psi}(y)\, u \notag \\
\mathcal{R}(z) &= \overline{v}\, \psi(z)  \quad \text{or} \quad \overline{\psi}(z)\, v
\end{align}
Here $u$ and $v$ are spinor operators, not depending on $y$ or $z$ and anticommuting with all the $\psi$ and $\overline{\psi}$ operators in our equations, as we assumed at the beginning of this section. 
\index{anticommute}%
For example take $u = \psi(w)$ where $w$ is a point far away outside the light-cones of both $y$ and $z$. We make a change in the Lagrangian by 
\begin{equation}
\delta_{\mathcal{Q}} \mathscr{L} = \epsilon \delta^{4}(x - y) \, \overline{\psi}(y) \, u
\end{equation}
The factor $u$ must be put in to make $\delta_{\mathcal{Q}} \mathscr{L}$ a bilinear expression, which is necessary for the Peierls method to be applicable. 
\index{Peierls!method}%
In fact only bilinear expressions have a physically observable meaning, and it is never meaningful under any circumstances to add together a term linear and a term bilinear in the field operators.

The changed field equations for $\psi$ and $\overline{\psi}$ are 
\begin{align}
\text{For}\; \overline{\psi}&: \qquad \text{No change} \notag \\
\text{For}\; \psi&: \quad \left( \sum_{\lambda} \gamma_{\lambda}\, \frac{\partial}{\partial x_{\lambda}} + \mu  \right) \psi - \frac{\epsilon}{\hbar c}\, \delta^{4}(x - y)\, u = 0
\end{align}
Thus $\delta_{\mathcal{Q}} \overline{\psi}(z) = 0 $ and $\epsilon  \delta_{\mathcal{Q}} \psi(z)$ satisfies (290). (Compare with (198).)

\noindent Hence $\delta_{\mathcal{Q}} \psi(z)$ is defined by the conditions
\begin{equation}
\begin{split}
\left( \sum_{\lambda} \gamma_{\lambda} \, \frac{\partial}{\partial x_{\lambda}} + \mu  \right)
\delta_{\mathcal{Q}} \psi(z) &= \frac{1}{\hbar c}\, \delta^{4}(z - y) \, u  \\
\delta_{\mathcal{Q}} \psi(z) &= 0 \quad \text{for $z_{0} < y_{0}$}
\end{split}
\end{equation}
From (291), $\delta_{\mathcal{Q}}\psi(z)$ is a c-number spinor. We write
\begin{equation}
\delta_{\mathcal{Q}}\psi(z) = - \frac{1}{\hbar c} \,S_{R}(z - y)\, u
\end{equation}
Then $S_{R}(x)$ is a c-number Dirac matrix function of x, satisfying 
\index{Dirac!matrices}%
\begin{equation}
\begin{split}
\left( \sum_{\lambda} \gamma_{\lambda}\, \frac{\partial}{\partial x_{\lambda}} + \mu  \right) S_{R}(x) &= - \delta^4(x)  \\
S_{R}(x) &= 0 \quad \text{for $x_{0} < 0$}
\end{split}
\end{equation}
and the $4 \times 4$ unit matrix $\mathbb{I}$ is understood on the right of (293).

If we have $\mathcal{R} = \overline{\psi}(z) \, v$ then $\delta \overline{\psi} = 0$, as before. If $\mathcal{R} = \overline{v} \, \psi(z)$ then
\begin{align}
\delta_{\mathcal{R}} \overline{\psi}(y) &= - \frac{1}{\hbar c} \, \overline{v} \, S_{A}(z - y) \notag \\
\delta_{\mathcal{R}} \psi(y) &= 0
\end{align}
where $S_{A}(x)$ is the Dirac matrix satisfying 
\index{Dirac!matrices}%
\begin{align}
\left( \sum_{\lambda} \gamma_{\lambda}\, \frac{\partial}{\partial x_{\lambda}} + \mu  \right) S_{A}(x) &= - \delta^4(x) \notag \\
S_{A}(x) &= 0 \quad \text{for $x_{0} > 0$}
\end{align}

Finally, if we had chosen $\mathcal{Q} = \overline{u} \, \psi(y)$ we should have had $\delta_{\mathcal{Q}} \psi(z) = 0$ in the same way.

Hence using the Peierls commutation law (194) with (292) and (294) we find 
\index{Peierls!formula}%
\begin{align}
\boldsymbol{[} \, \overline{v}\, \psi(z)\, , \, \overline{u} \, \psi(y) \, \boldsymbol{]} &= \boldsymbol{[} \, \overline{\psi}(z)\,v\, , \, \overline{\psi}(y)\,u \, \boldsymbol{]} = 0 \notag \\
\boldsymbol{[} \, \overline{v}\, \psi(z)\, , \, \overline{\psi}(y)\,u \, \boldsymbol{]} &= -i\overline{v} \left[ S_{A}(z - y) - S_{R}(z - y)  \right] u
\end{align}
These hold for every choice of $u$ and $v$ if we now choose $u$ and $v$ anti-commuting with all the $\psi$ and $\overline{\psi}$ operators. Therefore writing
\begin{equation}
S(x) = S_{A}(x) - S_{R}(x)
\end{equation}
we can write down the commutation rules for the operator components:
\begin{align}
\left\{ \psi_{\alpha}(z), \,\psi_{\beta}(y) \right\} &= \left\{ \overline{\psi}_{\alpha}(z), \,\overline{\psi}_{\beta}(y) \right\} = 0  \\
\left\{ \psi_{\alpha}(z), \,\overline{\psi}_{\beta}(y) \right\} &= - i S_{\alpha \beta} (z - y)
\end{align}
The invariant $S-$function satisfies by (293) and (295)
\begin{equation}
\left( \sum_{\lambda} \gamma_{\lambda}\, \frac{\partial}{\partial x_{\lambda}} + \mu  \right) S(x) = 0
\end{equation}
There are no simple formulae for the $S$-functions like (261) and (265) for the $D$-functions, in coordinate space. However in momentum representations the $S$-functions are equally simple. 
\index{momentum!representation}%

\subsection*{Momentum Representations}
\addcontentsline{toc}{subsection}{Momentum Representations}

Write 
\index{momentum!representation}%
\[
S_{R}(x) = \left( \sum_{\lambda} \gamma_{\lambda}\, \frac{\partial}{\partial x_{\lambda}} - \mu  \right) \Delta_{R}(x), \quad S_{A}(x) = \left( \sum_{\lambda} \gamma_{\lambda}\, \frac{\partial}{\partial x_{\lambda}} - \mu  \right) \Delta_{A}(x)
\]
\begin{equation}
S(x) = \left( \sum_{\lambda} \gamma_{\lambda}\, \frac{\partial}{\partial x_{\lambda}} - \mu  \right) \Delta(x)
\end{equation}
Then we have
\begin{align}
&\left( \Box^2 - \mu^2 \right) \Delta_{R}(x) = \left( \Box^2 - \mu^2 \right) \Delta_{A}(x) = - \delta^{4}(x) \notag \\
&\left( \Box^2 - \mu^2 \right) \Delta(x) = 0 	\qquad  \Delta(x) = \left( \Delta_{A} - \Delta_{R} \right)(x)
\end{align}
with the boundary conditions as before. And the $\Delta$-functions are exact analogues of the $D$-functions, the $D$-functions being the special case $\mu = 0$. Instead of (207) we have by making the formal substitution $k^2 \rightarrow k^2 + \mu^2$
\begin{equation}
\Delta_{R}(x) = \frac{1}{(2\pi)^4} \int_{+} e^{i k \cdot x}\, \frac{1}{k^2 + \mu^2} \, d^{\,4} k
\end{equation}
where the contour in the $k_{0}$-plane goes above the two poles at $k_{0} = \pm \sqrt{| \boldsymbol{k} |^2 + \mu^2}$. Similarly for (208). And instead of (210)
\begin{equation}
\Delta (x) = - \frac{i}{(2 \pi)^3} \int e^{i k \cdot x} \, \delta(k^2 + \mu^2) \, \epsilon(k) \, d^{\,4}k
\end{equation}
Hence using (301) and the notation (110) 
\begin{equation}
S(x) = \frac{1}{(2\pi)^3} \int e^{i k \cdot x} \, (\slashed{k} + i\mu) \delta(k^2 + \mu^2) \, \epsilon(k) \, d^{\,4}k
\end{equation}
Note the projection operator $\Lambda_{+}$ appearing here, as defined by (115) with the momentum $p = \hbar k$. 
\index{projection~operator}%
Thus the  $S$-function distinguishes automatically between the electron states $k_{0} = +\sqrt{| \boldsymbol{k} |^2 + \mu^2}$ and the positron states $k_{0} = -\sqrt{| \boldsymbol{k} |^2 + \mu^2}$ 
\index{positron!states}%

\subsection*{Fourier Analysis of Operators}
\addcontentsline{toc}{subsection}{Fourier Analysis of Operators}

We analyze $\psi_{\alpha}$ into Fourier components, written in a quite general form:
\index{Fourier!components}%
\begin{equation}
\psi_{\alpha}(x) = Q \int d^{\, 3} \boldsymbol{k} \left( \frac{\mu^2}{| \boldsymbol{k} |^2 + \mu^2}\right)^{1/4} \left\{ \sum_{u^{+}} u_{\alpha} e^{i \boldsymbol{k} \cdot \boldsymbol{x}} \, b_{ku} + \sum_{u^{-}} u_{\alpha} e^{-i \boldsymbol{k} \cdot \boldsymbol{x}} \, b_{ku} \right\}
\end{equation}
where, as in (211), the factor $\left(\mu^2 /(| \boldsymbol{k} |^2 + \mu^2)\right)^{1/4}$ only makes the notation simpler. The integration is over all 4-vectors $k$ with $k_{0} = +\sqrt{| \boldsymbol{k} |^2 + \mu^2}$. For each $k$, the sum $\sum_{u^{+}}$ is over the two spin-states $u$ satisfying, by (111), the equation
\begin{equation}
(\slashed{k} - i\mu) \, u = 0
\end{equation}
and the sum $\sum_{u^{-}}$ is over the two spin-states $u$ satisfying by (112)
\begin{equation}
(\slashed{k} + i\mu) \, u = 0
\end{equation}
the normalization being given by (106) and (113). The $b_{ku}$ are operators independent of $x$ and $\alpha$, whose properties are to be determined. 

Taking the adjoint to (306) we have 
\begin{equation}
\overline{\psi}_{\alpha}(x) = Q \int d^{\, 3} \boldsymbol{k} \left( \frac{\mu^2}{| \boldsymbol{k} |^2 + \mu^2}\right)^{1/4} \left\{ \sum_{u^{+}} b^{*}_{ku}\overline{u}_{\alpha} e^{-i \boldsymbol{k} \cdot \boldsymbol{x}} \,  + \sum_{u^{-}} b^{*}_{ku}\overline{u}_{\alpha} e^{i \boldsymbol{k} \cdot \boldsymbol{x}} \right\}
\end{equation}
Here the $b^{*}_{ku}$ are ordinary Hermitian conjugates of $b_{ku}$. 
\index{Hermitian!conjugate}%

Computing the anti-commutators (298), (299) from (306) and (309), and comparing the results with the momentum integral (305), using (115) and the properties of $\Lambda_{+}$ we have 
\index{momentum!integral}%
\begin{align}
\left\{ \, b_{ku}, \, b_{k^{\prime}v} \right\} &= \left\{ \, b^{*}_{ku}, \, b^{*}_{k^{\prime}v}\right\} = 0 \\
\left\{\,b_{ku}, \,b^{*}_{k^{\prime}v}\right\} &= \delta^{3}(\boldsymbol{k} - \boldsymbol{k}^{\prime})\, \delta_{uv}
\end{align}
and we find that the constant $Q$ in (306) and (309) is given by\endnote{
``is'' replaces ``being''}
\begin{equation}
Q = (2\pi)^{-3/2}
\end{equation}

\subsection*{Emission and Absorption Operators}
\addcontentsline{toc}{subsection}{Emission and Absorption Operators}

Let
\begin{equation}
E_{k} = \hbar c \sqrt{|\boldsymbol{k}|^2 + \mu^2}
\end{equation}
be the energy of an electron or a positron with momentum $\hbar k$. We apply the same argument which led to (215) for the Maxwell field. 
\index{Maxwell!field}%
From this it follows that 
\index{positron!states}%
\begin{align*}
b_{ku} \; &\text{for electron states $u$} \\
b^{*}_{ku} \; &\text{for positron states $u$}
\end{align*}
have matrix elements only for transitions from an initial state of energy $E_{1}$ to a final state of energy $E_{2}$, where 
\begin{equation}
E_{1} -  E_{2} = E_{k}
\end{equation}
And
\begin{align*}
b_{ku} \; &\text{for positron states $u$} \\
b^{*}_{ku} \; &\text{for electron states $u$}
\end{align*}
have matrix elements which are non-zero only when 
\begin{equation}
E_{2} -  E_{1} = E_{k}
\end{equation}
Thus we see as before that the field has the properties we require of a quantized field. It can carry energy only in discrete lumps of magnitude $E_{k}$ for each frequency $k$. And the energy can be carried by two kinds of excitation, which we have called electrons and positrons anticipating the later results of the theory. We see already that these two excitation states have particle properties, and that there are two kinds of particles. 

The absorption operators are 
\index{positron!absorption~and~emission~operators}%
\index{electron!absorption~and~emission~operators}%
\begin{align*}
b_{ku} \; &\text{for electrons} \\
b^{*}_{ku} \; &\text{for positrons}
\end{align*}
the emission operators are 
\begin{align*}
b_{ku} \; &\text{for positrons} \\
b^{*}_{ku} \; &\text{for electrons}
\end{align*}

\subsection*{Charge-Symmetrical Representation}
\addcontentsline{toc}{subsection}{Charge Symmetrical Representation}

We use the charge-conjugate field $\phi$ defined by (51), in order to put the whole theory into a form where there is complete symmetry between electrons and positrons. 
\index{charge~symmetry}%
This symmetry is known as the charge-symmetry of the theory. 

Let $k$ be given, and a spinor $u$ satisfying (308), representing a positron state. 
\index{positron!states}%
We represent the positron state alternatively by the charge conjugate spinor 
\begin{equation}
v = Cu^{+}
\end{equation}
which will satisfy (307) as $u$ does for electron states. We denote by
\[
b^{C}_{kv} = b^{*}_{ku}
\]
the absorption operator for the positron state $v$. Then instead of (306), (309) we may write the pair of equations
\begin{equation}
\psi_{\alpha}(x) = Q \int d^{\, 3} \boldsymbol{k} \left( \frac{\mu^2}{| \boldsymbol{k} |^2 + \mu^2}\right)^{1/4} \left\{ \sum_{u^{+}} u_{\alpha} e^{i \boldsymbol{k} \cdot \boldsymbol{x}} \, b_{ku} + \sum_{v^{+}} \left\{Cv^{+}\right\}_{\alpha} e^{-i \boldsymbol{k} \cdot \boldsymbol{x}} \, b^{*C}_{kv} \right\}
\end{equation}
\begin{equation}
\phi_{\alpha}(x) = Q \int d^{\, 3} \boldsymbol{k} \left( \frac{\mu^2}{| \boldsymbol{k} |^2 + \mu^2}\right)^{1/4} \left\{ \sum_{v^{+}} v_{\alpha} e^{i \boldsymbol{k} \cdot \boldsymbol{x}} \, b^{C}_{kv} + \sum_{u^{+}} \left\{Cu^{+}\right\}_{\alpha} e^{-i \boldsymbol{k} \cdot \boldsymbol{x}} \, b^{*}_{ku} \right\}
\end{equation}
The $\psi$ and $\phi$ fields are thus entirely symmetrical between positrons and electrons; $\phi$ could be taken as the starting-point and $\psi$ derived from it, just as easily as vice versa.

The commutation rules (311) become
\begin{align*}
\left\{ \, b_{ku}, \, b^{*}_{k^{\prime}u^{\prime}} \right\} &= \delta^{3}(\boldsymbol{k} - \boldsymbol{k}^{\prime})\, \delta_{uu^{\prime}}  \\
\left\{ \, b^{C}_{kv}, \,b^{*C}_{k^{\prime}v^{\prime}}\right\} &= \delta^{3}(\boldsymbol{k} - \boldsymbol{k}^{\prime})\, \delta_{vv^{\prime}} \tag{318a} \\
\left\{ \, b^{C}_{kv^{\prime}}, \,b^{*}_{ku} \right\} &= 0, \qquad \text{etc.} \notag
\end{align*}
These are also symmetrical between electron and positron. 

\subsection*{The Hamiltonian}
\addcontentsline{toc}{subsection}{The Hamiltonian}

The Hamiltonian $H$ has commutation rules with the emission and absorption operators as for the Maxwell field. 
\index{Maxwell!field}%
These rules come straight from the Heisenberg equations of motion for $\psi$ and $\phi$. For any electron state $u$ or positron $v$
\begin{equation}
\begin{split}
\boldsymbol{[} \, b_{ku}, \, H \, \boldsymbol{]} &= E_{k} b_{ku}	\qquad \qquad	\boldsymbol{[} \, b^{C}_{kv}, \, H \, \boldsymbol{]} = E_{k} b^{C}_{kv} \\
\boldsymbol{[} \, b^{*}_{ku}, \, H \, \boldsymbol{]} &= -E_{k} b^{*}_{ku}	\qquad \quad\,	\boldsymbol{[} \, b^{*C}_{kv}, \, H \, \boldsymbol{]} = -E_{k} b^{*C}_{kv} 
\end{split}
\end{equation}
Hence the Hamiltonian of the theory is
\index{Hamiltonian!Dirac~field}%
\begin{equation}
H = \int d^{\, 3} \boldsymbol{k} \, E_{k} \left\{ \sum_{u^{+}}b^{*}_{ku} b_{ku} + \sum_{v^{+}} b^{*C}_{kv} b^{C}_{kv} \right\}
\end{equation}
as one can verify at once by substituting into (319).  \\
The additive constant is again chosen so that $\Braket{H}_{o}$, the expectation value of $H$ in the vacuum state, is zero. This eliminates a possible arbitrary additive constant from $H$.

In (317), (318) and (320) there is complete symmetry between electrons and positrons.
\index{electron-positron!symmetry}%
The theory could just have well been built up from the positron as the fundamental particle, instead of from the electron.

But for practical calculations we shall generally not use (317), (318), (320). It is generally easier to work with the unsymmetrical form of the theory, with the fields $\psi$ and $\overline{\psi}$.

\subsection*{Failure of Theory with Commuting Fields}
\addcontentsline{toc}{subsection}{Failure of Theory with Commuting Fields}

Suppose we had taken the theory up to this point, only assuming $\psi$ and $\overline{\psi}$ to be ordinary commuting fields. Then $u$ and $v$ would be taken to be quantities commuting with all $\psi$ and $\overline{\psi}$, in the relations (296). Thus (298) and (299) would still be true only with commutators everywhere instead of anticommutators. 
\index{anticommute}%
Likewise (310) and (311). However in this symmetric representation, instead of (318a) we should have 
\begin{equation}
\begin{split}
\boldsymbol{[} \, b_{ku}, \,  b^{*}_{k^{\prime}u^{\prime}} \, \boldsymbol{]} &= \delta^{3}(\boldsymbol{k} - \boldsymbol{k}^{\prime})\, \delta_{uu^{\prime}} \\
\boldsymbol{[} \, b^{C}_{kv}, \,  b^{*C}_{k^{\prime}v^{\prime}} \, \boldsymbol{]} &= -\delta^{3}(\boldsymbol{k} - \boldsymbol{k}^{\prime})\, \delta_{vv^{\prime}} 
\end{split}
\end{equation}
And then the Hamiltonian instead of being given by (320) would have to be
\index{Hamiltonian!Dirac~field}%
\begin{equation}
H = \int d^{\, 3} \boldsymbol{k} \, E_{k} \left\{ \sum_{u^{+}} b^{*}_{ku} b_{ku} - \sum_{v^{+}} b^{*C}_{kv} b^{C}_{kv} \right\}
\end{equation}
So positrons  would actually be particles of negative energy, like the negative energy electrons of the one-electron theory. 
\index{positron!failure~of~commuting~fields}%
This is physically inadmissible.

Thus the use of anticommuting fields is the only thing which gives us a proper positive energy for the positrons. This is reasonable, because the intuitive Dirac hole theory  can only work by virtue of the Pauli exclusion principle, and the exclusion principle is a feature of anticommuting fields. 
\index{Dirac!hole~theory}%
\index{anticommute}%
\index{Pauli!exclusion~principle}%

\subsection*{The Exclusion Principle}
\addcontentsline{toc}{subsection}{The Exclusion Principle}
\index{Pauli!exclusion~principle}%

Take any creation operator $b^{*}_{ku}$. 
\index{creation!operator}%
As a special case of (310) we have identically
\begin{equation}
b^{*}_{ku} b^{*}_{ku} = 0
\end{equation}
Given any state $\Psi$, the result of creating \emph{two} electrons with frequency $k$ and spin $u$ in that state is $b^{*}_{ku} b^{*}_{ku} \Psi = 0$. Thus there are no states in which two electrons have the same momentum and spin. So we have the Pauli exclusion principle valid for both electrons and positrons. Also, an electron and a positron do not exclude each other. 

It is a very great success of the general field theory, that it has given us the Pauli principle automatically, not by special hypothesis as in the old particle theory of electrons.\endnote{
``by'' replaces ``be''} 
\index{Pauli!exclusion~principle}%

The most general state of the fields is described by specifying for each electron and positron state the number of particles occupying it. This number in each case can take just the two values 0 and 1. 

\subsection*{The Vacuum State}
\addcontentsline{toc}{subsection}{The Vacuum State}

The vacuum state $\Psi_{o}$ is defined by 
\begin{align*}
b_{ku} \Psi_{o} &= 0 \; \text{and thus} \qquad \Psi^{*}_{o}\, b^{*}_{ku} = 0 \; \text{for electron states $u$} \\
b^{*}_{ku} \Psi_{o} &= 0 \; \text{and thus} \qquad \Psi^{*}_{o}\, b_{ku} = 0 \; \text{for positron states $u$} \tag{323a}
\end{align*}
So the vacuum expectation value of products of emission and absorption operators are given by (311). We find using (323a)
\begin{align}
\Braket{b_{ku}b_{k^{\prime}v}}_{o} &= \Braket{b^{*}_{ku}b^{*}_{k^{\prime}v}}_{o} = 0 \notag \\
\Braket{b_{ku}b^{*}_{k^{\prime}v}}_{o} &= \Theta_{u} \, \delta^{3}(\boldsymbol{k} - \boldsymbol{k}^{\prime})\, \delta_{uv}  \\
\Braket{b^{*}_{ku}b_{k^{\prime}v}}_{o} &= (1 - \Theta_{u})\, \delta^{3}(\boldsymbol{k} - \boldsymbol{k}^{\prime})\, \delta_{uv} \notag
\end{align}
where
\[
\Theta_{u} = \begin{cases}
 = 1 &\text{for electron states $u$,} \\
 = 0 &\text{for positron states.}
 \end{cases}
\]
Hence by (306) and (309) the expectation value $\Braket{\psi_{\alpha}(z) \overline{\psi}_{\beta}(y)}_{o}$ is just the part of the anti-commutator $\{\psi_{\alpha}(z), \, \overline{\psi}_{\beta}(y)\}$ which contains positive frequencies $\exp [ ik \cdot (z - y) ]$, with $k_{o} > 0$. Thus, similarly to (221)
\begin{equation}
\Braket{\psi_{\alpha}(z) \overline{\psi}_{\beta}(y)}_{o} = -i S^{+}_{\alpha \beta} (z - y)
\end{equation}
\begin{equation}
S^{+}(x) = \frac{1}{(2\pi)^3} \int e^{i k \cdot x} \left( \slashed{k} + i \mu \right) \delta(k^2 + \mu^2)\, \Theta(k) \,d^{\,4} k
\end{equation}
where
\[
\Theta(x) = \begin{cases}
 = +1 &\text{for $x_{0} > 0$} \\
 = 0 &\text{for $x_{0} < 0$}
 \end{cases}
\]
writing as before
\begin{equation}
S^{+} = \tfrac{1}{2} \left(S - iS^{(1)} \right) \qquad \qquad S^{-} = \tfrac{1}{2} \left(S + iS^{(1)} \right) 
\end{equation}
\begin{equation}
\Braket{\overline{\psi}_{\beta}(y) \psi_{\alpha}(z)}_{o} = -i S^{-}_{\alpha \beta}(z -y)
\end{equation}
\begin{equation}
\Braket{ \boldsymbol{[}\, \psi_{\alpha}(z), \, \overline{\psi}_{\beta}(y) \, \boldsymbol{]}}_{o} = - S^{(1)}_{\alpha \beta}(z -y)
\end{equation}
\begin{equation}
S^{(1)}(x) = \frac{i}{(2\pi)^3} \int e^{i k \cdot x} \left( \slashed{k} + i \mu \right) \delta(k^2 + \mu^2) \,d^{\,4} k
\end{equation}
\[
S^{-}(x) = -\frac{1}{(2\pi)^3} \int e^{i k \cdot x} \left( \slashed{k} + i \mu \right) \delta(k^2 + \mu^2)\, \Theta(-k) \,d^{\,4} k \tag{330a}
\]

These results for the Dirac theory without electromagnetic interaction will be used a lot when we come to the complete quantum electrodynamics with both Dirac and Maxwell fields quantized.
\index{Dirac!electron~theory}%
Meanwhile, we should say a little about the theory of quantized Dirac particles in a given c-number Maxwell field. 
\index{Maxwell!field}%

\section*{III. Field Theory of Dirac Electron in External Field}
\addcontentsline{toc}{section}{III. Field Theory of Dirac Electron in External Field}

\hspace*{3ex}Lagrangian
\begin{equation}
\mathscr{L} = \mathscr{L}_{D} - ie \overline{\psi} \slashed{A}^{e} \psi
\end{equation}
\hspace*{3ex}Field equations
\begin{equation}
\left\{ \sum_{\lambda} \gamma_{\lambda} \left( \frac{\partial}{\partial x_{\lambda}} + \frac{ie}{\hbar c} A_{\lambda}^{e}\right) + \mu \right\} \psi = 0
\end{equation}
\begin{equation}
\sum_{\lambda} \left( \frac{\partial \overline{\psi} }{ \partial x_{\lambda}} - \frac{ie}{\hbar c} A_{\lambda}^{e} \overline{\psi} \right) \gamma_{\lambda} - \mu \overline{\psi} = 0
\end{equation}
These equations are still \emph{linear}, the $A_{\mu}^{e}$ being given functions of position. This makes the theory still simple. 

\subsection*{Covariant Commutation Rules}
\addcontentsline{toc}{subsection}{Covariant Commutation Rules}

Because of the linearity, a change $\overline{v} \,\psi(z)$ in $\mathscr{L}$ makes no difference to the field equation for $\psi(y)$. Therefore for every two space-time points $y$ and $z$ we still have, as in (298), 
\begin{equation}
\{ \psi_{\alpha}(z), \, \psi_{\beta}(y) \} = \{ \overline{\psi}_{\alpha}(z), \, \overline{\psi}_{\beta}(y) \} = 0
\end{equation}

Beyond this point not much can be done with the theory for general time-dependent potentials $A_{\mu}^{e}$. In practice when we have time-dependent  $A_{\mu}^{e}$ we always use perturbation theory starting from free field formalism, assuming the  $A_{\mu}^{e}$ to be small, or else use special tricks for particular problems. 

The important practical cases in which the  $A_{\mu}^{e}$ are not small are always those in which the  $A_{\mu}^{e}$ are \emph{time-independent} in a particular Lorentz system. 
\index{Lorentz!system}%
Examples: electrons bound in atoms by static Coulomb forces, electrons moving in constant macroscopic electric and magnetic fields. 
\index{Coulomb~potential}%

So we assume  $A_{\mu}^{e}$ =  $A_{\mu}^{e}(r)$ time-independent. We also assume the  $A_{\mu}^{e}$ to be physically well-behaved so that the stationary eigenvalue equation
\begin{equation}
E_{n} \psi_{n} = \left\{ -e \Phi + \sum_{j=1}^{3} \left(-i \hbar c \frac{\partial}{\partial x_{j}} + e A_{j}^{e} \right)\alpha^{j}  + mc^2 \beta \right\} \psi_{n}
\end{equation}
regarded as an equation for the c-number Dirac wave-function $\psi_{n}(r)$ has a complete set of eigenfunctions $\psi_{n}$ with eigenvalues $E_{n}$; the spectrum may be either discrete or continuous or mixed. 
\index{Dirac!wave~function}%
Equation (335) is derived from (332) by substituting into it the particular function
\begin{equation}
\psi =  \psi_{n}(r) \exp \left\{-i\frac{E_{n}}{\hbar} t \right\}
\end{equation}

We assume further the potentials to be such that the eigenfunctions $\psi_{n}$ separate clearly into two classes, the $\psi_{n+}$ with positive $E_{n}$ and the $\psi_{n-}$ with negative $E_{n}$. This is again true for all physically occurring potentials, though it would fail for the Coulomb field of a point nucleus with charge $Z > 137$.
\index{Coulomb~potential}%

The time-independent potentials make the problem essentially non-covariant, and so we shall use the non-covariant notations in developing the theory. We write the equations as if all the levels $n$ are \emph{discrete}, thus $\sum_{n}$ means a sum over discrete levels plus an integral over continuous levels suitably normalized. We are now chiefly interested in discrete levels, and so we need not bother to write the formulae for normalizing the continuous levels explicitly. This makes the picture apparently simpler than the free-particle theory, where the normalization of continuous levels was done at every stage accurately; the simplicity is only apparent, because we shall just be ignoring the complications arising from continuous levels.

The general solution of the field equations (332) is
\begin{equation}
\psi(r, t) =  \sum_{n} b_{n}\psi_{n}(r) \exp \left\{-i\frac{E_{n}}{\hbar} t \right\}
\end{equation}
where the $b_{n}$ are operators independent of $r$ and $t$, and $E_{n}$ can be both positive and negative.

\noindent By (334) we have
\[
\{ b_{m}, \, b_{n} \} = 0 \qquad \qquad \{ b^{*}_{m}, \, b^{*}_{n} \} = 0
\]
where $b^{*}_{m}$ is the Hermitian conjugate of $b_{m}$. 
\index{Hermitian!conjugate}%
We take the $\psi_{n}$ normalized so that
\begin{equation}
\int  \psi^{*}_{m}(r) \, \psi_{n}(r) \, d^{\,3}\boldsymbol{r} = \delta_{n m}
\end{equation}
i.e.in the usual non-covariant way. Note especially in (339) that we are treating all levels as if they were discrete; this can be done for example by enclosing our whole system in a finite box.

We want still to find the continuous rules between $\psi$ and $\psi^{*}$, or between $b_{n}$ and $b^{*}_{m}$.\endnote{
In v.2, the Hermitian conjugate $b^{*}_{m}$ lacked the asterisk.}
Suppose we add to the Lagrangian (331)
\begin{equation}
\delta \mathscr{L}(r, t) = \epsilon  \delta(t - t_{o}) \,\psi^{*}(r, t_{o})\, \psi_{n}(r)\, u
\end{equation}
where $u$ is an operator anticommuting with $\psi$ and $\psi^{*}$ as in (288). 
\index{anticommute}%
This produces a change in the field equation for  $\psi$ which now becomes 
\begin{equation}
\left\{ \sum_{\lambda} \gamma_{\lambda} \left( \frac{\partial}{\partial x_{\lambda}} + \frac{ie}{\hbar c} A_{\lambda}^{e}\right) + \mu \right\} \psi = \frac{\epsilon}{\hbar c} \, \delta(t - t_{o}) \beta \psi_{n}(r) u
\end{equation}
So the change $\delta \psi$ produced in $\psi$ by the addition of $\delta \mathscr{L}$ satisfies (341) with the initial condition  $\delta \psi(r, t) = 0$ for $t < t_{o}$. (Compare with (198) and (290).)  Now the solution of (341) will obviously be of the form
\begin{equation}
\delta \psi = a(t) \psi_{n}(r)
\end{equation}
where $a(t)$ is a function of $t$ only, because the right side of the linear equation also has this form. Substituting (342) into (341) and using (335) we have
\begin{equation}
\left( i \hbar \frac{\partial}{\partial t} - E_{n} \right) a(t) = - \epsilon \delta(t - t_{o}) u
\end{equation}
and hence
\begin{equation}
\delta \psi = \frac{i \epsilon}{\hbar} \Theta(t - t_{o}) \psi_{n}(r) \exp \left\{-i \frac{E_{n}}{\hbar} (t - t_{o}) \right\} u
\end{equation}
as we can verify using
\[
\frac{d}{dt} \Theta(t - t_{o}) = \delta(t - t_{o})
\]
Integrating (340) over space-time gives by (339)
\begin{equation}
c \iint \delta \mathscr{L}(r, t) \,d^{\,3}\boldsymbol{r} \, dt = \epsilon \,c \, b^{*}_{n} \exp \left\{i \frac{E_{n}}{\hbar} t_{o} \right\} u
\end{equation}

For $t > t_{o}$, (343) is the change produced in $\psi(r,t)$ by adding (344) to the action integral. Hence by the Peierls commutation rule, using (193), (343) and (344), 
\index{Peierls!formula}%
\[
\boldsymbol{[} \, b^{*}_{n}u, \psi(r,t) \, \boldsymbol{]} = - \psi_{n}(r) \exp \left\{-i \frac{E_{n}}{\hbar} t  \right\}
\]
and therefore
\begin{equation}
\{b^{*}_{n}, \psi(r,t) \} = \psi_{n}(r) \exp \left\{-i \frac{E_{n}}{\hbar} t \right\}
\end{equation}
because by assumption the $u$'s anticommute with the $\psi$'s. The time $t_{o}$ no longer appears in (345), checking the consistency of the method. 
\index{anticommute}%

Multiplying (345) by $\psi^{*}_{n}(r^{\prime}) \exp \{ -i E_{n}t^{\prime}/\hbar \}$ and summing over $n$, we have\endnote{
The argument of the $\exp$ function originally had the factor $(t - t_{o})$; this has been replaced by the factor $(t - t^{\prime})$.}
\begin{equation}
\{ \psi_{\alpha}(r,t), \, \psi^{*}_{\beta}(r^{\prime}, t^{\prime}) \} =  \sum_{n} \psi_{n \alpha}(r) \psi^{*}_{n \beta}(r^{\prime})\exp \left\{-i \frac{E_{n}}{\hbar} (t - t^{\prime}) \right\}
\end{equation}
This is the general commutation rule which reduces to (299) in the special case of free particles. 

Multiplying (345) by $\psi^{*}_{m}(r)$ and integrating over $r$, we have 
\begin{equation}
\{ b_{m}, \, b^{*}_{n} \} = \delta_{n m}
\end{equation}
which is identical to (311) in the case of free particles when normalizations are properly treated.

\subsection*{The Hamiltonian}
\addcontentsline{toc}{subsection}{The Hamiltonian}

As before, the $b_{n+}$ are absorption operators for electrons and the $b^{*}_{n-}$ absorption operators for positrons, only the electrons and positrons being defined by the bound wave-functions. The vacuum state $\Psi_{o}$ is given by 
\begin{equation}
b_{n+} \Psi_{0} = 0 \qquad \qquad b^{*}_{n-} \Psi_{0} = 0
\end{equation}
And the total Hamiltonian of the system, in order to give the correct commutators with the  $b_{n}$ and the $b^{*}_{n}$ and also to have zero expectation value in the vacuum, is
\index{Hamiltonian!Dirac~field!in~external~field}%
\begin{align}
H &= \sum_{n+} E_{n} b^{*}_{n} b_{n}  - \sum_{n-} E_{n} b_{n} b^{*}_{n} \\
    &= \sum_{n+} E_{n} b^{*}_{n} b_{n}  + \sum_{n-} | E_{n} | b_{n} b^{*}_{n} 
\end{align}

From this Hamiltonian it is clear that the system is just a superposition of non-interacting particle-states. In each particle-state independently there is a number of particles given by
\begin{align*}
N_{n} &= b^{*}_{n}b_{n}	\qquad \text{for electron states} \\
N_{n} &= b_{n}b^{*}_{n}	\qquad \text{for positron states}
\end{align*}
From the commutation rules (338) and (347)
\begin{equation}
N_{n}^{2} = N_{n}
\end{equation}
so each $N_{n}$ has just the two eigenvalues 0, 1. This describes exactly the physical situation in a many-electron atom where each atomic level may be full or empty independent of the others. 

If each $N$ is represented by the $(2 \times 2)$ diagonal matrix 
\begin{equation}
N_{n} = \left( \begin{matrix}
       0 & 0 \\
       0 & 1
	\end{matrix}  \right)
\end{equation}
then we have 
\begin{align}
b_{n+} &= \left( \begin{matrix}
           0  &  1 \\
           0  &  0
           \end{matrix} \right) \qquad \qquad
b^{*}_{n-} = \left( \begin{matrix}
           0  &  0 \\
           1  &  0
           \end{matrix} \right) \notag \\
 b_{n-} &= \left( \begin{matrix}
           0  &  0 \\
           1  &  0
           \end{matrix} \right) \qquad \qquad
b^{*}_{n-} = \left( \begin{matrix}
           0  &  1 \\
           0  &  0
           \end{matrix} \right) 
\end{align}
This gives an explicit matrix representation for the operators. Each of the states $n$ has its own two-valued row and column index. Thus for an atom with $M$ levels altogether the operators would be represented by $(2^{M} \times 2^{M})$ matrices.

Once having got the Hamiltonian (350) and the stationary states $\psi_{n}$ the theory of many-electron systems is completely straightforward. 
\index{Hamiltonian!Dirac~field}
We see that the levels of the hydrogen atom given by the one-electron Dirac theory are still exactly valid in this many-electron theory. 
\index{hydrogen~atom!energy~levels}%
Only now the Hamiltonian (350) having positive eigenvalues, the negative energy states give us no disasters. The positrons appear with positive energy, so that all the results of the Dirac theory are given to us simply and automatically.
\index{positron!positive~energy}%

\subsection*{Antisymmetry of the States}
\addcontentsline{toc}{subsection}{Antisymmetry of the States}

We know that in elementary quantum theory of many-electron systems we have to represent the system wave-functions by determinants of one-particle wave-functions, so as to make the system wave-functions always antisymmetric in the particle coordinates. We shall no longer need to make any such arbitrary choice of wave-functions in the field theory, all the results of the antisymmetry being given automatically by the theory.

For example, consider an atom with 2 electrons in states $\psi_{1}$ and $\psi_{2}$, all other states being empty. Then the state of the system is given by 
\begin{equation}
\Psi = b^{*}_{1}b^{*}_{2} \Psi_{o}
\end{equation}
where $ \Psi_{o}$ is the vacuum state. In (354) there is no arbitrariness; an interchange of the indices 1 and 2 will only change $\Psi$ into $-\Psi$ which means no physical change. Now consider a 2-particle interaction operator
\begin{equation}
V = \tfrac{1}{2} \iint d^{\, 3} \boldsymbol{r}_1 \, d^{\, 3} \boldsymbol{r}_2 \left\{ \psi^{*}(r_{1})  \psi(r_{1})\right\} V(r_{1} -  r_{2}) \left\{ \psi^{*}(r_{2})  \psi(r_{2}) \right\}
\end{equation}
For example, $V$ may be the Coulomb potential between two electrons, which is not included in the Lagrangian (331). 
\index{Coulomb~potential}%
The $\tfrac{1}{2}$ is put in so as to count each pair of points $r_{1}$, $r_{2}$ only once. We calculate the matrix element of $V$ for a transition from $\Psi$ to a state
\[
\Psi^{\prime} = b^{*}_{3}b^{*}_{4} \Psi_{o}
\]
where the 2 electrons are in two other states $\psi_{3}$ and $\psi_{4}$. This matrix element is 
\begin{equation}
M = \left( \Psi^{*}_{o}, b_{4}b_{3} V b^{*}_{1}b^{*}_{2} \Psi_{o} \right)
\end{equation}

Expanding $V$ by (337) into a sum of products of $b_{n}$ and $b^{*}_{n}$, a contribution to (356) will come only from the 4 terms in $V$ proportional to 
$b_{1}b_{2}b^{*}_{3}b^{*}_{4}$. Using the anticommutation rules, we have 
\index{anticommute}%
\begin{align}
\left( \Psi^{\prime*}, b_{1}b_{2} V b^{*}_{3}b^{*}_{4} \Psi \right) &= -1 \notag \\
\left( \Psi^{\prime*}, b_{1}b_{2} V b^{*}_{4}b^{*}_{3} \Psi \right) &= 1, \quad \text{etc.,}
\end{align}
Hence adding up the 4 terms
\begin{equation}
\begin{split}
M = \iint d^{\, 3} \boldsymbol{r}_1 \, d^{\, 3} \boldsymbol{r}_2 \,V(r_{1} -  r_{2}) 
 &\left\{ \{ \psi^{*}_{3}(r_{1})  \psi_{1}(r_{1}) \} \{ \psi^{*}_{4}(r_{2})  \psi_{2}(r_{2}) \} \right. \\
 &- \left. \{ \psi^{*}_{3}(r_{1})  \psi_{2}(r_{1}) \} \{ \psi^{*}_{4}(r_{2})  \psi_{2}(r_{2}) \} \right\}
\end{split}
\end{equation}
This is exactly the result, direct minus exchange interaction, which would be given by using antisymmetrized wave-functions.

The field theory thus gives the full force of the Fermi statistics for electrons. 
\index{Fermi!statistics}%
And we could have shown in the same way that it gives the Bose statistics for photons. 
\index{Bose~statistics}%

\subsection*{Polarization of the Vacuum}
\addcontentsline{toc}{subsection}{Polarization of the Vacuum}

Because of the possibility of exciting the vacuum by creating a positron-electron pair,  the vacuum behaves like a dielectric, just as a solid has dielectric properties in virtue of the possibility of its atoms being excited to excited states by Maxwell radiation. 
\index{positron-electron~pair}%
\index{Maxwell!radiation}%
This effect does not depend on the quantizing of the Maxwell field, so we calculate it using classical fields. 
\index{Maxwell!field!classical}%

Like a real solid dielectric, the vacuum is both non-linear and dispersive, i.e. the dielectric constant depends on the field intensity and on the frequency. And for sufficiently high frequencies and field intensities it has a complex dielectric constant, meaning it can absorb energy from the Maxwell field by real creation of pairs.  
\index{dielectric~constant}%
\index{creation}

We calculate the dielectric constant only in the linear region, i.e.assuming weak fields. The critical field for this problem is
\begin{equation}
E_{c} = \frac{m^{2}c^{3}}{e \hbar} \approx 10^{16} \; \frac{\text{Volts}}{\text{cm}} \text{\qquad (from $eE \cdot \dfrac{\hbar}{mc} \approx mc^2$)}
\end{equation}
and in fact the linear theory is good enough for almost all problems. The important case where it is \emph{not} good enough is the propagation of photons through the intense Coulomb field round a heavy nucleus like lead. 
\index{Coulomb~potential}%
Then the non-linearity produces a scattering of photons which is small but has been detected experimentally by Wilson. \cite{Wilson53} 
\index{Wilson, Robert~R.}%
\index{scattering!photon,~and~vacuum~polarization}%

We calculate the dispersive effects exactly, i.e. with no restriction on the frequency. Since the treatment is linear, the imposed Maxwell field may be supposed to be given by the potentials of a plane wave whose amplitude  increases slowly with time
\index{Maxwell!field}%
\index{amplitude}%
\begin{equation}
A_{\mu}^{e}(x) = e_{\mu} \exp \{ iq \cdot x + \delta_{o} x_{o} \}
\end{equation}

Here $e$ and $q$ are given vectors, $\delta_{o}$ is a small positive number. This exponentially increasing amplitude  is put in so that the potential $A_{\mu}^{e}$ effectively acts only for a finite time before any given time at which observations may be made. 
\index{amplitude}%
This enables us to fix the initial conditions of the problem unambiguously. At the end of the calculation we shall go to the limit $\delta_{o} = 0$.

The vacuum polarization is the effect of the fluctuations of the quantized electron-positron field on a given Maxwell field. 
\index{vacuum~polarization}%
\index{electron-positron!field}%
The Lamb shift  is the effect of the fluctuations of the quantized Maxwell field  on a given electron. 
\index{Lamb~shift}%
\index{Maxwell!field!quantized}%
The two effects are just opposites of each other, the roles of the two fields being interchanged. Thus we can treat the vacuum polarization now conveniently with the theory of the quantized electron field alone. The treatment will be relativistic and so more correct than the treatment given for the Lamb shift. 
\index{Lamb~shift}%
Later, to have a complete theory of both effects, we shall quantize both fields together and consider the reaction of each on the other.

Historically the electron self-energy (Lamb shift) and the vacuum polarization were the two problems on which the theory broke down because of divergences. 
\index{self-energy!and~Lamb~shift}%
\index{self-energy!electron}%
\index{vacuum~polarization}%
Schwinger showed that the vacuum polarization could be calculated and was finite, if one used the same kind of renormalization  idea that made the Lamb shift also finite.
\index{vacuum~polarization!Schwinger~calculation}%
\index{renormalization!and~vacuum~polarization}%
\index{Lamb~shift}%

The electron field operator $\psi_{H}$ in the field (360) satisfies (332). Here $\psi_{H}$ is the operator in the Heisenberg representation. Now a solution of (332) correct to first order in $A_{\mu}^{e}$ is 
\begin{equation}
\psi_{H}(x) = \psi(x) + \frac{ie}{\hbar c} \int dx^{\prime} \, S_{R}(x - x^{\prime}) \slashed{A}^{e}(x^{\prime}) \psi (x^{\prime})
\end{equation}
Here $S_{R}$ is given by (293), (301),  and (303), and $\psi(x)$ is a solution of the free-field equation (286). In fact $\psi(x)$ is the field operator of the interaction representation, when the effects of $A_{\mu}$ are represented in the wave-function instead of in the operators. Using the retarded potential in (361) means that the unperturbed states are specified in the past, as the initial states upon which $A_{\mu}^{e}$ later produces effects. Thus the vacuum state defined by (323a) is the state in which initially no electrons or positrons are present. This is the state which we wish to study, and we call it $\Psi_{o}$. 

Using the interaction representation operators $\psi(x)$, $\Psi_{o}$ is the vacuum state and remains so for all time; the physical state is initially $\Psi_{o}$ but does not remain $\Psi_{o}$. Using the Heisenberg operators $\psi_{H}(x)$, $\Psi_{o}$ is the physical state for all time; it is initially the vacuum state  but does not remain so. In the remote past as $x_{0} \rightarrow - \infty$, because $S_{R}$ is a retarded potential $\psi_{H}(x)$ and $\psi(x)$ become identical. 

The expression (361) is useful because we know how to calculate matrix elements of $\psi(x)$ from the state $\Psi_{o}$, whereas the matrix elements of $\psi_{H}$ do not have any simple form. We also need the adjoint equation
\begin{equation}
\overline{\psi}_{H}(x) = \overline{\psi}(x) + \frac{ie}{\hbar c} \int dx^{\prime}\,\overline{\psi}(x^{\prime}) \slashed{A}^{e}(x^{\prime}) S_{A}(x^{\prime} - x)
\end{equation}
where $S_{A}(x)$ is given by (295).

\noindent The total current operator to first order in $A_{\mu}$ is 
\begin{align}
j_{\mu H}(x) &= -iec\; \overline{\psi}_{H}(x) \gamma_{\mu} \psi_{H}(x) \notag \\
      &=  j_{\mu}(x) + \frac{e^2}{\hbar} \int d^{\, 4}x^{\prime} \left\{ \overline{\psi}(x) \gamma_{\mu}S_{R}(x - x^{\prime}) \slashed{A}^{e}(x^{\prime})
            \psi(x^{\prime}) +  \overline{\psi}(x^{\prime}) \slashed{A}^{e}(x^{\prime}) S_{A}(x^{\prime} - x) \gamma_{\mu} \psi(x) \right\}
\end{align}
Here
\begin{equation}
j_{\mu}(x) = -iec \; \overline{\psi}(x) \gamma_{\mu} \psi(x) 
\end{equation}
is the current operator of the interaction representation. The vacuum expectation value
\begin{align}
\left(\Psi^{*}_{o}\; j_{\mu}(x)\; \Psi_{o} \right) &= \Braket{j_{\mu}(x)}_{o} \notag \\
 &= -iec \; \Braket{\sum\nolimits_{\alpha , \beta} \overline{\psi}_{\beta}(x) (\gamma_{\mu})_{\beta \alpha} \psi_{\alpha}(x)}_{o} \notag \\
 &= -iec \; \sum_{\alpha , \beta} (\gamma_{\mu})_{\beta \alpha} \Braket{ \overline{\psi}_{\beta}(x)\psi_{\alpha}(x)}_{o}
\end{align}
is given by (328)
\begin{align}
\Braket{j_{\mu}(x)}_{o} &= - ec \; \text{Spur} \left\{ \gamma_{\mu} S^{-}(0) \right\} \notag \\
 &= \frac{ec}{(2\pi)^3} \int d^{\, 3} \boldsymbol{k} \, \delta(k^2 + \mu^2)\, \Theta(-k) \;\text{Spur}\{\gamma_{\mu} \left[ \slashed{k} - i \mu \right] \} \notag \\
 &= \frac{4ec}{(2\pi)^3} \int d^{\, 3} \boldsymbol{k} \, \delta(k^2 + \mu^2)\, \Theta(-k)\;k_{\mu}
\end{align}
This is a highly divergent integral and is mathematically meaningless. This is one of the difficulties of the theory about which one can argue for a long time. 

However there is no doubt that correct physical results are obtained by putting $\Braket{j_{\mu}(x)}_{o} = 0$ simply. There are two good reasons for doing this.

(1) Physical. $\Braket{j_{\mu}(x)}_{o}$, being the expectation value of the charge-current in the vacuum in the absence of all external fields, is known to be zero experimentally. Therefore if we calculated $\Braket{j_{\mu}(x)}_{o}$ and found it were not zero we should simply define the current operator to be $j_{\mu} - \Braket{j_{\mu}}_{o}$. With this definition the expectation value would become zero automatically. 

(2) Mathematical. $\Braket{j_{\mu}(x)}_{o}$ as we have calculated it is a vector, each component of which is a number quite independent of the coordinate system. There does not exist such a vector invariant under Lorentz transformations, except for this zero vector.
\index{Lorentz!transformations}%
Therefore $\Braket{j_{\mu}(x)}_{o} = 0$ is the only assumption we \emph{can} make which would keep the theory invariant.

This is a simple example of a method which has often to be used in quantum electrodynamics. 
\index{quantum~electrodynamics}%
When a calculation leads to a divergent integral or a mathematically indeterminate expression, we use physical arguments or arguments of Lorentz invariance to find a definite value for the quantity we are unable to calculate. 
\index{Lorentz!invariance}%
This is the reason for the great success of the covariant formulation of electrodynamics introduced by Schwinger. 
\index{Schwinger!covariant~electrodynamics}%

So using this principle we have by (328) 
\begin{equation}
\Braket{j_{\mu H}(x)}_{o} = -\frac{ie^2}{\hbar} \int d^{\, 4} x^{\prime}\; \text{Spur}\{\slashed{A}^{e}(x^{\prime}) S^{-}(x^{\prime} - x) \gamma_{\mu} S_{R} (x - x^{\prime}) + \slashed{A}^{e}(x^{\prime}) S_{A} (x^{\prime} - x) \gamma_{\mu}S^{-}(x - x^{\prime}) \}
\end{equation}
We use the momentum representation (303) for $S_{R}$. 
\index{momentum!representation}%
But instead of taking the contour along the real axis for $k_{0}$ we can use a path of integration running along a straight line parallel to the real axis a distance $\delta_{o}$ above it. That gives the momentum representation
\begin{equation}
e^{-\delta_{0}} S_{R}(x) = \frac{i}{(2\pi)^4} \int e^{i k \cdot x} \frac{\slashed{k} + i \slashed{\delta} + i \mu}{(k + i\delta)^2  + \mu^2} d^{\, 4}k
\end{equation}
where $\delta_{0}$ is any positive real number, $\delta$ is the vector with components $(0, 0, 0, \delta_{0})$ and the integral is along the real axis. The poles of (368) in the $k_{0}$ plane are displaced away from the real axis and so the integrand is free from singularities on the path of integration. Similarly
\begin{equation}
e^{+\delta_{0}} S_{A}(x) = \frac{i}{(2\pi)^4} \int e^{i k \cdot x} \frac{\slashed{k} - i \slashed{\delta} + i \mu}{(k - i\delta)^2  + \mu^2} d^{\, 4}k
\end{equation}
When using (368) and (369) we shall usually make $\delta_{0} \rightarrow 0$ after carrying out the integrations, so that the convergence factors $e^{\pm \delta_{0} x_{0}}$ will tend to 1 for every finite $x$.

Thus the momentum representation of (367) will be 
\index{momentum!representation}%
\[
\begin{split}
\Braket{j_{\mu H}(x)}_{o} = &- \frac{e^2}{\hbar} \frac{1}{(2\pi)^{7}} \int d^{\,4}x^{\prime} \, \iint d^{\,4}k_{1}\,d^{\,4}k_{2} \exp\{iq\cdot x^{\prime} + i (k_{1} - k_{2}) \cdot (x^{\prime} - x) + \delta_{0} x_{0} \} \times \\
& \times \left\{
            \text{Spur} \{ \slashed{e} (\slashed{k}_{1} + i \mu) 
            \gamma_{\mu} (\slashed{k}_{2} + i \slashed{\delta} + i \mu) \} 
            \frac{\delta(k_{1}^{2} + \mu^{2}) \, \Theta(- k_{1})}{(k_{2} + i \delta)^2 + \mu^2}
  \right. \\
& +  	  \left. \text{Spur} \{ \slashed{e} (\slashed{k}_{1} - i \slashed{\delta} + i \mu)  
	  \gamma_{\mu} (\slashed{k}_{2} + i \mu)  \} 
          \frac{\delta(k_{2}^{2} + \mu^{2})\, \Theta(- k_{2})}{(k_{1} - i \delta)^2 + \mu^2} \right\}
\end{split}
\]
The integration of $x^{\prime}$ is immediate and gives $(2\pi)^{4} \delta^{4}(k_{1} - k_{2} + q)$. Hence 
\begin{equation}
\begin{split}
\Braket{j_{\mu H}(x)}_{o} = & - \frac{e^2}{(2\pi)^{3}\hbar} e^{iq\cdot x + \delta_{0}x_{0}}  \times \\
&\times \int d^{\,4}k \,\left\{\text{Spur} \{ \slashed{e} (\slashed{k} + i \mu) \gamma_{\mu} (\slashed{k} + \slashed{q} +  i \slashed{\delta} + i \mu) \} 
            		\frac{\delta(k^{2} + \mu^{2}) \, \Theta(- k)}{(k + q + i \delta)^2 + \mu^2} \right. \\
& +  \left. \text{Spur} \{ \slashed{e} (\slashed{k} - i \slashed{\delta} + i \mu) \gamma_{\mu} (\slashed{k} + \slashed{q} + i \mu)  \} 
          \frac{\delta\{(k + q)^{2} +  \mu^{2}\}\, \Theta(- k - q)}{(k - i \delta)^2 + \mu^2} \right\}
\end{split}
\end{equation}
Now consider the function\endnote{
The function $F_{\nu}(k)$ was written as a scalar, $F(k)$. This is misleading; the right-hand side is a vector function, because it is linear in $\gamma_{\nu}$. So $F(k)$ was promoted to $F_{\nu}(k)$.}
\begin{equation}
F_{\nu} (k) = \text{Spur}\{ \slashed{e} (\slashed{k} + i \mu) \gamma_{\nu}(\slashed{k} + \slashed{q} + i \slashed{\delta} + i \mu) \} \frac{1}{(k^2 + \mu^2) [(k + q + i \delta)^2 + \mu^2]}
\end{equation}
This has poles at 4 points in the $k_{0}$ plane
\begin{equation}
k_{0} = \pm \sqrt{ | \boldsymbol{k} |^2 + \mu^2}  \qquad \qquad k_{0} = - q_{0} - i\delta_{0} \pm \sqrt{ |\boldsymbol{k} + \boldsymbol{q} |^2 + \mu^2}
\end{equation}
The integral in the expression (370) is just the sum of the residues at the two points\endnote{
The phrase ``integral in the'' was inserted.}
\begin{equation}
k_{0} = - \sqrt{ | \boldsymbol{k} |^2 + \mu^2}  \qquad \qquad k_{0} = - q_{0} - i\delta_{0} - \sqrt{ |\boldsymbol{k} + \boldsymbol{q} |^2 + \mu^2}
\end{equation}
\begin{center}
\includegraphics{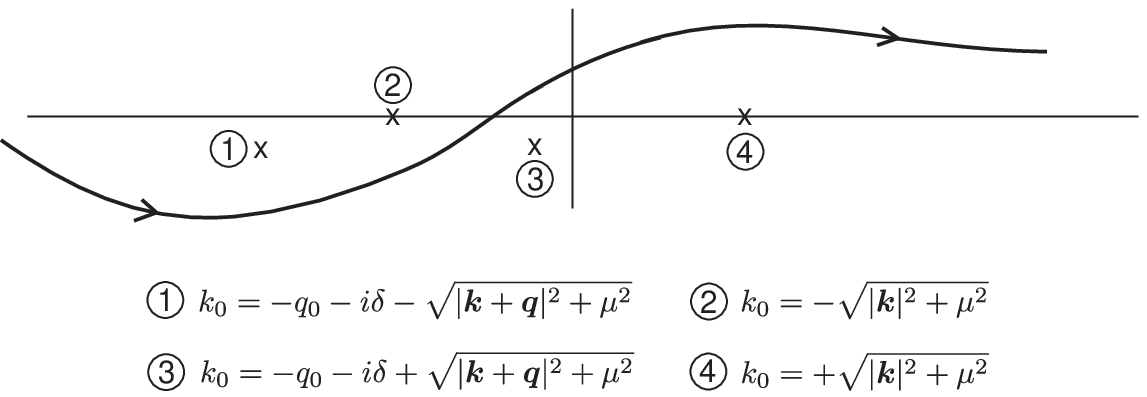}
\end{center}
Hence
\begin{equation}
\Braket{j_{\nu H} (x)}_{o} = \frac{ie^2}{(2\pi)^{4} \hbar} e^{i q \cdot x + \delta_{0}x_{0}} \int_{C} F_{\nu}(k)\, dk
\end{equation}
where $\int_{C}$ means a contour integral in the $k_{0}$ plane drawn as shown in the diagram, going from $-\infty$ to $+\infty$ below the two poles (373) and above the other two poles, and including the upper semicircle of the $k_{0}$ plane at infinity. So long as $\delta_{0} > 0$ the poles are always clearly separated from each other. Now the calculation really starts, with the evaluation of this integral (374). This will be typical of all the calculations which have to be done in quantum electrodynamics using the modern methods. 
\index{quantum~electrodynamics}%

\subsection*{Calculation of Momentum Integrals}
\addcontentsline{toc}{subsection}{Calculation of Momentum Integrals}
\index{momentum!integral}%

Write $J_{\nu} = \int_{C} F_{\nu}(k)\, dk$. Then $J_{\nu}$ is a vector function of the variables $\mu$ (the mass of the electron) and 
\begin{equation}
Q = q + i \delta
\end{equation}
 $J_{\nu}$ is certainly an analytic function of $\mu$ if $\delta > 0$, and it is an analytic function of $Q$ if $\mu$ is large enough so that the poles (373) always lie to the left of the imaginary axis and the other two poles (372) to the right. Therefore we can calculate  $J_{\nu}$ for large values of $\mu$ and $\delta = 0$ so that $Q = q$. 
 
To simplify (371) we take $\delta = 0$ and use Feynman's formula (\emph{Phys. Rev.} \textbf{76} (1949) 785) 
\index{Feynman!integration formula}%
\begin{equation}
\frac{1}{ab} = \int_{0}^{1} dz \,  \frac{1}{[az + b(1 - z)]^2}
\end{equation}
which we can get at once simply from the definite integral of the RHS.\endnote{
``definite'' replaces ``indefinite''}
Hence
\begin{align}
J_{\nu} &= \int_{0}^{1} dz \int_{C} dk \; \text{Spur} \{ \slashed{e} (\slashed{k} + i \mu) \gamma_{\nu} (\slashed{k} + \slashed{q} + i\mu) \} \frac{1}{\{k^2 + \mu^2 + z(2 k \cdot q + q^2) \}^2} \notag \\
&= \int_{0}^{1} dz \int_{C} dk \; \text{Spur} \{ \slashed{e} (\slashed{k} - z\slashed{q}  + i \mu) \gamma_{\nu} (\slashed{k} + (1-z)\slashed{q} + i\mu) \} \frac{1}{\{k^2 + \mu^2 + (z - z^2) q^2 \}^2}
\end{align}
In the last step a shift of origin was made in the $k$--integration replacing $k$ by $(k- zq)$. Again in (377) the poles in the $k_{0}$--plane for every $z$ are well separated by the imaginary axis, provided that $\mu$ is large. Evaluating the spur and dropping terms are odd in $k$, and using (33),\endnote{
In v.2, this reads ``Spur $\sum_{\mu} \gamma_{\mu} \gamma_{\nu} = 4$''}
Spur $\gamma_{\nu} = 0$, Spur $\gamma_{\mu} \gamma_{\nu} = 4\delta_{\mu \nu}$, we get
\begin{equation}
J_{\nu} = 4 \int_{0}^{1} dz \int_{C} dk \; \frac{e_{\nu}(- k^{2} - \mu^{2} + (z - z^{2}) q^2) + 2 (e \cdot k)\, k_{\nu} - 2 (z - z^{2}) (e \cdot q) \, q_{\nu}} {\{k^2 + \mu^2 + (z - z^2) q^2 \}^2}
\end{equation}
the odd terms going out because we can now take the $k_{0}$--integration straight up the imaginary axis from $-i\infty$ to $i\infty$ if we like. For the same reasons of symmetry we may replace
\[
e \cdot k \; k_{\nu} \quad \text{by} \quad \tfrac{1}{4} k^2 e_{\nu} \qquad \text{because} \quad  e \cdot k \; k_{\nu} = \sum_{\lambda} e_{\lambda} k_{\lambda} k_{\nu} \rightarrow e_{\nu} k_{\nu} k_{\nu} \rightarrow \tfrac{1}{4} e_{\nu} k^2
\]
in the numerator, and obtain finally
\begin{equation}
J_{\nu} = 4 \int_{0}^{1} dz \int_{C} dk \; \frac{e_{\nu}\{- \frac{1}{2}k^{2} - \mu^{2} + (z - z^{2}) q^2\} - 2 (z - z^{2}) (e \cdot q) \, q_{\nu}} {\{k^2 + \mu^2 + (z - z^2) q^2 \}^2}
\end{equation}

This integral is still badly divergent. So we again use a physical argument to give a definite value to the most divergent part of it. The current operator both in interaction and in Heisenberg representation must satisfy
\begin{equation}
\sum_{\nu} \frac{\partial j_{\nu}(x)}{\partial x_{\nu}} = 0
\end{equation}
identically. Therefore (374) gives (since we now take $\delta = 0$)
\begin{equation}
\sum_{\nu} q_{\nu} J_{\nu} = 0
\end{equation}
which gives the relation
\begin{equation}
\int_{0}^{1} dz \int_{C} dk \; \frac{- \frac{1}{2} k^{2} - \mu^{2} - (z - z^{2}) q^2}{\{k^2 + \mu^2 + (z - z^2) q^2 \}^2} \equiv 0
\end{equation}
This equation (382) is a guiding equation telling us that a certain divergent expression appearing in (379) is to be given the value zero in order to make physical sense. We are left with
\begin{equation}
J_{\nu} =  8 (q^{2} e_{\nu} - e \cdot q \; q_{\nu}) \int_{0}^{1} dz \, (z - z^{2}) \int_{C} \frac{dk}{\{k^2 + \mu^2 + (z - z^2) q^2 \}^2}
\end{equation}

For any positive $\Lambda$ the integral
\begin{equation}
I_{\Lambda} = \int_{C} \frac{dk}{(k^2 + \Lambda)^{3}}
\end{equation}
is convergent and can be evaluated by integrating for $k_{0}$ up the imaginary axis from $-i\infty$ to $+i\infty$. This gives (see the Appendix below) 
\begin{equation}
I_{\Lambda} = i \iiiint \frac{dk_{1} dk_{2} dk_{3} dk_{0}}{(k_{1}^2 + k_{2}^2 +k_{3}^2 +k_{0}^2 + \Lambda)^{3}} = 2\pi^2 i \int_{0}^{\infty} \frac{k^{3} \, dk}{(k^{2} + \Lambda)^{3}} = \pi^{2} i \int_{0}^{\infty} \frac{x \, dx}{(x + \Lambda)^{3}} = \frac{\pi^2 i}{2 \Lambda}
\end{equation}
Hence integrating with respect to $\Lambda$
\begin{equation}
\int_{C} dk \left\{ \frac{1}{(k^{2} + \Lambda)^2} - \frac{1}{(k^{2} + \mu^{2})^2} \right\} = \pi^{2} i \log \left( \frac{ \mu^2}{\Lambda} \right)
\end{equation}
this integral being also convergent. However
\[
\int_{C} \frac{dk}{(k^2 + \mu^2)^2 }
\]
is logarithmically divergent for large $k$. Its value is
\begin{equation}
2 i \pi^{2} \log \left( \frac{k_{\text{max}}}{\mu} \right) = 2 i \pi^2 R
\end{equation}
where $R$ is the logarithmic factor and is independent of $q$.

Using (386) and (387) in (383), with $\Lambda = \mu^{2} + (z - z^{2}) q^{2}$, we get
\begin{equation}
J_{\nu} =  8 \pi^{2} i (q^{2} e_{\nu} - e \cdot q \; q_{\nu}) \left\{ \tfrac{1}{3} R - \int_{0}^{1} dz \; (z - z^{2}) \log \left[ 1 + \frac{(z - z^{2}) q^{2}}{\mu^{2}}\right] \right\}
\end{equation}
This is the analytic formula for $J_{\nu}$ valid for large $\mu$ in which case the logarithm is real. We make the analytic continuation to small values of $\mu$ by writing $(q + i \delta)$ for $q$ in (388), treating $\delta_{0}$ as small and positive. Then $q^{2}$ becomes $q^{2} - 2i \delta q_{0}$ and the logarithm becomes 
\[
\log \left| 1 + \frac{(z - z^{2}) q^{2}}{\mu^{2}} \right| + \begin{cases} 
		\quad \quad 0 \qquad \; \, \text{for}  \; \dfrac{(z - z^{2}) q^{2}}{\mu^2} > -1 \\
		-i \pi \, \epsilon(q_{0}) \quad \text{for}  \; \dfrac{(z - z^{2}) q^{2}}{\mu^2} < -1
		\end{cases}
\]
Writing now $z$ for $4(z - z^{2})$ and using (374), we go to the limit $\delta_{0} = 0$ and find\endnote{
The bottom limit of 0 was added to the last integral sign. Note that the change of variable is easier to follow by first observing
\[
\int_{0}^{1} dz \, (z - z^{2}) f(z - z^{2}) = 2 \int_{0}^{1/2} dz \, (z - z^{2}) f(z - z^{2})
\]
because the expression $(z - z^{2})$ is symmetric about $z = \tfrac{1}{2}$.}
\begin{equation}
\begin{split}
\Braket{j_{\nu H} (x)}_{o} = -\frac{e^{2}}{2 \pi^{2}\hbar}& \left( q^{2}e_{\nu} - e \cdot q \; q_{\nu} \right) e^{i q \cdot x} \times \\
&\times \left\{ \tfrac{1}{3} R - \frac{1}{8} \int_{0}^{1} \frac{z \, dz}{\sqrt{1 - z}} \log \left| 1 + \frac{zq^{2}}{4 \mu^2} \right| + \frac{i \pi}{8} \, \epsilon(q_{0})  \int_{0}^{-4\mu^{2}/q^{2}} \frac{z \, dz}{\sqrt{1 - z}} \right\} 
\end{split} 
\end{equation}
the last term being zero except when
\begin{equation}
q^{2} < -4 \mu^{2}
\end{equation}
Now the external potential $A_{\nu}^{e}(x)$ is associated with a classical external charge-current density
\begin{align}
j_{\nu E} (x) &= - c \sum_{\lambda} \frac{\partial}{\partial x_{\lambda}} F_{\lambda \nu E} (x) \notag \\
                    &= - c \sum_{\lambda} \left\{ \frac{\partial^{2}}{\partial x_{\lambda}^{2}} A_{\nu}^{e}(x)  - \frac{\partial^{2}}{\partial x_{\nu} \partial x_{\lambda}} A_{\lambda}^{e}(x) \right\} \notag \\
                     &= c \left\{ q^{2} e_{\nu} - e \cdot q \; q_{\nu} \right\} e^{i q \cdot x}
\end{align}   
Hence (389) gives the final result, with $\alpha = \dfrac{1}{137} = \dfrac{e^{2}}{4 \pi \hbar c}$, (Heaviside units) 
\index{Heaviside~units}%
\begin{equation}
\Braket{j_{\nu H} (x)}_{o} =  -\alpha j_{\nu E} (x) \left\{ \frac{2}{3\pi} R - \frac{1}{4\pi} \int_{0}^{1} \frac{z \, dz}{\sqrt{1 - z}} \log \left| 1 + \frac{zq^{2}}{4 \mu^2} \right| 
+ \frac{i}{4} \, \epsilon(q_{0})  \int_{0}^{-4\mu^{2}/q^{2}} \frac{z \, dz}{\sqrt{1 - z}} \right\} \\
\end{equation}

\noindent \emph{Appendix} \\

The four dimensional volume element is (see (385)) $d\xi_{1}d\xi_{2}d\xi_{3}d\xi_{4} = 2 \pi^{2} r^{3}\, dr$ in four dimensional polar coordinates. To show this, we denote the surface of a $p$ dimensional unit sphere by $\omega$. Then the surface of a $p$ dimensional sphere of radius $R$ is $R^{p-1} \omega$, and hence the volume element in polar coordinates is $\omega R^{p-1} dR$.

To calculate the value of $\omega$, we compute the $p$ dimensional Laplace integral in Cartesian and polar coordinates. 
\index{Laplace~integral}%
\index{Gauss~integral|see{Laplace~integral}}%
We have
\[
J = \iint \dots \int \exp \left\{ - \sum_{i = 1}^{p} \xi_{i}^{2}\right\} \, d\xi_{1}d\xi_{2} \dots d\xi_{p} = \left( \sqrt{\pi} \right)^{p}
\]
On the other hand
\[
J = \omega \int_{0}^{\infty} e^{-\rho^{2}} \rho^{p -1} \, d\rho = \omega\, \frac{\Gamma(p/2)}{2}
\]
and thus by comparison
\[
\omega = \frac{2 \pi^{p/2}}{\Gamma(p/2)}
\]
For $p = 4$, we have $\Gamma(2) = 1$, $\omega = 2 \pi^{2}$. For $p = 3$, we have $\Gamma(\tfrac{3}{2}) = \dfrac{\sqrt{\pi}}{2}$, $\omega = \dfrac{2 \pi^{3/2}}{\sqrt{\pi}/2} = 4 \pi$, etc. 

\subsection*{Physical Meaning of the Vacuum Polarization}
\addcontentsline{toc}{subsection}{Physical Meaning of the Vacuum Polarization}

We now discuss the various physical effects arising from the calculation of 
\index{vacuum~polarization|(}%
\[
\Braket{j_{\nu H}(x)}_{o}
\]
\hspace*{3ex}1)\hspace*{1ex}The result is completely gauge-invariant. 
\index{gauge!invariance}%
This may be seen at once from (391): for if a gradient $\partial \Lambda/\partial x_{\nu}$ is added to $A_{\nu}^{e}$, there is no change in $j_{\nu E}$. \\
\hspace*{3ex}2)\hspace*{1ex}If the relation (382) had not been used in order to simplify (379), we should have had an addition to $\Braket{j_{\nu H}(x)}_{o}$ of the form $K^{\prime}e_{\nu} = K A_{\nu}^{e}(x)$ (see (360)) where $K$ is an indeterminate numerical factor containing the divergent integral standing on the left of (382). That is to say, an induced current proportional to the inducing potential. This would have destroyed the gauge-invariance of the result if $K \ne 0$. Therefore we can also use the physical requirement that the results must be gauge-invariant in order to  give the value zero to the indeterminate $K$. \\
\hspace*{3ex}3)\hspace*{1ex}The energy density of the vacuum arising from the polarization by the potentials $A_{\nu}^{e}(x)$ is
\begin{equation}
d(x) = - \frac{1}{2c} \sum_{\nu} A_{\nu}^{e}(x) \Braket{j_{\nu E}(x)}_{o}
\end{equation}
Thus the term $K A_{\nu}^{e}(x)$ would give an energy density
\begin{equation}
 - \frac{K}{2c} \sum_{\nu} A_{\nu}^{e}(x) A_{\nu}^{e}(x)
\end{equation}
associated with the electromagnetic potentials. This would give to the photon a finite rest-mass, and for this reason $K$ is often referred to as the ``self-energy of the photon''. 
\index{self-energy!photon}%
There is a lot of discussion in the literature of this photon self-energy. But since physical arguments lead us definitely to give $K$ the value zero, we have no choice but to say that the photon self-energy also is zero. This is of course the result that any consistent theory of electrodynamics must give. \\
\hspace*{3ex}4)\hspace*{1ex}The logarithmic divergence $R$ is a real divergence and cannot be given the value $0$ from physical arguments. However, it gives only an induced charge exactly proportional to the inducing external charge. It is never possible experimentally to separate the external charge from the proportional induced charge. Therefore in all measurements of the external charge, the measured charge will be not $j_{\nu E}(x)$ but 
\begin{equation}
j_{\nu R} (x) = \left(1 - \frac{2 \alpha}{3 \pi} R \right) j_{\nu E} (x)
\end{equation}
$j_{\nu R}$ here meaning ``renormalized charge''. 
\index{renormalization!charge}%
\index{fine~structure~constant}%
Thus the effect of the term $R$ in (392) is only change the unit in which the external charge is measured.  We write the results in terms of the observed external charge $j_{\nu R}$ instead of the unobservable $j_{\nu E}$, this change of units being called ``charge renormalization''.
\index{charge~renormalization}%
Note the similarity between charge and mass renormalization. In both cases a divergent effect is recognized as producing no observable phenomenon, because it only changes the value of one of the fundamental constants, in the one case the electron mass $m$ and in the other the unit of charge $e$. Since $m$ and $e$ are things which are observed directly, the divergent effects disappear completely when the results are written in terms of the observed $m$ and $e$. Thus (392) becomes 
\begin{equation}
\Braket{j_{\nu H} (x)}_{o} =  \alpha j_{\nu R} (x) \left\{\frac{1}{4\pi} \int_{0}^{1} \frac{z \, dz}{\sqrt{1 - z}} \log \left| 1 + \frac{zq^{2}}{4 \mu^2} \right|  - \frac{i}{4} \, \epsilon(q_{0})  \int_{0}^{-4\mu^{2}/q^{2}} \frac{z \, dz}{\sqrt{1 - z}} \right\} \\
\end{equation}
Everything here is now finite and observable.\\
\hspace*{3ex}5)\hspace*{1ex}When $A_{\nu}^{e}(x)$ is the potential of a pure radiation field without sources, $j_{\nu R} = 0$ and so there is no polarization. Thus for every photon or freely travelling wave, the vacuum behaves like a true vacuum; there are no dielectric effects  of any kind. 
\index{dielectric~constant}%
This agrees with the common-sense idea of the vacuum. The result is however only true so long as the polarization can be treated as linear; if non-linear effects were included then two beams of light crossing the same region would produce a polarization current, giving rise to a ``scattering of light by light''. The light-by-light scattering has been calculated, it is not zero but is much too small to be observable.  \\
\index{scattering!light~by~light}%

\hspace*{3ex}6)\hspace*{1ex}The induced current (396) consists of two components, the first in phase with the potential $A^{e}_{\nu}(x)$ and the second out of phase by $\pi/2$. Speaking of the vacuum as a circuit being driven by the potential $A^{e}_{\nu}(x)$, the first is an inductive effect, the second a resistive one. So only the second term gives an absorption of energy by the vacuum from the driving potentials.

We study the energy balance, remembering that the classical potentials $A^{e}_{\nu}(x)$ must always be real quantities like
\begin{equation}
A^{e}_{\nu}(x) = e_{\nu} \cos (q \cdot x)
\end{equation}
and we take without loss of generality $q_{0}$ positive. Then (396) gives, taking $e \cdot q = 0$, 
\begin{equation}
\Braket{j_{\nu H}(x)}_{o} = e_{\nu} \left\{ A \cos (q \cdot x) + B \sin (q \cdot x) \right\}
\end{equation}
where $A$ and $B$ are real and
\begin{equation}
B = \tfrac{1}{4} \alpha c q^{2} \int_{0}^{-4 \mu^{2}/q^{2}} \frac{z \, dz}{\sqrt{1 - z}}
\end{equation}
The energy supplied by the potentials to the vacuum per unit volume and per unit time is
\begin{align}
E &= - \frac{1}{c} \sum_{\nu} \Braket{j_{\nu H} (x)}_{o} \frac{\partial A_{\nu}^{e}(x)}{\partial t} \notag \\
    &= - q_{0} \sum_{\nu} e_{\nu}^{2} \left[  A \sin (q \cdot x) \cos (q \cdot x) + B \sin ^{2} (q \cdot x) \right]
\end{align}
From (400) one sees that the in-phase current gives no net absorption of energy, while the out-of-phase current gives a mean energy supply per unit time
\begin{equation}
\overline{E} = - \tfrac{1}{2} q_{0}e^{2} B = - \frac{\alpha c e^2 q^{2} q_{0}}{8} \int_{0}^{-4 \mu^{2}/q^{2}} \frac{z \, dz}{\sqrt{1 - z}}
\end{equation}
If $q$ does not satisfy (390), i.e.if
\begin{equation}
q_{0} < \sqrt{4\mu^2 + | \boldsymbol{q} |^2}
\end{equation}
then $B = 0$ and $E = 0$, there is not enough energy in the vibrations of the field to create a real positron-electron pair whose rest-mass alone requires $2 mc^{2}$, given that the field supplies in the interaction a momentum $\hbar k$ together with the energy $\hbar c q_{0}$.
\index{creation!pair}%
\index{electron-positron!pair}%

However if (390) is satisfied, then there is enough energy for creation of real pairs, each real pair carrying energy $\hbar c q_{0}$. 
\index{creation!pair}%
Since $q$ is time-like and $e \cdot q = 0$, $e$ is space-like and $(e^{2}) > 0$. This can be seen as follows:
\[
e \cdot q = 0 = \boldsymbol{e} \cdot \boldsymbol{q} - e_{0}q_{0}
\]
Since $q$ is timelike, we can use a Lorentz transformation  which makes $\boldsymbol{q} = 0$; then, of course, $q_{0} \ne 0$. 
\index{Lorentz!transformations}%
But then we must have $e_{0} = 0$, which means that $e$ is spacelike. Therefore $\overline{E} > 0$, verifying that the potentials can never extract energy from the vacuum. And we have the probability per unit volume and per unit time that the potential (397) will create a real pair 
\index{probability!pair~creation}%
\begin{equation}
w = \frac{\overline{E}}{\hbar c q_{0}} = - \frac{\alpha (e^{2})(q^{2})}{8 \hbar} \int_0^{-4 \mu^{2}/q^{2}} \frac{z \, dz}{\sqrt{1 - z}}
\end{equation}
The result could of course have been found more easily by elementary methods. The thing I want to stress here is that the elementary real pair-creation processes are necessarily tied up with the less elementary vacuum-polarization effect given by the $A$ term in (398), and the $A$ term will exist whether or not real pair production is possible. 
\index{creation!pair}%
The situation is quite the same as the tying up of the elementary line-width effect in atomic spectra with the less elementary line-shifts; we discussed these effects in detail earlier. Therefore we have quite as much reason to take seriously the vacuum-polarization effect given by the in-phase current in (396), as we have to take seriously the Lamb shift itself. 
\index{Lamb~shift}%
Because physicists were unwilling to take these two effects seriously, physics was held up for quite a number of years.

\subsection*{Vacuum Polarization For Slowly Varying Weak Fields. The Uehling Effect.}
\addcontentsline{toc}{subsection}{Slowly Varying Weak Fields. The Uehling Effect.}

Let now the external potential $A^{e}_{\nu}(x)$ be not only weak but also slowly varying in both space and time, i.e. let it be a superposition of Fourier components (360) with
\index{Fourier!components}%
\begin{equation}
|q^{2} | \ll \mu^{2}
\end{equation}
Then by (390) the second term in (396) is zero\endnote{
The original read ``Then by (390) in (396) the second term is zero''.}
and the logarithm may be expanded in terms of $(q^{2}/\mu^{2})$. Keeping only the term of order $q^{2}$
\[
\Braket{j_{\nu H}(x)}_{o} = \alpha \, \frac{q^{2}}{16 \pi \mu^{2}} \, j_{\nu R} (x) \int_0^{1} \frac{z^{2} \, dz}{\sqrt{1 - z}} = \frac{\alpha q^{2}}{15 \pi \mu^{2}}\,  j_{\nu R}(x)
\]
But in each Fourier component of $j_{\nu R}(x)$, operating with the D'Alembertian $\Box^{2}$ gives a factor $(-q^{2})$. 
\index{d'Alembertian~operator}%
Hence the result, independent of the Fourier decomposition, valid for slowly varying fields
\index{Fourier!components}%
\begin{equation}
\Braket{j_{\nu H}(x)}_{o} = - \frac{\alpha}{15 \pi \mu^{2}} \, \left\{ \Box^{2} j_{\nu R}(x) \right\}
\end{equation}

Consider the effect of (405) in the case of the hydrogen atom. 
\index{hydrogen~atom!vacuum~polarization}%
The proton\endnote{
``proton'' replaces ``photon''}
is represented by the static charge-density $\rho_{\text{P}}(r)$, and this induces a charge in the vacuum whose density is
\begin{equation}
\rho_{\text{IN}}(r) = - \frac{\alpha}{15 \pi \mu^{2}} \nabla^{2} \rho_{\text{P}}(r)
\end{equation}

The electrostatic potential of the proton is thus $V(r) + V_{\text{IN}}(r)$, where 
\begin{align*}
\nabla^{2}V(r) &= - \rho_{\text{P}}(r) \\
\nabla^{2}V_{\text{IN}}(r) &= - \rho_{\text{IN}}(r)  = \frac{\alpha}{15 \pi \mu^2}  \nabla^{2} \rho_{\text{P}}(r)
\end{align*}
and hence
\begin{equation}
V_{\text{IN}}(r) = + \frac{\alpha}{15 \pi \mu^2} \rho_{\text{P}}(r)
\end{equation}
Thus for a point proton, the potential added to the Coulomb potential  by vacuum-polarization is
\index{Coulomb~potential}%
\begin{equation}
V_{\text{IN}}(r) = + \frac{\alpha e}{15 \pi \mu^2} \delta^{3}(\boldsymbol{r}) 
\end{equation}
And the change in the energy of a state of the hydrogen atom with wave-function $\psi(r)$ is
\begin{equation}
\Delta E_{P} = - \frac{\alpha e^{2}}{15 \pi \mu^{2}} | \psi(0) |^{2} = - \tfrac{1}{5} \left\{ \frac{e^{4} \hbar}{12 \pi^{2} m^{2} c^{3}} | \psi(0) |^2 \right\}
\end{equation}
This is just the same as the formula for the Lamb shift,  (284), with $(- 1/5)$ instead of the logarithm. 
\index{Lamb~shift}%
Thus it is a factor 40 smaller than the Lamb shift, and in the opposite direction, $-27$ megacycles in the total of 1062. Still the experiments are good enough to show that the effect is there all right. 

The result (409) was calculated many years ago by Uehling \cite{Uehling35} using old-fashioned methods. 
\index{Uehling, E. A.}%
\index{vacuum~polarization|)}%

\section*{IV. Field Theory Of Dirac And Maxwell Fields In Interaction}
\addcontentsline{toc}{section}{IV. Field Theory Of Dirac And Maxwell Fields In Interaction}

\subsection*{The Complete Relativistic Quantum Electrodynamics}
\addcontentsline{toc}{subsection}{The Complete Relativistic Quantum Electrodynamics}

We now take the combined system of Dirac and Maxwell fields in interaction, and make out of it a relativistic quantum theory using the methods we have already developed.
\index{Dirac!field!interacting~with~Maxwell~field}%
This will then be the complete theory of quantum electrodynamics, applicable to all problems in which electrons, positrons and photons are concerned.
\index{quantum~electrodynamics}%
We also include in the theory a classical Maxwell field, which acts on the electrons and positrons and represents the effects of external charges such as protons which may happen to be present.
\index{Maxwell!field!external~classical}%

Lagrangian  
\index{Lagrangian~density!quantum~electrodynamics}%
\begin{equation}
\mathscr{L} = \mathscr{L}_{D} + \mathscr{L}_{M} - ie \overline{\psi} \slashed{A} \psi - ie \overline{\psi} \slashed{A}^{e} \psi
\end{equation}
Here we use $A_{\nu}(x)$ for the Maxwell potential operators and $A_{\nu}^{e}(x)$ for the potentials of the classical external field.
\index{Maxwell!potentials}%

Field equations
\begin{equation}
\left\{ 
\sum_{\lambda} \gamma_{\lambda} 
\left\{ \frac{\partial}{\partial x_{\lambda}} + \frac{ie}{\hbar c} \left(A_{\lambda} + A_{\lambda}^{e} \right) \right\} + \mu 
\right\} \psi = 0
\end{equation}
\begin{equation}
\sum_{\lambda} \left\{ \frac{\partial}{\partial x_{\lambda}} - \frac{ie}{\hbar c} \left(A_{\lambda} + A_{\lambda}^{e} \right) \right\} \overline{\psi} \gamma_{\lambda} - \mu \overline{\psi} = 0
\end{equation}
\begin{equation}
\Box^{2} A_{\nu} = i e \overline{\psi} \gamma_{\nu} \psi
\end{equation}
(See (384).)\endnote{
The potentials had a subscript $\nu$ and the gamma matrices a subscript $\lambda$.}

These equations are \emph{non-linear}. And so there is no possibility of finding the general commutation rules of the field operators in closed form. We cannot find any solutions of the field equations, except for the solutions which are obtained as formal power series expansions in the coefficient $e$ which multiplies the non-linear interaction terms. It is thus a basic limitation of the theory, that it is in its nature a perturbation theory stating from the non-interacting fields as an unperturbed system. Even to write down the general commutation laws of the fields, it is necessary to use a perturbation theory of this kind.

Since the perturbation theory treatment is forced on us from the beginning, it is convenient not to set up the theory in the Heisenberg representation but to use the interaction representation. 
\index{interaction~representation|(}%
The IR is just designed for a perturbation theory in which the radiation interaction is treated as small. In the IR the commutation rules can be obtained simply in closed form, and so the theory can be set up with a minimum of trouble.

There are two different interaction representations which we can use. The first may be called the \emph{Bound Interaction Representation}. 
\index{bound~interaction~representation}%
It is exactly the representation we used in discussing the radiation from an atom in the non-relativistic theory. We take all field operators to have the time-dependence of the Heisenberg operators in the theory of the free Maxwell field and the electron field interacting with external potential, only the interaction between the two fields being omitted. 
\index{Maxwell!field}%
Thus the field equations in the BIR are (332), (333), and 
\begin{equation}
\Box^{2} A_{\nu} = 0
\end{equation}
The wave-function $\Phi(t)$ in the BIR satisfies the Schr\"{o}dinger equation 
\index{Schr\"{o}dinger!equation}%
\begin{equation}
i \hbar \frac{\partial \Phi}{\partial t} = H_{R}(t)  \Phi
\end{equation}
\begin{equation}
H_{R}(t) = ie \int \overline{\psi}(r, t) \slashed{A}(r, t) \psi(r, t) d^{\,3} \boldsymbol{r}
\end{equation}
This $H_{R}(t)$ is just the difference between the Hamiltonians of the theories with and without the radiation interaction. 
\index{Hamiltonian!with~and~without~radiation}%
Because no derivatives of field operators occur in $H_{R}$, the difference is just minus the difference between the corresponding Lagrangians, and so has the simple form given by (416). (Compare with (243).)

Using the BIR we can discuss the radiation of light by an atom, as it was done before, but now treating the atom relativistically. In fact this representation must be used, as soon as we wish to calculate effects accurately enough to require exact Dirac wave-functions  for the unperturbed atomic states. 
\index{Dirac!wave~function}%
However, the BIR is not convenient to use, because the commutation rules for the electron field are given by (346) and are still too complicated for all but the simplest problems. Therefore we shall use the BIR only when we are compelled to, and then usually only in the final stages of a problem. Generally we can do the main part of the work, with the major calculations, in the second type of interaction representation.

\subsection*{Free Interaction Representation}
\addcontentsline{toc}{subsection}{Free Interaction Representation}
\hspace{3ex}Here we take all field operators $\overline{\psi}$, $\psi$ and $A_{\mu}$ to satisfy the free field equations (286) and (414). The commutation rules are then also given by the free-field formulae (203) and (298), (299).  The wave-function satisfies the Schr\"{o}dinger equation 
\index{Schr\"{o}dinger!equation}%
\begin{equation}
i \hbar \frac{\partial \Phi}{\partial t} = \left\{ H^{e}(t) + H_{R}(t) \right\} \Phi
\end{equation}

\begin{equation}
H^{e}(t) = ie \int \overline{\psi}(\boldsymbol{r}, t) \slashed{A}^{e}(\boldsymbol{r}, t) \psi(r, t) d^{\,3} \boldsymbol{r}
\end{equation}
with $H_{R}$ formally again given by (416). But here $H_{R}$ is not the same operator as in (415) because of the time-dependence of the $\overline{\psi}$ and $\psi$ in the two cases.

This FIR is the interaction representation normally used in quantum electrodynamics, and will be called henceforth the Interaction Representation or IR simply. 
\index{quantum~electrodynamics}%
\index{free~interaction~representation}%
It is very well suited to relativistic calculations because it makes the field-commutators and expectation values invariant functions. Thus the calculations can be explicitly and formally invariant, even when the potentials $A_{\nu}^{e}$ are given in a special Lorentz frame as in the hydrogen atom.
\index{Lorentz!frame}%

It was Schwinger and Feynman  who first discovered the importance of making calculations formally invariant when using a relativistic theory. 
\index{Schwinger, Julian}%
\index{Feynman, Richard~P.}%
They made this discovery in completely different ways, characteristically. Feynman simply found that the calculations become much easier and simpler when done in a way which did not conceal the invariance of the theory. This is still true, in fact the main reason why we can tackle now more difficult problems than we could 10 years ago is just that the calculations with the new methods are so much shorter. But the greater and essential advantage of the covariant calculations, pointed out by Schwinger, is that they enable the separation of finite observable effects from infinite renormalization terms to be made in a clear and unambiguous way. 
\index{renormalization!advantages~of~covariant~calculation}%
We had an example of this in the treatment we gave of vacuum polarization,  where the covariant type of calculation was used. 
\index{vacuum~polarization}%
The divergent term (382) could be separated clearly from (379), because of the way in which (379) depends formally on the vectors $e_{\nu}$ and $q_{\nu}$. If the calculation had been done in a non-covariant way we could not have used (381) as we did.

So now we shall apply the covariant methods, working in the IR, to solve a number of the standard problems of electrodynamics, in increasing order of difficulty.
\index{interaction~representation|)}%

%


\newpage

\pagestyle{fancy}
\fancyhead{}
\lhead{\emph{\MakeUppercase{Free Particle Scattering Problems}}}
\chead{}
\rhead{\thepage}
\lfoot{}
\cfoot{}
\rfoot{}

\chapter*{Free Particle Scattering Problems}
\addcontentsline{toc}{chapter}{Free Particle Scattering Problems}

In this extensive class of problems we are interested in calculating the over-all transition matrix element $M$ between an initial state $A$ and a final state $B$, $A$ and $B$ being specified by assigning the spins and momenta of the free particles which are present in these states. The scattering process is supposed to occur as follows: The free particles which are specified by state $A$ in the remote past, converge and interact, and other free particles emerge or are created in the interaction and finally constitute the state $B$ in the remote future. 
\index{create}%
We wish to calculate the matrix element $M$ for this process, without studying the equations of motion or investigating the behavior of the system at intermediate times while the interaction is in progress.

The unperturbed states $A$ and $B$ are supposed to be states of free particles without interaction and are therefore represented by constant state-vectors $\Phi_{A}$ and $\Phi_{B}$ in the interaction representation. The actual initial and final states in a scattering problem will consist of particles each having a self-field with which it continues to interact even in the remote future and past, hence $\Phi_{A}$ and $\Phi_{B}$ do not accurately represent the initial and final states. However, so long as we are using perturbation theory and not including the higher order effects arising form the self-fields of the particles, it is consistent to use constant $\Phi_{A}$ and $\Phi_{B}$ representing bare particles without radiation interaction. Even when self-field effects are considered, it turns out that the bare-particle $\Phi_{A}$ and $\Phi_{B}$ can still be used, although in this case some careful justification for it is needed. 

The matrix element $M$ is
\begin{equation}
M = (\Phi_{B}^{*} U \Phi_{A})
\end{equation}
Here $U\Phi_{A}$ is the state which is obtained at $t = + \infty$ by solving the equation of motion (417) with the initial condition $\Phi = \Phi_{A}$ at $t = -\infty$. $U$ can easily be written down as a perturbation expansion in the operators $H^{e}$ and $H_{R}$, 
\begin{equation}
\begin{split}
U = 1 &+ \left(- \frac{i}{\hbar} \right) \int_{-\infty}^{\infty} dt_{1} \, \left\{H^{e}(t_{1}) + H_{R}(t_{1}) \right\} \\
          &+ \left(- \frac{i}{\hbar} \right)^2  \int_{-\infty}^{\infty} dt_{1} \, \int_{-\infty}^{\infty} dt_{2} \,  \left\{H^{e}(t_{1}) + H_{R}(t_{1}) \right\}  \left\{H^{e}(t_{2}) + H_{R}(t_{2}) \right\}  + \dots
\end{split}
\end{equation}
\begin{equation}
= \sum_{n = 0}^{\infty} \left(- \frac{i}{\hbar} \right)^{n} \frac{1}{n!} \int_{-\infty}^{\infty} dt_{1} \, \dots  \int_{-\infty}^{\infty} dt_{n} \, P\left\{ \left\{H^{e}(t_{1}) + H_{R} (t_{1})\right\} \dots  \left\{H^{e}(t_{n}) + H_{R} (t_{n})\right\} \right\}
\end{equation}
Here the $P$ signifies a chronological product,  the factors in it being multiplied not in the order in which they are written but in the order of the times $t_{1}, t_{2}, \dots, t_{n}$, the factors with later times standing to the left of those with earlier times. 
\index{chronological~product}%
This accounts for the factor $1/n!$ after we change all the limits to cover the whole range from $-\infty$ to $+\infty$. The operator $U$ is generally referred to as ``the S-matrix''. 
\index{S@\emph{S}-matrix}%

Before discussing the general analysis of the series expansion (421), we shall use it to solve some standard problems.

\section*{A. M\o ller Scattering of Two Electrons} 
\addcontentsline{toc}{section}{A. M\o ller Scattering of Two Electrons}
\index{Moller@M\o ller~scattering}%

We have in the initial state $A$ two electrons in states $(p_{1}u_{1})(p_{2}u_{2})$ and in the final state $B$ two electrons $(p_{1}^{\prime}u_{1}^{\prime})(p_{2}^{\prime}u_{2}^{\prime})$. The electron $(p_{1}u_{1})$ is given by the one-particle wave-function
\begin{equation}
u_{1}e^{i p_{1} \cdot x}
\end{equation}
normalized by $(\overline{u}_{1}u_{1}) = 1$. With the wave function normalized in this way, (422) is just the matrix element of the operator $\psi(x)$ between the vacuum state and the state containing electron 1. We can see this from
\[
\psi(x) = \sum_{p, u} b_{p u} u e^{i p \cdot x}
\]
where $\{ b_{pu}, b^{*}_{p^{\prime} u^{\prime}} \} = \delta_{p p^{\prime}} \delta_{u u^{\prime}}$. Then $ (\Phi_{o}^{*}, \psi(x) \Phi_{pu}) = (\Phi_{o}^{*}, b_{pu} \Phi_{pu})ue^{i p \cdot x} = (\Phi_{o}^{*},\Phi_{o})ue^{i p \cdot x} = ue^{i p \cdot x} $. \\
Thus we consider the states 1, 2 and $1^{\prime}, \; 2^{\prime}$ as if they were discrete states, the $\psi$ operator being given by the expansion (337). It would also be possible to use the continuous-state expansion (306) for $\psi$, but then the normalization of the initial and final states would have to be considered afresh. Since we fixed the normalization (472) when we previously derived the M\o ller formula (144), we shall now stick to it. 
\index{Moller@M\o ller~scattering|(}%
\index{scattering!M\o ller|(}%

We shall calculate in the Born approximation as before, thus keeping only the term $n = 2$ in (421) which gives the matrix element $M$ proportional to $e^2$. In this problem the external potential $A^{e}$ is zero. The term $n = 2$ in (421) is 
\index{Born~approximation}%
\begin{equation}
U_{2} = \frac{+e^{2}}{2 \hbar^{2} c^{2}} \iint dx_{1} \, dx_{2} \, P \left\{ \overline{\psi}(x_{1}) \slashed{A}(x_{1}) \psi(x_{1}), \overline{\psi}(x_{2}) \slashed{A}(x_{2}) \psi(x_{2}) \right\}
\end{equation}
the integration going over all space-time. To obtain the matrix element $M = (\Phi^{*}_{B} U_{2} \Phi_{A})$ we only have to replace according to (377)
\begin{equation}
\psi(x_{i}) = u_{1} e^{i p_{1} \cdot x_{i}} b_{1} + u_{2} e^{i p_{2} \cdot x_{i}} b_{2} + u^{\prime} e^{i p^{\prime}_{1} \cdot x_{i}} b^{\prime}_{1} + u^{\prime}_{2} e^{i p^{\prime}_{2} \cdot x_{i}} b^{\prime}_{2} \quad \text{by} \quad  u_{1} e^{i p_{1} \cdot x_{i}} b_{1} + u_{2} e^{i p_{2} \cdot x_{i}} b_{2}
\end{equation}
and
\[
\overline{\psi}(x_{i}) = \overline{u}_{1} e^{-i p_{1} \cdot x_{i}} b^{*}_{1} + \overline{u}_{2} e^{-i p_{2} \cdot x_{i}} b^{*}_{2} + \overline{u}^{\prime}_{1} e^{-i p^{\prime}_{1} \cdot x_{i}} b^{*\prime}_{1} + \overline{u}^{\prime}_{2} e^{-i p^{\prime}_{2} \cdot x_{i}} b^{*\prime}_{2} \quad \text{by} \quad  \overline{u}^{\prime}_{1} e^{-i p^{\prime}_{1} \cdot x_{i}} b^{*\prime}_{1} + \overline{u}^{\prime}_{2} e^{-i p^{\prime}_{2} \cdot x_{i}} b^{*\prime}_{2} 
\]
because we only absorb 1 and 2 and only create 1$^{\prime}$ and 2$^{\prime}$. Then we want to pick out the coefficient of
\index{creation!pair}%
\begin{equation}
(b^{*\prime}_{1} b_{1})(b^{*\prime}_{2} b_{2})
\end{equation}
in the resulting expansion. There are no photons in the initial and final states and so the vacuum expectation value is taken for the Maxwell potential operators. 
\index{Maxwell!potentials}%
We thus find, taking account of the fact that the $b$ and $\overline{b}$ anticommute with each other as in the derivation of (358), the result 
\index{anticommute}%
\begin{equation} 
\begin{split}
M = \sum_{\mu , \lambda} \frac{e^{2}}{\hbar^{2} c^{2}} & \iint dx_{1} \, dx_{2} \, \left\{ \exp \left[ i(p_{1} - p^{\prime}_{1}) \cdot x_{1} + i  (p_{2} - p^{\prime}_{2}) \cdot x_{2} \right] (\overline{u}^{\prime}_{1} \gamma_{\lambda} u_{1})(\overline{u}^{\prime}_{2} \gamma_{\mu} u_{2}) \right. \\
& \left. -\exp \left[ i(p_{1} - p^{\prime}_{2} \cdot x_{1} +  i (p_{2} - p^{\prime}_{1}) \cdot x_{2} \right] (\overline{u}^{\prime}_{2} \gamma_{\lambda} u_{1})(\overline{u}^{\prime}_{1} \gamma_{\mu} u_{2}) \right\}  \Braket{P \{A_{\lambda}(x_{1}), A_{\mu}(x_{2}) \} }_{o}
\end{split}
\end{equation}

The expectation-value of the chronological product brings in an important new function, 
\index{chronological~product}%
\begin{equation}
\Braket{P \{A_{\lambda}(x_{1}), A_{\mu}(x_{2}) \} }_{o} = \tfrac{1}{2} \hbar c \, \delta_{\lambda \mu} \, D_{F}(x_{1} - x_{2})
\end{equation}
where $F$ stands for Feynman. Also called by Stueckelberg $D^{c}$, $c$ for causality.  \cite{Stueckelberg46} 
\index{Stueckelberg~$D^{c}$}%
\index{Stueckelberg, E. C. G.}%
\index{Feynman!propagator|see{Feynman, $D_{F},S_{F},\Delta_{F}$}}%
\index{Feynman!$D_{F}$|(}%

\subsection*{Properties of the $D_{F}$ Function}
\addcontentsline{toc}{subsection}{Properties of the $D_{F}$ Function}
\hspace*{3ex}Since 
\begin{equation}
P \{A_{\lambda}(x_{1}), A_{\mu}(x_{2}) \} = \tfrac{1}{2} \{A_{\lambda}(x_{1}), A_{\mu}(x_{2}) \} +  \tfrac{1}{2} \epsilon(x_{1} - x_{2}) \boldsymbol{[}\,A_{\lambda}(x_{1}), A_{\mu}(x_{2})\,\boldsymbol{]}
\end{equation}
we have by (203) and (205) 
\begin{align}
D_{F}(x) &= D^{(1)} (x) + i \epsilon (x) D(x) = \frac{1}{2 \pi^{2}} \left[\frac{1}{x^{2}} - i \pi \delta(x^2) \right] \notag \\
	      &= D^{(1)} (x) - i \left\{ D_{A}(x) + D_{R}(x) \right\} \notag \\
	      &= D^{(1)} (x) - 2i \overline{D}(x)
\end{align}	       
according to (228). Obviously $D_{F}$ is an even function. It has the property that asymptotically as $x_{0} \rightarrow \infty$ in the future, $D_{F} = 2iD^{+}$ contains only positive frequencies, while as $x_{0} \rightarrow -\infty$  in the past, $D_{F} = - 2iD^{-}$ contains only negative frequencies. See Fierz, \emph{Helv.\ Phys.\ Acta} \textbf{23} (1950) 731 for a full discussion.
\index{Fierz, Markus}%

Thus $D_{F}$ is the potential arising from a point source disturbance at the origin, when all the potential travelling out in the future represents particles created, and all the potential travelling in from\endnote{
``from'' replaces a second ``in''}
the past represents particles to be absorbed, all particles having positive energy. 
\index{create}%
It is thus the potential which maintains the correct causal time-sequence of events, and in this way it was discovered by Stueckelberg. 
\index{Stueckelberg, E. C. G.}%
But the definition (427) is easier to understand and  use. 

The momentum representation of $D_{F}$ is 
\index{momentum!representation}%
\begin{equation}
D_{F}(x) =  \frac{-2i}{(2 \pi)^{4}} \int_{F} e^{i k \cdot x} \, \frac{d^{\, 4} k}{k^{2}}
\end{equation}
The contour of integration here is along the real axis, below the pole at $k_{0} = - | \boldsymbol{k} |$ and above the pole at $k_{0} = + | \boldsymbol{k} |$, in the $k_{0}$ plane:
\begin{center}
\includegraphics{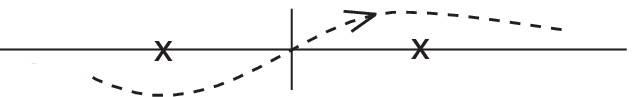}
\end{center}
We can see this using (429), (207), (208), and comparing (210), (226) and (209) one also knows that
\begin{center}
\includegraphics{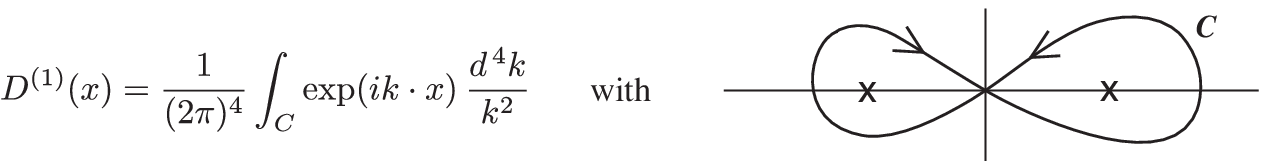}
\end{center}
(430) is always known as a ``Feynman integral''. We can also write 
\index{Feynman!contour~integral}%
\begin{equation}
D_{F}(x) = \frac{-2i}{(2 \pi)^{4}} \int  e^{i k \cdot x} \, \frac{d^{\, 4} k}{k^{2} - i\epsilon}
\end{equation}
Here the integration is along the real axis for all four components of  $k$. $\epsilon$ is a small positive real number and the limit $\epsilon \rightarrow 0$ is supposed to be taken in (431) after the integration is done. Before going to the limit, the $\epsilon$ term just shifts the poles off the real axis, the pole $| \boldsymbol{k} |$ going done and the pole $- | \boldsymbol{k} |$ going up, and so the integral is well-defined and non-singular.
\index{Feynman!$D_{F}$|)}%

\subsection*{The M\o ller Formula, conclusion} 
\addcontentsline{toc}{subsection}{The M\o ller Formula -- Conclusion}

Using (427) and (431) in (426), the integrations over $x_{1}$ and $x_{2}$ can be done at once, giving a $\delta$-function involving $k$, and the $k$ integration can then also be done immediately. The result is
\begin{equation}
M_{2} = \sum_{\lambda} \frac{-ie^{2}}{\hbar c} (2\pi)^{4} \delta^{4} (p_{1} + p_{2} - p^{\prime}_{1} - p^{\prime}_{2}) \left\{ \frac{(\overline{u}^{\prime}_{1}\gamma_{\lambda}u_{1})(\overline{u}^{\prime}_{2}\gamma_{\lambda}u_{2})}  {(p_{1} - p^{\prime}_{1})^{2} - i \epsilon} - \frac{(\overline{u}^{\prime}_{2}\gamma_{\lambda}u_{1})(\overline{u}^{\prime}_{1}\gamma_{\lambda}u_{2})}  {(p_{1} - p^{\prime}_{2})^{2} - i \epsilon} \right\}
\end{equation}
Now $p_{1}$ and $p^{\prime}_{1}$ are both electron momentum 4-vectors and hence $(p_{1} - p^{\prime}_{1})$ is a space-like vector and its square cannot be zero. Hence we can go directly to the limit by putting $\epsilon = 0$ in (432). This gives us the M\o ller formula (144) allowing for a difference in units in $p$ and $e$.

It is clear that the formula comes very directly out of (423), once we know the momentum expansion (431) of the $D_{F}$ function. And we shall find that things are equally simple in other free-particle scattering problems. 
\index{Moller@M\o ller~scattering|)}%
\index{scattering!M\o ller|)}%

\subsection*{Electron-Positron Scattering}
\addcontentsline{toc}{subsection}{Electron-Positron Scattering} 
\index{electron-positron!scattering|(}%

Exactly the same formula (432) also gives the matrix element of scattering of an electron by a positron. We must assume that the electron is initially in state 1 given by 
\begin{equation}
u_{1}e^{i p_{1} \cdot x}
\end{equation}
and finally in state $1^{\prime}$ by
\begin{equation}
u_{1}^{\prime}e^{i p_{1}^{\prime} \cdot x}
\end{equation}
But now the initial state of the positron is given by the wave-function
\begin{equation}
\overline{u}_{2}^{\prime}e^{-i p_{2}^{\prime} \cdot x}
\end{equation}
and the final state by
\begin{equation}
\overline{u}_{2}e^{-i p_{2} \cdot x}
\end{equation}
using the negative-energy electron wave-functions and \emph{not} the charge-conjugate functions to represent the positron. The correctness of (435) and (436) is clear since $b_{2}$ is the \emph{emission} operator and $b_{2}^{\prime}$ the \emph{absorption} operator for this positron.

The second term in (432) now represents not an ordinary exchange effect but a special short-range scattering due to a virtual annihilation of the positron and electron. 
\index{annihilation}%
This term has been observed experimentally by measuring the fine-structure constant of positronium. (M.\ Deutsch and E.\ Dulit, \emph{Phys.\ Rev.} \textbf{84} (1951) 601, (Nov.\ 1, 1951).) 
\index{positronium}%
\index{Deutsch, Martin}%
\index{Dulit, Everett}%
\index{electron-positron!scattering|)}%

\section*{B. Scattering of a Photon by an Electron. The Compton Effect. Klein-Nishina Formula.}                                                                                                                            
\addcontentsline{toc}{section}{B. Scattering of a Photon by an Electron -- The Compton Effect -- The Klein-Nishina Formula.}
\index{electron-photon~scattering|(}%
\index{Compton!effect|(}%
\index{scattering!Compton|(}%

Again we use the same operator $U_{2}$ given by (423). We only have to calculate its matrix element $M_{2}$ between an initial state $A$ and a final state $B$, where $A$ consists of an electron with wave-function
\begin{equation}
ue^{i p \cdot x}
\end{equation}
and a photon with potentials given by 
\begin{equation}
A_{\mu} = e_{\mu} e^{i k \cdot x}
\end{equation}
and $B$ consists of the electron in state
\begin{equation}
u^{\prime}e^{i p^{\prime} \cdot x}
\end{equation}
and the photon with potentials
\begin{equation}
A_{\mu} = e^{\prime}_{\mu} e^{i k^{\prime} \cdot x}
\end{equation}

The operator $A_{\lambda}(x_{1})$ appearing in (423) contains both photon emission and absorption components according to (211), and so does $A_{\mu}(x_{2})$. Hence the matrix element $M_{2}$ is a sum of contributions; we can either take $e_{\lambda} e^{i k \cdot x_{1}}$ from $A_{\lambda}(x_{1})$ and\endnote{
The factor $x_{2}$ had a superfluous superscript ``$^\prime$''.}
 $e_{\mu}^{\prime} e^{i k^{\prime} \cdot x_{2}}$ from $A_{\mu}(x_{2})$ or vice versa. Likewise the electron can be absorbed by $\psi(x_{2})$ or emitted again by $\overline{\psi}(x_{1})$, or vice versa. Thus altogether we find for $M_{2}$, after taking into account that the whole expression is symmetrical in $x_{1}$ and $x_{2}$, 
\begin{equation}
\begin{split}
M_{2} = \frac{e^{2}}{\hbar^{2}c^{2}} \sum_{\lambda , \mu , \alpha , \beta} \; \iint dx_{1} \, dx_{2}\; &\left\{ \exp(i p \cdot x_{2} - i p^{\prime} \cdot x_{1}) \left(\overline{u}_{1}^{\prime} \gamma_{\lambda}\right)_{\alpha} \Braket{\epsilon(x_{1} - x_{2}) P \left\{\psi_{\alpha}(x_{1}), \overline{\psi}_{\beta}(x_{2}) \right\}}_{o} \left( \gamma_{\mu} u\right)_{\beta} \right\} \times \\
          & \left\{e_{\lambda} e^{\prime}_{\mu} \exp(i k \cdot x_{1} - i k^{\prime} \cdot x_{2}) + e_{\mu} e^{\prime}_{\lambda} \exp(i k \cdot x_{2} - i k^{\prime} \cdot x_{1}) \right\}
\end{split}
\end{equation}
Note that with anticommuting fields the expression $\epsilon(x_{1} - x_{2}) P \left\{\psi_{\alpha}(x_{1}), \overline{\psi}_{\beta}(x_{2}) \right\} $ is a relativistic invariant whereas the $P$-product by itself is not. Thus in analogy with (427) we write 
\index{anticommute}%
\index{Feynman!$S_{F}$|(}%
\begin{equation}
\Braket {\epsilon(x_{1} - x_{2}) P \left\{\psi_{\alpha}(x_{1}), \overline{\psi}_{\beta}(x_{2}) \right\} }_{o} = - \tfrac{1}{2} S_{F \alpha \beta} (x_{1} - x_{2})
\end{equation}
where $S_{F}$ is a new invariant function. Since
\begin{equation}
\epsilon(x_{2} - x_{1}) P \left\{ \psi_{\alpha}(x_{1}), \overline{\psi}_{\beta}(x_{2}) \right\} = \tfrac{1}{2} \boldsymbol{[}\, \psi_{\alpha}(x_{1}), \overline{\psi}_{\beta}(x_{2}) \, \boldsymbol{]} + \tfrac{1}{2} \epsilon(x_{2} - x_{1})  \left\{ \psi_{\alpha}(x_{1}), \overline{\psi}_{\beta}(x_{2}) \right\}
\end{equation}
we have by (299) and (329)
\begin{equation}
S_{F}(x) = S^{(1)} + i \epsilon(x) S(x)
\end{equation}
in exact analogy to (429). We write also 
\index{Feynman!$\Delta_{F}$|(}%
\begin{equation}
S_{F}(x) = \sum_{\lambda} \left( \gamma_{\lambda} \frac{\partial}{\partial x_{\lambda}} - \mu \right) \Delta_{F}(x)
\end{equation}
and we find from (444) the momentum representations 
\index{momentum!representation}%
\begin{equation}
\Delta_{F}(x) = \frac{-2i}{(2 \pi)^{4}} \int e^{i k \cdot x}\frac{d^{\,4}k}{k^{2} + \mu^{2} - i \epsilon}
\end{equation} 
\index{Feynman!$\Delta_{F}$|)}%
\begin{equation}
S_{F}(x) = \frac{2}{(2 \pi)^{4}} \int e^{i k \cdot x} \left( \frac{\slashed{k} + i \mu}{k^{2} + \mu^{2} - i \epsilon} \right)\, d^{\,4}k
\end{equation}
We can also write (447) conveniently as 
\begin{equation}
S_{F}(x) = \frac{2}{(2 \pi)^{4}} \int_{F} e^{i k \cdot x} \frac{1}{\slashed{k} - i \mu}\, d^{\,4}k
\end{equation}
Here the Dirac matrix in the denominator means that we have to multiply above and below by $(\slashed{k} + i \mu)$ in order to evaluate the integral. 
\index{Dirac!matrices!in~denominator}%
Thus (448) is not a real simplification of (447), only it saves writing. The Feynman integration in (448) is defined as a contour integral exactly as in (430). 
\index{Feynman!contour~integral}%
\index{Feynman!$S_{F}$|)}%

Now substituting from (448) into (441), just like for (432)
\begin{equation}
M_{2} = \frac{-e^{2}}{\hbar^{2}c^{2}} (2 \pi)^{4} \delta^{\,4}(p + k - p^{\prime} - k^{\prime})\, \overline{u}^{\prime} \left[ \slashed{e} \frac{1}{\slashed{p} - \slashed{k}^{\prime} - i \mu} \slashed{e}^{\prime} + \; \slashed{e}^{\prime} \!\frac{1}{\slashed{p} + \slashed{k} - i \mu} \slashed{e}  \right] u
\end{equation}
Again the denominators $(p - k)^{2} + \mu^{2}$ can never vanish so the $\epsilon$ can be put equal to zero in (447). In fact if the electron is initially at rest 
\[
(p - k^{\prime})^{2} + \mu^{2} = (p^{2} + \mu^{2}) + k^{\prime \, 2} - 2 p \cdot k^{\prime} = 2 p_{0} k_{0}^{\prime} = 2 \mu k_{0}^{\prime}
\]
and similarly
\begin{equation}
(p + k)^{2} + \mu^{2} = - 2 \mu k_{0}
\end{equation}
because $k^{\prime \, 2} = 0$ and $p^{2}= - \mu^{2}$; $\boldsymbol{p} = 0$ because the electron is at rest.
\begin{equation}
M_{2} = \frac{-e^{2}}{2\hbar^{2}c^{2} \mu} (2 \pi)^{4} \delta^{\,4}(p + k - p^{\prime} - k^{\prime}) \, \overline{u}^{\prime} \left[\frac{1}{k_{0}^{\prime}} \slashed{e} (\slashed{p} - \slashed{k}^{\prime} + i \mu)\slashed{e}^{\prime} - \frac{1}{k_{0}} \; \slashed{e}^{\prime}(\slashed{p} + \slashed{k} + i \mu) \slashed{e}  \right] u
\end{equation}
Now we can simplify (451) further. Since the photon is not polarized in time, $e_{4} = e_{4}^{\prime} = 0$; since the electron is at rest, $\boldsymbol{p} = 0$. Hence $e \cdot p = 0$, and thus $\slashed{e} \slashed{p} = - \slashed{p} \slashed{e} + 2 \mathbb{I} \, e \cdot p = - \slashed{p} \slashed{e}$, i.e. $\slashed{p}$ and $\slashed{e}$ anticommute. This plus the fact that $u$ is a spin state of momentum $\hbar p$, i.e.
\index{anticommute}%
\begin{equation}
(\slashed{p} - i\mu)u = 0
\end{equation}
mean that the term $\slashed{p} + i \mu$ in (451) can be omitted. Thus we get
\begin{equation}
M_{2} = \frac{e^{2}}{2 \hbar^{2} c^{2} \mu} (2 \pi)^{4} \delta^{\,4}(p + k - p^{\prime} - k^{\prime}) \, \overline{u}^{\prime} \left[ \slashed{e} \frac{\slashed{k}^{\prime}}{k_{0}^{\prime}} \slashed{e}^{\prime} + \slashed{e}^{\prime} \frac{\slashed{k}}{k_{0}} \slashed{e} \right] u
\end{equation}

\subsection*{Calculation of the Cross-Section}
\addcontentsline{toc}{subsection}{Calculation of  Cross Section}

We write as in (145)
\begin{equation}
M_{2} = K(2\pi)^{4} \delta^{\,4}(p + k - p^{\prime} - k^{\prime})
\end{equation}
Then the scattering probability per unit volume and per unit time for the single final state is as before 
\index{probability!scattering}%
\begin{equation}
w_{\delta} = c |K |^{2} (2\pi)^{4} \delta^{\,4}(p + k - p^{\prime} - k^{\prime})
\end{equation}
The number of final states for the electron is
\begin{equation}
\frac{1}{(2 \pi)^{3}} \left( \frac{mc^{2}}{E^{\prime}}  \right) \, dp_{1}^{\prime} \, dp_{2}^{\prime} \, dp_{3}^{\prime}
\end{equation}
The photon with potentials (440), when $e_{\mu}^{\prime}$ is a space-like vector having $(e_{\mu}^{\prime})^2 = 1$ is normalized to one particle per volume $\hbar c/2k_{0}^{\prime}$. This can be seen at once, comparing (440) with (211) and (214), and taking into account the difference of $(2 \pi)^{3}$ between the continuous and discrete normalizations. Hence the number of final states for the photon is 
\begin{equation}
\frac{1}{(2 \pi)^{3}} \left( \frac{\hbar c}{2 k_{0}^{\prime}} \right) \, dk_{1}^{\prime}\, dk_{2}^{\prime} \, dk_{3}^{\prime}
\end{equation}
The total transition probability is thus 
\index{probability!scattering}%
\begin{equation}
w = c | K |^{2} \frac{1}{(2 \pi)^{2}} \left( \frac{mc^{2}}{E^{\prime}}\; \frac{\hbar c}{2 k_{0}^{\prime}} \right) \frac{dk_{1}^{\prime}\, dk_{2}^{\prime} \, dk_{3}^{\prime}}{d(p_{0}^{\prime} + k_{0}^{\prime})}
\end{equation}
We write this as a probability for scattering the \emph{photon} with frequency $k_{0}^{\prime}$ into a unit solid angle $d\Omega$. 
\index{probability!scattering}%
\index{angle!solid}%
Then using momentum conservation, we have 
\index{momentum!conservation}%
\begin{equation}
\begin{split}
\frac{dp_{0}^{\prime}}{dk_{0}^{\prime}} = \, & \frac{p_{1}^{\prime}\, dp_{1}^{\prime} + \dots}{p_{0}^{\prime} \, dk_{0}^{\prime}} = - \frac{p_{1}^{\prime} \, dk_{1}^{\prime} + \dots}{p_{0}^{\prime}\, dk_{0}^{\prime}} = 
- \frac{p_{1}^{\prime} k_{1}^{\prime} + \dots + p_{3}^{\prime} k_{3}^{\prime}}{p_{0}^{\prime} k_{0}^{\prime}} \\
\phantom{k} \\
&\frac{dk_{0}^{\prime}}{d(p_{0}^{\prime} + k_{0}^{\prime})} = \frac{p_{0}^{\prime} k_{0}^{\prime}}{- (p^{\prime} \cdot k^{\prime})} = \frac{p_{0}^{\prime} k_{0}^{\prime}}{- p \cdot k} = \frac{p_{0}^{\prime} k_{0}^{\prime}}{p_{0}  k_{0}}
\end{split}
\end{equation}
Hence finally 
\begin{equation}
w = c | K |^{2} \frac{1}{(2 \pi)^{2}} \, \frac{\hbar c}{2 k_{0}} k_{0}^{\prime 2}\, d\Omega
\end{equation}
The differential cross-section  for scattering the photon of frequency $k_{0}$ into solid angle  $d\Omega$ is then 
\index{cross-section!differential,~photon}%
\index{angle!solid}%
\[
\sigma = \frac{w V_{1} V_{2}}{c}
\]
where $V_{1}$ is according to (149) the electron normalization volume $\dfrac{mc^2}{E} = 1$ and $V_{2}$ is the photon volume, $V_{2} = \dfrac{\hbar c}{2k_{0}}$. Thus
\begin{equation}
\sigma = \left( \frac{\hbar c}{4 \pi k_{0}} \right)^{2} |K|^{2} k_{0}^{\prime 2} d \Omega
\end{equation}
\begin{equation}
K = \frac{e^{2}}{2 \hbar^{2} c^{2} \mu} \; \overline{u}^{\prime} \left[ \slashed{e} \frac{\slashed{k}^{\prime}}{k_{0}^{\prime}} \slashed{e}^{\prime} +  \slashed{e}^{\prime} \frac{\slashed{k}}{k_{0}} \slashed{e} \right] u
\end{equation}
This gives the cross-section for a known electron spin in the initial and final states.

\subsection*{Sum over Spins}
\addcontentsline{toc}{subsection}{Sum Over Spins}

Experimentally we do not observe the electron spins. Hence we observe only the cross-section $\overline{\sigma}$ obtained by averaging $\sigma$ over the two spin states $u$ and summing over the two spin states $u^{\prime}$. The summing and averaging we do by the method of projection operators according to (109), (114). 
\index{projection~operator}%

\noindent Here are some rules for spurs and dagger operators:
\index{Dirac!matrices!spur~theorems}%
\index{spur~theorems}%
\[
\xi \; \text{is a Dirac matrix in general.}
\]
1)\hspace*{2ex}Spur $\left( \xi^{(1)} \xi^{(2)} \dots \,\xi^{(2k - 1)} \right) = 0$, i.e.the spur of an odd number of factors is 0.\\

\noindent 2)\hspace*{2ex}Spur $\left( \xi^{(1)} \xi^{(2)} \dots \, \xi^{(2k)} \right)$ =  Spur $ \left\{ P\left( \xi^{(1)} \xi^{(2)} \dots \, \xi^{(2k)} \right) \right\} $ where $P$ is any cyclic permutation. \\

This is clear, because any cyclic permutation consists of steps of the form 
\[
\left( \xi^{(1)} \xi^{(2)} \dots \, \xi^{(s)} \right) \xi^{(m)} \rightarrow \xi^{(m)}\left( \xi^{(1)} \xi^{(2)} \dots \, \xi^{(s)} \right)
\]
and for two square matrices\endnote{
$A$, $B$ added for clarity}
$A$, $B$
\[
\text{Spur}\, AB = \sum_{i, j} a_{ij} b_{ji} = \text{Spur}\, BA
\]
3)\hspace*{2ex}Spur  $\left( \xi^{(1)} \xi^{(2)} \dots \,\xi^{(2k - 1)} \xi^{(2k)} \right)$ =  Spur $\left( \xi^{(2k)} \xi^{(2k - 1)}  \dots \, \xi^{(2)} \xi^{(1)} \right)$ \\

\noindent To show this, it is enough to assume that all $\xi^{(i)}$ are different; by the commutation rules of the $\gamma$'s we can always reduce the product to this form. Then, since each inversion (of neighbors) brings in a minus sign, and since there are an even number of inversions, we get our rule immediately. \\

\noindent 4)\hspace*{2ex}$\slashed{a} \slashed{b} = - \slashed{b} \slashed{a} + 2\, \mathbb{I}\, (a \cdot b)$ \\

In particular
\[
\slashed{e} \slashed{e} = \mathbb{I} \, (e \cdot e) \qquad \qquad \slashed{e} \slashed{k} = - \slashed{k} \slashed{e} \qquad \qquad \slashed{e}^{\prime} 
\slashed{k}^{\prime} = - \slashed{k}^{\prime} \slashed{e}^{\prime}
\]

\noindent 5)\hspace*{2ex}By 2), one can cyclically permute a product of dagger operators without changing its spur. \\

\noindent Now we can proceed to evaluate the sum over spins. We have
\begin{align}
\tfrac{1}{2} \sum_{u} \sum_{u^{\prime}} | K |^{2} &= -\frac{e^{4}}{8 \hbar^{4} c^{4} \mu^{2}} \sum_{u} \sum_{u^{\prime}} \left\{ \overline{u}^{\prime}
 \left( \slashed{e} \frac{\slashed{k}^{\prime}}{k_{0}^{\prime}} \slashed{e}^{\prime} + \slashed{e}^{\prime} \frac{\slashed{k}}{k_{0}} \slashed{e} \right) u \right\}
 \left\{\overline{u}\left( \slashed{e} \frac{\slashed{k}}{k_{0}} \slashed{e}^{\prime} + \slashed{e}^{\prime} \frac{\slashed{k}^{\prime}}{k_{0}^{\prime}} \slashed{e} \right) u^{\prime} \right\} \notag \\
 &= \frac{e^{4}}{32 \hbar^{4} c^{4} \mu^{4}} \, \text{Spur} \, \left\{ ( \slashed{p} + i \mu) \left( \slashed{e} \frac{\slashed{k}}{k_{0}} \slashed{e}^{\prime} + \slashed{e}^{\prime} \frac{\slashed{k}^{\prime}}{k_{0}^{\prime}} \slashed{e} \right) (\slashed{p}^{\prime} + i \mu) \left( \slashed{e} \frac{\slashed{k}^{\prime}}{k_{0}^{\prime}} \slashed{e}^{\prime} + \slashed{e}^{\prime} \frac{\slashed{k}}{k_{0}} \slashed{e} \right) \right\}
 \end{align}
 Now $\dfrac{\slashed{k}}{k_{0}} = i \beta + \dfrac{k_{1}\gamma_{1} + k_{2}\gamma_{2} + k_{3}\gamma_{3}}{k_{0}} = i \beta + \gamma_{k}$ , say. \\

\noindent Similarly $\dfrac{\slashed{k}^{\prime}}{k_{0}^{\prime}} = i \beta + \gamma_{k^{\prime}}$. \\

\noindent Since $\slashed{p}$ anticommutes with $\slashed{e}$, $\slashed{e}^{\prime}$, $\gamma_{k}$ and $\gamma_{k^{\prime}}$, (compare with remark after (451)) we may write (463) in the form 
\index{anticommute}%
\[
\frac{e^{4}}{32 \hbar^{4} c^{4} \mu^{4}} \, \text{Spur} \, \left[ \left\{ \left( \slashed{e} \frac{\slashed{k}}{k_{0}} \slashed{e}^{\prime} + 
\slashed{e}^{\prime} \frac{\slashed{k}^{\prime}}{k_{0}^{\prime}} \slashed{e} \right) (i \mu - \slashed{p}) + 4 \mu (e \cdot e^{\prime})\right\}(\slashed{p}^{\prime} + i \mu) \left\{ \slashed{e} \frac{\slashed{k}^{\prime}}{k_{0}^{\prime}} \slashed{e}^{\prime} + \slashed{e}^{\prime} \frac{\slashed{k}}{k_{0}} \slashed{e}\right\} 
\right]
\]
because, using 4), and again compare with remark after (451),\endnote{
The words after ``because'', ``using 4) \dots (451)'' were added.}
\[
\slashed{p} \slashed{e} \frac{\slashed{k}}{k_{0}} \slashed{e}^{\prime} = - \slashed{e} \slashed{p} \frac{\slashed{k}}{k_{0}} \slashed{e}^{\prime} = + \slashed{e} \frac{\slashed{k}}{k_{0}} \slashed{p} \slashed{e}^{\prime} - 2 \slashed{e} \frac{k \cdot p}{k_{0}} \slashed{e}^{\prime} = - \slashed{e} \frac{\slashed{k}}{k_{0}} \slashed{e}^{\prime} \slashed{p} + 2 \mu\slashed{e} \slashed{e}^{\prime}
\]
and similarly\endnote{
In the second edition, the intermediate calculation was wrong; however, the conclusion was correct. It was rewritten up to Eq.\ (464). See the trace theorems Eq.\ (585) \emph{et seq.} Also, note the identity
\[
\text{Sp} \, (\slashed{a}_{1}\slashed{a}_{2}\slashed{a}_{3}\slashed{a}_{4}) = (a_{1} \bold{\cdot}\, a_{2})(a_{3} \bold{\cdot}\, a_{4})  - (a_{1} \bold{\cdot}\, a_{3})(a_{2} \bold{\cdot} \,a_{4}) + (a_{1} \bold{\cdot}\, a_{4})(a_{2} \bold{\cdot} \, a_{3})
\]}
\[
\slashed{p} \slashed{e}^{\prime} \frac{\slashed{k}^{\prime}}{k_{0}^{\prime}} \slashed{e} = - \slashed{e}^{\prime} \frac{\slashed{k}^{\prime}}{k_{0}^{\prime}} \slashed{e} \slashed{p} + 2 \mu \slashed{e}^{\prime} \slashed{e}
\]
so adding the terms gives
\[- \left( \slashed{e} \frac{\slashed{k}}{k_{0}} \slashed{e}^{\prime} + 
\slashed{e}^{\prime} \frac{\slashed{k}^{\prime}}{k_{0}^{\prime}} \slashed{e} \right)\slashed{p} +  2 \mu \{\slashed{e}, \slashed{e}^{\prime}\}
= - \left( \slashed{e} \frac{\slashed{k}}{k_{0}} \slashed{e}^{\prime} + 
\slashed{e}^{\prime} \frac{\slashed{k}^{\prime}}{k_{0}^{\prime}} \slashed{e} \right)\slashed{p} + 4 \mu (e \cdot e^{\prime})
\]
Now since $\slashed{k} \slashed{k} = \slashed{k}^{\prime}\slashed{k}^{\prime} = \slashed{p} \slashed{p} + \mu^{2} = 0$ (for the photons because they are on the lightcone, for the electron because $p^{2} = - \mu^{2}$), we get for $|K|^{2}$
\begin{equation}
\frac{e^{4}}{32 \hbar^{4} c^{4} \mu^{4}} \, \text{Spur} \left[ 4 \mu ( e \cdot e^{\prime}) (\slashed{p}^{\prime} + i \mu) (\slashed{e} \frac{\slashed{k}^{\prime}}{k_{0}^{\prime}} \slashed{e}^{\prime} + \slashed{e}^{\prime} \frac{\slashed{k}}{k_{0}} \slashed{e}) + (\slashed{e} \frac{\slashed{k}^{\prime}}{k_{0}^{\prime}} \slashed{e}^{\prime} \slashed{e} \frac{\slashed{k}}{k_{0}} \slashed{e}^{\prime} + 
 \slashed{e}^{\prime} \frac{\slashed{k}}{k_{0}} \slashed{e} \slashed{e}^{\prime} \frac{\slashed{k}^{\prime}}{k_{0}^{\prime}} \slashed{e})
 (i \mu - \slashed{p})(\slashed{k} - \slashed{k}^{\prime}) \right]
\end{equation}
because
\[
(i \mu - \slashed{p})(\slashed{p}^{\prime} + i \mu) = i \mu(\slashed{p}^{\prime} - \slashed{p}) - \mu^{2} - \slashed{p} \slashed{p}^{\prime} = i \mu(\slashed{k} - \slashed{k}^{\prime}) - \mu^{2} - \slashed{p}( \slashed{p} - \slashed{k}^{\prime} + \slashed{k}) = (i \mu - \slashed{p}) (\slashed{k} - \slashed{k}^{\prime})
\]

We consider the second part of (464) first:
\begin{align*}
\text{Spur} & \left[ \slashed{e} \frac{\slashed{k}^{\prime}}{k_{0}^{\prime}} \slashed{e}^{\prime} \slashed{e} \frac{\slashed{k}}{k_{0}} \slashed{e}^{\prime} \left\{ - \slashed{p} (\slashed{k} - \slashed{k}^{\prime}) -  (\slashed{k} - \slashed{k}^{\prime})\slashed{p} \right\} \right] = 2 p_{0} (k_{0} - k_{0}^{\prime}) \; \text{Spur} \left[  \slashed{e} \slashed{e}^{\prime}  \frac{\slashed{k}^{\prime}}{k_{0}^{\prime}} \frac{\slashed{k}}{k_{0}}  \slashed{e} \slashed{e}^{\prime} \right] \\
&= 2 \mu (k_{0} - k_{0}^{\prime}) \; \text{Spur} \left[ - 2(e \cdot e^{\prime}) \left(\frac{\slashed{k}^{\prime}}{k_{0}^{\prime}} \slashed{e}^{\prime} \frac{\slashed{k}}{k_{0}} \slashed{e} \right) - \frac{\slashed{k}^{\prime}}{k_{0}^{\prime}} \frac{\slashed{k}}{k_{0}} \right] \\
&= -8 \mu \frac{k_{0} - k_{0}^{\prime}}{k_{0} k_{0}^{\prime}}\, (k \cdot k^{\prime}) + 4 \mu (e \cdot e^{\prime}) \; \text{Spur} \left[ -\slashed{k} \slashed{e}
 \frac{\slashed{k}^{\prime}}{k_{0}^{\prime}} \slashed{e}^{\prime} + \slashed{k}^{\prime}\slashed{e}^{\prime} \frac{\slashed{k}}{k_{0}} \slashed{e} \right]
 \end{align*}
since
\[
\text{Sp} \left[\slashed{e} \slashed{e}^{\prime} \slashed{k}^{\prime} \slashed{k}\slashed{e}\slashed{e}^{\prime}\right] = \text{Sp} \left[\slashed{e}\slashed{e}^{\prime}\slashed{e} \slashed{e}^{\prime} \slashed{k}^{\prime} \slashed{k}\right] = \text{Sp} \left[- \slashed{e}\slashed{e}\slashed{e}^{\prime} \slashed{e}^{\prime} \slashed{k}^{\prime} \slashed{k} + 2\, \mathbb{I}\, \slashed{e}\slashed{e}^{\prime} \slashed{k}^{\prime} \slashed{k} (e \cdot e^{\prime}) \right] = \text{Sp} \left[ - \slashed{k}^{\prime} \slashed{k} - 2 (e \cdot e^{\prime}) \slashed{k}^{\prime} \slashed{e}^{\prime} \slashed{k} \slashed{e} \right]
\]
Hence altogether (464) becomes
\[
\frac{e^{4}}{32 \hbar^{4} c^{4} \mu^{4}} \left\{ - 8 \mu \, \frac{k_{0} - k_{0}^{\prime}}{k_{0}k_{0}^{\prime}}\,(k \cdot k^{\prime}) +  4 \mu (e \cdot e^{\prime}) \, 
\text{Spur} \left[ (\slashed{p} - \slashed{k}^{\prime}) \slashed{e} \frac{\slashed{k}^{\prime}}{k_{0}^{\prime}} \slashed{e}^{\prime} + 
(\slashed{p} + \slashed{k}) \slashed{e}^{\prime} \frac{\slashed{k}}{k_{0}} \slashed{e} \right] \right\}
\]
But $\slashed{k} \slashed{e} \slashed{k} = \slashed{k}^{\prime} \slashed{e}^{\prime} \slashed{k}^{\prime} = 0$ and 
\[
(p^{\prime} - p)^{2} = (k - k^{\prime})^{2} = k^{2} + k^{\prime 2} - 2 k \cdot k^{\prime} = - 2 k \cdot k^{\prime}
\]
\[
(p^{\prime} - p)^{2} =  p^{\prime 2} + p^{2}  - 2  p^{\prime} \cdot p = 2 \mu^{2} + 2 \mu p_{0}^{\prime} =  2 \mu^{2} + 2 \mu (- \mu + k_{0} - k_{0}^{\prime}) = 2 \mu (k_{0} - k_{0}^{\prime})
\]
Hence $k \cdot k^{\prime}  = - \mu (k_{0} - k_{0}^{\prime})$ and then
\begin{align*}
\text{Sp} &\left[ (\slashed{p} - \slashed{k}^{\prime}) \slashed{e} \frac{\slashed{k}^{\prime}}{k_{0}^{\prime}} \slashed{e}^{\prime} + 
(\slashed{p} + \slashed{k}) \slashed{e}^{\prime} \frac{\slashed{k}}{k_{0}} \slashed{e} \right] = \text{Sp} \left[\slashed{p} \slashed{e} \frac{\slashed{k}^{\prime}}{k_{0}^{\prime}} \slashed{e}^{\prime} -  \slashed{k}^{\prime} \slashed{e}^{\prime} \frac{\slashed{k}^{\prime}}{k_{0}^{\prime}} \slashed{e} + \slashed{p} \slashed{e}^{\prime} \frac{\slashed{k}}{k_{0}} \slashed{e} + \slashed{k}\slashed{e} \frac{\slashed{k}}{k_{0}} \slashed{e}^{\prime} \right] \\
& = \text{Sp} \left[\slashed{p} \slashed{e} \frac{\slashed{k}^{\prime}}{k_{0}^{\prime}} \slashed{e}^{\prime}+ \slashed{e} \frac{\slashed{k}}{k_{0}} \slashed{e}^{\prime}\slashed{p}\right] = \text{Sp} \left[\slashed{p} \slashed{e} \frac{\slashed{k}^{\prime}}{k_{0}^{\prime}} \slashed{e}^{\prime} + \slashed{p} \slashed{e} \frac{\slashed{k}}{k_{0}} \slashed{e}^{\prime} \right]
\end{align*}
and so (464) is
\begin{equation}
\frac{e^{4}}{8 \hbar^{4} c^{4} \mu^{4}} \left\{  \frac{2 \mu^{2}(k_{0} - k_{0}^{\prime})^{2}}{k_{0}k_{0}^{\prime}} +  2 \mu (e \cdot e^{\prime}) \, 
\text{Spur} \left[ \slashed{p}\slashed{e} i \beta \slashed{e}^{\prime} \right] \right\} = \frac{e^{4}}{4 \hbar^{4} c^{4} \mu^{2}} \left\{  \frac{(k_{0} - k_{0}^{\prime})^{2}}{k_{0}k_{0}^{\prime}} +  4 (e \cdot e^{\prime})^{2}\right\}
\end{equation}
Hence by (461) the cross-section averaged over electron spins is 
\[
\overline{\sigma} = \frac{e^{4}k_{0}^{\prime 2} \, d\Omega}{64 \pi^{2} \hbar^{2} c^{2} \mu^{2} k_{0}^{2}} \left\{  \frac{(k_{0} - k_{0}^{\prime})^{2}}{k_{0}k_{0}^{\prime}} +  4 (e \cdot e^{\prime})^{2}\right\}
\]
The classical electron radius is 
\index{electron!classical~radius}%
\[
r_{o} = \frac{e^{2}}{4\pi mc^{2}} = \frac{e^{2}}{4 \pi \hbar c \mu}
\]
Hence
\begin{equation}
\overline{\sigma} = \tfrac{1}{4} r_{o}^{2} \, d\Omega \left( \frac{k_{0}^{\prime}}{k_{0}} \right)^{2} \left\{  \frac{(k_{0} - k_{0}^{\prime})^{2}}{k_{0}k_{0}^{\prime}} +  4 \cos^{\,2} \!\phi \right\}
\end{equation}
where $\phi$ is the angle between the polarizations of the incident quantum $k_{0}$ and the emitted quantum $k_{0}^{\prime}$.
\index{angle}%
 
This is the famous \emph{Klein-Nishina formula}. 
\index{Klein-Nishina~formula}%
\index{Compton!effect|)}%
\index{scattering!Compton|)}%
\index{cross-section!Klein-Nishina}%

To put $\overline{\sigma}$ explicitly as a function of the scattering angle $\theta$,  we must use the equations\endnote{
The third equation lacked a subscript ``0'' on the variable $k_{0}^{\prime}$; the fourth equation lacked a superscript ``$^\prime$'' on the variable  $k_{0}^{\prime}$.}
\index{angle!scattering}%
\begin{align*}
&k \cdot k^{\prime} = - \mu (k_{0} - k_{0}^{\prime}) \\
& k \cdot k^{\prime} = | \boldsymbol{k} | |\boldsymbol{k}^{\prime} | \cos \theta - k_{0} k_{0}^{\prime} = k_{0} k_{0}^{\prime} (\cos \theta - 1) \\
& k_{0} k_{0}^{\prime} (1 - \cos \theta) = \mu (k_{0} - k_{0}^{\prime}) \\
& \frac{k_{0}}{k_{0}^{\prime}} = 1 + (1 - \cos \theta) \frac{k_{0}}{\mu}
\end{align*}
Put 
\[
\epsilon = \frac{k_{0}}{\mu} = \left( \frac{\text{Photon energy}}{mc^{2}} \right)
\]
Then
\begin{equation}
\overline{\sigma} =  \tfrac{1}{4} r_{o}^{2} \, d\Omega\;  \frac{\left(\dfrac{(1 - \cos \theta)^{2} \epsilon^{2}}{1 + \epsilon(1 - \cos \theta)} + 4 \cos^{\,2} \!\phi\right)}{\left[1 + \epsilon (1 - \cos \theta)\right]^{2} }
\end{equation}
Thus for large $\epsilon$ the scattered photons are mainly unpolarized and concentrated in the forward direction.

For small $\epsilon$ (non-relativistic problem) we have simply\endnote{
Eq.\ (468) lacked a label. The word ``simply'' was inserted.}
\begin{equation}
\overline{\sigma} = r_{o}^{2}  \cos^{\,2} \!\phi \, d\Omega
\end{equation}
the classical result. Summing over the two polarizations of the photon $k^{\prime}$ and averaging over all polarizations of $k$, this gives the cross-section for all polarizations
\begin{equation}
\overline{\overline{\sigma}} = \tfrac{1}{2} r_{o}^{2} (1 + \cos^{\,2}\! \theta) \, d\Omega
\end{equation}
We get this by evaluating $\tfrac{1}{2} \sum_{e} \sum_{e^{\prime}} (e \cdot e^{\prime})^{2}$. First we have to sum over the two polarization directions of photon $k^{\prime}$. This summation for three directions would give
\[
\sum (e \cdot e^{\prime})^{2} = e^{2} = 1
\]
Hence for the two directions perpendicular to $k^{\prime}$ we can write
\[
\sum_{e^{\prime}} (e \cdot e^{\prime})^{2} = 1 - (\boldsymbol{e} \cdot \Hat{\boldsymbol{k}}^{\prime})^2
\]
Now we perform the other summation over the two polarization directions of photon $k$, using the same argument. This gives
\[
\sum_{e} \sum_{e^{\prime}} (e \cdot e^{\prime})^{2} = \sum_{e} \left[ 1 - (\boldsymbol{e} \cdot \Hat{\boldsymbol{k}}^{\prime})^2 \right] = 2 - \left[  \Hat{\boldsymbol{k}}^{\prime\; 2} - ( \Hat{\boldsymbol{k}} \cdot \Hat{\boldsymbol{k}}^{\prime})^{2} \right] = 1 + \cos^{2} \! \theta
\]
This with the averaging factor of $\tfrac{1}{2}$ gives (469).

The total cross-section then is
\begin{equation}
\sigma = \tfrac{8}{3} \pi r_{o}^{2}
\end{equation}
This non-relativistic scattering given by (468) -- (470) is called \emph{Thomson scattering}.\endnote{
``Thomson'' replaces ``Thompson''.}
\index{Thomson~scattering}%
\index{electron-positron!scattering|)}%
\index{scattering!Thomson}%
\index{cross-section!Thomson~scattering}%

\section*{C. Two Quantum Pair Annihilation}
\addcontentsline{toc}{section}{C. Two-Quantum Pair Annihilation} 
\index{electron-positron!annihilation|(}%
\index{annihilation!pair}%

Consider a process by which an electron in the state $(p, u)$ and a positron associated with the wave-function (439) are both annihilated,  with the emission of two photons given by the potentials (438) and (440). 
\index{annihilate}%
The positron momentum-energy 4-vector is then $(- \hbar p^{\prime})$ so we write $p_{+} = - p^{\prime}$. The positron spinor in the charge-conjugate representation is $v = Cu^{\prime \,+}$.

This annihilation process will again be effected by the operator $U_{2}$ given by (423). And the matrix element for the transition is exactly as before given by an expression identical with (449) except that $k$ is now replaced by $-k$, namely
\index{annihilation}%
\begin{equation}
\begin{split}
M_{2} &= - \frac{e^{2}(2 \pi)^{4}}{\hbar^{2} c^{2}} \delta^{\,4}( p + p_{+} - k - k^{\prime})\; \overline{u}^{\prime} \left\{ \slashed{e} \frac{1}{\slashed{p} - \slashed{k}^{\prime} - i \mu} \slashed{e}^{\prime} + \slashed{e}^{\prime} \frac{1} {\slashed{p} - \slashed{k} - i \mu} \slashed{e} \right\} u \\
& = K (2\pi)^{4} \delta^{\,4}( p + p_{+} - k - k^{\prime})
\end{split}
\end{equation}

We consider the probability for this process for an electron and positron \emph{both at rest}.
\index{probability!annihilation}%
The result will then apply to the decay of a positronium atom, where the velocities are only of the order of $\alpha c$ and can be treated as zero to a good approximation. \\
Then
\begin{equation}
\begin{split}
p &= p_{+} = (0, 0, 0, i \mu) \\
k_{0} &= k_{0}^{\prime} = \mu
\end{split}
\end{equation}
As in (453) we have
\begin{equation}
K = \frac{e^2}{2 \hbar^{2}c^{2}\mu^{2}}\overline{u}^{\prime} (\slashed{e} \slashed{k}^{\prime} \slashed{e}^{\prime} + \slashed{e}^{\prime} \slashed{k} \slashed{e}) u
\end{equation}
The decay probability per unit volume per unit time into a solid angle $d\Omega$ for \emph{one} of the photons is 
\index{angle!scattering}%
\[
w = c|K|^{2} \frac{1}{(2 \pi)^{2}} \left( \frac{\hbar c}{2 \mu} \right)^{2} \frac{dk_{1}dk_{2} dk_{3}}{d(k_{0} + k_{0}^{\prime})}
\]
(because $k_{0} = \mu$ here) in analogy to (458). But now $d(k_{0} + k_{0}^{\prime}) = 2 dk_{0}^{\prime}$ and so 
\begin{equation}
w =  c|K|^{2} \frac{1}{(2 \pi)^{2}} \frac{1}{8} \hbar^{2} c^{2} \, d\Omega
\end{equation}

For \emph{parallel} polarizations, $e = e^{\prime}$, and 
\[
(\slashed{e} \slashed{k}^{\prime} \slashed{e}^{\prime} + \slashed{e}^{\prime} \slashed{k} \slashed{e}) = - (\slashed{k}^{\prime} + \slashed{k}) = - 2i \mu \beta
\]
But $\beta$ has zero matrix element between the spin-states $u$ and $u^{\prime}$, which are positive and negative-frequency states both of zero momentum. Hence for parallel polarizations
\begin{equation}
w = 0
\end{equation}
For \emph{perpendicular} polarizations, take coordinate axes 1 along $e$, 2 along $e^{\prime}$, and 3 along $k$. Then
\begin{equation}
(\slashed{e} \slashed{k}^{\prime} \slashed{e}^{\prime} + \slashed{e}^{\prime} \slashed{k} \slashed{e})  = \mu \{ \gamma_{1}(-\gamma_{3} + i \beta) \gamma_{2} + \gamma_{2}(\gamma_{3} + i \beta) \gamma_{1} \} = 2 \mu \gamma_{1} \gamma_{2} \gamma_{3}
\end{equation}
Hence for perpendicular polarizations\endnote{
The first spinor $u$ lacked a bar; $\overline{u}$ replaces $u$.}
\begin{align*}
\overline{u}[\slashed{e} \slashed{k}^{\prime} \slashed{e}^{\prime} + \slashed{e}^{\prime} \slashed{k} \slashed{e}]u &= 2 \mu v^{T}C \gamma_{4}\gamma_{1}\gamma_{2}\gamma_{3} u =  -2 \mu i v^{T} \gamma_4 \sigma_{2} u \\
&= 2 \mu v^{T} \left[ \begin{matrix} 
                                      0 & -1 & 0 & 0 \\
                                      1 & 0 & 0 & 0 \\
                                      0 & 0 & 0 & 1 \\
                                      0 & 0 & -1 & 0 \\
                                  \end{matrix} \right] u \\
                                  &= \begin{cases}
					\;\;\; 0 &\text{when spins $u$ and $v$ are parallel}\\
					2 \mu\sqrt{2}  &\text{when spins $u$ and $v$ are antiparallel}
				\end{cases}
\end{align*}
We get this latter result by observing that for antiparallel spins the initial wave function is 
\[
\psi = \frac{1}{\sqrt{2}} \left( \begin{matrix}
                                                1 \\
                                                0 \\
                                                \end{matrix} \right) 
                                     \left( \begin{matrix}
                                                0 \\
                                                1 \\
                                              \end{matrix} \right) 
                                         -
           \frac{1}{\sqrt{2}} \left( \begin{matrix}
                                                0 \\
                                                1 \\
                                                 \end{matrix} \right)
                                     \left( \begin{matrix}
                                               1 \\
                                                0 \\
                                              \end{matrix} \right) 
\]
(neglecting the ``small components'') and therefore 
\begin{align*}
v^{T} \left[ \begin{matrix}
                    0 & -1 \\
                    1 &  0 \\
                  \end{matrix}
          \right] u  &=  \frac{1}{\sqrt{2}} \; (0 \;  1) \left[ \begin{matrix} 
                                                                0 & -1 \\
                                                                1 &  0 \\
                                                               \end{matrix} \right]
                                                               \left( \begin{matrix}
                                                               1 \\
                                                               0 \\
                                                               \end{matrix} \right) 
                                                                   -
                                 \frac{1}{\sqrt{2}} \; (1 \;  0) \left[ \begin{matrix} 
                                                                0 & -1 \\
                                                                1 &  0 \\
                                                               \end{matrix} \right]
                                                               \left( \begin{matrix}
                                                               0 \\
                                                               1 \\
                                                               \end{matrix} \right) \\
                            &\phantom{0}\\
                            &= \frac{1}{\sqrt{2}} \left[ (1 \;  0)    \left( \begin{matrix}
                                                               1 \\
                                                               0 \\
                                                               \end{matrix} \right)  -   (0 \; -1)  \left( \begin{matrix}
                                                               0 \\
                                                               1 \\
                                                               \end{matrix} \right) \right] = \frac{2}{\sqrt{2}} = \sqrt{2}                     
\end{align*}

This is one place where charge-conjugate spinors are useful and necessary! \\
\index{spinors}%

Summarizing, we find for electron and positron with spins parallel, in triplet state, the 2-photon decay is forbidden. This selection rule is in fact an exact one for positronium in the ground state 1$s$ triplet. 
\index{positronium}%
Only 3-photon decay can occur and this makes the lifetime $\sim 1100$ times longer. For electron and positron in singlet state, the 2-photon decay occurs with the photons always polarized perpendicularly to each other. The probability for the decay, integrating (474) over a solid angle $2 \pi$ since the photons are indistinguishable, is 
\index{angle!solid}%
\begin{equation}
w = \frac{\hbar^{2} c^{3}}{16 \pi} 2 |K|^{2} = \frac{2 e^{4}}{\hbar^{2}c \mu^{2} 8 \pi} = 4 \pi c r_{o}^{2}
\end{equation}
Formula (477) is for electron and positron normalized to one particle per unit volume. If the density of the electron probability relative to the position of the positron is $\rho$, then the mean annihilation life-time will be\endnote{
The expression for $r_{o}$ was added.}
(the ``classical electron radius'' $r_{o} = e^{2}/(4\pi mc^{2})$ in Heaviside units)
\index{classical~electron~radius}%
\index{r@$r_{o}$|see{classical~electron~radius}}%
\index{annihilation!life-time}%
\begin{equation}
\tau = \frac{1}{4 \pi c r_{o}^{2} \rho}
\end{equation}
For the positronium 1$s$ singlet state
\index{Bohr~radius}%
\[
\rho = \frac{1}{8 \pi a_{o}^{3}}		\qquad \qquad a_{o} = \text{Bohr radius} \, = 137^{2} r_{o}
\]
\begin{equation}
\tau = 2 \times 137^{4} \times \frac{a_{o}}{c} = 2 \times 137^{5} \times \frac{\hbar}{mc^{2}} \approx 1.2 \times 10^{-10} \, \text{sec.}
\end{equation}

For slowly-moving electrons and positrons with relative velocity $v$, the annihilation cross-section according to (477) will be
\index{annihilation!cross-section}%
\index{cross-section!annihilation}%
\begin{equation}
4 \pi r_{o}^{2} \left(\frac{c}{v} \right) \qquad \qquad \text{singlet state,}
\end{equation}
proportional to $\dfrac{1}{v}$ just like neutron cross-sections  at low (thermal) energies. 
\index{cross-section!neutron}%
\index{electron-positron!annihilation|)}%

\section*{D. Bremsstrahlung and Pair Creation in the Coulomb Field of an Atom}
\addcontentsline{toc}{section}{D. Bremsstrahlung and Pair Creation in the Coulomb Field of an Atom}
\addtocontents{toc}{\protect\newpage}
 
We consider these two important processes together. Given an external potential $A_{\mu}^{e}$ representing the Coulomb field,  the processes are:\\  
\index{Coulomb~potential}%
\index{creation!pair}%

\emph{Bremsstrahlung}\\
\hspace*{7ex}Electron $(pu) \rightarrow$ Electron $(p^{\prime}u^{\prime})$ + Photon $(k^{\prime} e^{\prime})$ \\
                 
\emph{Pair-creation}\\
\hspace*{7ex} Photon $(k^{\prime} e^{\prime}) \rightarrow$ Electron $(pu)$ + Positron $(p_{+}^{\prime}u^{\prime})$ \\
    
\noindent We treat not only the photon $(ke)$ but also the potential $A^{e}$ in \emph{Born approximation.} This is valid so long as 
\index{Born~approximation}%
\begin{align}
\text{Potential energy} \, &\times \, \text{time of transit} \ll \hbar  \notag \\
\text{or}  \qquad \qquad \; \frac{Ze^{2}}{4\pi r} &\times \frac{r}{v} \ll \hbar \notag \\
\text{or}  \qquad \quad \! \frac{Ze^{2}}{4 \pi \hbar v} =&\; \frac{Z}{137} \frac{c}{v} \ll 1 \quad     
\end{align}
The treatment will only be good for \emph{relativistic velocities} $v \sim c$, and for \emph{lighter atoms} $Z \ll 137$. In fact for heavy atoms ($Z = 82$ for lead) and $v \sim c$ the error from the Born approximation is about 10\%. 
\index{Born~approximation}%

The processes arise in the Born approximation just from the term linear in $A_{\mu}$ and linear in $A_{\mu}^{e}$ in (421). This term is 
\index{Born~approximation}%
\begin{equation}
U_{1} = \frac{e^{2}}{\hbar^{2}c^{2}} \iint dx_{1}\, dx_{2}\, P \left\{ \overline{\psi}(x_{1}) \slashed{A}(x_{1}) \psi(x_{1}), \overline{\psi}(x_{2}) \slashed{A}^{e}(x_{2})\psi(x_{2}) \right\}
\end{equation}
The factor $\tfrac{1}{2}$ in (423) is now missing, otherwise everything is the same as before. We suppose $A_{\mu}^{e}(x_{2})$ is a superposition of Fourier components 
\index{Fourier!components}%
\begin{equation}
A_{\mu}^{e}(x_{2}) = \frac{1}{(2 \pi)^{4}} \int dk \, f(k) \,e_{\mu} e^{i k \cdot x_{2}}
\end{equation}
where $f(k)$ is a known function of $k$. For a static Coulomb field all the vectors $k$ appearing in (483) have zero 4$^{\text{th}}$ component, and $e_{\mu}$  is the constant vector $(0, 0, 0, i)$. 
\index{Coulomb~potential} %
We calculate the matrix element $M_{1}$ for bremsstrahlung  or pair creation with $A_{\mu}^{e}$ given by the pure Fourier component (438) ; the results are then to be afterwards superposed to give the actual potential according to (483). 
\index{bremsstrahlung}%
\index{Fourier!components}%

For bremsstrahlung the formula for $M_{1}$ is (449), identically the same as for the Compton effect, or integrating over the frequency $k$  
\index{bremsstrahlung}%
\index{Compton!effect}%
\begin{equation}
M_{1} = - \frac{e^{2}}{\hbar^{2}c^{2}} f(p^{\prime} + k^{\prime} - p) \, \overline{u}^{\prime} \left\{ \slashed{e}\frac{1}{\slashed{p} - \slashed{k}^{\prime} - i \mu}\slashed{e}^{\prime} +\, \slashed{e}^{\prime} \frac{1}{\slashed{p}^{\prime} + \slashed{k}^{\prime} - i \mu}  \slashed{e} \right\} u
\end{equation}
The factor 2 difference between (482) and (423) is just compensated by the fact that the photon $k^{\prime}e^{\prime}$ can be emitted by two operators $A_{\mu}$ in (423) and by only one in (482). The bremsstrahlung cross section is then calculated by squaring (484) and integrating over $k^{\prime}$ and $p^{\prime}$ with appropriate normalization factors. 
\index{bremsstrahlung}%
\index{cross-section!bremsstrahlung}%
For the details see Heitler's book \textsection 17.\endnote{
In the 3$^{\text{rd}}$ edition of Heitler's book, see \textsection 25.}
\index{Heitler, Walter}%

For pair creation the same formula (449) gives the matrix element $M_{1}$, allowing for the fact that the roles of the particles are now changed around so that electron $(pu)$ instead of $(p^{\prime}u^{\prime})$ is created etc. 
\index{creation!pair}%
\index{create}%
Thus
\begin{equation}
M_{1} = - \frac{e^{2}(2 \pi)^{4})}{\hbar^{2}c^{2}} \delta^{\,4}(k + k^{\prime} - p - p_{+}) \, \overline{u} \left\{ \slashed{e}\frac{1}{\slashed{k}^{\prime} - \slashed{p}_{+} - i \mu}\slashed{e}^{\prime} +\, \slashed{e}^{\prime} \frac{1}{\slashed{k} - \slashed{p}_{+}  - i \mu}  \slashed{e} \right\} u^{\prime}
\end{equation}
and integrating over the components of the potential
\begin{equation}
M_{1} = - \frac{e^{2}}{\hbar^{2}c^{2}}\, f(p + p_{+} - k^{\prime}) \, \overline{u} \left\{ \slashed{e}\frac{1}{\slashed{k}^{\prime} - \slashed{p}_{+} - i \mu}\slashed{e}^{\prime} +\, \slashed{e}^{\prime} \frac{1}{\slashed{p} - \slashed{k}^{\prime} - i \mu}  \slashed{e} \right\} u^{\prime}
\end{equation}
For the cross-section calculation see again Heitler, \textsection 20.\endnote{
In the 3$^{\text{rd}}$ edition of Heitler's book, see \textsection 26.}
\index{Heitler, Walter}%
 
%

 
\newpage

\pagestyle{fancy}
\fancyhead{}
\lhead{\emph{\MakeUppercase{General Theory of Free Particle Scattering}}}
\chead{}
\rhead{\thepage}
\lfoot{}
\cfoot{}
\rfoot{}

\chapter*{General Theory of Free Particle Scattering}
\addcontentsline{toc}{chapter}{General Theory of Free Particle Scattering}
\hspace{3ex}We have shown how (421) leads to matrix elements for standard scattering processes, from which cross-sections can be calculated. In each case we used only the term $n = 2$ in (421), which happened to be the lowest term giving a contribution to these processes. The higher terms $n = 4, \, 6, \, \dots$ will also give contributions to the matrix elements for these processes, contributions which are collectively called ``radiative corrections''. It turns out that the results without radiative corrections agree with the experimental scattering cross-sections in all cases. 
\index{cross-section!experimental}%
The experiments are never accurate to better than a few percent, and the radiative corrections are always smaller than the lowest-order terms by at least one power of $(e^{2}/4 \pi \hbar c) = (1/137)$. 
\index{radiative~corrections!scattering}%
Thus the study of the radiative corrections for scattering processes will not lead to any directly observable effects.

Nevertheless we shall work out a method of calculating the high-order terms of (421). This method turns out to be simplest and easiest to explain when we are discussing scattering problems. Incidentally we shall see what the radiative corrections to scattering look like, and we shall learn something about the nature of radiative corrections in general. Finally at the end we shall be able to use the method of calculation in order to find the radiative corrections to the motion of an electron in a hydrogen atom, which is the case in which these small effects can be accurately observed, but where the pure scattering theory is not directly applicable.
\index{hydrogen~atom!radiative~corrections~to~electron~motion}%
\index{radiative~corrections!electron~motion~in~hydrogen~atom}%

To avoid unnecessary complications we suppose there is no external field $A^{e}$. Problems in which there is an external field, so long as it can be treated in the Born approximation,  can always be simply related to problems without external field, just as the bremsstrahlung  matrix element (484) is related to the Compton effect (449). 
\index{Born~approximation}%
\index{bremsstrahlung}%
\index{Compton!effect}%
When there is no external field the matrix element for any scattering process is
\begin{equation}
M = \left( \Phi^{*}_{B} S \Phi_{A} \right)
\end{equation}
\begin{equation}
S = \sum_{n = 0}^{\infty} \left( \frac{e}{\hbar c} \right)^{n} \frac{1}{n!} \int \dots \int dx_{1} \dots dx_{n} P\left\{ \overline{\psi}\slashed{A}\psi(x_{1}), \dots, \overline{\psi}\slashed{A}\psi(x_{n}) \right\}
\end{equation}
The operators in (488) are field-operators of the interaction representation, the integrations over the points $x_{1}, \dots x_{n}$ extend over all space-time, and the initial and final states $A$ and $B$ are entirely arbitrary.
\index{interaction~representation}%

We wish to calculate the matrix element $M$ of $S$ for a particular scattering process, in which the states $A$ and $B$ are specified by enumerating the particles present in the two states. We must now take properly into account the fact that the particles in states $A$ and $B$, although well separated and no longer interacting with one another, are real particles interacting with their self-fields and with the vacuum-fluctuations of the fields in their neighborhood. Thus $A$ and $B$ are really time-dependent states in the interaction representation and will not be given by time-independent vectors $\Phi_{A}$ and $\Phi_{B}$, except in the lowest-order approximation. 
\index{interaction~representation}%
(See page 81.) Let $\Psi_{B}(t)$ be the actual time-dependent state-vector of the state $B$ in the IR. We are not interested in the dependence of $\Phi_{B}(t)$ on $t$. In an actual scattering experiment the particles in state B are observed in counters or photographic plates or cloud-chambers and the time of their arrival is not measured precisely. Therefore it is convenient to use for $B$ not the state-function $\Psi_{B}(t)$ but a state-function $\Phi_{B}$ which is \emph{by definition} the state-function describing a set of bare particles without radiation interaction, the bare particles having the same momenta and spins as the real particles in the state $B$. In the IR the state-function $\Phi_{B}$ is time-independent. The question is only, what is the connection between $\Psi_{B}(t)$ and $\Phi_{B}$?

Suppose $t_{B}$ to be a time so long in the future after the scattering process is over, that from $t_{B}$ to $+\infty$ the state $B$ consists of separated outward-traveling particles. Then the relation between $\Psi_{B}(t)$ and $\Phi_{B}$ is simple. We imagine a fictitious world in which the charge $e$ occurring in the radiation interaction decreases infinitely slowly (adiabatically)  from its actual value at time $t_{B}$ to zero at time ($+\infty$). In the fictitious world, the state $\Psi_{B}(t_{B})$ at time $t_{B}$ will grow into the bare-particle state $\Phi_{B}$ at time $+\infty$. 
\index{adiabatic}%
Thus
\begin{equation}
\Phi_{B} = \Omega_{2}(t_{B}) \Psi_{B}(t_{B})
\end{equation}
where\endnote{
The first potential lacked a slash; $\slashed{A}$ replaces $A$.}
\begin{equation}
\Omega_{2}(t_{B}) = \sum_{n = 0}^{\infty} \left( \frac{e}{\hbar c} \right)^{n} \frac{1}{n!} \int_{t_{B}}^{\infty}\! \dots \,\int_{t_{B}}^{\infty} dx_{1} \dots \, dx_{n} \, P\left\{ \overline{\psi}\slashed{A} \psi(x_{1}), \dots , \overline{\psi}\slashed{A} \psi(x_{n})\right\} g_{B}(t_{1}) \dots g_{B}(t_{n})
\end{equation}
and $g_{B}(t)$ is a function decreasing adiabatically from the value 1 at $t = t_{B}$ to zero at $t = \infty$. Similarly, when $t_{A}$ is a time so far in the past that the state $A$ consists of separated converging particles from $t = - \infty$ to $t = t_{A}$ we have
\index{adiabatic}%
\begin{equation}
\Psi_{A}(t_{A}) = \Omega_{1}(t_{A}) \Phi_{A}
\end{equation}
\begin{equation}
\Omega_{1}(t_{A}) = \sum_{n = 0}^{\infty} \left( \frac{e}{\hbar c} \right)^{n} \frac{1}{n!} \int_{-\infty}^{t_{A}}\! \dots \, \int_{-\infty}^{t_{A}} dx_{1} \dots \, dx_{n} \, P\left\{ \overline{\psi}\slashed{A} \psi(x_{1}), \dots , \overline{\psi}\slashed{A} \psi(x_{n})\right\} g_{A}(t_{1}) \dots g_{A}(t_{n})
\end{equation}
where $g_{A}(t)$ is a function increasing adiabatically from $t = -\infty$ to $t = t_{A}$.

The scattering matrix element between states $A$ and $B$ is given exactly by
\index{adiabatic}%
\begin{equation}
M = \left(\Psi^{*}_{B}(t_{B}) S_{t_{A}}^{t_{B}} \Psi_{A}(t_A) \right)
\end{equation}
\begin{equation}
S_{t_{A}}^{t_{B}} = \sum_{n = 0}^{\infty} \left( \frac{e}{\hbar c} \right)^{n} \frac{1}{n!} \int_{t_{A}}^{t_{B}}\! \dots \, \int_{t_{A}}^{t_{B}} P \left\{ \overline{\psi}\slashed{A} \psi(x_{1}), \dots , \overline{\psi}\slashed{A} \psi(x_{n})\right\}\, dx_{1} \dots \, dx_{n}
\end{equation}
Of course (493) is independent of the times $t_{A}$ and $t_{B}$. When  $t_{A}$ and $t_{B}$ are chosen so far in the past that (489) and (491) are satisfied, then (493) may be written in the form (487), where now 
\begin{equation}
\begin{split}
S &= \Omega_{2}(t_{B})S_{t_{A}}^{t_{B}}\Omega_{1}(t_{A})\\
   &= \sum_{n = 0}^{\infty} \left( \frac{e}{\hbar c} \right)^{n} \frac{1}{n!} \int_{-\infty}^{\infty}\! \dots \, \int_{-\infty}^{\infty} \, dx_{1} \dots \, dx_{n}\, P \left\{ \overline{\psi}\slashed{A} \psi(x_{1}), \dots , \overline{\psi}\slashed{A} \psi(x_{n})\right\} g(t_{1}) \dots g(t_{n})
\end{split}
\end{equation}
and $g(t)$ is a function increasing adiabatically from 0 to 1 for $-\infty < t < t_{A}$, equal to 1 for $t_{A} \le t \le t_{B}$, and decreasing adiabatically from 1 to 0 for $t_{A} < t < \infty$. 
\index{adiabatic}%
Thus we come to the important conclusion that formula (487) for the matrix element is correct, using the bare particle state-functions $\Phi_{A}$ and $\Phi_{B}$, provided that formula (488) for $S$ is interpreted by putting in the slowly varying cut-off function $g(t_{i})$ to make the integrals converge at $t_{i} = \pm \infty$. The cut-off functions are to be put in as they appear in (495), and then $S$ is defined as the limit to which (495) tends as the rate of variation of $g(t)$ is made infinitely slow.

The main practical effect of this limiting process in the definition of $S$ is to justify us in throwing away all terms in the integrals which oscillate finitely at $t_{i} = \pm \infty$. There are however certain cases in which the integral (488) is in a more serious way ambiguous due to bad convergence at $t_{i} = \pm \infty$. In these cases the cut-off functions have to be kept explicitly until a late stage of the calculations before going to the limit $g(t) = 1$. In all cases, if the limiting process is done in this way, the matrix element $M$ is obtained correctly and unambiguously.

The use of bare-particle wave-functions $\Phi_{A}$ and $\Phi_{B}$ in (487) is thus justified. This makes the calculation of $M$ in principle simple. It is only necessary to pick out from (488) the terms which contain the right combination of elementary emission and absorption operators to annihilate the particles in $A$ and to create  those in $B$. 
\index{annihilate}%
\index{create}%
We shall next describe a general method of systematically picking out those terms, which is due originally to Feynman. 
\index{Feynman!quantization}%
It was first explained in published form by G.\ C\. Wick, \emph{Phys.\ Rev.} \textbf{80} (1950) 268. 
\index{Wick, Gian~Carlo}%
Feynman and Wick have applied the method only to chronologically ordered products such as appear in (488). However the method applies in the same way to all products whether chronological or not, and we shall describe it in full generality.

\section*{The Reduction of an Operator to Normal Form}
\addcontentsline{toc}{section}{Reduction of an Operator to Normal Form}

Given any operator $\mathcal{O}$ which is a product of field operators, for example
\begin{equation}
\mathcal{O} = \overline{\psi}(x_{1})\slashed{A}(x_{1}) \psi(x_{1}) \overline{\psi}(x_{2}) \slashed{A}(x_{2}) \psi(x_{2})
\end{equation}
we want to pick out the matrix element of $\mathcal{O}$ for a transition between states $A$ and $B$ in which there is a known distribution of bare particles. 
\index{distribution}
For example $A$ may be a state with only one electron in state 1, and $B$ a state with one electron in state 2. Then we wish to pick out from (496) terms in which there appear the operators $b_{1}$ and $b^{*}_{2}$. In order to pick out all such terms systematically, we make a complete analysis of $\mathcal{O}$ into a sum of terms $\mathcal{O}_{n}$, each $\mathcal{O}_{n}$ being a sum of products of emission and absorption operators in which all emission operators stand to the left of all absorption operators.\endnote{
The phrase ``in which all emission operators stand to the left of all absorption operators'' was lost in the transition from the first edition to the second.}.
Any operator in which the emission and absorption operators are arranged in this special way is called ``Normal''\endnote{
Nowadays called ``normal order'', this ordering arises in connection with \emph{Wick's Theorem}:
\begin{center} (time ordered operators) = (normal ordered operators) + (all contractions) \end{center}
the contractions being equal to the propagators $S_{F}$, $D_{F}$, and so on.} 
\cite{Dyson51a}.
\index{normal~form|(}%
The $\mathcal{O}_{n}$ will be called the ``Normal constituents'' of $\mathcal{O}$.  Once $\mathcal{O}$ has been analyzed in this way, then we find the matrix element simply by taking the coefficient of $b^{*}_{2}b_{1}$ in the expansion $\sum \mathcal{O}_{n}$. No other term in the expansion can give any contribution to the matrix element. In $\mathcal{O}$ itself there might appear a term such as 
\begin{equation}
b^{*}_{2}b_{3}b^{*}_{3}b_{1}
\end{equation}
which would give a contribution to the matrix element, since the operator $b_{3}^{*}$ could create a particle in an intermediate state 3 which the operator $b_{3}$ would then annihilate.
\index{create}%
\index{annihilate}%
The expansion of $\mathcal{O}$ into normal constituents eliminates all terms such as (497) and replaces them by sums of normal products with numerical coefficients. Thus using the anticommutation rule for $b_{3}$ and $b^{*}_{3}$, (497) becomes replaced by 
\index{anticommute}%
\begin{equation}
Ab^{*}_{2}b_{1} - b^{*}_{2}b^{*}_{3}b_{3}b_{1}
\end{equation}
where $A$ is a numerical coefficient.  The second term in (498) gives no contribution to the matrix element. 

It is clear that by using the commutation rules of the operators in this way, every $\mathcal{O}$ can be written as a sum of normal products, and that the analysis leads to a unique expansion of $\mathcal{O}$. But we do not need to go through the tedious algebra of using the commutation rules, instead we can write down the normal constituents $\mathcal{O}_{n}$ directly by following simple rules.

First, we define the notation $N(\mathcal{Q})$, where $\mathcal{Q}$ is any product of emission and absorption operators, to be the product obtained by simply rearranging the factors of $\mathcal{Q}$ in a normal order, irrespective of the commutation rules, with a factor $(-1)$ if the rearrangement involves an odd permutation of the electron-positron operators. Similarly if $\mathcal{Q}$ is any sum of products, $N(\mathcal{Q})$ is defined by rearranging factors in each term of the sum in the same way. Thus we have for example (see (211))
\begin{equation}
N\!\left(A_{\lambda}(x) A_{\mu}(y) \right) = A_{\lambda}^{+}(x) A_{\mu}^{+}(y) +A_{\lambda}^{-}(x) A_{\mu}^{-}(y)  + A_{\lambda}^{-}(x) A_{\mu}^{+}(y)  + A_{\mu}^{+}(y) A_{\lambda}^{+}(x) 
\end{equation}
where $A_{\mu}^{+}(x)$ is the positive-frequency part of $A_{\mu}(x)$, i.e. the part containing absorption operators. Observe that the order of factors in the first two products in (499) is immaterial, only the third and fourth products have their order fixed by the condition of being normal. Similarly (see (306) and (309))
\begin{equation}
N\! \left( \psi_{\alpha}(x) \overline{\psi}_{\beta}(y)\right) = \psi_{\alpha}^{\,+}(x) \overline{\psi}_{\beta}^{\,+}(y) + \psi_{\alpha}^{\,-}(x) \overline{\psi}_{\beta}^{\,-}(y) + \psi_{\alpha}^{\,-}(x) \overline{\psi}_{\beta}^{\,+}(y) - \overline{\psi}_{\beta}^{\,-}(y) \psi_{\alpha}^{\,+}(x)
\end{equation}
With this notation, every product of \emph{two} field operators can immediately be written as a sum of normal constituents. Using the commutation rules (213) and the vacuum expectation values given by (219), (220), 
\begin{equation}
A_{\lambda}(x) A_{\mu}(y) = \Braket{A_{\lambda}(x) A_{\mu}(y)}_{o} + N\!\left(A_{\lambda}(x) A_{\mu}(y) \right)
\end{equation}
Similarly, using (310), (311), (324),
\begin{equation}
\psi_{\alpha}(x) \overline{\psi}_{\beta}(y) = \Braket{\psi_{\alpha}(x) \overline{\psi}_{\beta}(y)}_{o} + N\!\left(\psi_{\alpha}(x) \overline{\psi}_{\beta}(y) \right)
\end{equation}
And in fact for \emph{any} two field operators $\mathcal{P}, \mathcal{Q}$ we have
\begin{equation}
\mathcal{PQ} = \Braket{\mathcal{PQ}}_{o} + N\! \left( \mathcal{PQ} \right)
\end{equation}
provided $\mathcal{P}$ and $\mathcal{Q}$ are both linear in emission and absorption operators. The proof of (503) has in effect been done by proving (501) and (502), because these include all of the possible products of two boson or two fermion operators, and (503) is trivial for the product of one boson and one fermion operator, because they commute. Equations (501) -- (503) are operator identities, and hold whether or not the physical problem is directly concerned with the vacuum state of the fields. In fact we could if we wished define the ``vacuum expectation values'' as the functions appearing in (501) -- (503) and so avoid speaking about the vacuum state at all.

Next we shall state the generalization of the rule (503) to any product $\mathcal{O}$ of field operators, for example the $\mathcal{O}$ given by (496). We define a  ``factor pairing'' of $\mathcal{O}$ by picking out from $\mathcal{O}$ a certain even number of factors, either all or none or any intermediate number, and associating them together in pairs. For the product $\mathcal{PQ}$ there are only two factor-pairings, either we choose the pair $\mathcal{PQ}$ together or we choose no pairs at all. To each factor-pairing $n$ corresponds to a normal constituent $\mathcal{O}_{n}$ obtained as follows: For each pair of factors $\mathcal{PQ}$ which is paired in $n$, $\mathcal{O}_{n}$ contains the numerical factor $\Braket{\mathcal{PQ}}_{o}$, the order of $\mathcal{P}$ and $\mathcal{Q}$ being maintained as it was in $\mathcal{O}$. The unpaired factors $\mathcal{R}_{1}\mathcal{R}_{2} \dots \mathcal{R}_{m}$ in $\mathcal{O}$ appear in $\mathcal{O}_{n}$ rearranged in normal form. Thus the complete form of $\mathcal{O}_{n}$ is
\begin{equation}
\mathcal{O}_{n} = \pm \Braket{\mathcal{P}\mathcal{Q}}_{o}\Braket{\mathcal{P}^{\prime}\mathcal{Q}^{\prime}}_{o}\dots N\! \left(\mathcal{R}_{1}, \mathcal{R}_{2},  \dots \mathcal{R}_{m}\right)
\end{equation}
the sign in front being $+$ or $-$ according to the even or odd character of the permutation of the electron-positron operators from the order in which they are written in $\mathcal{O}$ to the order in which they are written in (504). With this definition of (504) of $\mathcal{O}_{n}$, we have the following theorem:
\[
\text{\emph{Every operator product $\mathcal{O}$ is identically equal to the sum of the $\mathcal{O}_{n}$ obtained from all its factor-pairings.}}
\]
This theorem gives the decomposition of $\mathcal{O}$ into its normal constituents. Equations (501) -- (503) are just special cases of it. Clearly non-zero $\mathcal{O}_{n}$ are only obtained when each pair of factors is either a $\overline{\psi}$ and a $\psi$ operator or two $A_{\mu}$ operators. We shall therefore suppose that the factor-pairings are always restricted in this way. 

The proof of the theorem is very simple, by induction on $m$, the number of factors in $\mathcal{O}$. The theorem is true when $m = 1$ or 2, so we need only prove it true for $m$ assuming it true for $m-2$. Let then $\mathcal{O}^{\prime}$ be a product of $(m-2)$ factors. First we show that the theorem is true for
\begin{equation}
\mathcal{O} = (\mathcal{PQ} \pm \mathcal{QP}) \mathcal{O}^{\prime}
\end{equation}
where $\mathcal{P}$ and $\mathcal{Q}$ are field operators and the plus sign appears only if $\mathcal{P}$ and $\mathcal{Q}$ are both electron-positron operators. In fact, the normal constituents of $\mathcal{PQO}^{\prime}$ and of $(\pm \mathcal{QPO^{\prime}})$ will be identical, so long as $\mathcal{P}$ and $\mathcal{Q}$ are not paired together. Therefore the sum of the normal constituents of $\mathcal{O}$ reduces to 
\begin{equation}
\sum \mathcal{O}_{n} = \left\{ \Braket{\mathcal{PQ}}_{o} \pm \Braket{\mathcal{QP}}_{o} \right\} \sum \mathcal{O}^{\prime}_{n}
\end{equation}
But $\sum \mathcal{O}^{\prime}_{n} = \mathcal{O}^{\prime}$, and 
\begin{equation}
\Braket{\mathcal{PQ}}_{o} \pm \Braket{\mathcal{QP}}_{o} = (\mathcal{PQ} \pm \mathcal{QP})
\end{equation}
this being a number and not an operator. Therefore (506) gives $\sum \mathcal{O}_{n} = 0$, and the theorem is proved for $\mathcal{O}$ given by (505). Next let $\mathcal{O}$ be any product of $m$ factors. Then by using the commutation relations we can write
\begin{equation}
\mathcal{O} = N(\mathcal{O}) + \Sigma
\end{equation}
where $\Sigma$ is a sum of terms of the form (505). The theorem is true for each term (505) and so is true for $\Sigma$. The theorem is trivially true for $N(\mathcal{O})$, because $\Braket{\mathcal{PQ}}_{o} = 0$ for every pair of factors  $\mathcal{P}$, $\mathcal{Q}$ in the order in which they occur in  $N(\mathcal{O})$, and so all normal constituents (504) of  $N(\mathcal{O})$ are zero except the constituent  $N(\mathcal{O})$ itself. Therefore the theorem holds for every  $\mathcal{O}$ given by (508), and this completes the proof. 
\index{normal~form|)}%

\section*{Feynman Graphs}
\addcontentsline{toc}{section}{Feynman Graphs}

We use a method of Feynman to enumerate the possible factor-pairings of $\mathcal{O}$. Each pairing is pictured in a diagram or graph $G$. 
\index{Feynman!graph}%
$G$ consists of a certain number of vertices with lines joining them. The vertices represent simply the different field-points at which the factors of $\mathcal{O}$ operate. Thus for $\mathcal{O}$ given by (496) each $G$ has the two vertices $x_{1}$, $x_{2}$. The lines in $G$ are either dotted, representing photon operators, or undotted, representing electron-positron operators. The rules for drawing these lines are the following: \\

\noindent \hspace*{3ex} 1)\hspace*{2ex}For each factor-pair $\overline{\psi}(x) \psi(y)$, an undotted line is drawn in $G$ running from $x$ to $y$, the direction being marked by an arrow in the line. \\

\noindent \hspace*{3ex} 2)\hspace*{2ex}For each unpaired factor $\overline{\psi}(x)$, an undotted line is drawn running from $x$ out of the diagram, the other end of the line being free and not being a vertex of $G$. \\

\noindent \hspace*{3ex} 3)\hspace*{2ex}For each unpaired factor $\psi(y)$, an undotted line is drawn running into $y$, the other end of the line being free. \\

\noindent \hspace*{3ex} 4)\hspace*{2ex}For each factor-pair $A_{\mu}(x)A_{\nu}(y)$, a dotted line joins $x$ and $y$. \\

\noindent \hspace*{3ex} 5)\hspace*{2ex}For unpaired factor $A_{\mu}(x)$, a dotted line is drawn with one end of $x$ and the other end free. \\

\noindent \hspace*{3ex} 6)\hspace*{2ex}Every undotted line has a definite direction marked by an arrow. A dotted line has no direction and no arrow. \\

In general, we must allow factor-pairings in which two operators at the same field-point are paired together. This will give a line in $G$ with both ends at the same point. However in the case of operators such as (496) or more generally (488), a pair of factors taken from the same point will always give rise to a factor 
\begin{equation}
\Braket{j_{\mu}(x)}_{o} = - iec \Braket{\overline{\psi}(x) \gamma_{\mu} \psi(x)}_{o}
\end{equation}
in the corresponding normal constituents (504). 
\index{normal~form}%
We saw in the discussion following (366) that the vacuum expectation value (509) is zero, the operators being IR operators. Therefore factor-pairings in which two factors at the same field-point are paired, in the analysis of quantum-electrodynamical operators such as (488), always give zero contributions. So we may add to our list of rules for the construction of $G$: \\

\noindent \hspace*{3ex} 7)\hspace*{2ex}Lines joining a point to itself are forbidden. \\

The possible factor pairings of (496) are then represented by the following $G$'s:
\vspace*{0.2in}
\begin{center}
\scalebox{1}{\includegraphics{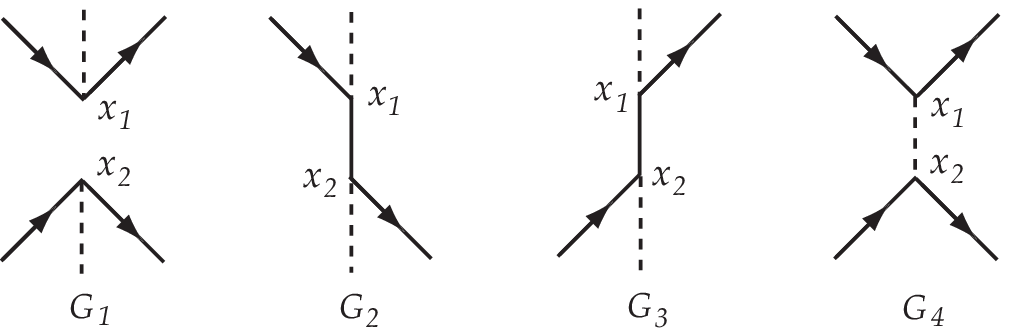}}
\end{center}

\vspace*{0.2in}
\begin{center}
\scalebox{1}{\includegraphics{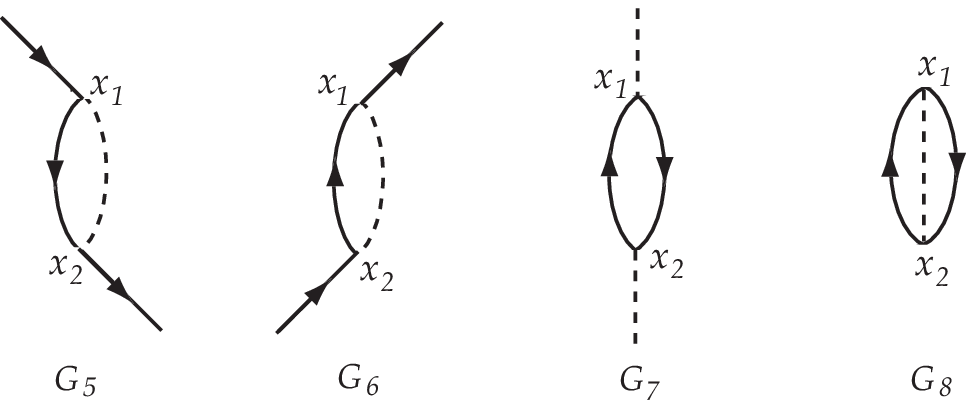}}
\end{center}

Corresponding to these $G$'s there will be just 8 normal constituents of $\mathcal{O}$, which we call $\mathcal{O}_{1} \dots \mathcal{O}_{8}$. 
\index{normal~form}%
These normal constituents are, in their respective order:
\begin{align*}
\mathcal{O}_{1} &= \pm N \!\left\{ \overline{\psi}(x_{1}) \slashed{A}(x_{1}) \psi(x_{1}) \overline{\psi}(x_{2}) \slashed{A}(x_{2}) \psi(x_{2}) \right\} \phantom{A^{A^{A}}}\\
\mathcal{O}_{2} &= \pm \Braket{ \overline{\psi}(x_{1}) \psi(x_{2})}_{o}  \, N \!\left\{ \slashed{A}(x_{1}) \psi(x_{1}) \overline{\psi}(x_{2}) \slashed{A}(x_{2})  \right\} \phantom{A^{A^{A}}}\\
\mathcal{O}_{3} &= \pm \Braket{ \overline{\psi}(x_{2}) \psi(x_{1})}_{o}  \, N \!\left\{\overline{\psi}(x_{1}) \slashed{A}(x_{1}) \slashed{A}(x_{2})\psi(x_{2})  \right\} \phantom{A^{A^{A}}}\\
\mathcal{O}_{4} &= \pm \Braket{\slashed{A}(x_{1}) \slashed{A}(x_{2})}_{o}  \, N \!\left\{\overline{\psi}(x_{1}) \psi(x_{1}) \overline{\psi}(x_{2}) \psi(x_{2})  \right\} \phantom{A^{A^{A}}}\\
\mathcal{O}_{5} &= \pm \Braket{ \overline{\psi}(x_{1}) \psi(x_{2})}_{o} \Braket{\slashed{A}(x_{1}) \slashed{A}(x_{2})}_{o} \, N \! \left\{\overline{\psi}(x_{2}) \psi(x_{1}) \right\} \phantom{A^{A^{A}}} \\
\mathcal{O}_{6} &= \pm \Braket{ \overline{\psi}(x_{2}) \psi(x_{1})}_{o} \Braket{\slashed{A}(x_{1}) \slashed{A}(x_{2})}_{o} \, N \! \left\{\overline{\psi}(x_{1}) \psi(x_{2}) \right\} \phantom{A^{A^{A}}} \\
\mathcal{O}_{7} &= \pm \Braket{ \overline{\psi}(x_{1}) \psi(x_{2})}_{o} \Braket{ \overline{\psi}(x_{2}) \psi(x_{1})}_{o} \, N \! \left\{ \slashed{A}(x_{1}) \slashed{A}(x_{2}) \right\} \phantom{A^{A^{A}}} \\
\mathcal{O}_{8} &= \pm \Braket{ \overline{\psi}(x_{1}) \psi(x_{2})}_{o} \Braket{ \overline{\psi}(x_{2}) \psi(x_{1})}_{o}  \Braket{\slashed{A}(x_{1}) \slashed{A}(x_{2})}_{o} \phantom{A^{A^{A}}}
\end{align*}
This type of process to which $\mathcal{O}_{i}$ gives matrix elements can be seen at once by looking at the \emph{external} lines of $G_{i}$, i.e. the lines which have one end free. Thus, for M\o ller scattering only $G_{4}$ contributes. 
\index{Moller@M\o ller~scattering}%
For Compton scattering only $G_{2}$ and $G_{3}$. 
\index{Compton!scattering}%
And for a transition from a one-electron atom to a one-electron state, which requires an operator of the form $b^{*}_{2}b_{1}$, and $G_{5}$ and $G_{6}$ will contribute.

A $\psi$ operator not only annihilates electrons but also creates positrons. 
\index{annihilate}%
\index{create}%
And a $\overline{\psi}$ not only creates electrons but annihilates positrons. 
\index{create}%
\index{annihilate}%
Thus an undotted external line with the arrow pointing in represents \emph{either} an electron in the initial state \emph{or} a positron in the final state. And an undotted external line with the arrow pointing out represents either an outgoing electron or an incoming positron. Dotted external lines likewise represent a photon either in the initial or the final state, because $A_{\mu}$ can both annihilate and create a photon. 
\index{annihilate}%
\index{create}%
Thus $G_{4}$ will give not only electron-electron scattering, but also electron-positron scattering. $G_{2}$ and $G_{3}$ give not only the Compton effect,  but also two-quantum annihilation of a positron-electron pair, and also the inverse process of pair-creation  by the collision of two photons. 
\index{Compton!effect}%
\index{annihilation!pair}%
\index{creation!pair}%

We have introduced the Feynman graphs simply as a convenient pictorial way of visualizing the analysis of an operator into its normal constituents. 
\index{Feynman!graph}%
\index{normal~form}%
The graphs are just diagrams drawn on the paper. But according to Feynman\endnote{
In Schwinger's QED anthology.}
, ``Space-time Approach to Quantum Electrodynamics'', \emph{Phys.\ Rev.} \textbf{76} (1949) 769, 
\index{quantum~electrodynamics}%
the graphs are more than this. He regards the graphs as a picture of an actual process which is occurring physically in space-time. Thus $G_{2}$ represents an electron and a photon coming together and interacting at the space-time point $x_{1}$, where the photon is absorbed; then the electron propagates through space-time along the line $x_{1}x_{2}$, until at $x_{2}$ it emits a photon, and the electron and photon then travel out along the outgoing lines at $x_{2}$. According to Feynman an internal undotted line running from $x_{1}$ to $x_{2}$ represents an electron propagating from $x_{1}$ to $x_{2}$ if the time $x_{2}$ is later than the time $x_{1}$, and represents a positron propagating from $x_{2}$ to $x_{1}$ if the time $x_{2}$ is earlier. In this sense a positron can be regarded as precisely equivalent to an electron which happens to propagate itself backwards instead of forwards in time.

The space-time picture of Feynman is a perfectly consistent and reasonable one. It gives a correct account of everything that happens, including pair creation and annihilation and all the other phenomena connected with positrons. It is in fact equivalent mathematically to the field-theory treatment we follow in this course.
\index{Feynman!quantization}%
\index{creation!pair}%
\index{annihilation!pair}%

The disadvantage of the Feynman theory is that it is constructed as a particle theory. 
\index{Feynman!quantization}%
The fact that there are many particles, indistinguishable from each other and obeying quantum statistics, has to be put into the theory as a special assumption. And the equations of motion of the particles become quite complicated when interactions between several particles, not to mention vacuum-polarization effects, are included. Thus the logical basis of the Feynman theory is much less simple than that of the field theory, where everything follows from general principles once the form of the Lagrangian is chosen.

In this course we follow the pedestrian route of logical development, starting from the general principles of quantizations applied to covariant field equations, and deriving from these principles first the existence of particles and later the results of the Feynman theory. Feynman by the use of imagination and intuition was able to build a correct theory and get the right answers to problems much quicker than we can. It is safer and better for us to use the Feynman space-time pictures not as the basis for our calculations but only as a help in visualizing the formulae which we derive rigorously from the field-theory. In this way we have the advantages of the Feynman theory, its concreteness and its simplification of calculations, without its logical disadvantages.

\section*{Feynman Rules of Calculation}
\addcontentsline{toc}{section}{Feynman Rules of Calculation}

The Feynman rules  of calculation arise when we analyze into normal  constituents a \emph{chronologically ordered} operator such as (488).
\index{Feynman!rules|(}%
\index{normal~form}%
\index{chronological~product}%
In this case the vacuum expectation values in (504) are always taken for pairs of operators which are already chronologically ordered. Hence the numerical factors in (504) are all either 
\begin{equation}
\Braket{P \!\left(A_{\lambda}(x), A_{\mu}(y) \right) }_{o} = \tfrac{1}{2}\, \hbar c\,  D_{F}(x - y) \, \delta_{\lambda \mu}
\end{equation}
or
\begin{equation}
\epsilon(x - y) \Braket{ P\! \left( \psi_{\alpha}(x), \overline{\psi}_{\beta}(y) \right) }_{o} = - \tfrac{1}{2} \, S_{F \alpha \beta} (x - y)
\end{equation}
using (427) and (442). The factor $\epsilon$ is put into (511) so that the $\pm$ sign still characterizes the permutation of electron-positron operators in going from the order \emph{as written} in (504) to the order as written in $\mathcal{O}$. For the same reason, we shall follow Wick and generally use for chronological products the notation
\index{Wick, Gian~Carlo}%
\index{chronological~product}%
\begin{equation}
T\!\left( \mathcal{R}_{1} \mathcal{R}_{2} \dots \mathcal{R}_{n} \right) = \pm P\!\left( \mathcal{R}_{1} \mathcal{R}_{2} \dots \mathcal{R}_{n} \right)
\end{equation}
where the sign is plus or minus according to the even or odd character of the permutation of electron-positron operators involved in going from the written order to the chronological order in (512). Hence in particular we have
\begin{equation}
\begin{split}
T\! \left( A_{\lambda}(x), A_{\mu}(y) \right\} &= P\! \left\{ A_{\lambda}(x), A_{\mu}(y) \right) \\
T\! \left( \psi_{\alpha}(x), \overline{\psi}_{\beta}(y) \right) &= \epsilon(x - y) \, P\! \left( \psi_{\alpha}(x), \overline{\psi}_{\beta}(y) \right)
\end{split}
\end{equation}
and for every set of field operators $ \mathcal{R}_{1} \mathcal{R}_{2} \dots \mathcal{R}_{n}$ the quantity (512) is a relativistic invariant although the $P$-product by itself is not. In (488) itself the $P$-product may be written as a $T$-product, the sign in (512) in this case always being plus.

The rules for writing down the normal constituents of (488) are therefore extremely simple. We are generally only interested in those normal constituents which give matrix elements for some specified type of scattering process. 
\index{normal~form}%
Then the rules are \\

\noindent \hspace*{3ex}1)\hspace*{1ex}Draw all the graphs which have the right set of external lines corresponding to the particles absorbed and emitted in the process considered. Each graph $G$ will have the same external lines, but the number of vertices and of internal lines will vary from graph to graph. We shall always calculate only up to some definite order $N$ in the series (488), and so we draw only graphs with not more than $N$ vertices. The total number of such graphs is finite. Each vertex in each graph must have precisely 3 lines ending at it, one incoming electron line, one outgoing electron line, and one photon line. \\

\noindent \hspace*{3ex}2)\hspace*{1ex}To each graph $G$ with $n$ vertices corresponds one normal constituent $S_{G}$ of $S$. \\

\noindent \hspace*{3ex}3)\hspace*{1ex}Choosing a particular $G$, write down the $n^{\text{th}}$ term $S_{n}$ of the series (488) and pair off the factors of $S_{n}$ as indicated by $G$. Replace each factor-pair $A_{\lambda}(x) A_{\mu}(y)$ by (510), and replace each factor-pair $\psi_{\alpha}(x) \overline{\psi}_{\beta}(y)$ by (511). Apply an $N$-ordering to the remaining unpaired factors of $S_{n}$, and multiply the whole expression by $(\pm 1)$ following the rule given for equation (504). The result of applying these operations to $S_{n}$ is the normal constituent $S_{G}$. \\
\index{contraction~of~field~operators}%

If we wish to calculate the matrix element for the scattering process, then we have only to add one more rule to the three already given. \\

\noindent \hspace*{3ex}4)\hspace*{1ex}In each $S_{G}$, substitute for the unpaired operators the wave-functions of the absorbed and emitted particles, for example writing (437) for $\psi(x)$ when an electron $(p, u)$ is absorbed, and writing (438) for $A_{\mu}(x)$ when a photon $(k, e)$ is absorbed. These substitutions may sometimes be made in more than one way (for example in the Compton effect when the absorbed and emitted photon may be assigned in two ways to the two unpaired photon operators.) 
\index{Compton!effect}%
In such cases the substitutions are to be made in all possible ways, and the results added together, taking account of Fermi statistics by putting in a minus sign when two electron or positron wave-functions are interchanged. \\

The rules (1)--(4) constitute the Feynman rules for calculating the matrix elements of all processes in electrodynamics. According to Feynman they have an intermediate concrete interpretation. Thus (510) is the probability amplitude for a photon emitted at $x$ with polarization $\lambda$ to propagate itself and arrive at $y$ with polarization $\mu$, plus the amplitude for a photon having been emitted at $y$ to arrive at $x$. 
\index{amplitude}%
\index{amplitude}%
And (511) is the amplitude for an electron emitted at $y$ to arrive at $x$, plus the amplitude for a positron emitted at $x$ to arrive at $y$, with the assigned spins $\alpha$ and $\beta$. In this way the matrix element is just the probability amplitude for the succession of events, interactions and propagations, that are depicted in the vertices and lines of $G$. The total probability amplitude for a process is just the sum of the amplitudes derived by from the various graphs $G$ which contribute to the process.

The Feynman rules of calculation take their most practical form when we use the momentum representations (430) and (448) for the $D_{F}$ and $S_{F}$ functions, carry out the integrations over the points $x_{1} \dots x_{n}$, and so obtain the matrix elements as integrals of rational functions in momentum-space. In this way for example the simple matrix elements (432) and (449) were obtained. 
\index{momentum!integral}%

In the momentum space integral for $S_{G}$, there will appear
\begin{align}
&\text{\hspace*{3ex}(1)\hspace*{3ex}A factor}\; \frac{1}{k^{2}} \; \text{corresponding to each internal photon line of $G$,} \\
&\text{\hspace*{3ex}(2)\hspace*{3ex}A factor}\; \frac{1}{\slashed{k} - i \mu} \; \text{corresponding to each internal electron line of $G$,} \\
&\text{\hspace*{3ex}(3)\hspace*{3ex}A factor}\; (2 \pi)^{4} \delta^{\,4}(k_{1} + k_{2} + k_{3})  \\
&\text{\hspace*{9ex}corresponding to each internal photon line of $G$ at which the 3 lines associated with} \notag \\
&\text{\hspace*{9ex}momenta ($k_{1}, k_{2},k_{3}$) meet. This factor arises from the integration over the space-time} \notag \\
&\text{\hspace*{9ex}position of the vertex.} \notag
\end{align}

In addition to these factors there will be numerical factors and Dirac matrices $\gamma_{\alpha}$ arising from the particular form of $S_{n}$. 
\index{Dirac!matrices}%
In practice it is easiest not to write down the $S_{G}$ directly in momentum space, but to use the rules (1) -- (4) to obtain formulae in configuration space with the right numerical constants, and then transform to momentum space by (430) and (448). 
\index{Feynman!rules|)}%

We shall now show how these general methods work by calculating in detail \emph{the} historic problem, the second-order radiative correction to the scattering of an electron by a weak external potential. 
\index{radiative~corrections!scattering!electron~by~a~weak~potential}%
This problem has been first satisfactorily treated by Schwinger,\endnote{
In Schwinger's anthology.}
\emph{Phys.\ Rev.} \textbf{76} (1949) 790. Schwinger's paper is outstandingly difficult to read, and I hope you will find my treatment at least slightly easier. But the problem is in its nature complicated and cannot be done without some fairly heavy mathematics. Once the calculations are done for this problem of scattering, it turns out that the results can be used without much further trouble for the relativistic calculation of the Lamb shift too. The scattering and Lamb shift problems are very closely related: in both cases one is calculating the second-order radiative corrections to the motion of an electron, only in one case the electron is in a high continuum state so that the external field can be treated as weak, in the other case the electron is in a discrete state and the potential must be treated as strong. 
\index{Schwinger!difficult~to~read}%
\index{Lamb~shift}%

\section*{The Self-Energy of the Electron}
\addcontentsline{toc}{section}{The Self-Energy of the Electron}

Before we can study the effect of radiation interaction on an electron scattered by an external potential, we must first consider the effect of the radiation interaction on a single free electron in the absence of external potentials. Let the free electron be given initially in the state $(pu)$. The effect of  the radiation interaction acting alone is given by the scattering matrix (488). If the initial state is $\Phi_{A}$, then the final state, reached after the radiation interaction has been acting for an infinitely long time, will be $S\Phi_{A}$. Now $S$ has matrix elements only for transitions which conserve momentum and energy. Starting from a one-electron state, it is impossible to make a transition to a many-particle state, for example by emitting one or more photons, while conserving momentum and energy. Therefore the only non-zero matrix elements of $S$ from the state $\Phi_{A}$ will be given by (487), where $\Phi_{B}$ is also a one-electron state. In $\Phi_{B}$ let the electron have the momentum and spin $(p^{\prime} u^{\prime})$.  

We consider radiative effects only up to the second order. The term of order 1 in (488) gives transitions only with emission an absorption of photons, and hence gives no contribution to the transition $\Phi_{A} \rightarrow \Phi_{B}$. Therefore we may write simply
\begin{equation}
S = 1 + U_{2}
\end{equation}
with $U_{2}$ given by (423). We have to calculate the matrix element $M_{2}$ of $U_{2}$ between the states $(pu)$ and $(p^{\prime}u^{\prime})$. 

To write down $M_{3}$ we use the Feynman rules. 
\index{Feynman!rules}%
\index{contraction~of~field~operators}%
The factor-pairings of $U_{2}$ are represented in the 8 graphs on pp 98-99. Of these only $G_{5}$ and $G_{6}$ contribute to $M_{2}$, and they contribute equally since the integral (423) is symmetrical in the variables $x_{1}$ and $x_{2}$. The normal constituent  of $U_{2}$ arising from $G_{5}$ and $G_{6}$ is, 
\index{normal~form}%
using (510) and (511)\endnote{
Both the time ordering brackets lacked a right bracket. These were added.}
\begin{equation}
\begin{split}
U_{2N} &= \sum_{\lambda, \mu} \frac{e^{2}}{\hbar^{2} c^{2}} \iint dx_{1}\, dx_{2} \, N\!\left(
 \overline{\psi}(x_{1}) \gamma_{\lambda} \Braket{T\!\left(\psi(x_{1}),\overline{\psi}(x_{2})\right)}_{o} \gamma_{\mu} \psi(x_{2})\right)\Braket{T\!\left( A_{\lambda}(x_{1}), A_{\mu}(x_{2})\right) }_{o} \\
&= -\frac{e^{2}}{4\hbar c} \sum_{\lambda} \iint dx_{1}\, dx_{2} \, N\!\left(\overline{\psi}(x_{1}) \gamma_{\lambda} S_{F}(x_{1} - x_{2}) \gamma_{\lambda} \psi(x_{2})\right) D_{F}(x_{2} - x_{1})
\end{split}
\end{equation}
To obtain $M_{2}$ from (518) we substitute for $\psi(x_{2})$ and $\overline{\psi}(x_{1})$ the wave-functions of the initial and final states, and use the momentum integrals (430), (448). Then the integration over $x_1$ and $x_2$  can be carried out and we find 
\begin{equation}
\begin{split}
M_{2} &= \sum_{\lambda} \frac{ie^{2}}{\hbar c}  \int_{F} \int_{F} dk_{1}\, dk_{2}\, \left( \overline{u}^{\prime} \gamma_{\lambda} \frac{1}{\slashed{k}_{1} - i \mu} \gamma_{\lambda} u \right) \frac{1}{k_{2}^{2}} \, \delta(k_{1} - k_{2} - p^{\prime}) \, \delta(k_{2} - k_{1} + p) \\
&= \sum_{\lambda} \frac{ie^{2}}{\hbar c} \, \delta(p - p^{\prime}) \int_{F} dk \, \left( \overline{u}^{\prime} \gamma_{\lambda} \frac{1}{\slashed{k}+ \slashed{p} - i\mu} \gamma_{\lambda} u \right) \frac{1}{k^{2}}
\end{split}
\end{equation}
We consider the Dirac operator 
\index{Dirac!equation}%
\begin{equation}
\Sigma(p) = \sum_{\lambda} \int_{F} dk\, \left(\gamma_{\lambda} \frac{1}{\slashed{k} + \slashed{p} - i \mu} \gamma_{\lambda} \right) \frac{1}{k^{2}}
\end{equation}
appearing in (519). Since $(p, u)$ are the momentum and spin of a real electron, we may use the relations
\begin{equation}
p^{2} + \mu^{2} = 0, \qquad (\slashed{p} - i \mu) u = 0
\end{equation}
when we evaluate $\Sigma(p)$ in (519). So using (376), (585) and following the same method that was used in evaluating (377)\endnote{
Unlike Dyson, Moravcsik cited Eq.\ (585) as well as Eq.\ (376). In Eq.\ (585) are Dirac matrix identities which establish the equality between the first two integrals in Eq.\ (522). Logically these identities should have been introduced before Chapter 6, but nothing prevents a reader making use of a ``forward'' reference.}
\begin{equation}
\begin{split}
\Sigma(p) &= \int_{F} dk \, \sum_{\lambda} \frac{\gamma_{\lambda} (\slashed{k} + \slashed{p} + i\mu) \gamma_{\lambda}}{k^{2}(k^2 + 2 p \cdot k)} = \int_{F} dk \, \frac{4 i \mu - 2 \slashed{k} - 2 \slashed{p}}{k^{2}(k^{2} + 2 p \cdot k)} \\
&= 2 \int_{0}^{1} dz\, \int_{F} dK \, \frac{i \mu - \slashed{k}}{[k^{2} + 2 z p \cdot k]^{2}} = 2 \int_{0}^{1} \int_{F} dk \,\frac{(i \mu - \slashed{k} + z\slashed{p})}{[k^{2}- z^{2}p^{2}]^{2}} \\
&= 2 \int_{0}^{1} dz \, \int_{F} dk \frac{i\mu(1 + z)}{[k^{2} + z^{2}\mu^{2}]^{2}}
\end{split}
\end{equation}
where we changed the origin of the $k$ integration by the replacement $k \rightarrow k - zp$ and we eliminated the odd terms. Using (386) and introducing the logarithmic divergence $R$ again according to (387), 
\begin{equation}
\Sigma(p) = 2 \int_{0}^{1} dz \, i \mu (1 + z) \{ 2 i \pi^{2} (R - \log z) \} = - \pi^{2} \mu \, [6R + 5] = - 6 \pi^{2} \mu R^{\prime}
\end{equation}
Thus $\Sigma(p)$ is a logarithmically divergent constant, depending only on the electron mass and independent of the state of the electron. The difference $5/6$ between $R$ and $R^{\prime}$ is of course not significant. Substituting (523) into (519) gives the value of $M_{2}$
\begin{equation}
M_{2} = -6\pi^{2} i \frac{e^{2}\mu}{\hbar c} R^{\prime} \delta(p - p^{\prime})\, (\overline{u}^{\prime} u)
\end{equation}
Thus $U_{2}$ does not give any transitions between different one-electron states. It has only the diagonal matrix elements given by (524) . 

Now (524) has precisely the correct relativistic form to be identified with a pure self-energy  effect. 
\index{self-energy!electron}%
Suppose that in consequence of the radiation interaction the mass of a real electron is
\begin{equation}
m = m_{o} + \delta m
\end{equation}
where $m_{o}$ is the mass of the bare electron without interaction and $\delta m$ is the electromagnetic contribution to the mass. 
\index{electron!bare}%
The mass-change $\delta m$ would be represented by a term 
\begin{equation}
\mathscr{L}_{S} = - \delta m \,c^{2} \, \overline{\psi} \psi
\end{equation}
in the Lagrangian density (410). This would produce an interaction energy 
\begin{equation}
H_{S}(t) = \delta m c^{2} \int \overline{\psi}(r, t) \psi(r, t) \, d^{\,3}\boldsymbol{r}
\end{equation}
in the Schr\"{o}dinger equation (415), and finally a contribution 
\index{Schr\"{o}dinger!equation}%
\begin{equation}
U_{S} = - i \frac{\delta m c}{\hbar} \int \overline{\psi}\psi(x) \,dx
\end{equation}
in the scattering matrix (421) or (488).

The matrix element of (528) between the states $(pu)$ and $(p^{\prime} u^{\prime})$ is 
\begin{equation}
M_{S} = - i \frac{\delta m c}{\hbar} (2\pi)^{4}\, \delta(p -p^{\prime}) \, (\overline{u}^{\prime}u)
\end{equation}
This is identical with (524) if we identify the self-mass $\delta m$ by the equation
\begin{equation}
\delta m = \frac{3}{8\pi^{2}} \frac{e^{2}m}{\hbar c} R^{\prime} = \frac{3 \alpha}{2 \pi} R^{\prime} m
\end{equation}
For all one-electron matrix elements, $U_{2}$ is identical with $U_{S}$. That is to say, the whole effect of the radiation interaction upon a free electron is to change its mass by the amount (530). This is a most satisfactory conclusion. It means that an electron with its self-field still has the correct relationship between momentum and energy for a relativistic particle, only the value of the rest-mass being changed by the self-field. It was always one of the central difficulties of the classical electron theory, that a classical extended electron did not have the right relativistic behavior.
  
The self-mass $\delta m$ is an unobservable quantity. The observed mass of an electron is $m$, and neither $m_{o}$ nor $\delta m$ can be measured separately. Thus it is unsatisfactory that $\delta m$ appears in the scattering matrix $S$ which is supposed to represent the results of experiments.

The reason why $\delta m$ still appears explicitly is just that we have \emph{not} used the observed mass $m$ in defining the initial and final states of the system. We defined these states as states of a free electron with the bare mass $m_{o}$. 
\index{bare~mass}%
Wherever we used the letter $m$ for the electron mass in the theory up to this point, in fact it was an inconsistency of notation and we meant by $m$ the mass of a \emph{bare} electron. 
\index{electron!bare}%

It is much better not to change the notation, but to keep the notation and change the interpretation, so that $m$ everywhere in the theory is now intended to mean the mass of a real electron. In particular, we set up the interaction representation operators with the real electron mass $m$, and the initial and final states of scattering problems are defined as free particles with the correct mass $m$. 
\index{interaction~representation}%
With this changed interpretation, the whole of the theory up to this point is correct, except that in $\mathscr{L}_{D}$ which appears in the Lagrangian (410) of quantum electrodynamics, and in the field equations (411), (412) satisfied by the Heisenberg operators, the bare mass $m_{o}$ must be used instead of $m$. 
\index{quantum~electrodynamics}%
\index{bare~mass}%
We prefer to keep the observed mass $m$ in $\mathscr{L}_{D}$, and correct for it by writing instead of (410)\endnote{
The subscript on the second $\mathscr{L}$ was originally ``$_{O}$''. It has been replaced with a subscript ``$_{D}$''.}
\begin{equation}
\mathscr{L} = \mathscr{L}_{D} + \mathscr{L}_{M} - ie \overline{\psi} \slashed{A} \psi - ie \overline{\psi} \slashed{A}^{e} \psi  - \mathscr{L}_{S}
\end{equation}
with $\mathscr{L}_{S}$ given by (526). The radiation interaction becomes then 
\begin{equation}
H_{R}(t) - H_{S}(t) = H^{I}(t)
\end{equation}
with $H_{R}$ given by (416) and $H_{S}$ by (527). After making the changes (537) and (532), the whole theory becomes consistent with the interpretation that $m$ is everywhere the observed electron mass. 

In particular, one result of (532) is that for one-electron states the scattering operator $S$ becomes 
\begin{equation}
S = 1 + U_{2} - U_{S}
\end{equation}
instead of (517), keeping only terms of order $e^{2}$.  The matrix elements of $(U_{2} - U_{S})$ for one-electron states are all zero. Thus, if we use the correct mass $m$ in defining the states of an electron, \emph{there are no longer any observable effects of the radiation interaction on the motion of a free electron.} This shows that the mass-renormalization, the procedure of inserting the term $(- \mathscr{L}_{S})$ in (531), is consistent and is likely to give sensible results. 
\index{renormalization!mass}%

\section*{Second-Order Radiative Corrections to Scattering}
\addcontentsline{toc}{section}{Second-Order Radiative Corrections to Scattering}

Let an electron be scattered from the initial state $(pu)$ to the final state $(p^{\prime}u^{\prime})$ by the external potential
\begin{equation}
A_{\mu}^{e}(x) = \frac{1}{(2\pi)^{4}} \int e^{i q \cdot x} \, e_{\mu}(q) \, dq
\end{equation}
At the same time the electron is interacting with the quantized Maxwell field with the interaction (532), since we suppose the initial and final states to be defined with the observed mass of a free electron. 
\index{Maxwell!field!quantized}%
The scattering matrix element $M$ is then given by (419), with $U$ given by (421) after replacing each $H_{R}$  by $H^{I}$ according to (532). 

We treat $A_{\mu}^{e}$ in the linear Born approximation. 
\index{Born~approximation}%
Thus we keep only terms of order 0 and 1 in $A_{\mu}^{e}$. The terms of order 0 give the effects of the radiation interaction alone; as we have seen, these effects are zero for an initial state consisting of a single electron.

The scattering matrix is thus given effectively by the terms of order 1 in  $A_{\mu}^{e}$ taken from (421), namely
\begin{equation}
U = \sum_{n = 0}^{\infty} \left( \frac{-i}{\hbar} \right)^{n} \frac{1}{n!} \int \dots \int dt \, dt_{1}  \dots dt_{n}\, P \left\{ H^{e}(t), H^{I}(t_{1}), \dots, H^{I}(t_{n}) \right\}
\end{equation}
We shall calculate radiative effects only up to the second order in the radiation interaction. Since $\delta m$ is itself of second order, this means that we go to second order in $H_{R}$ and to first order in $H_{S}$. Thus
\begin{align}
U &= U_{0} + U_{1} + U_{2} + U_{2}^{\prime} \\
U_{0} &= \frac{e}{\hbar c} \int dx \, \overline{\psi} \slashed{A}^{e} \psi(x) \\
U_{1} &= \frac{e^{2}}{\hbar^{2} c^{2}} \iint dx \, dx_{1} P \left\{ \overline{\psi} \slashed{A}^{e} \psi(x),  \overline{\psi} \slashed{A} \psi(x_{1}) \right\} \\
U_{2} &= \frac{e^{3}}{2 \hbar^{3} c^{3}} \iiint dx \, dx_{1} \, dx_{2} \, P \left\{ \overline{\psi} \slashed{A}^{e} \psi(x),  \overline{\psi} \slashed{A} \psi(x_{1}), \overline{\psi} \slashed{A} \psi(x_{2}) \right\}\\
U_{2}^{\prime} &= \frac{ie \, \delta m}{\hbar^{2}} \iint dx \, dx_{1} \,P \left\{ \overline{\psi} \slashed{A}^{e} \psi(x),  \overline{\psi} \psi(x_{1}) \right\}
\end{align}
The matrix element we wish to calculate is then correspondingly 
\begin{equation}
M = M_{0} + M_{1} + M_{2} + M_{2}^{\prime}
\end{equation}
The wave-functions of initial and final states are
\begin{equation}
u e^{i p \cdot x} \qquad \qquad u^{\prime} e^{i p^{\prime} \cdot x}
\end{equation}
Then by (534) we have
\begin{equation}
M_{0} = \frac{e}{\hbar c} (u^{\prime} \slashed{e} u)
\end{equation}
where $q$ is the constant vector
\begin{equation}
q = p^{\prime} - p
\end{equation}
and 
\begin{equation}
e_{\mu} = e_{\mu}(q)
\end{equation}

The operator $U_{1}$ gives transitions from a one-electron state only to states consisting of an electron and a photon. This is just the bremsstrahlung process, scattering of the electron with real photon emission, and the matrix element for it is given by (484). 
\index{bremsstrahlung}%
In any scattering experiment, this process will of course go on at the same time as the scattering without radiation. Experimentally, the scattering with photon emission will only be separable from the radiationless scattering if the emitted photon has an energy greater than some limit $\Delta E$, roughly equal to the energy resolution with which the energy of the electron can be measured. The scattering with emission of soft quanta (low frequency, $k^{\prime}$ small) will always be included in the radiationless scattering cross-section. 
\index{cross-section!non-radiative}%
Therefore we shall be interested in the value of $M_{1}$ for a final state consisting of the electron $(p^{\prime}u^{\prime})$ and a photon with potentials (440), in the case where $k^{\prime}$ is so small as to be negligible in comparison with $p$, $p^{\prime}$ and $q$. In this case (484) gives 
\begin{equation}
M_{1} = \frac{e^{2}}{\hbar^{2} c^{2}} \left[ \frac{p \cdot e^{\prime}}{p \cdot k^{\prime}} - \frac{p^{\prime} \cdot e^{\prime}}{p^{\prime} \cdot k^{\prime}} \right] (\overline{u}^{\prime} \slashed{e} u) = \frac{e}{\hbar c} \left[ \frac{p \cdot e^{\prime}}{p \cdot k^{\prime}} - \frac{p^{\prime} \cdot e^{\prime}}{p^{\prime} \cdot k^{\prime}} \right] M_{0}
\end{equation}
where we used (521) and rule (4) on page 87. 

We now come to the calculation of $M_{2}$, the matrix element of (539) between the states (542). This is the main part of the problem. To do it we use the Feynman rules. 
\index{Feynman!rules}%
There are just 9 graphs giving contributions to $M_{2}$, namely
\begin{center}
\includegraphics{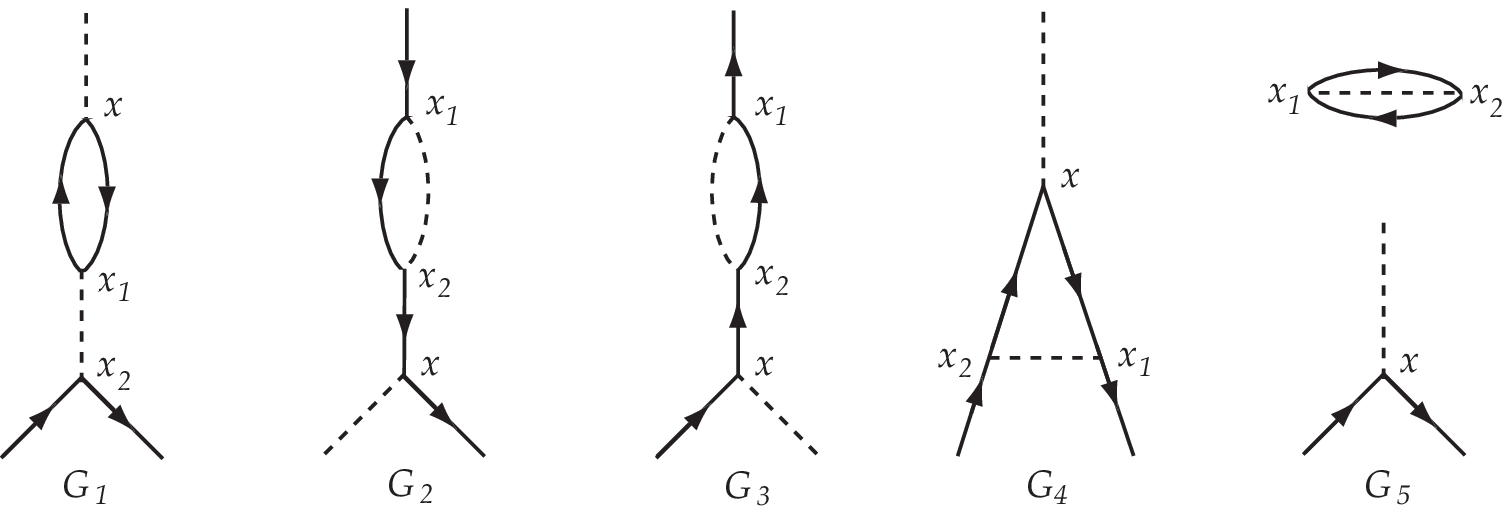}
\end{center}
and $\{G_{6}, G_{7}, G_{8}, G_{9}\}$ obtained by interchanging the labels $(x_{1}, x_{2})$ in $\{G_{1}, G_{2}, G_{3}, G_{4}\}$, respectively. We can see this in  the following way: The process we are interested in calls for one external photon line and two external electron lines. Hence $\slashed{A}^{e}(x)$, one $\overline{\psi}$ and one $\psi$ must be unpaired. Therefore the two $\slashed{A}$'s are always coupled. The free electron lines can be $\overline{\psi}_{0}, \psi_{1}$; $\overline{\psi}_{1}, \psi_{0}$; $\overline{\psi}_{1}, \psi_{1}$; $\overline{\psi}_{1}, \psi_{2}$ and four more cases obtained by the substitution $1 \leftrightarrow 2$. In each case, the rest is uniquely determined by the rules. The ninth case is the one with external electron lines $\overline{\psi}_{0}, \psi_{0}$. 

The effect of $\{G_{6}, G_{7}, G_{8}, G_{9}\}$ is only to double the contribution from $\{G_{1}, G_{2}, G_{3}, G_{4}\}$, since (539) is symmetrical in the variables $x_{1}$ and $x_{2}$. Also $G_{5}$ will give only a numerical phase factor multiplying $M_{0}$, the phase factor being the same for all final states; it is in fact precisely the phase-shift between the initial and final states due to the self-energy of the vacuum. 
\index{self-energy!vacuum}%
Similar phase shift factors would also multiply the contributions of $G_{1}$, $G_{2}$, etc., coming from disconnected graphs in higher order contributions. In this sense therefore, we can consider $G_{5}$ to be really the contribution of $U_{0}$ with one of the many possible disconnected additions. A numerical phase factor of this kind, the same for all final states, is entirely unobservable and without physical meaning, since it can be cancelled by changing the phase of all wave-functions by the same amount. Hence we can always ignore graphs such as $G_{5}$ having a disconnected part without external lines. There remain to be considered only $\{G_{1}, G_{2}, G_{3}, G_{4}\}$. 

Using the Feynman rules, the contribution of $G_{1}$ to $M_{2}$ is (with the factor 2 from $G_{6}$)
\begin{equation}
\begin{split}
M_{21} = - \frac{e^{3}}{\hbar^{3}c^{3}} 
\iiint dx\, dx_{1}\, dx_{2} \, \sum_{\mu} \text{Spur} &\left\{ \slashed{A}^{e}(x) \Braket{T\{\psi(x), \overline{\psi}(x_{1})\}}_{o} \gamma_{\mu} \Braket{T\{\psi(x_{1}), \overline{\psi}(x)\}}_{o} \right\} \times \\
& \times \overline{\psi}(x_{2}) \Braket{T\{A_{\mu}(x_{1}), \slashed{A}(x_{2})\}}_{o} \psi(x_{2})
\end{split}
\end{equation}
where the spur appears because of the contraction according to rule 3, and the minus sign comes from the change in order of $\overline{\psi}$ and $\psi$ factors between (539) and (547). 
\index{contraction~of~field~operators}%
By (510) and (511) 
\[
M_{21} = - \frac{e^{3}}{8\hbar^{2}c^{2}} 
\iiint dx\, dx_{1}\, dx_{2} \, \sum_{\mu} \text{Spur} \left\{ \slashed{A}^{e}(x) S_{F}(x - x_{1}) \gamma_{\mu}S_{F}(x_{1} - x) \right\} D_{F}(x_{1} - x_{2}) \overline{\psi}(x_{2}) \gamma_{\mu} \psi(x_{2})
\]
Hence, using the momentum integrals (430), (448), (534), (542) and carrying out the integration over $(x, x_{1}, x_{2})$,
\begin{equation}
\begin{split}
M_{21} &= \frac{ie^{3}}{(2 \pi)^{4} \hbar^{2} c^{2}}
 \iiiint dk_{1} \, dk_{2} \, dk_{3} \, dq\, \sum_{\mu} \text{Spur} \left\{ \slashed{e}(q) \frac{1}{\slashed{k}_{1} - i\mu} \gamma_{\mu} \frac{1}{\slashed{k}_{2} - i\mu} \right\} \frac{1}{k_{3}^{2}}\,(\overline{u}^{\prime} \gamma_{\mu} u) \, \times \\
& \qquad \qquad \qquad \qquad \qquad \times \delta(q + k_{1} - k_{2})\, \delta(-k_{1} + k_{2} + k_{3})\, \delta(-k_{3} - p^{\prime} + p) \\
&= \frac{ie^{3}}{(2 \pi)^{4} \hbar^{2} c^{2}} \int_{F} dk \sum_{\mu} \text{Spur} \left\{ \slashed{e} \frac{1}{\slashed{k} - i\mu} \gamma_{\mu} \frac{1}{\slashed{k} + \slashed{q} -i\mu} \right\} \frac{1}{q^{2}} \,(\overline{u}^{\prime} \gamma_{\mu} u) \\
&= \frac{ie^{3}}{(2 \pi)^{4} \hbar^{2} c^{2}} \sum_{\mu}  \frac{1}{q^{2}} \,(\overline{u}^{\prime} \gamma_{\mu} u) \, J_{\mu}
\end{split}
\end{equation}
with $q$ given by (544) and\endnote{
Again, a subscript $\mu$ has been appended to the function $F(k)$ to make it a Lorentz vector. See note 31 at Eq.\ (371).}
\begin{equation}
J_{\mu} = \int_{F} F_{\mu} (k)\, dk
\end{equation}
the function $F_{\mu}(k)$ being identical with (371) for $\delta = 0$. Note that (549) is a Feynman integral, which is precisely the same thing as the contour integral (374) with the contour drawn in the diagram. 
\index{Feynman!contour~integral}%
The effect of the $\epsilon$ in (431) is just equivalent to the contour $C$. Hence\endnote{
The last curly bracket was missing; it has been added.} 
using (388) for $J_{\mu}$,
\begin{equation}
M_{21} = - \frac{e^{3}}{2 \pi^{2} \hbar^{2} c^{2}}\, (\overline{u}^{\prime} \slashed{e} u) \left\{ \tfrac{1}{3} R - \int_{0}^{1} (z - z^{2}) \log \left(1 + (z - z^{2}) \frac{q^{2}}{\mu^{2}} \right) dz \right\}
\end{equation}
where we have also dropped the term $q_{\mu}$ in(388), since 
\begin{equation}
(\overline{u}^{\prime} \slashed{q} u) = \{ \overline{u}^{\prime}(\slashed{p}^{\prime} - i \mu) u\} -  \{ \overline{u}^{\prime}(\slashed{p} - i \mu) u \} = 0
\end{equation}
Writing $\alpha = \dfrac{e^{2}}{4 \pi \hbar c}$, (550) becomes
\index{fine~structure~constant}%
\begin{equation}
M_{21} = \alpha M_{0} \left \{ - \frac{2}{3\pi} R + \frac{2}{\pi} \int_{0}^{1} (z - z^{2}) \log \left(1 + (z - z^{2}) \frac{q^{2}}{\mu^{2}} \right) dz \right\}
\end{equation}
This is just the scattering which is produced by the charge-current density induced in the vacuum by the potential $A_{\mu}^{e}$ according to (392). As before, the term in $R$ is unobservable since it can never be separated experimentally from the simple scattering 
$M_{0}$ to which it is proportional. The observed external potential, measured in any way whatever, will not be $A_{\mu}^{e}$ but $A_{\mu}^{e}(1 - \tfrac{2 \alpha}{3 \pi} R)$, which we may call the ``renormalized external potential''. 
\index{renormalization!and~external~potential}%
Hence in terms of the observed $A_{\mu}^{e}$ the total contribution from $G_{1}$ will be 
\begin{equation}
M_{21} = \frac{2 \alpha}{\pi} M_{0}  \int_{0}^{1} (z - z^{2}) \log \left(1 + (z - z^{2}) \frac{q^{2}}{\mu^{2}} \right) dz 
\end{equation}
The integral will be in general complex as before. But for small $q$ it will be real, and neglecting terms of order higher than $q^{2}$, we have 
\begin{equation}
M_{21} = \frac{2 \alpha}{\pi} \frac{q^{2}}{\mu^{2}} M_{0}  \int_{0}^{1} (z - z^{2})^{2}  dz = \frac{\alpha}{15 \pi} M_{0} \frac{q^{2}}{\mu^{2}}
\end{equation}

Next we consider the contribution to $M_{2}$ from $G_{2}$. This is
\begin{align}
M_{22} &=  \frac{e^{3}}{8\hbar^{2}c^{2}} 
\iiint dx\, dx_{1}\, dx_{2} \, \sum_{\lambda} \overline{\psi}(x) \slashed{A}^{e}(x) S_{F}(x - x_{2}) \gamma_{\lambda}S_{F}(x_{2} - x_{1})\gamma_{\lambda} \psi(x_{1})D_{F}(x_{1} - x_{2}) \\
&= - \frac{ie^{3}}{(2 \pi)^{4} \hbar^{2} c^{2}}
 \iiiint dk_{1} \, dk_{2} \, dk_{3} \, dq\, \sum_{\lambda} \left\{\overline{u}^{\prime} \slashed{e}(q)\, \frac{1}{\slashed{k}_{1} - i\mu}\, \gamma_{\lambda}\,\frac{1}{\slashed{k}_{2} - i\mu}\, \gamma_{\lambda} u \right\} \frac{1}{k_{3}^{2}} \times  \notag \\
& \qquad \qquad \qquad \qquad \qquad \times \delta(k_{1} + q - p^{\prime})\, \delta(k_{2} - k_{1} - k_{3})\, \delta(k_{3} + p - k_{2}) \phantom{A_{A_{A_{A_{A_{A}}}}}}\\
&= - \frac{ie^{3}}{(2 \pi)^{4} \hbar^{2} c^{2}} \sum_{\lambda} \int_{F} dk\, \left(\overline{u}^{\prime} \slashed{e} \frac{1}{\slashed{p} - i\mu}\, \gamma_{\lambda} \,\frac{1}{\slashed{k} + \slashed{p} - i\mu} \,\gamma_{\lambda} u \right) \frac{1}{k^{2}} \phantom{A^{A^{A^{A^{A^{A}}}}}} \phantom{A_{A_{A_{A_{A_{A}}}}}}\\
&= -\frac{ie^{3}}{(2 \pi)^{4} \hbar^{2} c^{2}}\left(\overline{u}^{\prime} \slashed{e} \frac{1}{\slashed{p} - i \mu} \, \Sigma(p)\, u \right) \phantom{A^{A^{A^{A^{A^{A}}}}}}
\end{align}
with $\Sigma(p)$ given by(520). 

Before discussing $\Sigma(p)$ we must look at the factor $\dfrac{1}{\slashed{p} - i\mu}$ which appears in (558). This factor is 
\begin{equation}
\frac{\slashed{p} + i\mu}{p^{2} + \mu^{2}}
\end{equation}
But since $p$ is the  momentum vector of a real electron, $p^{2} + \mu^{2} = 0$ and the factor (559) is singular. This means that the integrals over $x_{1}$ and $x_{2}$ are really divergent and not merely finitely oscillating at $t = \pm \infty$, and the transformation into momentum integrals is not allowable. Eq. (558) as it stands is strictly meaningless.

This is the place where we have to take explicitly into our calculations the slowly-varying cut-off functions $g(t_{i})$ appearing in (495), which are put in for the purpose of defining unambiguously the initial and final states of the problem. So we write instead of (555)
\begin{equation}
M_{22} =  \frac{e^{3}}{8\hbar^{2}c^{2}} 
\iiint dx\, dx_{1}\, dx_{2} \, \sum_{\lambda} \overline{\psi}(x) \slashed{A}^{e}(x) S_{F}(x - x_{2}) \gamma_{\lambda}S_{F}(x_{2} - x_{1})\gamma_{\lambda} \psi(x_{1})D_{F}(x_{1} - x_{2}) g(t_{1})g(t_{2})
\end{equation}
Here the $g(t_{1})g(t_{2})$ factors are attached to the radiation interaction operating at $x_{1}$ and $x_{2}$. It is supposed that the time $T$ over which $g(t)$ varies appreciably  is long compared with the duration of the scattering process. Let the Fourier integral  representation of $g(t)$ be 
\index{Fourier!integral}%
\begin{equation}
\begin{split}
g(t) &= \int_{-\infty}^{\infty} G(\epsilon_{0}) e^{-i \epsilon_{0} c t} \, d\epsilon_{0}  \\
       &=  \int_{-\infty}^{\infty} G(\epsilon_{0}) e^{i \epsilon \cdot x} \, d\epsilon_{0}
\end{split}
\end{equation}
where $\epsilon_{0}$ is a real variable and $\epsilon$ is the vector
\begin{equation}
\epsilon = (0, 0,0, \epsilon_{0})
\end{equation}
We have the normalization 
\begin{equation}
g(0) = \int_{-\infty}^{\infty} G(\epsilon_{0})\, d\epsilon_{0}  = 1
\end{equation}
and we suppose that $G(\epsilon_{0})$ is ``almost'' a $\delta$-function, that is to say a function which is large only for values of $\epsilon_{0}$ in a range of about $(cT)^{-1}$ on either side of zero. Substituting (561) into (560), we obtain instead of (558) the corrected formula 
\begin{equation}
M_{22} = - \frac{ie^{3}}{(2 \pi)^{4} \hbar^{2}c^{2}} \iint G(\epsilon_{0})\,G(\epsilon^{\prime}_{0}) \, d\epsilon_{0} \, d\epsilon_{0}^{\prime} \, \left\{ \overline{u}\, \slashed{e}(q - \epsilon - \epsilon^{\prime}) \, \frac{1}{\slashed{p} + \slashed{\epsilon} + \slashed{\epsilon}^{\prime} - i\mu} \Sigma(p + \epsilon) \, u \right\}
\end{equation}
In (564) the inadmissible factor (559) is replaced by something finite and mathematically well-defined. There will be a singularity in the integration of (564) over $\epsilon_{0}$, but this is an ordinary pole and the integration over $\epsilon_{0}$ will give a well-determined result when taken as a Feynman integral.
\index{Feynman!contour~integral}%
We have as $T \rightarrow \infty$ and $\epsilon_{0}$ and $\epsilon_{0}^{\prime} \rightarrow 0$
\begin{equation}
\frac{1}{\slashed{p} + \slashed{\epsilon} + \slashed{\epsilon}^{\prime} - i \mu} = \frac{\slashed{p} + \slashed{\epsilon} + \slashed{\epsilon}^{\prime} + i\mu}{2 p \cdot (\epsilon + \epsilon^{\prime}) + (\epsilon + \epsilon^{\prime})^{2}} \sim - \frac{\slashed{p} + i\mu}{2 p_{0} (\epsilon_{0} + \epsilon_{0}^{\prime})}
\end{equation}
Hence when evaluating $\Sigma(p + \epsilon)$ we need retain only terms of order zero and one in $\epsilon_{0}$; the terms of order two and higher are negligible, because even when multiplied by (565) they tend to zero as $T \rightarrow \infty$. 

Keeping only terms of order zero and one in $\epsilon$, $\Sigma(p + \epsilon)$ becomes 
\begin{equation}
\Sigma( p + \epsilon) = \Sigma(p) - \sum_{\alpha} \epsilon_{\alpha} I_{\alpha}(p)
\end{equation}
\begin{equation}
I_{\alpha}(p) = \int_{F} dk \, \sum_{\lambda} \left( \gamma_{\lambda} \, \frac{1}{\slashed{k} + \slashed{p} -i\mu} \, \gamma_{\alpha} \, \frac{1}{\slashed{k} + \slashed{p} -i\mu} \gamma_{\lambda} \right) \frac{1}{k^{2}}
\end{equation}
Here we have used the identity
\begin{equation}
\frac{1}{A + B} = \frac{1}{A} - \frac{1}{A} B \frac{1}{A} + \frac{1}{A} B \frac{1}{A} B \frac{1}{A} - \dots
\end{equation}
which is valid for any two operators $A$ and $B$, commuting or not, if the series on the RHS converges in some sense. This can be seen at once by multiplying through by $A + B$; then the condition becomes $\left( B/A \right)^{n}  \rightarrow 0$ in some sense.

In (564) we may use the conditions (521), and these give for $\Sigma(p)$ the constant value (523). The integral $I_{\alpha}(p)$ is like $\Sigma(p)$ logarithmically divergent for large $k$, and it is also logarithmically divergent for small $k$, which $\Sigma(p)$ is not. We shall not attempt to evaluate $I_{\alpha}(p)$ mathematically. From general principles of covariance we can say what is its form as a function of $p$. For general $p$ not satisfying (521), $I_{\alpha}(p)$ is a Dirac matrix  transforming like a vector under Lorentz transformations, and therefore must be of the form
\index{Dirac!matrices}%
\index{Lorentz!transformations}%
\begin{equation}
I_{\alpha}(p) = F_{1}(p^{2}) \gamma_{\alpha} + F_{2}(p^{2}) (\slashed{p} - i \mu) \gamma_{\alpha} + F_{3}(p^{2}) \gamma_{\alpha} (\slashed{p} - i\mu) + F_{4}(p^{2}) (\slashed{p} - i \mu)\gamma_{\alpha}(\slashed{p} - i\mu)
\end{equation}
where $F_{1}, \dots, F_{4}$ are functions of the scalar $p^{2}$.  Hence using (521) and (523), we see that in (564) we may put
\begin{equation}
\Sigma(p + \epsilon) = - 6 \pi^{2} \mu R^{\prime} - I_{1} \slashed{\epsilon} - I_{2} (\slashed{p} - i\mu) \slashed{\epsilon}
\end{equation}
where $I_{1}$ and $I_{2}$ are new absolute constants, and in particular 
\begin{equation}
I_{1} = F_{1}(- \mu^{2})
\end{equation}
But in (564) the term 
\[
\left( \frac{1}{\slashed{p} + \slashed{\epsilon} + \slashed{\epsilon}^{\prime} -i \mu} \right) ( \slashed{p} - i\mu) \slashed{\epsilon} = \slashed{\epsilon} - \frac{1}{(\slashed{p} + \slashed{\epsilon} + \slashed{\epsilon}^{\prime} -i \mu)} (  \slashed{\epsilon} +  \slashed{\epsilon}^{\prime})  \slashed{\epsilon}
\]
is of order $\epsilon$ and tends to zero as $T \rightarrow 0$. This term may be dropped, and then (564) becomes\endnote{
The times symbol $\times$ was inserted.}
\begin{equation}
\begin{split}
M_{22} = - \frac{ie^{3}}{(2 \pi)^{4} \hbar^{2}c^{2}} \iint & G(\epsilon_{0})\,G(\epsilon^{\prime}_{0}) \, d\epsilon_{0} \, d\epsilon_{0}^{\prime} \, \times \\
& \left\{ \overline{u}\, \slashed{e}(q - \epsilon - \epsilon^{\prime}) \, \frac{1}{\slashed{p} + \slashed{\epsilon} + \slashed{\epsilon}^{\prime} - i\mu} (- 6 \pi^{2} \mu R^{\prime} - I_{1} \slashed{\epsilon}) \, u \right\}
\end{split}
\end{equation}
Note that if $I_{\alpha}(p)$ given by (567) were to be evaluated assuming that $p^{2}+ \mu^{2} = 0$ and $\slashed{p} - i \mu = 0$ operating both to the left and to the right, instead of only to the right as in (521), the result obtained would be just 
\begin{equation}
I_{\alpha}(p) = I_{1} \gamma_{\alpha}
\end{equation}
This is a convenient definition of $I_{1}$ for future reference.

Now it is clear that the term $R^{\prime}$ in $M_{22}$ represents some kind of effect of the electron self-energy, which ought not to be observable. 
\index{self-energy!electron}%
We may expect that this term will be cancelled by the term $M_{2}^{\prime}$ arising from the self-energy correction $H_{S}$ in (532). 
\index{self-energy!electron}%
This is all the more plausible, because the graph $G_{2}$ on page 106 contains as a part the graph $G_{5}$ on page 99 which represents the self-energy of a free electron. We now turn to the calculation of $M_{2}^{\prime}$.

$M_{2}^{\prime}$ is the sum of two contributions arising from the two graphs shown below.
\begin{center}
\includegraphics{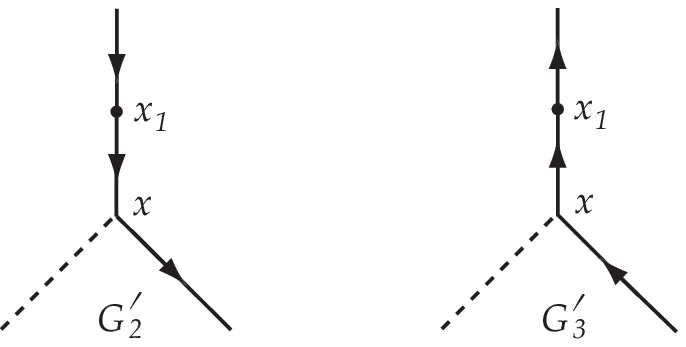}
\end{center}
From $G_{2}^{\prime}$ the contribution is (compare with (528))
\begin{equation}
M_{22}^{\prime} = - \frac{ie \,\delta m}{2 \hbar^{2}} \iint dx\, dx_{1}\, \overline{\psi}(x) \slashed{A}^{e} S_{F}(x - x_{1})\, \psi(x_{1})
\end{equation}
Like (555) this integral does not oscillate but diverges at $t_{1} = \pm \infty$. Therefore we must take explicitly into account the cut-off factor multiplying the radiation interaction. At time $t_{1}$ the radiation interaction $H_{R}(t_{1})$ will carry the cut-off factor $g(t_{1})$. But the self-energy $\delta m$ at time $t_{1}$ is an effect of second order in $H_{R}$, and therefore becomes multiplied by $\left[g(t_{1})\right]^{2}$ if $g(t_{1})$ varies slowly enough. In the definition of the scattering matrix element (487), the cut-off factors $g(t)$ were introduced in order to represent the initial and final states by simple bare-particle wave-functions in an unambiguous way. We now require that the bare-particles' wave-functions should always have the same mass $m$ as a real electron. This is achieved if we add the term $(-H_{S})$ to the radiation interaction appearing in (495), each $H_{S}(t_{i})$ being multiplied by the cut-off factor $\left[ g(t_{i}) \right]^{2}$ so long as we are dealing only with second order terms in $\delta m$. If we were calculating effects to fourth order in $e$, then the fourth order part of $\delta m$ should be multiplied by $\left[ g(t_{i}) \right]^{4}$, and so on.

The effect of the cut-off factors is therefore to replace (574) by 
\begin{equation}
M_{22}^{\prime} = - \frac{ie\, \delta m}{2 \hbar^{2}} \iint dx\, dx_{1}\, \overline{\psi}(x) \slashed{A}^{e} S_{F}(x - x_{1})\, \psi(x_{1})\left[ g(t_{1}) \right]^{2}
\end{equation}
Using (561) and performing the integrations as before, this becomes
\begin{equation}
M_{22}^{\prime} = - \frac{ie\, \delta m}{2 \hbar^{2}} \iint G(\epsilon_{0}) G(\epsilon_{0}^{\prime})\, d\epsilon_{0} \, d\epsilon_{0}^{\prime} \, \left\{\overline{u}^{\prime} \slashed{e}(q - \epsilon - \epsilon^{\prime}) \left( \frac{1}{\slashed{p} + \slashed{\epsilon} + \slashed{\epsilon}^{\prime} - i\mu} \right) u \right\}
\end{equation}
In virtue of (530), this term does precisely cancel the term in $R^{\prime}$ in (572), as was to be expected. 

To simplify the term in $I_{1}$ in (572), we may replace $\slashed{\epsilon}$ by $\tfrac{1}{2} ( \slashed{\epsilon} + \slashed{\epsilon}^{\prime} )$ since the integrand is otherwise symmetrical between $\epsilon$ and $\epsilon^{\prime}$. And using (521) we may replace this in turn by $\tfrac{1}{2} (\slashed{p} +  \slashed{\epsilon} + \slashed{\epsilon}^{\prime} - i \mu)$. This cancels the denominator of (572) precisely. After the denominator is cancelled, the expression is non-singular and we may go to the limit $T \rightarrow \infty$, using (563) to perform the integration over $\epsilon_{0}$ and $\epsilon_{0}^{\prime}$. Since it is assumed that the external potential is of a limited duration not tending to infinity with $T$, the factor $\slashed{e}(q - \epsilon - \epsilon^{\prime})$ is a continuous function of $\epsilon + \epsilon^{\prime}$ and tends to $\slashed{e}(q)$ as $T \rightarrow \infty$.\endnote{
This sentence read formerly ``\dots the factor $\epsilon(q - \epsilon - \epsilon^{\prime})$ is a continuous function of ($\epsilon + \epsilon^{\prime}$) and tends to $\slashed{\epsilon}(q)$ \dots''} 
Hence in the limit $T \rightarrow \infty$ we have
\begin{equation}
M_{22}+ M_{22}^{\prime} = \frac{ie^{3}}{(2 \pi)^{4} \hbar^{2} c^{2}} \,\frac{1}{2}\, I_{1}\, ( \overline{u}^{\prime} \slashed{\epsilon} u ) = \frac{i \alpha}{(2 \pi)^{3}} \,I_{1}\, M_{o}
\end{equation}
The graphs $G_{3}$ and $G_{3}^{\prime}$ give an exactly equal contribution. Hence
\begin{equation}
M_{22} + M_{23} + M_{22}^{\prime} + M_{23}^{\prime} = \frac{i \alpha}{4\pi^{3}} \, I_{1} \, M_{o}
\end{equation}
It turns out that $I_{1}$ is pure imaginary, and the factor multiplying $M_{o}$ in (578) is real and negative.

What is the physical interpretation of the divergent term (578)? It is just a divergent constant multiple of $M_{o}$, like the charge-renormalization term in (552). 
\index{renormalization!charge}%
So one is tempted at first to call it an additional charge-renormalization effect. But this cannot be right, because the whole charge-renormalization was calculated in (392) and the result agreed with (552). In fact (578) has a much more elementary interpretation. When the electron arrives at the point $x$ of the external potential where it undergoes scattering, there will be a certain probability $P$ that it will have previously emitted and not yet reabsorbed a photon, as pictured in the Feynman graph $G_{4}$.
\index{Feynman!graph}%
There will be a probability $(1 - P)$ that it will arrive at $x$ not accompanied by a photon, as pictured in $G_{2}$ or $G_{3}$.

Consider then the contribution $M_{NP}$ to the matrix element $M$, produced by scattering processes in which the electron arrives at $x$ with no photon present. In the zero-order approximation $M_{NP} = M_{o}$ simply. But in the second-order approximation we must take account of the reduced probability that the electron comes to $x$ without a photon present; this is done by multiplying the wave-function of the electron in both initial and final states by the factor 
\index{probability!scattering!reduced}%
\begin{equation}
(1 - P)^{1/2}
\end{equation}
Therefore in second-order approximation 
\begin{equation}
M_{NP} = (1 -P)M_{o}
\end{equation}
Since in second order
\begin{equation}
M_{NP} = M_{o} + M_{22} + M_{22}^{\prime} + M_{23} + M_{23}^{\prime},
\end{equation}
the result (578) agrees with (580) provided that we identify $P$ with
\index{fine~structure~constant}%
 \begin{equation}
P = - \frac{i\alpha}{4 \pi^{3}}\, I_{1}
\end{equation}

The factor (579) represents a renormalization of the amplitude of the wave-function, and for this reason (578) is usually called a ``wave-function renormalization'' effect. 
\index{renormalization!wave~function}%
But this does not mean that the term (578) is to be removed by a process analogous to mass or charge renormalization. No difficulties arise if we simply retain (578) as it stands. Finally it will cancel out against a term $(+ PM_{o})$ which comes from $G_{4}$. 

The contribution from $G_{4}$ to $M$ is, with a factor 2 to allow for $G_{9}$, using the Feynman rules, 
\index{Feynman!rules}%
\begin{equation}
\begin{split}
M_{24} &= \frac{e^{3}}{8 \hbar^{2} c^{2}} \iint dx\, dx_{1} \, dx_{2} \sum_{\lambda} \overline{\psi}(x) \gamma_{\lambda} S_{F}(x_{1}- x)\slashed{A}^{e}(x) S_{F}(x - x_{2}) \gamma_{\lambda} \psi(x_{2}) D_{F}(x_{2} - x_{1}) \\
&= - \frac{i e^{3}}{(2 \pi)^{4} \hbar^{2}c^{2}} \left( \overline{u}^{\prime} \Lambda(p, p^{\prime}) u \right)
\end{split}
\end{equation}
where 
\begin{equation}
\Lambda(p, p^{\prime}) = \int_{F} dk \, \sum_{\lambda} \left( \gamma_{\lambda} \frac{1}{\slashed{k} + \slashed{p}^{\prime} - i \mu} \slashed{e} \frac{1}{\slashed{k} + \slashed{p} - i \mu}  \gamma_{\lambda} \right) \frac{1}{k^{2}}
\end{equation}
There is no singular factor in (584) such as we had in (558). To perform the sum over $\lambda$ in (584), we use the table 
\index{Dirac!matrices!spur~theorems}%
\index{spur~theorems}%
\index{trace~theorems|see{spur~theorems}}%
\begin{equation}
\begin{split}
\sum_{\lambda} \gamma_{\lambda} \gamma_{\lambda} &= 4 \\
\sum_{\lambda} \gamma_{\lambda} \slashed{a} \gamma_{\lambda} &= -2 \slashed{a} \\
\sum_{\lambda} \gamma_{\lambda} \slashed{a} \slashed{b} \gamma_{\lambda} &= 4 (a \cdot b) \\
\sum_{\lambda} \gamma_{\lambda} \slashed{a} \slashed{b} \slashed{c} \gamma_{\lambda} &= -2 \slashed{c} \slashed{b} \slashed{a}
\end{split}
\end{equation}
valid for any vectors $a$, $b$, $c$. These formulae can be deduced from the following recursive formula: \\

\noindent Denote $\slashed{q}_{(n)} = \slashed{q}_{1} \slashed{q}_{2} \dots \slashed{q}_{n}$ where $q_{i}$ are arbitrary vectors, and $\chi_{n} = \sum_{\lambda} \gamma_{\lambda} \slashed{q}_{(n)} \gamma_{\lambda}$, $\chi_{o} = 4$. Then we have
\begin{align*}
\chi_{n+1} &= \sum_{\lambda} \gamma_{\lambda} \slashed{q}_{(n)} \slashed{q}_{n+1} \gamma_{\lambda} = \sum_{\lambda} \sum_{\mu} \gamma_{\lambda} \slashed{q}_{(n)} \gamma_{\mu} \gamma_{\lambda} (q_{n+1})_{\mu}= \sum_{\lambda} \sum_{\mu} \gamma_{\lambda} \slashed{q}_{(n)} \left[ 2 \delta_{\lambda \mu} - \gamma_{\lambda} \gamma_{\mu} \right] (q_{n+1})_{\mu} \\
&= 2 \slashed{q}_{n+1} \slashed{q}_{(n)} - \chi_{n} \slashed{q}_{n+1}
\end{align*}
which then gives (585) for $n = 1, 2, 3$. Thus
\begin{equation}
\Lambda(p, p^{\prime}) = -2 \int_{F} dk \, \frac{(\slashed{k} + \slashed{p}) \slashed{e} (\slashed{k} + \slashed{p}^{\prime} ) - 2i \mu ( 2 e \cdot k + e \cdot p  + e \cdot p^{\prime}) - \mu^{2} \slashed{e}}{k^{2} \left[k^{2} + 2 k \cdot p^{\prime} \right]\left[ k^{2} + 2 k \cdot p \right]}
\end{equation}
In (583) we may use the relations
\begin{equation}
p^{2} + \mu^2 = p^{\prime 2} + \mu^2 = 0 \qquad \qquad (\slashed{p} - i \mu)u = 0 \qquad \qquad \overline{u}^{\prime} (\slashed{p}^{\prime} - i \mu) = 0
\end{equation}
We also assume that the external potential satisfies the Lorentz condition 
\index{Lorentz!gauge~condition}%
\index{gauge!condition}%
\begin{equation}
\sum_{\lambda} \frac{\partial A_{\mu}^{e}}{\partial x_{\mu}} = 0 \qquad \text{so that} \qquad e \cdot q = 0
\end{equation}

To evaluate (586) we use the 3-variable generalization of (376)
\begin{equation}
\frac{1}{abc} = 2 \int_{0}^{1} dx \, \int_{0}^{1}x\, dy \, \frac{1}{\left[a(1 - x) + bxy + cx(1-y) \right]^{3}}
\end{equation}
which one can verify at once by direct integration. We write
\begin{equation}
\begin{split}
p_{y} &= py + p^{\prime}(1 - y) \\
p_{y}^{2} &= [-(p^{\prime} - p)y + p^{\prime}]^{2} = q^{2}y^{2} - (2p^{\prime 2} - 2 p \cdot p^{\prime}) y - \mu^{2} = q^{2}y^{2} - (p^{\prime 2} - 2 p \cdot p^{\prime} + p^{2}) - \mu^{2} \\
&= - \mu^{2} - (y - y^{2})q^{2}
\end{split}
\end{equation}

Then changing the origin of the $k$-integration by the substitution $k \rightarrow k - xp_{y}$ (586) and (589) give
\begin{equation}
\Lambda(p, p^{\prime}) = - 4 \iint x \, dx \, dy  \int_{F} dk \, \frac{(\slashed{k} - x\slashed{p}_{y} 
+ \slashed{p}) \slashed{e} (\slashed{k}  - x\slashed{p}_{y} + \slashed{p}^{\prime} ) - 2i \mu e \cdot ( 2 k - 2x p_{y} +  p+ p^{\prime}) - \mu^{2} \slashed{e}}{\left[k^{2} - x^{2}p_{y}^{2} \right]^{3}}
\end{equation}
In (591) we may drop terms which are odd in $k$. Also using (587) and (588) we may put 
\begin{equation}
e \cdot p = e \cdot p^{\prime} = e \cdot p_{y} = i \mu \slashed{e} + \tfrac{1}{2} \slashed{e} \slashed{q}
\end{equation}
\begin{equation}
\begin{split}
(\slashed{p} - x \slashed{p}_{y}) \slashed{e} (\slashed{p}^{\prime} - x \slashed{p}_{y}) &= \left\{(1 - x) i\mu- (1 - xy)\slashed{q} \right\} \slashed{e} \left\{(1 - x) i\mu + (1 - x + xy) \slashed{q} \right\} \\
&= - (1 - x)^{2} \mu^{2} \slashed{e} + (1- x) i\mu \slashed{e} \slashed{q} (2 - x) + (1 - xy)(1 - x + xy) q^{2} \slashed{e}
\end{split}
\end{equation}
Collecting terms from (592), (593), we have
\begin{equation}
\Lambda(p, p^{\prime}) = - 4 \iint x \, dx \, dy  \int_{F} dk \, \frac{\slashed{k}\slashed{e}\slashed{k} + (1 - xy)(1 - x+ xy)q^{2} \slashed{e} - (x - x^{2}) i\mu \slashed{e}\slashed{q} + (2 - 2x - x^{2}) \mu^{2}\slashed{e}}{\left[k^{2} + x^{2}\left(\mu^{2} + (y - y^{2})\,q^{2}\right) \right]^{3}}
\end{equation}

Now we saw earlier that the integral (567), evaluated using the conditions $p^{2} + \mu^{2} = 0, \slashed{p} - i \mu = 0$, has the value (573). Comparing (567) with (584), this implies that
\begin{equation}
\Lambda(p, p) = \Lambda(p^{\prime}, p^{\prime}) = I_{1}\, \slashed{e}
\end{equation}
when the conditions (587) with $p^{\prime} = p$ are assumed to hold. Thus $(I_{1} \slashed{e})$ is just the value of (594) for $p^{\prime} = p$. Adding together (583) and (578), this gives
\begin{equation}
M_{2T} = M_{24} + M_{22} + M_{22}^{\prime} + M_{23} + M_{23}^{\prime} = - \frac{ie^{3}}{(2\pi)^{4}\hbar^{2}c^{2}} \left(\overline{u}^{\prime} \Lambda_{c}(p, p^{\prime}) u \right)
\end{equation}
\begin{equation}
\begin{split}
\Lambda_{c}(p, p^{\prime}) &= \Lambda(p, p^{\prime}) - \tfrac{1}{2} \left\{ \Lambda(p, p) + \Lambda(p^{\prime}, p^{\prime}) \right\} \\
&= - 4 \iint x \, dx \, dy \left\{ \int_{F} dk \, \left[ \slashed{k}\slashed{e}\slashed{k} +  (2 - 2x - x^{2}) \mu^{2}\slashed{e} \right] \left( \frac{1}{\left[ k^{2} + x^{2}\left(\mu^{2} + (y - y^{2})\,q^{2}\right) \right]^{3}} - \frac{1}{\left[k^{2} + x^{2}\mu^{2}\right]^{3}} \right) \right.\\
&+ \left. \left[ (1 - xy)(1 - x + xy) q^{2} \slashed{e} - (x - x^{2})i\mu \slashed{e} \slashed{q}\right] \int_{F} dk \, \frac{1}{\left[ k^{2} + x^{2}\left(\mu^{2} + (y - y^{2})\,q^{2}\right) \right]^{3}} \right\}
\end{split}
\end{equation}
The $k$-integrals in (597) are now convergent. Thus the effect of the ``wave-function renormalization'' term (578) is just to cancel out the part of $M_{24}$ which is independent of $q$ and divergent at high frequencies.
\index{renormalization!wave~function}%

To evaluate (597) we use (385) and (386). First, in the term $\slashed{k}\slashed{e}\slashed{k}$ we may replace $k_{\mu} k_{\nu}$ by $\tfrac{1}{4} \delta_{\mu \nu} k^{2}$ because of the symmetry of the integral in $k$-space. So we may write, using (585)
\begin{equation}
\slashed{k}\slashed{e}\slashed{k} = \sum_{\alpha} \tfrac{1}{4} k^{2} \gamma_{\alpha} \slashed{e} \gamma_{\alpha} = - \tfrac{1}{2} k^{2} \slashed{e}
\end{equation}
Then by (385) and (386)
\begin{equation}
\begin{split}
\int_{F}& dk \, k^{2} \left\{ \frac{1}{\left[k^{2}+ \Lambda\right]^{3}} - \frac{1}{\left[k^{2}+ \Lambda^{\prime}\right]^{3}} \right\} = \\
&= \int_{F} dk \, k^{2} \left\{ \frac{1}{\left[k^{2}+ \Lambda\right]^{2}} - \frac{1}{\left[k^{2}+ \Lambda^{\prime}\right]^{2}} - \frac{\Lambda}{\left[k^{2}+ \Lambda\right]^{3}} +   \frac{\Lambda^{\prime}}{\left[k^{2}+ \Lambda^{\prime}\right]^{3}}\right\} = \pi^{2}i \log \frac{\Lambda^{\prime}}{\Lambda}
\end{split}
\end{equation}
and (597) becomes
\begin{equation*}
\begin{split}
\Lambda_{c}(p, p^{\prime}) = 2 \pi^{2} i \iint x\, dx\, dy \, & \left\{ - \slashed{e} \log \left[ 1 + (y - y^{2}) \frac{q^{2}}{\mu^{2}}\right] \right. \\
&+ \left. \frac{\{ x - 1 + 2 (y - y^{2}) (1 - x - x^{2}) \} q^{2} \slashed{e} + (x - x^{2}) i\mu \slashed{e} \slashed{q}}{x^{2}(\mu^{2} + (y - y^{2}) \,q^{2}} \right\}
\end{split}
\end{equation*}
Integrating the logarithmic term by parts with respect to $y$,
\begin{equation}
\Lambda_{c}(p, p^{\prime}) = - 2\pi^{2}i \int_{0}^{1}\int_{0}^{1} dx \, dy \, \frac{1}{x \left[\mu^{2} + (y - y^{2}) \, q^{2} \right]} \left\{ \left[(1  - x)(1 - 2y + 2y^{2}) + x^{2} y \right] q^{2} \slashed{e} - (x - x^{2}) i \mu \slashed{e} \slashed{q} \right\}
\end{equation}

When $q^{2} < - 4 \mu^{2}$, the external potential is able to create real pairs, and the denominator in (600) has poles in the range of the $y$-integration. 
\index{creation!pair}%
In this case the Feynman rule of adding a term $(-i \epsilon)$ to $\mu^{2}$, where $\epsilon$ is an infinitesimal positive real number, will give an unambiguous determination of the integral.
\index{Feynman!$i\epsilon$~prescription}%
Just as in the case of the vacuum polarization  formula (389), the integral will split into a real part and an imaginary part describing the effects of the real pairs created. 
\index{vacuum~polarization}%
\index{creation!pair}%
We shall not discuss these effects of real pairs since they are not practically important. So we assume $q^{2} > -4 \mu^{2}$.

In (600) there are no longer any divergences arising from large $k$. But (584) has a logarithmic divergence at small $k$ which appears in (600) as a divergence in the $x$-integration arising from the factor $(1/x)$. This last remaining divergence must now be examined in detail. It is the famous ``Infra-Red Catastrophe''. 
\index{infra-red~divergence}%

To discover the physical meaning of the $x$-divergence, we consider what would be the effect on our calculations if the Maxwell field were somehow modified so that all field oscillations with wave-numbers satisfying 
\index{Maxwell!field!modified}%
\begin{equation}
|\boldsymbol{k}| \ge r
\end{equation}
were present as usual, while all oscillations not satisfying (601) were simply absent or incapable of being excited. We suppose $r$ to be a constant small compared with $m$, $p$, $p^{\prime}$ and $q$. Thus photons will exist only if their energies exceed
\begin{equation}
\Delta E = \hbar c r
\end{equation}
In the modified Maxwell theory the $D_{F}$ function will still be given by the integral (431), the $k_{1}, k_{2}, k_{3}$ integrations being limited by (601), and the $k_{0}$ integration being taken as usual along the whole real axis from $-\infty$ to $+\infty$. 
\index{Maxwell!electromagnetic~theory}%
Let $\Lambda^{r}(p, p^{\prime})$ and $\Lambda^{r}_{c}(p, p^{\prime})$ be the integrals which replace $\Lambda(p, p^{\prime})$ and $\Lambda_{c}(p, p^{\prime})$ when the Maxwell field is modified. We calculate the differences $(\Lambda - \Lambda^{r})$ and $(\Lambda_{c} - \Lambda^{r}_{c})$, considering these integrals only in the limit of small $r$, neglecting all terms which tend to zero with $r$. This means that we may neglect terms containing either $k$ or $x$ as a factor in the numerator of integrals such as (591) or (594).

In (583) there is just one factor $D_{F}$. Thus $ \Lambda^{r}(p, p^{\prime})$ is obtained from (584) simply by restricting the $k_{1}, k_{2}, k_{3}$ integration by (601). We can now follow the reduction of (584) to the form (594), except that we do not shift the origin of the $k$-integration through $(xp_{y})$ since this would disturb the condition (601). Dropping terms in the numerator having $k$ or $x$ as a factor, this gives the result
\begin{equation}
\Lambda(p, p^{\prime}) - \Lambda^{r}(p, p^{\prime}) = - 4 \iint x\, dx\, dy\, (q^{2} + 2 \mu^{2}) \slashed{e} \int_{F} \frac{dk}{\left[k^{2} + 2 xk \cdot p_{y} \right]^{3} }
\end{equation}
Hence by (597)
\begin{equation}
\begin{split}
&\Lambda_{c}(p, p^{\prime}) - \Lambda_{c}^{r} (p, p^{\prime}) = \\
&= -4 \iint x\, dx\, dy\, \slashed{e} \int_{F} dk \, \left\{ \frac{q^{2} + 2 \mu^{2}}{\left[ k^{2} + 2 x k \cdot p_{y} \right]^{3}} - \frac{\mu^{2}}{\left[ k^{2} + 2 x k \cdot p \right]^{3}} - \frac{\mu^{2}}{\left[ k^{2} + 2 x k \cdot p^{\prime} \right]^{3}} \right\}
\end{split}
\end{equation}
The integral (604), with the integration extended over the whole $k$-space, would give, using (385), (587), and the substitutions $k \rightarrow k - xp_{y}, k \rightarrow  k - xp, k \rightarrow xp^{\prime}$, respectively in the three integrals, 
\begin{equation}
- 2 \pi i \iint dx \, dy \, \frac{(1 - 2y + 2y^{2}) \, q^{2}}{x( \mu^{2} + (y - y^{2}) \, q^{2} )} \slashed{e}
\end{equation}
which is just the divergent part of (600). But the integration in (604) actually extends over $k$ not satisfying (601). Therefore subtracting (604) from (600) and using (605), we find for $\Lambda_{c}^{r}(p, p^{\prime})$ the final result
\begin{equation}
\begin{split}
\Lambda_{c}^{r}(p, p^{\prime}) &= - 2\pi^{2} i \int_{0}^{1} dy\, \frac{1}{\mu^{2} + (y - y^{2}) \, q^{2}} \left\{ (-1 + \tfrac{5}{2} y - 2 y^{2})\, q^{2} \slashed{e} - \tfrac{1}{2}i \mu \slashed{e} \slashed{q} \right\} \\
&-4 \iint x\, dx\, dy\, \slashed{e} \int_{F} dk \, \left\{ \frac{q^{2} + 2 \mu^{2}}{\left[ k^{2} + 2 x k \cdot p_{y} \right]^{3}} - \frac{\mu^{2}}{\left[ k^{2} + 2 x k \cdot p \right]^{3}} - \frac{\mu^{2}}{\left[ k^{2} + 2 x k \cdot p^{\prime} \right]^{3}} \right\}
\end{split}
\end{equation}
This integral is completely convergent for every finite $r$, the $k$-integration being restricted to $k$ satisfying (601). The formula (606) is exact except for terms which tend to zero with $r$. 

To evaluate the $k$ integral in (606) for general $p$ and $p^{\prime}$ is possible but tedious. So we shall do it only in the case of non-relativistic velocities, when 
\begin{equation}
|\boldsymbol{p}| \ll \mu, \qquad\qquad |\boldsymbol{p}^{\prime}| \ll \mu, \qquad\qquad | \boldsymbol{q} | \ll \mu 
\end{equation}
where $|\boldsymbol{p}|$ means $\sqrt{p_{1}^{2} + p_{x}^{2} + p_{3}^{2}}$, the magnitude of the space-like part of the 4-vector $p$. In addition to (607) we still assume $r$ small compared to $q$, $p$, $p^{\prime}$.

We consider the integral 
\begin{equation}
K = \int_{0}^{1} \int_{0}^{1} x\, dx\, dy\, \int_{F} dk \, \frac{1}{\left[k^{2} + 2x k \cdot p_{y} \right]^{3}}
\end{equation}
integrated over $k$ satisfying (601), and evaluate it including terms of order $|\boldsymbol{p}|^{2}$,  $q^{2}$,  $|\boldsymbol{p}^{\prime}|^{2}$, but neglecting higher terms. Integrating over $k_{0}$ only, we have for any positive $b$
\begin{equation}
\int_{F} dk_{0} \, \frac{1}{\left[k^{2} + 2 ak_{0} + b \right]} = i \int_{- \infty}^{\infty} \frac{dk_{0}}{\left[ |\boldsymbol{k}|^{2} + k_{0}^{2} + 2 ia k_{0} + b\right]} = i \pi \frac{1}{\sqrt{|\boldsymbol{k}|^{2} + a^{2} + b}}
\end{equation}
Differentiating (609) twice with respect to $b$, 
\begin{equation}
\int_{F} \frac{dk_{0}}{\left[k^{2}+ 2ak_{0} + b\right]^{3}} = \frac{3i\pi}{8} \left\{ |\boldsymbol{k}|^{2} +a^{2} + b \right\}^{-5/2}
\end{equation}
Hence\endnote{
A sentence, ``Here, $(k \cdot p_{y})_{3}$ denotes the scalar product of the space-like parts of the vectors $k$ and $p_{y}$.'', was deleted, because the expression $\boldsymbol{k} \cdot \boldsymbol{p}_{y}$ is self-explanatory. In both the first and second editions, scant attention was paid to three-vectors; sometimes an overhead arrow was used, but these were very few. In this typed version care has been taken to represent three-vectors with bold type, thus: $(A_{x}, A_{y}, A_{z}) = \boldsymbol{A}$.}
\begin{equation}
\begin{split}
K &= \iint x\, dx\, dy\, \frac{3i\pi}{8} \int_{|\boldsymbol{k}| > r} d^{\,3} \boldsymbol{k} \left\{ |\boldsymbol{k}+ x\boldsymbol{p}_{y}|^{2} + x^{2} \left(\mu^{2}   +(y - y^{2}) \, q^{2} \right) \right\}^{-5/2} \\
&= \iint x\, dx\, dy\, \frac{3i\pi}{8} \int_{|\boldsymbol{k}| > r} d^{\,3} \boldsymbol{k} \left\{ 
\left(|\boldsymbol{k}|^{2} + x^{2}\mu^{2}  \right)^{-5/2}  - \tfrac{5}{2}\left(2x \boldsymbol{k} \cdot \boldsymbol{p}_{y} + x^{2} | \boldsymbol{p}_{y} |^{2}  + 
x^{2}(y - y^{2}) \, q^{2}  \right)   \right. \times \\
& \left. \qquad \qquad \qquad \qquad \qquad \qquad \quad \;\left( |\boldsymbol{k}|^{2} + x^{2} \mu^{2} \right)^{-7/2} + \tfrac{35}{8}\, 4x^{2} (\boldsymbol{k} \cdot \boldsymbol{p}_{y})^{2} \left( |\boldsymbol{k}|^{2} + x^{2} \mu^{2} \right)^{-9/2} \right\} \\
&= \iint x\, dx\, dy\, \frac{3i\pi^{2}}{2} \int_{r}^{\infty} k^{2}\, dk \, \left\{ \left(k^{2} + x^{2}\mu^{2}  \right)^{-5/2}  - \tfrac{5}{2}\, x^{2} \left( |\boldsymbol{p}_{y}|^{2} + (y - y^{2}) \, q^{2}  \right) \left(k^{2} + x^{2}\mu^{2}  \right)^{-7/2}  \right. \\
& \left. \qquad \qquad \qquad \qquad \qquad \qquad \quad \;\ + \tfrac{35}{6}\, x^{2} k^{2} |\boldsymbol{p}_{y}|^{2} \left(k^{2} + x^{2}\mu^{2}  \right)^{-9/2} \right\}
\end{split}
\end{equation}

We can now carry out the integrations over $x$ and $y$, using
\begin{equation}
\int_{0}^{1} |\boldsymbol{p}_{y}|^{2} dy = \boldsymbol{p} \cdot \boldsymbol{p}^{\prime} + \tfrac{1}{3} q^{2}
\end{equation}
This gives 
\begin{equation}
\begin{split}
K = \frac{3i\pi^{2}}{2} \int_{r}^{\infty} k^{2} \, dk\, &
\left\{ \frac{1}{3\mu^{2}} \left( \frac{1}{k^{3}} - \frac{1}{(k^{2} +\mu^{2})^{3/2}} \right) - \frac{\boldsymbol{p} \cdot \boldsymbol{p}^{\prime} + \tfrac{1}{2} 
q^{2}}{3 \mu^{4}} \left(\frac{1}{k^{3}} - \frac{1}{(k^{2} +\mu^{2})^{3/2}}   \right. \right. \\
& \left. \left. - \frac{1}{2\mu^{2}(k^{2} + \mu^{2})^{5/2}} \right) + \tfrac{1}{6} k^{2} \left[ \boldsymbol{p} \cdot \boldsymbol{p}^{\prime} + \tfrac{1}{3} q^{2} \right] \frac{2}{\mu^{4}} \left( \frac{1}{k^{5}} - \frac{1}{(k^{2} + \mu^{2})^{5/2}} \right. \right. \\
& \left. \left. - \frac{5}{\mu^{2}} \frac{1}{(k^{2} + \mu^{2})^{7/2}} \right) \right\}
\end{split}
\end{equation}
The $k$-integration is now elementary, and after dropping terms which tend to zero with $r$ we have 
\begin{equation}
\begin{split}
K &= \frac{3i \pi^{2}}{2 \mu^{2}} \left\{ \tfrac{1}{3}\left(\log \frac{\mu}{2r} + 1 \right) - \frac{\boldsymbol{p} \cdot \boldsymbol{p}^{\prime} + \tfrac{1}{2} 
q^{2}}{ \mu^{2}} \left(\tfrac{1}{3} \log \frac{\mu}{2r} + \tfrac{1}{6} \right) \right. \\
& \qquad \qquad \qquad \left. + \; \tfrac{1}{6} \frac{\boldsymbol{p} \cdot \boldsymbol{p}^{\prime} + \tfrac{1}{3} q^{2}}{ \mu^{2}} \left( 2 \log \frac{\mu}{2r} +  \tfrac{5}{3} \right) \right\} 
\end{split}
\end{equation}
Putting $p = p^{\prime}$, $q = 0$ in (614), we find
\begin{equation}
K_{0} = \int_{0}^{1} \int_{0}^{1} x \, dx\, dy \, \int_{F} dk \, \frac{1}{[k^{2} + 2xk \cdot p]^{3}} =\frac{3i\pi^{2}}{2 \mu^{2}} \left\{ \tfrac{1}{3} \left( \log \frac{\mu}{2r} + 1 \right) + \tfrac{1}{9} \frac{|\boldsymbol{p}|^{2}}{\mu^{2}} \right\}
\end{equation}
Substituting (614) and (615) into (606) and dropping terms of higher order than $|\boldsymbol{p}|^{2}$, $|\boldsymbol{p}^{\prime}|^{2}$ and $q^{2}$,
\begin{equation}
\begin{split} 
\Lambda^{r}_{c}(p, p^{\prime}) &= -2\pi^{2}i \left\{ - \tfrac{5}{12}\, \frac{q^{2}}{\mu^{2}} \slashed{e} - \tfrac{1}{2}\, \frac{i}{\mu} \slashed{e} \slashed{q} \right\} 
 - 4 \slashed{e} \frac{3i \pi^{2}}{2 \mu^{2}} \left\{ \tfrac{1}{3}\,q^{2} \left( \log \frac{\mu}{2r} + 1 \right) - \tfrac{1}{18} \,q^{2} \left(2 \log \frac{\mu}{2r} + \tfrac{5}{3} \right) \right\} \\
&= -\tfrac{4}{3} \pi^{2} i \frac{q^{2}}{\mu^{2}} \slashed{e} \left\{ \log \frac{\mu}{2r} + \tfrac{11}{24} \right\} - \frac{\pi^{2}}{\mu} \slashed{e} \slashed{q}
\end{split}
\end{equation}
Using (596) and adding the contribution (554) from $G_{1}$, we find for the second-order terms in (541) the value
\begin{equation}
M_{2} + M_{2}^{\prime} = -  \frac{\alpha}{3 \pi} \left\{ \log \frac{\mu}{2r} + \tfrac{11}{24} - \tfrac{1}{5} \right\} \frac{q^{2}}{\mu^{2}}\, M_{0} + \frac{\alpha}{4\pi} \frac{ie}{mc^{2}} \left( \overline{u}^{\prime} \slashed{e} \slashed{q} u \right)
\end{equation}

\section*{The Treatment of Low-Frequency Photons. The Infra-Red Catastrophe.}
\addcontentsline{toc}{section}{Treatment of Low-Frequency Photons -- The Infra-Red Catastrophe.}
\index{infra-red~divergence|(}%

The second-order correction (617) to the scattering matrix element $M_{0}$ has been made convergent by only taking into account the effects of photons with energy greater than $\Delta E$ according to (603). When $\Delta E \rightarrow 0$ the correction diverges logarithmically, and this divergence must now be interpreted. 

In the approximation where $| \boldsymbol{p} |^{2}$ and $| \boldsymbol{p}^{\prime} |^{2}$ are small compared with $\mu^{2}$, (546) gives 
\begin{equation}
M_{1} =  \frac{e}{\hbar c \mu | \boldsymbol{k}^{\prime} |} \, (  q \cdot e^{\prime} ) M_{0}
\end{equation}
The total probability that an electron is scattered between the initial and final states (542), with the emission of a photon with potentials (440), summed over all photons with frequencies lying in the range\endnote{
``in'' inserted; the original read ``lying the range''}
\index{probability!scattering!with~one~emitted~photon}%
\begin{equation}
r_{1} < | \boldsymbol{k}^{\prime} | < r_{2}
\end{equation}
is therefore
\begin{equation}
\begin{split}
W_{R}(r_{1}, r_{2}) 
&= \int d^{\,3} \boldsymbol{k}^{\prime} \, \sum_{e^{\prime}} \frac{1}{(2 \pi)^{3}} \left( \frac{\hbar c}{2  | \boldsymbol{k}^{\prime} |} \right) | M_{1} |^{2} \\
&= \frac{e^{2}}{16 \pi^{3} \hbar c \mu^{2}}\, | M_{0} |^{2}\, \int d^{\,3} \boldsymbol{k}^{\prime} \, \frac{1}{ |\boldsymbol{k}^{\prime} |^{3}} \sum_{e^{\prime}} | q \cdot e^{\prime} |^{2} \\
&= \frac{\alpha}{\pi  \mu^{2}} \, | M_{0} |^{2}\, \int_{r_{1}}^{r_{2}} \frac{dk^{\prime}}{k^{\prime}}  \, \tfrac{2}{3} q^{2} \\
&= \frac{2 \alpha}{3 \pi} \left( \log \frac{r_{2}}{r_{1}} \right)  \frac{q^{2}}{\mu^{2}}\, | M_{0} |^{2}
\end{split}
\end{equation}
In (620) it is assumed that both $r_{2}$ and $r_{1}$ are frequencies small compared with $| \boldsymbol{q} |$. 

On the other hand, the probability that an electron is scattered between the states (542) without emitting a photon is given by 
\index{probability!scattering!with~no~emitted~photon}%
\begin{equation}
\begin{split}
W_{N} &= | M_{0} + M_{2} + M_{2}^{\prime} |^{2} \\
&= | M_{0} |^{2} + M_{0}^{*}\left(M_{2} + M_{2}^{\prime}\right) + \left(M_{2} + M_{2}^{\prime}\right)^{*} M_{0}
\end{split}
\end{equation}
neglecting terms of fourth order in the radiation interaction. If we consider in (621) the contribution only from virtual photons with frequencies in the range (619), then by (617) we have
\begin{equation}
\begin{split}
M_{2} + M_{2}^{\prime} &= -\frac{\alpha}{3 \pi} \left( \log \frac{r_{2}}{r_{1}} \right) \frac{q^{2}}{\mu^{2}}\, |M_{0}|^{2} \\
W_{N}(r_{1}, r_{2}) &=  |M_{0}|^{2} - \frac{2 \alpha}{3 \pi} \left( \log \frac{r_{2}}{r_{1}} \right) \frac{q^{2}}{\mu^{2}}\, |M_{0}|^{2}
\end{split}
\end{equation}
So  the contributions to (617) from low-frequency virtual photons serve only to compensate exactly the probability for scattering with emission of low-frequency real photons. 
\index{probability!scattering!infra-red}%
The non-radiative probability is decreased by the effects of low-frequency virtual photons, so that the total scattering probability, radiative plus non-radiative, is essentially independent of the presence of very low-frequency photons. The total scattering probability is thus a finite quantity free from any infra-red divergence.

To describe correctly the radiative corrections to scattering it is essential to specify the critical energy $\Delta E$ below which real photons will not be detected. 
\index{radiative~corrections!scattering|(}%
Ideally we suppose that in each scattering event a photon of energy greater than $\Delta E$ is detected with 100\% efficiency, a photon of energy less than $\Delta E$ with zero efficiency. Then the total observed probability for non-radiative scattering is given by (621), with $M_{2} + M_{2}^{\prime}$ given by (617) and 
\begin{equation}
\log \frac{\mu}{2r}   = \log \left( \frac{mc^{2}}{2 \Delta E} \right)
\end{equation}
This probability (621) will include the scatterings in which a photon is emitted with energy below the limit of detection. The formula (617) is valid so long as 
\index{probability!scattering!infra-red}%
\begin{equation}
r \ll |\boldsymbol{p}|,  |\boldsymbol{p}^{\prime}|,  |\boldsymbol{q}| \ll \mu
\end{equation}
The probability for radiative scattering (i.e. scattering with emission of a detectable photon) is given by (546).

It can be proved that this removal of the infra-red divergence by taking into account the existence of unobservable photons is quite general, and works equally well when $q$ is not small. Only then the evaluation of (608) is considerably more unpleasant. Also, the same argument removes all infra-red divergences, also when higher-order radiative corrections are considered, in which case we are concerned with the effects of the emission of two or more soft photons during the scattering process. 
\index{infra-red~divergence}%
For a general discussion of this question see Bloch and Nordsieck, \emph{Phys.\ Rev.} \textbf{52} (1937) 54. 
\index{Bloch, Felix}%
\index{Nordsieck, Arnold E.}%
\index{infra-red~divergence|)}%

%


\newpage

\pagestyle{fancy}
\fancyhead{}
\lhead{\emph{\MakeUppercase{Scattering By A Static Potential}}}
\chead{}
\rhead{\thepage}
\lfoot{}
\cfoot{}
\rfoot{}

\chapter*{Scattering By A Static Potential. Comparison With Experimental Results.}
\addcontentsline{toc}{chapter}{Scattering By A Static Potential}
\hspace{3ex}Consider the scattering of an electron by a time-independent electrostatic potential
\begin{equation}
V(r) = \frac{1}{(2\pi)^{3}} \int d^{\,3} \boldsymbol{q} \, V(\boldsymbol{q})\, e^{i \boldsymbol{q} \cdot \boldsymbol{r}}
\end{equation}
Then (543) gives the matrix element for scattering without radiative corrections (compare (625) and (534))
\begin{equation}
M_{0} = 2\pi i \frac{e}{\hbar c} (u^{\prime *} u) V(\boldsymbol{q}) \, \delta(q_{0})
\end{equation}
Non-radiative scattering occurs only between states for which
\begin{equation}
q_{0} = 0, \qquad \qquad | \boldsymbol{p} | =  | \boldsymbol{p}^{\prime} | 
\end{equation}
The cross-section for scattering between the states (542), per element of solid angle $d \Omega$ in the direction of $p^{\prime}$, is without radiative corrections
\index{cross-section!scattering~by~a~static~potential}%
\index{angle!solid}%
\begin{equation}
\sigma_{0} = \left( \frac{em}{2\pi \hbar^{2}} \right)^{2} | u^{\prime *} u |^{2} \, | V(\boldsymbol{q}) |^{2} d \Omega
\end{equation}
This follows immediately from (626) using (627), using the prescription derived on page 29, when we treated the M\o ller scattering the first time. 
\index{Moller@M\o ller~scattering}%
The procedure is as follows:\endnote{
In the first line, twice in the fourth line, and in the first appearance in the sixth line, the expression $| u^{\prime *} u  |$ lacked the exponent 2. These have been supplied.}
\[
w_{S} = \frac{c |M_{0}|^{2}}{2 \pi\, \delta(q_{0})} = 2 \pi \frac{e^{2}}{\hbar^{2}c}\, | u^{\prime *} u  |^{2} \, | V(\boldsymbol{q}) |^{2} \,  \delta(q_{0})
\]
\[
\rho \,dE =\frac{mc^{2}}{E} \, \frac{d^{\,3}\boldsymbol{p}}{(2\pi)^{3}} \qquad \qquad E\,dE = \hbar^{2}c^{2} p\,dp \qquad \qquad d^{\,3}\boldsymbol{p} = p^{2} dp \, d\Omega
\]
\[
\rho = \frac{mc^{2}}{E}\, \frac{p^{2}}{(2\pi)^{3}} \, \frac{dp}{dE} \, d\Omega = \frac{mp}{\hbar^{2} (2\pi)^{3}} \, d\Omega \qquad \qquad \delta(q_{0}) = \hbar c \, \delta(E)
\]
\[w = \frac{2 \pi e^{2}}{\hbar^{2}c} \,\hbar c \, \frac{mp}{\hbar^{2}(2\pi)^{3}} \, d\Omega\, | u^{\prime *} u |^{2} \, | V(\boldsymbol{q}) |^{2} = \frac{e^{2}mp}{(2\pi)^{2} \hbar^{3}} \, d\Omega \,| u^{\prime *} u |^{2} \, | V(\boldsymbol{q}) |^{2}
\]
\[
\sigma = \frac{wV}{v} \qquad \qquad V = \frac{mc^{2}}{E} \qquad \qquad v = \frac{c^{2}\hbar p}{E}
\]
\[
\sigma = \frac{e^{2}mp}{(2\pi)^{2} \hbar^{3}} \, \frac{mc^{2}}{E} \, \frac{E}{c^{2}\hbar p} \, d\Omega \,  | u^{\prime *} u |^{2} \, | V(\boldsymbol{q}) |^{2} =  \left( \frac{em}{2\pi\hbar^{2}} \right)^{2} \, d\Omega\, | u^{\prime *} u |^{2} \, | V(\boldsymbol{q}) |^{2}
\]
Summing over the final spin states, and averaging over the initial spin states, we have
\begin{align}
\tfrac{1}{2} \sum_{u} \sum_{u^{\prime}} | u^{\prime *} u |^{2} 
&= \tfrac{1}{2} \frac{1}{(2i \mu)^{2}} \, \text{Spur} \left\{ (\slashed{p} + i \mu) \gamma_{4} (\slashed{p}^{\prime} + i \mu)\gamma_{4} \right\} \notag \\
&= \frac{1}{2 \mu^{2}} \left\{ \mu^{2} + p_{0}p_{0}^{\prime}  + \boldsymbol{p} \cdot \boldsymbol{p}^{\prime} \right\} = \frac{1}{2 \mu^{2}} \left\{2 p_{0}^{2} -  \tfrac{1}{2} | \boldsymbol{q} |^{2} \right\} \\
&= \frac{p_{o}^{2}}{\mu^{2}} \left\{ 1 - \tfrac{1}{4} \frac{ | \boldsymbol{p} |^{2} - 2\boldsymbol{p} \cdot \boldsymbol{p}^{\prime} + | \boldsymbol{p}^{\prime} |^{2}}{p_{0}^{2}} \right\} = \frac{p_{o}^{2}}{\mu^{2}} \left( 1 - \beta^{2} \sin^{2} \frac{\theta}{2} \right) \notag
\end{align}
where $\theta$ is the angle between $p$ and $p^{\prime}$, and 
\index{angle}%
\begin{equation}
\beta = \frac{| \boldsymbol{p} |}{p_{0}} = \frac{v}{c}
\end{equation}
where $v$ is the velocity of the incident electron. Hence the cross-section  for an unpolarized electron beam is
\index{cross-section!unpolarized~electron~beam}%
\begin{equation}
\overline{\sigma}_{0} = \left( \frac{eE}{2 \pi \hbar^{2} c^{2}} \right)^{2} \left( 1 - \beta^{2} \sin^{2} \frac{\theta}{2} \right) \, | V(\boldsymbol{q}) |^{2} \, d\Omega
\end{equation}
where $E$ is the energy of the incident electron. 

The second-order radiative correction to $M_{0}$ is given by (617), which in this case becomes 
\begin{equation}
M_{2} + M_{2}^{\prime} = - \frac{\alpha}{3 \pi} \left\{ \log \frac{\mu}{2r} + \tfrac{11}{24} - \tfrac{1}{5} \right\} \frac{q^{2}}{\mu^{2}}\, M_{0} - \frac{\alpha}{2}\, \frac{e}{\hbar c \mu} \, (u^{\prime *} \slashed{q} u) V(\boldsymbol{q}) \, \delta(q_{0})
\end{equation}
This gives a second order correction to the cross-section $\sigma_{0}$ according to (621). The total non-radiative cross-section, for scattering without the emission of a photon of energy greater than $\Delta E$, becomes 
\index{cross-section!non-radiative|(}%
\begin{equation}
\begin{split}
\sigma_{N} &= \sigma_{0} + \sigma_{2N}   \\
&= \left( \frac{em}{2\pi \hbar^{2}} \right)^{2} | V(\boldsymbol{q}) |^{2} \, d\Omega  \left| \left\{1 - \frac{\alpha}{3\pi} \left(\log \frac{\mu}{2r} + \tfrac{11}{24} - \tfrac{1}{5}\right)\,\frac{q^{2}}{\mu^{2}} \right\} (u^{\prime *} u) + \frac{i\alpha}{4 \pi \mu} \, (u^{\prime *} \slashed{q} u) \right|^{2}
\end{split}
\end{equation}
Summing and averaging over the spin-states 
\begin{equation}
\begin{split}
\tfrac{1}{2} \sum_{u} \sum_{u^{\prime}} (u^{\prime *}\slashed{q} u)(u^{*} u^{\prime}) &= \frac{1}{2(2i\mu)^{2}} \, \text{Spur} \left\{ (\slashed{p} + i\mu) \gamma_{4} (\slashed{p}^{\prime} + i \mu) \gamma_{4} \slashed{q} \right\} \\
&= - \frac{1}{8\mu^{2}} \, \text{Spur} \left\{ i \mu \left( \slashed{p} \gamma_{4} \gamma_{4} \slashed{q} + \gamma_{4} \slashed{p}^{\prime} \gamma_{4}\slashed{q} \right) \right\} = - \frac{1}{8\mu^{2}} \, \text{Spur} \left\{i\mu ( \slashed{p}\slashed{p}^{\prime} - \slashed{p}\slashed{p} - \slashed{p}^{\prime}\slashed{p}^{\prime} + \slashed{p}^{\prime}\slashed{p}) \right\} \\
&= - \frac{1}{8\mu^{2}} \, \text{Spur} \left\{i\mu (\slashed{p}^{\prime} - \slashed{p})(\slashed{p}^{\prime} - \slashed{p}) \right\} = \frac{i}{2\mu}\, q^{2}
\end{split}
\end{equation}
Hence for an unpolarized electron beam the non-radiative cross-section is 
\index{cross-section!non-radiative|)}%

\begin{equation}
\overline{\sigma}_{N} = \left( 1 - \frac{2\alpha}{3\pi} \left( \log \frac{\mu}{2r} +  \tfrac{11}{24} - \tfrac{1}{5} \right) \frac{q^{2}}{\mu^{2}} \right) \overline{\sigma}_{0} - \left(\frac{em}{2\pi \hbar^{2}} \right)^{2} | V(\boldsymbol{q}) |^{2} \, d\Omega \, \frac{\alpha}{4\pi} \, \frac{q^{2}}{\mu^{2}}
\end{equation}
Since we are working only to order $q^{2}$ in the radiative corrections, the second term in (635) may be replaced by
\begin{equation}
-\frac{\alpha}{4\pi} \, \frac{q^{2}}{\mu^{2}} \, \overline{\sigma}_{0}
\end{equation}
and then we find 
\begin{equation}
\sigma_{N} = \left( 1 - \frac{2\alpha}{3\pi} \left( \log \frac{mc^{2}}{2 \Delta E} + \tfrac{5}{6} - \tfrac{1}{5}  \right)  \frac{q^{2}}{\mu^{2}} \right) \overline{\sigma}_{0}
\end{equation}
Formulae (628) and (631) are exact for electrons of any energy; while (632) and (637) are valid only for slow electrons, terms of order higher than $\alpha q^{2}$ being neglected. 

To remove the dependence of (637) on $r$ we must consider the cross-section for scattering with emission of a photon of frequency greater than $r$. Since the electron is now considered to be slow, the maximum possible energy of the photon is 
\begin{equation}
\hbar ck_{\text{max}} = E - mc^{2} \approx \frac{\hbar^{2}}{2m} | \boldsymbol{p} |^{2}
\end{equation}
and therefore for all possible photons the momentum $\hbar | \boldsymbol{k}^{\prime} |$ will be very small compared with the electron momentum $\hbar | \boldsymbol{p}|$.\endnote{
The modulus bars around $p$ were absent.} 
 Thus the recoil of the electron, arising from the momentum carried off by the photon, can always be neglected. The matrix element and probability for radiative scattering will be given by (618) and (620), even when the photon takes away a large fraction of the kinetic energy of the electron. 
\index{probability!scattering!radiative}%

We imagine a scattering experiment in which only the direction of the emerging electron is measured and not its energy. Then the radiative cross-section $\overline{\sigma}_{R}$ 
will measure the total probability of scattering the electron into a solid angle $d\Omega$ with emission of a photon having any frequency between the lower limit $r$ and the upper limit $k_{\text{max}}$ given by (638).
\index{cross-section!radiative|(}%
\index{angle!solid}%
The observed cross-section will be
\begin{equation}
\sigma_{T} = \overline{\sigma}_{N} + \overline{\sigma}_{R}
\end{equation}
with the same low-frequency cut-off $r$ in both $\overline{\sigma}_{N}$ and $\overline{\sigma}_{R}$. Thus $\sigma_{T}$ gives the cross-section for scattering into a given solid angle $d\Omega$ with or without photon emission. Being directly observed, $\sigma_{T}$ must be divergence-free and independent of $r$.
\index{angle!solid}%

In the radiative scattering process we may take the final momentum of the electron to be $\lambda \hbar p^{\prime}$ where $0 < \lambda < 1$ and $p^{\prime}$ satisfies (627). Instead of (627) the conservation of energy now gives by (638) 
\begin{equation}
\hbar | \boldsymbol{p} |^{2} (1 - \lambda^{2}) = 2mc |\boldsymbol{k}^{\prime} |
\end{equation}
According to (620), the probability for scattering the electron into a state $\lambda \hbar p^{\prime}$, with emission of a photon in any direction with frequency in the range $(k^{\prime}, k^{\prime}  + dk^{\prime})$ is
\index{probability!scattering!radiative}%
\begin{equation} 
w_{R}(k^{\prime}) = \frac{2 \alpha}{3 \pi} \frac{dk^{\prime}}{k^{\prime}}\, \frac{| \boldsymbol{p} - \lambda \boldsymbol{p}^{\prime} |^{2}}{\mu^{2}}\, |M_{0}|^{2}
\end{equation}
This corresponds to a differential cross-section 
\index{cross-section!differential}%
\begin{equation}
\sigma_{R}(k^{\prime}) = \frac{2 \alpha}{3 \pi} \frac{dk^{\prime}}{k^{\prime}}\, \frac{| \boldsymbol{p} - \lambda \boldsymbol{p}^{\prime} |^{2}}{\mu^{2}}\,\lambda \left(\frac{eE}{2\pi \hbar^{2} c^{2}} \right)^{2} |V(p - \lambda p^{\prime}) |^{2} \, d\Omega
\end{equation}
for scattering into the solid angle $d\Omega$, neglecting now the term in $\beta$ which appeared in (631) since (642) is itself of order $\alpha \beta^{2}$ and higher terms are neglected. 
\index{angle!solid}%
The factor $\lambda$ comes in from $\dfrac{p_{\text{final}}}{p^{\prime}}$. According to (640)
\begin{equation}
\frac{dk^{\prime}}{k^{\prime}} = - \frac{2 \lambda \, d\lambda}{1 - \lambda^{2}}
\end{equation}
Hence the radiative cross-section integrated over the quantum frequency is
\index{cross-section!radiative}%
\begin{equation}
\overline{\sigma}_{R} = \frac{2 \alpha}{3 \pi} \int_{0}^{\lambda_{m}} \frac{2 \lambda^{2} \, d\lambda}{1 - \lambda^{2}}\, \frac{| \boldsymbol{p} - \lambda \boldsymbol{p}^{\prime} |^{2}}{\mu^{2}}\,\left(\frac{eE}{2\pi \hbar^{2} c^{2}} \right)^{2} |V(p - \lambda p^{\prime}) |^{2} \, d\Omega
\end{equation}
where by (640) and (638) 
\begin{equation}
\lambda_{m} = \sqrt{1 - \frac{r}{k_{\text{max}}}} = \sqrt{1 - \frac{\Delta E}{T}}
\end{equation}
$T$ is the initial kinetic energy of the electron given by (638).

Now we can combine (637) and (644) to give by (639) and (629) 
\begin{equation}
\begin{split}
\sigma_{T} &= \left(1 - \frac{2\alpha}{3 \pi} \left( \log \frac{mc^{2}}{2T} + \tfrac{5}{6} - \tfrac{1}{5} \right) 4\beta^{2} \sin^{2} \frac{\theta}{2} \right) \overline{\sigma}_{0} \\
& +  \frac{2\alpha}{3 \pi} \left( \frac{e}{2 \pi \hbar c} \right)^{2} \, d\Omega \times \int_{0}^{1} \frac{2 \lambda \, d\lambda}{1 - \lambda^{2}}\, \left\{\lambda \, | \boldsymbol{p}-\lambda \boldsymbol{p}^{\prime} |^{2} \, |V(p - \lambda p^{\prime}) |^{2} - | \boldsymbol{p} - \boldsymbol{p}^{\prime} |^{2}\, |V(p - p^{\prime}) |^{2} \right\}
\end{split}
\end{equation}
Here we used the following trick: the integral over $\lambda$ blows up at $\lambda = 1$. Therefore we subtract from the numerator its value at $\lambda = 1$, which makes the integral behave decently and permits us to change the upper limit from $\lambda_{m}$ to $1$, for small $\Delta E$'s. We also have to add then the integrand with the numerator having $\lambda = 1$; this gives then a logarithmic term which combines with (637) to give the first part of (646). 

Formula (646) gives a result of the form
\begin{equation}
\sigma_{T} =\left(1 - \frac{8 \alpha}{3\pi} \beta^{2} \sin^{2} \frac{\theta}{2} \left\{\log \frac{mc^{2}}{2T} + f(\theta) \right\} \right) \overline{\sigma}_{0}
\end{equation}
where for low velocities $f(\theta)$ is independent of $T$ and of order 1 compared to the logarithm. For any special potential $f(\theta)$ can be calculated. 

From (647) we see that the observable radiative correction is not of order $\alpha$ but of order
\index{fine~structure~constant}%
\begin{equation}
\alpha \left(\frac{v}{c}\right)^{2} \log \left( \frac{c}{v} \right)
\end{equation}
which is much smaller if $v$ is not relativistic. Thus the correction \emph{cannot be observed at all} in a non-relativistic scattering experiment. In the relativistic region the effect is actually of order $\alpha$ as indicated by (647), but the correct formula is then enormously more complicated. 

The exact formulae in both non-relativistic and relativistic cases have been published by J.\ Schwinger, \emph{Phys.\ Rev.} \textbf{76} (1949) 790. %

An experimental test in the relativistic range is just on the limits of the possible. See Lyman, Hanson and Scott, \emph{Phys.\ Rev.} \textbf{84} (1951) 626. 
\index{Lyman, E. M.}%
\index{Hanson, A. O.}%
\index{Scott, M. B.}%
Scattering of 15 MeV electrons by nuclei were measured with very good energy-resolution, $\Delta E/E$ being 1 - 3 \%. In this case it was \emph{only} the non-radiative cross-section $\overline{\sigma}_{N}$ that was observed, and so the radiative correction given by the relativistic form of (637) becomes quite large. In fact the radiative correction in $\overline{\sigma}_{N}$ in the relativistic range becomes of the order
\begin{equation}
\alpha \left\{ \log \frac{\Delta E}{E} \right\} \left\{ \log \frac{E}{mc^{2}} \right\}
\end{equation}
according to Schwinger, \emph{Phys.\ Rev.} \textbf{76}, 813, Eq.\ (2.105) (with misprint $K$ for $k$, in my notation $\mu$.) In the conditions of the experiment of Lyman-Hanson-Scott (649) becomes of the order 5\% and was clearly observed, the experimental errors being $\sim 2\%$. However, (649) arises mainly from the low-energy virtual photons, with energies going down to $\Delta E$. What is observed is just the decrease in the non-radiative cross-section due to the competition of radiative scattering with energy-loss in the range $[ \Delta E, E ]$. Thus the measurement of (649) by LHS is actually only a very inaccurate measurement of the cross-section for bremsstrahlung, which could be observed much more accurately by observing the photons which are really emitted. 
\index{bremsstrahlung}%
\index{cross-section!bremsstrahlung}%
 
The theoretically interesting part of the radiative corrections is the part which is not just the effect of real bremsstrahlung. 
\index{bremsstrahlung}%
This part is given by the terms in $\overline{\sigma}_{N}$ which are of order 1 compared with the $\log (\Delta E/E)$ appearing in (649). For example, we would have to observe with sufficient accuracy to see the terms $(\tfrac{5}{6} - \tfrac{1}{5})$ in (637) if we wished to verify the theoretical radiative corrections at low velocities. In the relativistic range the ``genuine'' radiative corrections are of order 
\begin{equation}
\alpha \log \left( \frac{E}{mc^{2}} \right)
\end{equation}
instead of (649), i.e. about 2\% in the LHS experiment. To detect such effects is already possible, but to observe them accurately in a scattering experiment seems not very hopeful. 
\index{cross-section!radiative|)}%

This is all we can say at present about radiative corrections to scattering by an electrostatic potential.

\section*{A. The Magnetic Moment of the Electron}
\addcontentsline{toc}{section}{A. The Magnetic Moment of the Electron}
\index{electron!anomalous~magnetic~moment|(}%

The scattering by an electrostatic potential, the two terms in (617) were lumped together. Both gave contributions of the same order of magnitude, $\alpha (q^{2}/\mu^{2})$ in the cross section. What then is the meaning of the special form of the second term in (617)? This term has no infra-red divergence and therefore should be particularly simple to interpret experimentally.
\index{infra-red~divergence}%

Consider scattering of a slow electron by a slowly-varying\endnote{
``slowly-varying'' replaces ``slowly-carrying''} 
magnetic field. The potentials (534) can then be taken to be a pure vector potential, so that
\begin{equation}
e_{4}(q) = 0
\end{equation}
The matrix elements of $\gamma_{1}, \gamma_{2}, \gamma_{3}$ between positive energy electron states are of the order $(v/c)$. Hence $M_{0}$ given by (543) is of order $(v/c)$. The first term in (617) is thus of the order $\alpha (v/c)^{3}$ while the second is of order $\alpha (v/c)$. Therefore the second term in (617) is the main term in considering magnetic effects, and the first term can be ignored. The meaning of the second term must be a change in the \emph{magnetic properties} of a non-relativistic electron.

As we saw in discussing the Dirac equation, (Eq's (99) and (100)), an electron by virtue of its charge $(-e)$ behaves in non-relativistic approximation  as if it had a magnetic moment  
\index{Dirac!equation}%
\index{electron!magnetic~moment}%
\begin{equation}
M = - \frac{e\hbar}{2mc}
\end{equation}
This moment has an energy of interaction with an external Maxwell field $(\boldsymbol{E}, \boldsymbol{H})$ given by
\index{Maxwell!field!external~classical}%
\begin{equation}
H_{M} = - M (\boldsymbol{\sigma} \cdot \boldsymbol{H} - i \boldsymbol{\alpha} \cdot \boldsymbol{E} )
\end{equation}
the term which appears in the non-relativistic Schr\"{o}dinger equation (100). 
\index{Schr\"{o}dinger!equation}%

Now suppose that the electron possesses an additional magnetic moment $\delta M$ which does not arise from its charge. Such an additional moment is called  ``anomalous''.  To give the electron an anomalous moment, we only need to add arbitrarily a term proportional to (653) to the Hamiltonian. 
\index{Hamiltonian!and~anomalous~magnetic~moment}%
Comparing (654) with (97) and (98), we see that (653) is a relativistic invariant and can be written
\begin{equation}
H_{M} = \tfrac{1}{2} i M \sum_{\mu} \sum_{\nu} \sigma_{\mu \nu} F_{\mu \nu}
\end{equation}
Hence an anomalous magnetic moment $\delta M$ will be given to the electron if the term 
\begin{equation}
L_{M} = -\tfrac{1}{2} i \delta M \sum_{\mu} \sum_{\nu} \sigma_{\mu \nu} F_{\mu \nu}
\end{equation}
is added to the Lagrangian. This refers still to the one-electron Dirac equation.
\index{Dirac!equation}%
In the theory of the quantized Dirac field, the corresponding addition to the Lagrangian density (410) is 
\index{Lagrangian~density!inclusion~of~anomalous~magnetic~moment}%
\begin{equation}
\mathscr{L}_{M} = -\tfrac{1}{2} i \delta M \overline{\psi} \sum_{\mu} \sum_{\nu} \sigma_{\mu \nu} \psi  F_{\mu \nu}^{e}
\end{equation}
where it is supposed that the anomalous moment is interacting with the external Maxwell field. 
\index{Maxwell!field!external~classical}%
The addition (656) to the Lagrangian\endnote{
For ``Lagrangian'' read here ``Lagrangian density''. Field theorists, by an abuse of language, often say the first and mean the second.}
gives a relativistically invariant description of an anomalous moment.

Consider the effect of (656) on the scattering of an electron by the potentials (534). Treating the scattering in Born approximation and using (420), the contribution of (656) to the scattering matrix element is
\index{Born~approximation}%
\begin{equation}
U_{M} = \sum_{\mu, \nu} \frac{\delta M}{2\hbar c} \int \overline{\psi}(x) \sigma_{\mu \nu} \psi(x) F_{\mu \nu}^{e}(x)\, dx
\end{equation}
the integral being over all space-time. Using (542) for the initial and final electron wave-functions, and defining $q$, $e$ by (544) and (545), this matrix element becomes 
\begin{equation}
U_{M} = i \frac{\delta M}{2 \hbar c} \sum_{\mu, \nu} (\overline{u}^{\prime} \sigma_{\mu \nu} u) (q_{\mu} e_{\nu} - q_{\nu}e_{\mu}) = i \frac{\delta M}{2 \hbar c} \left[ \overline{u}^{\prime} (\slashed{q} \slashed{e} - \slashed{e} \slashed{q}) u \right]
\end{equation}
where we used $\gamma_{k} \gamma_{\ell} = i \sigma_{m}, \; k, \ell, m = (1, 2, 3) $ cyclically permuted. Since we have also assumed (588) we may write simply 
\begin{equation}
U_{M} = - i \frac{\delta M}{\hbar c} ( \overline{u}^{\prime} \slashed{e} \slashed{q} u)
\end{equation}

Now comparing the matrix element (659) with (617), we see that the magnetic effect of the second-order radiative correction to scattering is exactly described by saying that the electron has an anomalous magnetic moment $\delta M$ given by 
\begin{equation}
\delta M = - \frac{\alpha}{4\pi} \, \frac{e\hbar}{mc} = + \frac{\alpha}{2\pi} M
\end{equation}
This is the famous Schwinger correction to the electron magnetic moment, which we have now calculated.
\index{Schwinger, Julian}%
\index{electron!anomalous~magnetic~moment!Schwinger~correction}%
\index{fine~structure~constant}%
Not only for scattering but for all phenomena in the non-relativistic range, the magnetic part of the second-order radiative correction to the motion of an electron is equivalent simply to the anomalous magnetic moment (660).  

This anomalous moment  has been extremely accurately confirmed experimentally by\endnote{
The original citation lacked Prodell's name.} 
Kusch, Prodell and Koenig (\emph{Phys.\ Rev.} \textbf{83} (1951) 687) who find 
\index{Kusch, Polykarp}%
\index{Koenig, Seymour~H.}%
\index{Prodell, Albert~G.}%
\[
\frac{\delta M}{M} = 0.001145 \pm 0.000013
\]
The calculated value including a fourth-order $\alpha^{2}$ correction found by Karplus and Kroll (\emph{Phys. Rev.} \textbf{77} (1950) 536) is 
\index{Karplus, Robert}%
\index{Kroll, Norman~M.}%
\index{electron!anomalous~magnetic~moment|)}%
\index{fine~structure~constant}%
\[
\frac{\delta M}{M} = \frac{\alpha}{2\pi} - 2.973 \left( \frac{\alpha^{2}}{\pi^{2}} \right) = 0.0011454 
\]

\section*{B. Relativistic Calculation of the Lamb Shift}
\addcontentsline{toc}{section}{B. Relativistic Calculation of the Lamb Shift} 
\index{Lamb~shift}%

To make a correct relativistic calculation of the Lamb shift we have to repeat the treatment of line-shifts and line-widths which we gave earlier, only now using the relativistic theory for the atom. 
\index{Lamb~shift}%
So we should set up the equation of motion of atom plus radiation field in the \emph{Bound Interaction Representation}. 
\index{bound~interaction~representation}%
The equation of motion is then given by (245), (247), only with the $j_{\mu}$ operator now describing the system of a relativistic atom. The solution of (245) can then be found as in the non-relativistic case, using the known wave-functions of the stationary states of the atom. In this way the Lamb shift  was actually calculated by Lamb and Kroll\endnote{
In Schwinger's anthology.},
\emph{Phys.\ Rev.} \textbf{75} (1949) 388.
\index{Lamb~shift}%
\index{Kroll, Norman~M.}%
\index{Lamb, Willis~E.}%
However, in their calculations Lamb and Kroll had troubles with the subtraction of the divergent mass-renormalization effect. 
\index{renormalization!mass}%
Because the calculation was all in terms of the atomic wave-functions, it was not possible to keep using relativistically invariant notations all the way through. Hence the mass term could not be clearly separated from the remaining finite terms by its dependence on the particle momentum $p$, as it was separated for example in Eq. (566) during the calculation of radiative corrections to scattering. The final result of Lamb and Kroll was uncertain because of this difficulty in the mass separation.
\index{fine~structure~constant}%
They obtained the correct answer 1052 Mc but only by making use of the experimentally \emph{measured} value $\alpha/2\pi$ of the electron anomalous magnetic moment.\endnote{
The notation ``Mc'' is outmoded; usually this is written ``MHz.''}
\index{electron!anomalous~magnetic~moment}%
\index{radiative~corrections!scattering|)}%

Learning from the calculation of the radiative corrections to scattering, we see that to make a clear separation of the mass renormalization from observable effects, we must arrange the calculations so that the separation is done for a particle of momentum $p$ in a \emph{variable} Lorentz system.
\index{Lorentz!system}%
Then by varying the Lorentz system we can vary $p$ and identify the mass term unambiguously as the expression which has the correct dependence on $p$. To work in a variable Lorentz system, we must work in a representation which is independent of the Lorentz system, so that the calculations are formally invariant. The only convenient invariant representation is the \emph{Free Interaction Representation}.
\index{free~interaction~representation}%

Hence our program is to set up the equation of motion first in the \emph{Free Interaction Representation}, then carry through a relativistically invariant calculation to identify and cancel the divergent renormalization effects unambiguously. After this we must transform to the \emph{Bound Interaction Representation} for the final calculation of the line-shift.  
\index{bound~interaction~representation}%
This two-stage calculation is absolutely necessary in order to get the right answers. It was Schwinger's invention. 
\index{bound~interaction~representation!Schwinger~invention}%

In the FIR the equation of motion is 
\begin{equation}
i \hbar \frac{\partial \Psi}{\partial t} = \left( H^{e}(t) + H^{I}(t) \right) \Psi
\end{equation}
with $H^{I}$ given by (532) including the mass-renormalization term $H_{S}$. The first stage in the solution of (661) is to write
\begin{equation}
\Psi(t) = \Omega_{1} \Phi(t)
\end{equation}
where $\Omega_{1}(t)$ is defined as in (492), only with $e\overline{\psi}\slashed{A}\psi$ replaced by $[e\overline{\psi}\slashed{A}\psi + i \,\delta m\, c^{2} \overline{\psi} \psi]$, and the function $g_{A}(t)$ is as before supposed to tend to the limiting form $g_{A}(t) \rightarrow 1$ at the end of the calculation. The operator $\Omega_{1}(t)$ satisfies 
\begin{equation}
i \hbar \frac{\partial\, \Omega_{1}(t)}{\partial t} = H^{I}(t) \, \Omega_{1}(t)
\end{equation}
for any value of $t$ not in the remote past, so that we may take $g_{A}(t) = 1$. Hence the equation of motion for $\Phi(t)$ derived from (661) is just
\begin{equation}
i \hbar \frac{\partial \Phi}{\partial t} = H_{T}(t) \Phi
\end{equation}
\begin{equation}
H_{T}(t) = \left( \Omega_{1}(t)\right)^{-1} H^{e}(t) \, \Omega_{1}(t)
\end{equation}
The covariant part of the calculation, which is done in the FIR, is just the evaluation of this transformed Hamiltonian $H_{T}(t)$.
\index{Hamiltonian!transformed}%

\subsection*{Covariant Part of the Calculation}
\addcontentsline{toc}{subsection}{Covariant Part of the Calculation}

Let us write
\begin{equation}
H_{F}(t) = \Omega_{2}(t) \, H^{e}(t) \, \Omega_{1}(t)
\end{equation}
with $\Omega_{2}$ given by (490). Then
\begin{equation}
H_{T}(t) = \left( \Omega_{2}(t) \Omega_{1}(t) \right)^{-1} H_{F}(t) = S^{-1}H_{F}(t)
\end{equation}
where $S$, given by (495), is the scattering matrix defined in the absence of the external potential $A_{\mu}^{e}$. We shall now restrict our attention to systems in which only one electron is actually present. We saw by Eq. (533) that  $S$ applied to a one-electron state is equivalent to the unit operator, i.\ e.g.  $S$ produces no real scattering or shift of phase in one-particle states. Therefore in discussing the hydrogen atom  we may simply omit the $S^{-1}$ in (667) and write
\index{hydrogen~atom!Lamb~shift|(}%
\begin{equation}
H_{T}(t) = H_{F}(t)
\end{equation}

Now $H_{F}(t)$ is just the term involving $H^{e}(t)$ in the series expansion (421). In fact we have, by direct multiplication of the series for $\Omega_{1}$ and $\Omega_{2}$, 
\begin{equation}
H_{F}(t) = \sum_{n = 0}^{\infty} \left( \frac{-i}{\hbar} \right)^{n} \frac{1}{n!} \int \dots \int dt_{1} \, dt_{2} \, \dots \,dt_{n} \, P\left\{ H^{e}(t), H^{I}(t_{1}), \dots, H^{I}(t_{n}) \right\}
\end{equation}
the damping functions $g(t_{i})$ being always understood when they are not written explicitly. Now (667) is directly related to the operator $U$ given by (535), namely
\begin{equation}
U = - \frac{i}{\hbar} \int_{- \infty}^{\infty} H_{F}(t) \, dt
\end{equation}
We write
\begin{equation}
H_{T}(t) = H^{e}(t) + H_{T1}(t) + H_{T2}(t)
\end{equation}
expanding $H_{T}$ in powers of the radiation interaction just as $U$ was expanded in (536)--(540). But the matrix elements of $\left(U_{2} + U_{2}^{1} \right)$ referring to one-electron transition have already been calculated and are given by (617) assuming non-relativistic velocities for the electron. This enables us to write down immediately a formula for the operator $H_{T2}$, valid for one-electron transitions at non-relativistic velocities. In (617) each factor $q_{\lambda}$ may be replaced by $(-i \partial / \partial x_{\lambda})$ operating on the potentials (534). Then (617) becomes 
\begin{equation}
\begin{split}
U_{2} + U_{2}^{\prime} &= \frac{1}{\mu^{2}} \, \frac{\alpha}{3 \pi} \, \left\{ \log \frac{\mu}{2r} + \tfrac{11}{24} - \tfrac{1}{5} \right\} \left( \frac{e}{\hbar c} \right) \int \overline{\psi} (\Box^{2} \slashed{A} ) \psi(x) \, dx \\
&+ \frac{\alpha}{4 \pi} \, \frac{e}{mc^{2}} \int \overline{\psi} \sum_{\lambda} \frac{\partial \slashed{A}}{\partial x_{\lambda}} \gamma_{\lambda} \psi(x) \, dx
\end{split}
\end{equation}
And using (670) and (668) we deduce 
\begin{equation}
\begin{split}
H_{T2} &= \frac{ie}{\mu^{2}} \, \frac{\alpha}{3 \pi} \left\{ \log \frac{\mu}{2r} + \tfrac{11}{24} - \tfrac{1}{5} \right\} \int \overline{\psi} (\Box^{2} \slashed{A}) \psi(x) \, d^{\,3} \boldsymbol{x} \\
&+ \frac{\alpha}{4\pi} \, \frac{ie\hbar}{mc} \int \overline{\psi} \sum_{\lambda} \frac{\partial \slashed{A}}{\partial x_{\lambda}} \gamma_{\lambda} \psi(x) \, d^{\,3} \boldsymbol{x}
\end{split}
\end{equation}
Specializing now to the case of a time-independent electrostatic potential given by 
\begin{equation}
A_{4} = i \varphi(r),  \qquad \qquad V = - e \varphi ,
\end{equation}
\begin{equation}
H^{e}(t) = \int V(r) ( \psi^{*} \psi ) \, d^{\,3} \boldsymbol{r},
\end{equation}
we find\endnote{
``$\alpha$'' is the fine-structure constant, $\tfrac{1}{137.036}$; ``$\boldsymbol{\alpha}$'' is the Dirac matrix. Originally ``$\boldsymbol{\alpha} \boldsymbol{\cdot} \nabla$'' was rendered as ``$\alpha \cdot \text{grad}$''.}
\index{Dirac!matrices}%
\begin{equation}
\begin{split}
H_{T2} &= \frac{\alpha}{3\pi \mu^{2}} \left\{ \log \frac{\mu}{2r} + \tfrac{11}{24} - \tfrac{1}{5} \right\} \int (\nabla^{2} V)(\psi^{*} \psi) d^{\,3}\boldsymbol{r}\\
&- \frac{i \alpha}{4\pi \mu} \, \int \psi^{*} \gamma_{4} (\boldsymbol{\alpha}  \boldsymbol{\cdot}\nabla V) \psi \, d^{\,3}\boldsymbol{r}
\end{split}
\end{equation}
The calculation of $H_{T2}$, which is the main part of the Lamb shift calculation, can thus be taken over directly from the scattering calculation. 
\index{Lamb~shift}%
In particular, the elimination of divergent renormalization effects does not have to be considered afresh. 
\index{renormalization!elimination~of~divergent~effects}%
Once (676) has been derived, everything is finite and we are permitted to carry out the rest of the calculation in a non-covariant way. Note however that the low-energy photon frequency cut-off $r$ still appears in (676). We expect that this dependence on $r$ will finally disappear when the effects of $H_{T1}$ are considered, the same cut-off being used in both $H_{T1}$ and $H_{T2}$. 

We next turn to the evaluation of $H_{T1}$, which is related to $U_{1}$ in the same way as $H_{T2}$ to $(U_{2} + U_{2}^{1})$. According to (484) the matrix element of $U_{1}$ for a one-electron transition between the states (542), with emission of the photon (440), is \endnote{
The equation labeling was faulty. In the second edition, what is here labeled (676a) was a \emph{second} (676), and what is here labeled (677a) had no label at all. It might have been all right to leave Eq.\ (677a) unlabeled, except that in the first edition, \emph{both} Eq.\ (677) \emph{and} Eq.\ (677a) are labeled (677)! This is a compromise.}
\[
M_{1} = - \frac{e^{2}}{\hbar^{2}c^{2}} \overline{u}^{\prime} \left\{ \slashed{e} \frac{1}{\slashed{p} - \slashed{k}^{\prime} - i\mu} \slashed{e}^{\prime} + \slashed{e}^{\prime} \frac{1}{\slashed{p}^{\prime} + \slashed{k}^{\prime} - i\mu} \slashed{e} \right\} u  \tag{676a}
\]
where now
\begin{equation}
e_{\mu} = e_{\mu}(p^{\prime} + k^{\prime} - p)
\end{equation}
is given by the Fourier expansion (534). 
\index{Fourier!expansion}%
Note that we cannot yet use the simple form (546) for $M_{1}$, because we do not know that $k^{\prime} \ll p, q$ for the photons which will be important in this problem. Using the Dirac equation satisfied by $u$ and $u^{\prime}$, we may write without approximations using (587) and rule 4 on page 87,
\index{Dirac!equation}%
\[
M_{1} = - \frac{e^{2}}{2\hbar^{2}c^{2}}\, \overline{u}^{\prime} \left\{ \frac{\slashed{e} \slashed{k}^{\prime} \slashed{e}^{\prime} - 2 ( p \cdot e^{\prime} )\slashed{e}}{p \cdot k^{\prime}} + \frac{\slashed{e}^{\prime} \slashed{k}^{\prime} \slashed{e} + 2 ( p^{\prime} \cdot e^{\prime}) \slashed{e}}{p^{\prime} \cdot k^{\prime}} \right\} u	 \tag{677a}
\]
Since $p$ and $p^{\prime}$ are assumed non-relativistic we may write
\[
p \cdot k^{\prime} = p^{\prime} \cdot k^{\prime} = - \mu k_{0}^{\prime}
\]
and then
\begin{equation}
M_{1} = \frac{e^{2}}{2 \hbar^{2} c^{2} \mu k_{0}^{\prime}} \, \overline{u}^{\prime} \left\{ 2\left( (p^{\prime} - p) \cdot e^{\prime} \right) \slashed{e} + \slashed{e}^{\prime} \slashed{k}^{\prime} \slashed{e} + \slashed{e} \slashed{k}^{\prime} \slashed{e}^{\prime} \right\} u
\end{equation}
Now because we shall be considering only an electrostatic potential (674), $\slashed{e}$ is a multiple of $\gamma_{4}$ simply. Then if $k_{3}^{\prime}$ is the space-like part of the vector $k^{\prime}$, we have
\[
\slashed{e}^{\prime} \slashed{k}^{\prime}_{3} \, \slashed{e} + \slashed{e} \slashed{k}^{\prime}_{3}\, \slashed{e}^{\prime} = \slashed{e} \left\{  \slashed{e}^{\prime} \slashed{k}^{\prime}_{3} + \slashed{k}^{\prime}_{3} \, \slashed{e}^{\prime} \right\} = 0
\]
Therefore 
\begin{equation}
\slashed{e}^{\prime} \slashed{k}^{\prime}_{3} \, \slashed{e} + \slashed{e} \slashed{k}^{\prime}_{3}\, \slashed{e}^{\prime} = i k^{\prime}_{0} \, \slashed{e}^{\prime} \, (2 \gamma_{4} \slashed{e} )
\end{equation} 
Now this term (679) is small compared to the other term in (678), because $\slashed{e}^{\prime}$ involves the matrices $\gamma_{1},\gamma_{2},\gamma_{3}$ while $\slashed{e}$ involves $\gamma_{4}$, and the matrix elements of $\gamma_{1},\gamma_{2},\gamma_{3}$ for non-relativistic transitions are small, of the order of $(v/c)$. The term (679) in fact describes magnetic radiation, whereas the other term in (678) gives electric. The electric term gives an effect of the order of the Lamb shift. 
\index{Lamb~shift}%
Hence in our approximation we may neglect the magnetic term and write 
\begin{equation}
M_{1} =  \frac{e^{2}}{\hbar^{2}c^{2}\mu k_{0}^{\prime}} \left( (p^{\prime} - p) \cdot e^{\prime} \right) ( \overline{u}^{\prime} \slashed{e} u )
\end{equation}
which is the same result as we should have obtained from (546). 

Let $Z_{A}(x)$ be the Hertzian vector corresponding to the radiation field potentials $A_{\lambda}(x)$, defined by
\begin{equation}
A_{\lambda}(x) = \frac{d}{dt} Z_{\lambda}(x)
\end{equation}
Then the matrix element of $Z_{\lambda}(x)$ for emitting the photon with potentials (440) is
\begin{equation}
Z_{\lambda}(x) = \frac{1}{ick_{0}^{\prime}} e^{\prime}_{\lambda} e^{-i k^{\prime} \cdot x}
\end{equation}
Compare with (422)\endnote{
The comparison with Eq.\ (422) is not obvious. Perhaps Eq.\ (438) was meant?}.

\noindent Thus the operator $U_{1}$ which has the matrix element (680) may be written\endnote{
The integrand was originally written 
\[
\overline{\psi} \left( Z \boldsymbol{\cdot} \text{grad} \right) \slashed{A}^{e} \psi(x)
\]
Since $Z_{\lambda}$ is a Lorentz vector, the gradient must likewise be. So ``grad'' here must be $\partial_{\lambda}$, not $\nabla$. One goes from (682) to (680) by an integration by parts; since there are in (680) apparently the dot product of two 4-vectors, this supports the identification here of grad $= \partial_{\lambda}$.}
\begin{equation}
U_{1} = \frac{e^{2}}{\hbar^{2}c\mu} \int dx\, \overline{\psi} (Z \cdot \partial ) \slashed{A}^{e} \psi(x)
\end{equation}
Using (670) and specializing by means of (674), this gives
\begin{equation}
H_{T1} = \frac{e}{\hbar \mu} \int \psi^{*} (Z \cdot \partial V) \psi \,d^{\,3} \boldsymbol{r}
\end{equation}
This completes the evaluation of $H_{T}$.

\subsection*{Discussion and the Nature of the $\Phi$-Representation}
\addcontentsline{toc}{subsection}{Discussion of the Nature of the $\Phi$-Representation}

To understand the effect of the transformation (662), we observe that if $\Psi(t)$ is the state of one real electron in the absence of an external field, then $\Phi(t)$ will be independent of  $t$ and will represent one ``bare'' electron with the same momentum as the real electron. 
\index{electron!bare}%
In an actual hydrogen atom,  we may consider the state $\Psi(t)$ to a very good approximation as a superposition of states of a single real free electron; then $\Phi(t)$ is a superposition of states of a single bare electron  with the same distribution  of momenta.
\index{hydrogen~atom!Lamb~shift}
\index{electron!bare}
Thus by the transformation from $\Psi$ to $\Phi$ we have eliminated the radiation field surrounding the electron, all remaining effects of this radiation field being contained in the operator $H_{T}$. 

It is essential at this point to make sure that in the $\Phi$ representation the field-operators are still free-particle operators, with the correct equations of motion for operators in the FIR. Thus the transformation (662) is only a transformation from one set of variables to another within the FIR, and does not take us out of the FIR. This point was never explained properly by Schwinger in his papers, although he no doubt understood it himself. 
\index{Schwinger, Julian}%

Let then $Q(x)$ be a field-operator of the $\Psi$ representation. Being a FIR operator, $Q(x)$ satisfies
\begin{equation}
i \hbar \frac{dQ}{dt} = \boldsymbol{[} Q, H_{0} \boldsymbol{]}
\end{equation}
where $H_{0}$ is the Hamiltonian of the Dirac and Maxwell fields without interaction. 
\index{Hamiltonian!Dirac~and~Maxwell~fields}%
In the $\Phi$ representation the corresponding field-operator is 
\index{Dirac!field}%
\begin{equation}
Q^{\prime}(x) = \left( \Omega_{1}(t) \right)^{-1} Q(x) \, \Omega_{1}(t)
\end{equation}
Now $\Omega_{1}(t)$ is given by (492) where we now take the limit $g_{A} = 1$. The operators appearing in (492) are all FIR operators satisfying equations of motion of the form (685). When the integrations in (492) are carried out, the integrated terms will still have the same time variation given by (685), except for those terms which correspond to transitions in which energy is conserved. The energy-conserving matrix elements will have an explicit linear dependence on $t$ after integration, which is not in accordance with (685). Therefore we conclude that the equation of motion
\begin{equation}
i \hbar \frac{d\Omega_{1}}{dt} = \boldsymbol{[} \Omega_{1}, H_{0} \boldsymbol{]}
\end{equation}
is valid for all matrix elements of $\Omega_{1}$ which are not diagonal in the FIR. The same equation of motion is satisfied by $\left(\Omega_{1}(t) \right)^{-1}$ with the same condition. Now we have seen that $\Omega_{1}(t)$ has no matrix elements diagonal in $H_{0}$ which give transitions either from or into one-particle states. Hence (687) holds for all matrix elements in which either initial or final state is a one-particle state. 

Combining (687) and the corresponding equation for $\left(\Omega_{1} \right)^{-1}$ with (685) and (686), we have 
\begin{equation}
i \hbar \frac{dQ^{\prime}}{dt} = \boldsymbol{[} Q^{\prime}, H_{0} \boldsymbol{]}
\end{equation}
This equation (688) is valid for all matrix elements between one-particle states. Thus we can conclude that so long as we confine attention to a one-electron system, (688) is satisfied by all field operators of the $\Phi$ representation, and so the $\Phi$ representation is still within the FIR.

When we consider systems containing more than one particle, then $\Omega_{1}(t)$ will have an explicit dependence on time in addition to (687). Then the $\Phi$ representation would no longer belong to the FIR. And this is physically reasonable, because in many-electron systems it is not in general possible to transform away the radiation interaction completely, the radiation interaction by itself giving rise to real effects such as M\o ller scattering which we should not wish to transform away. 
\index{Moller@M\o ller~scattering}%

\subsection*{Concluding Non-Covariant Part of the Calculation}
\addcontentsline{toc}{subsection}{Concluding Non-Covariant Part of the Calculation}

Having established that the $\Phi$-representation in which (664) holds is the FIR, we now proceed at once to transform to the BIR in which we shall finish the calculation of the Lamb shift. 
\index{Lamb~shift}%
To transform to the BIR, we write
\begin{equation}
\Phi(t) = e^{iH_{0}t/\hbar} e^{-i\left\{H_{0} + H^{e} \right\}t/\hbar}\, \Phi^{\prime}(t)
\end{equation}
The new wave-function $\Phi^{\prime}(t)$ thus satisfies
\begin{equation}
i \hbar \frac{\partial \Phi^{\prime}}{\partial t} = \left\{H_{T1}(t) + H_{T2}(t) \right\} \Phi^{\prime}
\end{equation}
where $H_{T1}$ and $H_{T2}$ are given by (684) and (676), only now the $\psi^{*}$ and $\psi$ operators have the time-variation of the Dirac field in the external potential $V$.
\index{Dirac!field!in~external~potential}%

To solve (690) we may now use precisely the method that we used for solving (245) in the non-relativistic treatment. There are only two differences, (i) we have now the extra term $H_{T2}$, and (ii) the form of $H_{T1}$ is different form (247).

Since we are working only to second order in the radiation interaction and $H_{T2}$ is already of that order, $H_{T2}$ is to be treated only as a first-order perturbation. Then $H_{T2}$ will have no effect on the line-width $\Gamma$, and will contribute to the line-shift $\Delta E$ just the expectation-value of (676) in the state $\psi_{0}$ of the atom, namely
\begin{equation}
\Delta E_{2} = \frac{\alpha}{3 \pi \mu^{2}} \left\{ \log \frac{\mu}{2r} + \tfrac{11}{24} - \tfrac{1}{5} \right\} \int (\nabla^{2} V) |\psi_{0}|^{2} \, d^{\,3}\boldsymbol{r} - \frac{i\alpha}{4 \pi \mu} \int \psi^{*}_{0} \gamma_{4} ( \boldsymbol{\alpha} \cdot \nabla V) \psi_{0}\; d^{\,3} \boldsymbol{r}
\end{equation}

The effect of the change from (247) to (684) is that the matrix element $j_{\mu}^{k}(n \; m)$ given by (256) now becomes everywhere replaced by
\begin{equation}
J_{\mu}^{k}(n \; m) = \frac{ie}{\hbar \mu | \boldsymbol{k} |}  \int \psi^{*}_{n} \, \frac{\partial V}{\partial x_{\mu}} \, e^{-i \boldsymbol{k} \cdot \boldsymbol{r}} \, \psi_{m} \, d^{\, 3}\boldsymbol{r}
\end{equation}
We see this by comparing (247) and (684), noting that 
\[
\boldsymbol{j}(r, t) \leftrightarrow \frac{ec}{\hbar \mu} \psi^{*} \psi \nabla V\, \int dt
\]
or 
\[
\int \boldsymbol{j}^{S}(r) \,  e^{-i \boldsymbol{k} \cdot \boldsymbol{r}} d^{\, 3}\boldsymbol{r} \leftrightarrow \frac{iec}{\hbar \mu | \boldsymbol{k} | c} \int \psi^{*} \psi \, \nabla V \,  e^{-i \boldsymbol{k} \cdot \boldsymbol{r}}\, d^{\, 3}\boldsymbol{r}
\]
In the previous calculation we used a non-relativistic dipole approximation for $j_{\mu}$ which gave according to (272)
\begin{equation}
j_{\mu}^{k}(n \; m) = + \frac{ie \hbar}{m} \int \psi^{*}_{n} \frac{\partial \psi}{\partial x_{\mu}} d^{\, 3}\boldsymbol{r}
\end{equation}
We shall again use a dipole approximation and drop the exponential factor in (692). Then taking for the atom the non-relativistic Hamiltonian 
\index{Hamiltonian!non-relativistic~atom}%
\begin{equation}
H = \frac{p^{2}}{2m} + V
\end{equation}
the difference between (692) and (693) becomes 
\begin{equation}
J_{\mu}^{k}(n \; m) - j_{\mu}^{k}(n \; m) = \frac{ie}{\hbar \mu | \boldsymbol{k} |} \left\{ \int \psi^{*}_{n} \frac{\partial \psi}{\partial x_{\mu}} d^{\, 3}\boldsymbol{r} \right\} (E_{m} - E_{n} - h c | \boldsymbol{k} |)
\end{equation}
where we used $ \boldsymbol{[} \,p^{2}, p_{\mu}\, \boldsymbol{]} = 0 $ and $\int p_{\mu}\psi^{*}_{n}\, p^{2} \psi_{m} \, d^{\, 3}\boldsymbol{r}  = - \int \psi^{*}_{n}\, p_{\mu}p^{2} \psi_{m}\, d^{\, 3}\boldsymbol{r}$. This difference vanishes for transitions in which energy is conserved. Hence the value of $\Gamma$ given by (262) is unaffected by the change from $j$ to $J$. The value $\Gamma$ calculated previously is still valid in the relativistic theory, except for very small effects from magnetic radiation which we have neglected. 

Using (695) we have the simple relation between $j_{\mu}$ and $J_{\mu}$,
\begin{equation}
J_{\mu}^{k}(n \; m) = j_{\mu}^{k}(n \; m) \, \frac{E_{m} - E_{n}}{hc | \boldsymbol{k} |}
\end{equation}
Using formula (261) with $J$ substituted for $j$, the contribution from $H_{T1}$ to the line-shift becomes instead of (273)
\begin{equation}
\Delta E_{1} = - \frac{e^{2}}{6 \pi^{2}m^{2} \hbar c^{3}} \int_{r}^{\infty} \frac{dk}{k} \, \sum_{n} \frac{(E_{n} - E_{0})^{2} | p_{n0} |^{2} }{E_{n} - E_{0} + hc | \boldsymbol{k} |}
\end{equation}
The integral is now convergent at high frequencies and only divergent at low frequencies where the cut-off $r$ is now required to make it finite. The shift (697) would be zero for a free particle, and so there is no question of subtracting away a mass-renormalization term  as we did from (273); in the relativistic treatment the mass-subtraction was already done long before this stage of the calculation was reached.
\index{renormalization!mass}%

Integrating (697) directly over $k$, and taking $r$ to be small compared with $(E_{n} - E_{0})$, we find
\begin{equation}
\Delta E_{1} = - \frac{e^{2}}{6 \pi^{2}m^{2} \hbar c^{3}}  \sum_{n} (E_{n} - E_{0})^{2}\, | p_{n0} |^{2} \log \frac {|E_{n} - E_{0}|} {hcr}
\end{equation}
This is precisely the non-relativistic line-shift (278) with $r$ substituted for $K$. Defining $(E - E_{0})_{\text{av}}$ by (279) and using (281), we have
\begin{equation}
\Delta E_{1} =  \frac{\alpha}{3 \pi \mu^{2}}  \left\{\log \frac {hcr}{(E - E_{0})_{\text{av}}} \right\} \int (\nabla^{2} V) \, | \psi_{0} |^{2}\,d^{\, 3}\boldsymbol{r}
\end{equation}
This combines with (691) to give for the total level shift
\begin{equation}
\Delta E 
=  \frac{\alpha}{3 \pi \mu^{2}}  \left\{\log \frac {mc^{2}}{2(E - E_{0})_{\text{av}}} + \tfrac{11}{24} - \tfrac{1}{5} \right\} \int (\nabla^{2} V) \, | \psi_{0} |^{2}\,d^{\,3}\boldsymbol{r} - \frac{i\alpha}{4 \pi \mu} \int \psi_{0}^{*} \gamma_{4} (\boldsymbol{\alpha}  \boldsymbol{\cdot} \nabla V) \psi_{0}\, d^{\, 3}\boldsymbol{r}
\end{equation}
a result which is completely divergence-free and independent of $r$.

The second term of (700) represents the effect of the anomalous magnetic moment  of the electron on the energy levels. 
\index{electron!anomalous~magnetic~moment}%
It therefore gives a spin-dependent shift which modifies slightly the fine-structure which arises from the Dirac magnetic moment. To evaluate this term we use the Dirac equations (see (38)). 
\begin{equation}
mc^{2} \gamma_{4} \psi_{0} = (E_{0} - V) \psi_{0} + i \hbar c (\boldsymbol{\alpha} \boldsymbol{\cdot} \nabla) \psi_{0}
\end{equation}
\begin{equation}
mc^{2} \psi_{0}^{*} \gamma_{4} =  \psi_{0}^{*}(E_{0} - V) - i \hbar c (\nabla \psi_{0}^{*} \boldsymbol{\cdot} \boldsymbol{\alpha})
\end{equation}
Using both (701) and (702) in turn in the second term of (700) and adding the results using $\alpha^{i}\gamma_{4} + \gamma_{4}\alpha^{i} = 0$ the terms in $(E_{0} - V)$ cancel and we find\endnote{
In the second to last line, the term  $i \boldsymbol{\sigma} \boldsymbol{\cdot} ( \nabla \psi_{0} \times \nabla V )$ was written with a dot product between the two gradients, rather than a cross product.} 
\begin{equation}
\begin{split} 
&2mc^{2} \int \psi^{*}_{0} \gamma_{4} (\boldsymbol{\alpha} \boldsymbol{\cdot} \nabla V) \psi_{0}\, d^{\, 3}\boldsymbol{r} \\
&= - i \hbar c \int \left\{ ( \nabla \psi^{*}_{0} \boldsymbol{\cdot} \boldsymbol{\alpha})(\boldsymbol{\alpha}  \boldsymbol{\cdot} \nabla V) \psi_{0}  
 + \psi^{*}_{0} (\boldsymbol{\alpha} \boldsymbol{\cdot} \nabla V) (\boldsymbol{\alpha} \boldsymbol{\cdot} \nabla\psi_{0})\right\}\, d^{\, 3}\boldsymbol{r}
\\
 &=- i \hbar c \int \left\{ ( \nabla \psi^{*}_{0} \boldsymbol{\cdot} \boldsymbol{\sigma})(\boldsymbol{\sigma} \boldsymbol{\cdot} \nabla V) \psi_{0}  
 + \psi^{*}_{0} (\boldsymbol{\sigma} \boldsymbol{\cdot} \nabla V) (\boldsymbol{\sigma} \boldsymbol{\cdot} \nabla\psi_{0})\right\}\, d^{\, 3}\boldsymbol{r}
 \\
 &= -i \hbar c \int \left\{ \psi_{0} \left[\nabla \psi^{*}_{0} \boldsymbol{\cdot} \nabla V + i \boldsymbol{\sigma} \boldsymbol{\cdot} ( \nabla \psi^{*}_{0} \times \nabla V ) \right]  +  \psi^{*}_{0} \left[ \nabla\psi_{0} \boldsymbol{\cdot}  \nabla V  +  i \boldsymbol{\sigma} \boldsymbol{\cdot} ( \nabla \psi_{0} \times \nabla V ) \right]\right\}\,d^{\, 3}\boldsymbol{r} \\
 &= i \hbar c \int \left\{ + (\nabla^{2} V) \psi^{*}_{0} \psi_{0} - 2i \psi^{*}[\boldsymbol{\sigma} \boldsymbol{\cdot} (\nabla V 
 \times \nabla)]\psi_{0}\right\}\,d^{\, 3}\boldsymbol{r} 
\end{split}
\end{equation}
Here we used $\alpha^{i} = \epsilon \sigma_{i}$ (see p. 20), $\epsilon^{2} = \mathbb{I}$, and the formula
\[
(\boldsymbol{\sigma}\cdot \boldsymbol{B}) (\boldsymbol{\sigma} \cdot \boldsymbol{C}) = (\boldsymbol{B} \cdot \boldsymbol{C}) + 
i (\boldsymbol{\sigma} \cdot \boldsymbol{B} \times \boldsymbol{C})
\]
See Dirac, \emph{The Principles of Quantum Mechanics}, third edition, p.\  263. 
\index{Dirac, P. A. M.}%

Now suppose $V$ is a central potential, a function of $r$ only. Then
\begin{equation}
\nabla V \times \nabla = \frac{1}{r} \frac{dV}{dr} \left( \boldsymbol{r} \times \nabla \right) =
\frac{1}{r} \frac{dV}{dr} \left( \frac{i}{\hbar} \boldsymbol{L} \right)
\end{equation}
with $\boldsymbol{L}$, the orbital angular momentum, is given by (39). In this case (700) becomes 
\index{angular~momentum}%
\begin{equation}
\begin{split}
\Delta E &= \frac{\alpha}{3\pi \mu^{2}} \left\{ \log \frac {mc^{2}}{2(E - E_{0})_{\text{avg}}} + \tfrac{5}{6} - \tfrac{1}{5} \right\} \int (\nabla^{2} V)\, 
|\psi_{0}|^{2} \,d^{\, 3}\boldsymbol{r} \\
&+ \frac{\alpha}{4\pi \mu^{2} \hbar} \int \psi^{*}_{0} \left( \frac{1}{r} \frac{dV}{dr}\right) (\boldsymbol{\sigma} \cdot \boldsymbol{L})\, \psi_{0} \,d^{\, 3}\boldsymbol{r}
\end{split}
\end{equation}
In the non-relativistic theory of the hydrogen atom, the quantum number $j$ given by (72) is related to the operator $(\boldsymbol{\sigma} \cdot \boldsymbol{L})$ by
\begin{equation}
\frac{1}{\hbar} (\boldsymbol{\sigma} \cdot \boldsymbol{L}) = \begin{cases}
\ell, &j = \ell + \tfrac{1}{2} \\
-\ell - 1, &j = \ell - \tfrac{1}{2}
\end{cases}
\end{equation}
Therefore for the hydrogen atom 
\begin{equation}
\Delta E = \frac{\alpha e^{2}}{3 \pi \mu^{2}} \left\{ \log \frac{mc^{2}}{2(E - E_{0})_{\text{avg}}} + \tfrac{5}{6} - \tfrac{1}{5} \right\} \, |\psi_{0}(0)|^{2} 
+ \frac{\alpha e^{2}}{16 \pi^{2} \mu^{2}}\, q \,\int \frac{1}{r^{3}} \, |\psi_{0}|^{2} d^{\, 3}\boldsymbol{r}
\end{equation}
where $q$ is the coefficient (706).

For $s$-states we have $q = 0$ and so the shift reduces to (see (284))
\begin{equation}
\Delta E = \frac{8\alpha^{3}}{3 \pi} \frac{1}{n^{3}}\, \text{Ry}\,  \left\{ \log \frac{mc^{2}}{2(E - E_{0})_{\text{avg}}} + \tfrac{5}{6} - \tfrac{1}{5} \right\}
\end{equation}
for the state with principal quantum number $n$. For all other states the term in $ |\psi_{0}(0)|^{2}$ is zero, and the shift depends only on the integral
\begin{equation}
\overline{\left( \frac{1}{r^{3}} \right)} = \int \frac{1}{r^{3}} \, |\psi_{0}|^{2} d^{\, 3}\boldsymbol{r}
\end{equation}
The value of (709) is given by Bethe, \emph{Handbuch der Physik}, Vol. 24/1, p. 286, Eq. (3.26) 
\cite{BetheSalpeter57}. It is
\begin{equation}
\overline{ \left( \frac{1}{r^{3}} \right) } =  \frac{1}{\ell (\ell + \tfrac{1}{2})(\ell + 1) n^{3} a_{o}^{3} }
\end{equation}
where $a_{o}$ is the Bohr radius of the hydrogen atom. 
\index{Bohr~radius}%
Hence the shift for states with $\ell \ne 0$ becomes 
\begin{equation}
\Delta E = \frac{\alpha^{3}}{2\pi} \frac{1}{n^{3}}\, \text{Ry}\, \frac{1}{(\ell + \tfrac{1}{2})(\ell + 1)} \qquad \;\text{for $j = \ell + \tfrac{1}{2}$}
\end{equation}
\begin{equation}
\Delta E = -\frac{\alpha^{3}}{2\pi} \frac{1}{n^{3}}\, \text{Ry}\, \frac{1}{\ell(\ell + \tfrac{1}{2})} \qquad \qquad \text{for $j = \ell - \tfrac{1}{2}$}
\end{equation}
For the relative displacement of the 2$s$ and 2$p_{1/2}$ levels, which in the Dirac theory were degenerate, we have finally by subtracting (712) from (708)
\begin{equation}
\Delta E = \frac{\alpha^{3}}{3\pi} \, \text{Ry} \left\{\log \frac{mc^{2}}{2(E - E_{0})_{\text{avg}}} + \tfrac{5}{6} - \tfrac{1}{5} + \tfrac{1}{8} \right\} = 1051 \, \text{Megacycles}
\end{equation}
\index{hydrogen~atom!Lamb~shift|)}
\subsection*{Accuracy of the Lamb Shift Calculation}
\addcontentsline{toc}{subsection}{Accuracy of the Lamb Shift Calculation}

With the relativistic calculation of the Lamb shift which we have done, this course comes to an end.
\index{Lamb~shift|(}%
In this calculation we have met and seen how to overcome all the problems of mass and charge renormalization. 
\index{renormalization!mass~and~charge}%
We can say we now have a workable quantum electrodynamics  which will give finite and unambiguous values for all observable quantities.
\index{quantum~electrodynamics}%

This calculation of the Lamb shift was of course not exact. The two most important errors were

(i) using non-relativistic wave-functions and the dipole radiation approximation in evaluating the effects of $H_{T1}$;

(ii) neglecting the finite mass of the proton. \\
To correct these errors, very long calculations have been done. In connection with (i), Baranger \cite{BarangerBetheFeynman53} 
 has calculated the effect of using relativistic theory in the treatment of $H_{T1}$ and he finds the observed shift increases by 7 megacycles.
\index{Baranger, Michael}%
\index{Bethe, Hans~A.}%
\index{Feynman, Richard~P.}%
The effects of (ii) are being looked at by Salpeter \cite{Salpeter52} but are not greater than 1 - 2 megacycles at most.
\index{Salpeter, Edwin~E.}%
In addition we have not considered 

(iii) effects of fourth order in the radiation interaction. These are being looked at by Kroll  and others \cite{BarangerDysonSalpeter52}; they are certainly less than 1 megacycle.
\index{Kroll, Norman~M.}%
\index{Baranger, Michael}%
\index{Dyson, Freeman~J.}%
\cite{Salpeter53}%

Therefore the theoretical value of the Lamb shift now stands at 1058 $\pm$ 2 megacycles. There is no clear discrepancy between this and the experimental value 1062 $\pm$ 5 though a discrepancy may be found when the experiments and the theory are further cleaned up.
\index{Lamb~shift|)}%

\begin{center}
\textsc{The End}
\end{center}
%


\chapter*{Typist's Afterword}
\addcontentsline{toc}{chapter}{Typist's Afterword} 

\pagestyle{fancy}
\fancyhead{}
\lhead{}
\chead{}
\rhead{\thepage}
\lfoot{}
\cfoot{}
\rfoot{}

Both Kaiser's admirable \emph{Drawing Theories Apart} \cite{Kaiser05} and Schweber's masterful \emph{QED and the Men Who Made It} \cite{Schweber94} refer frequently to the famous lectures on quantum electrodynamics given by Freeman Dyson at Cornell University in 1951. Two generations ago, graduate students (and their professors) wishing to learn the new techniques of QED passed around copies of Dyson's Cornell lecture notes, then the best and fullest treatment available. Textbooks appeared a few years later, e.\ g.\ by Jauch \& Rohrlich \cite{Jauch55} and Schweber \cite{Schweber61}, but interest in Dyson's notes has never fallen to zero. Here is what the noted theorist E. T. Jaynes wrote in an unpublished article \cite{Jaynes84} on Dyson's autobiographical \emph{Disturbing the Universe}, 1984:

\begin{quote}
But Dyson's 1951 Cornell course notes on Quantum Electrodynamics were the original basis of the teaching I have done since. For a generation of physicists they were the happy medium: clearer and better motivated than Feynman, and getting to the point faster than Schwinger. All the textbooks that have appeared since have not made them obsolete. Of course, this is to be expected since Dyson is probably, to this day, best known among the physicists as the man who first explained the unity of the Schwinger and Feynman approaches.
\end{quote}

As a graduate student in Nicholas Kemmer's department of theoretical physics (Edinburgh, Scotland) I had heard vaguely about Dyson's lectures (either from Kemmer or from my advisor, Peter Higgs) and had read his classic papers \cite{Dyson49A}, \cite{Dyson49B} in Schwinger's collection \cite{Schwinger58}. It never occurred to me to ask Kemmer for a copy of Dyson's lectures which he almost certainly had.

My interest in the legendary notes was revived thirty years later by the Kaiser and Schweber books. Within a few minutes Google led to scans of the notes \cite{Dyson51b} at the Dibner Archive (History of Recent Science \& Technology) at MIT, maintained by Karl Hall, a historian at the Central European University in Budapest, Hungary. He had gotten permission from Dyson to post scanned images of the Cornell notes. Through the efforts of Hall, Schweber and Babak Ashrafi these were uploaded to the Dibner Archive. To obtain a paper copy would require downloading almost two hundred images, expensive in time and storage. Was there a text version? Had anyone had retyped the notes? Hall did not know, nor did further searching turn anything up. I volunteered to do the job. Hall thought this a worthwhile project, as did Dyson, who sent me a copy of the second edition, edited by Michael J. Moravcsik. (This copy had originally belonged to Sam Schweber.) Dyson suggested that the second edition be retyped, not the first. Nearly all of the differences between the two editions are Moravcsik's glosses on many calculations; there is essentially no difference in text, and (\emph{modulo} typos) all the labeled equations are identical.

Between this typed version and Moravcsik's second edition there are few differences; all are described in the added notes. (I have also added references and an index.) About half are corrections of typographical errors. Missing words or sentences have been restored by comparison with the first edition; very infrequently a word or phrase has been deleted. A few changes have been made in notation. Intermediate steps in two calculations have been corrected but change nothing. Some notes point to articles or books. No doubt new errors have been introduced. Corrections will be welcomed! The young physicists will want familiar terms and notation, occasionally changed from 1951; the historians want \emph{no} alterations. It was not easy to find the middle ground.

I scarcely knew \LaTeX\  before beginning this project. My friend (and Princeton '74 classmate) Robert Jantzen was enormously helpful, very generous with his time and his extensive knowledge of \LaTeX. Thanks, Bob. Thanks, too, to Richard Koch, Gerben Wierda and their colleagues, who have made \LaTeX\  so easy on a Macintosh. George Gr\"{a}tzer's textbook \emph{Math into \LaTeX} was never far from the keyboard. No one who types technical material should be ignorant of  \LaTeX.

This project would never have been undertaken without the approval of Prof.\ Dyson and the efforts of Profs.\ Hall, Schweber and Ashrafi, who made the notes accessible. I thank Prof.\ Hall for his steady encouragement through the many hours of typing. I thank Prof.\ Dyson both for friendly assistance and for allowing his wonderful lectures to become easier to obtain, to be read with pleasure and with profit for many years to come.

\begin{flushright}
David Derbes \\
Laboratory Schools\\
University of Chicago\\
loki@uchicago.edu\\
11 July 2006
\end{flushright}

%


\newpage

\renewcommand{\enotesize}{\normalsize} 
\renewcommand{\notesname}{}
\chapter*{Notes}
\addcontentsline{toc}{chapter}{Notes} 

\pagestyle{fancy}
\fancyhead{}
\lhead{}
\chead{}
\rhead{\thepage}
\lfoot{}
\cfoot{}
\rfoot{}

\theendnotes

%


\newpage
\pagestyle{fancy}
\fancyhead{}
\lhead{}
\chead{}
\rhead{\thepage}
\lfoot{}
\cfoot{}
\rfoot{}

\renewcommand{\bibname}{References}

%

\newpage
\addcontentsline{toc}{chapter}{Index}
\printindex

\end{document}